\documentclass[twocolumn,pra,floatfix,longbibliography,showpacs]{revtex4-2}

\usepackage{amsmath}
\usepackage{amsfonts}
\usepackage{amsbsy}
\usepackage{xcolor}
\usepackage{soul}
\usepackage{graphicx}
\usepackage{epsfig}
\usepackage{hyperref}
\usepackage{txfonts}
\usepackage{mathtools}
\usepackage{multirow}
\usepackage{dsfont}
\hypersetup{
    unicode=true, 
    plainpages=false,
    colorlinks=true,
    linkcolor=blue,
    citecolor=blue,
    filecolor=blue,
    urlcolor=blue
}
\usepackage{bbold}

\newcommand{\pol}{\hat{\bf e}}
\newcommand{\rv}{{\bf r}}

\newcommand{\qv}{{\bf q}}
\newcommand{\kv}{{\bf k}}
\newcommand{\pv}{{\bf p}}
\newcommand{\eo}{\epsilon_0}
\newcommand{\beq}{\begin{equation}}
\newcommand{\eeq}{\end{equation}}
\newcommand{\bea}{\begin{eqnarray}}
\newcommand{\eea}{\end{eqnarray}}
\newcommand{\BEQAL}{\begin{align}}
\newcommand{\EEQAL}{\end{align}}
\newcommand{\EQREF}[1]{Eq.~(\ref{#1})}

\newcommand{\comment}[1]{{}}

\newcommand{\<}{\langle}
\renewcommand{\>}{\rangle}
\renewcommand{\(}{\left(}
\renewcommand{\)}{\right)}
\renewcommand{\[}{\left[}
\renewcommand{\]}{\right]}

\newcommand{\Pc}{{\cal P}}

\newcommand{\commentout}[1]{{}}

\newcommand{\Jop}{\ensuremath{\hat{J}}}

\newcommand{\bE}{{\bf E}}

\newcommand{\cbE}{\boldsymbol{\mathbf{\cal E}}}

\newcommand{\h}{\hat}

\newcommand{\vecHat}[2]{\hat{\mathbf{#1}}_{\mathrm{#2}}}

\newcommand{\cbEL}{\boldsymbol{\mathbf{\cal E}}}

\newcommand{\rAtom}{\mathfrak{r}}
\newcommand{\rAtomBold}{\boldsymbol{\mathfrak{r}}}

\newcommand{\boldvec}[1]{\ensuremath{\boldsymbol{\mathbf{#1}}}}
\newcommand{\spvec}[1]{\ensuremath{\boldvec{#1}}}

\newcommand{\Dvhat}{\hat{{\spvec{D}}}}

\newcommand{\bra}[1]{\ensuremath{\left\langle #1 \right\vert}}
\newcommand{\ket}[1]{\ensuremath{\left\vert #1 \right\rangle}}
\newcommand{\unitvec}[1]{\hat{\mathbf{{#1}}}}

\newcommand{\radKernel}{\ensuremath{\mathsf{G}}}

\newcommand{\psihat}[1]{\hat{\psi}_{#1}}
\newcommand{\psihatdag}[1]{\hat{\psi}_{#1}^{\dag}}

\begin{document}
\title{Cooperative quantum-optical planar arrays of atoms}
\author{Janne Ruostekoski}
\affiliation{Department of Physics, Lancaster University, Lancaster, LA1 4YB, United Kingdom}

\begin{abstract}
Atomic planar arrays offer a novel emerging  quantum-optical many-body system in which light mediates strong interactions between the atoms. The regular lattice structure provides a cooperatively enhanced light-matter coupling and allows for increased control and harnessing of these interactions. In subwavelength arrays, coherent scattering of incident light beams can be highly collimated in the forward and backward direction, resembling one-dimensional light propagation without the need for waveguides, fibers, or resonators. The atomic planar arrays share common features with fabricated metasurfaces, formed by thin nanostructured films that have shown great promise in manipulating and structuring classical light.
Here we describe theoretical methods commonly employed to analyze the cooperative responses of atomic arrays and explore some recent developments and potential future applications of planar arrays as versatile quantum interfaces between light and matter.
\end{abstract}

\date{\today}
\maketitle

\section{Introduction}

Resonant emitters play a crucial role in optical devices for both classical and quantum technologies that rely on interfaces between light and matter. In this research, there is a natural tendency towards using cold samples for better coherence properties and dense samples due to the critical parameter of resonance optical depth.
As refined media, cold atomic ensembles provide a quantum platform for light-matter interfaces that underpin a wide range of quantum technologies from atomic clocks~\cite{Bothwell19} to quantum information processing~\cite{HAM10}. At high densities, however, atoms exhibit strong light-mediated resonant dipole-dipole (DD) interactions that can lead to resonance broadening, shifts, and dephasing. 
Traditionally, these are considered unwanted phenomena that are challenging to control and pose limitations in applications.

Due to the light-mediated DD interactions, cold atoms at densities with many atoms per cubic optical resonance wavelength exhibit cooperative responses to light, characterized by collective excitations with their own resonance linewidths and line shifts. When the linewidth is below the single-particle
linewidth, excitations are called subradiant, whereas in the opposite case, they are super-radiant~\cite{Dicke54}. While super-radiance has been extensively studied for many years~\cite{GrossHarochePhysRep1982}, experiments on subradiance were, until a few years ago, limited to systems of two or a few particles~\cite{DeVoe,Hettich,McGuyer,Takasu,Lovera,Frimmer}.
More recent experiments have shown small long-lived fractions of disordered atom clouds~\cite{Guerin_subr16,Guerin2023} and spatially-extended subradiant mode resonances in resonator arrays~\cite{Jenkins17}.

Atoms confined in regular arrays have emerged as pristine systems where cooperative light-mediated interactions can be controlled and harnessed.
Giant subradiance was observed in the transmission of light through a two-dimensional (2D) subwavelength planar optical lattice, with  the spectral resonance narrowing below the fundamental quantum limit set by a single atom~\cite{Rui2020}\footnote{Subradiant decay in lattices was also observed in Ref.~\cite{Ferioli21}}. Unlike conventional free-space optical media that rely on long propagation distances in their ability to control light, planar atomic arrays strongly interact with light within a single ultrathin atomic layer. 
Highly collimated coherent emission from planar arrays results in effectively 1D light propagation~\cite{Shahmoon,Facchinetti16,Facchinetti18,Guimond2019,Javanainen19}, with
significantly enhanced optical cross sections compared with similar disordered ensembles.

The light-matter interface with atomic planar arrays exhibits an intriguing analogy with artificially fabricated metasurfaces~\cite{Yu14,Chen2016_review,Luo18,Qiu21_review}, which have garnered considerable attention in nanophotonics in recent years with their unique abilities to manipulate light. 
However, reaching the quantum regime in nanostructured metasurfaces remains a formidable challenge~\cite{Solntsev21}, while quantum-optical control of atoms is commonplace~\cite{WallsMilburn}. 

There has been rapid progress in theoretical understanding of light propagation in dense and cold atomic ensembles. Improved computing resources have enabled even atom-by-atom simulations for the most miniature samples. These simulations go far beyond the traditional analysis of light propagation that has relied on a continuous description of the atomic medium, characterized by macroscopic electrodynamics quantities such as susceptibility or refractive index.

In this Perspective, we provide a review of the relevant theoretical background and describe  commonly employed methods for analyzing optical responses. We present and highlight selective results that shed light on the optical behaviors of these arrays. Our focus is on atoms trapped in regular planar geometries, while 1D chains of atoms typically analyzed in the context of waveguides and optical fibers are mostly outside the scope of this paper. 
For interested readers, recent review articles provide coverage of these systems~\cite{Sheremet23,Chang18,Reitz22}.

\section{Theoretical models} 

\subsection{Light interacting with atoms}
\label{sec:basic}

In the analysis of light interacting with closely spaced atoms, we utilize a nonrelativistic Hamiltonian formalism of electrodynamics in the dipole approximation.
The quantum-optical interaction between atoms with light is expressed in the \emph{length gauge}, obtained through the Power-Zienau-Woolley transformation~\cite{PowerZienauPTRS1959,Woolley1971a,CohenT}. 
We write the Hamitonian in terms of  the electric displacement $\Dvhat(\rv)$ which serves as the fundamental dynamical variable for light in the length gauge~\cite{CohenT}. Our focus is on incident light 
with the dominant frequency $\omega=c |{\bf k}|=ck=2\pi c/\lambda$ and wavevector ${\bf k}$. To facilitate the rotating wave approximation, we introduce positive and negative frequency components, $\Dvhat=\Dvhat{}^++\Dvhat{}^-$, with $\Dvhat{}^+=[\Dvhat{}^-]^\dagger$. For the quantized electromagnetic field, the mode frequency and the photon
annihilation and creation operators are denoted by $\omega_q$, $\hat a_q$, and $\hat a^\dagger_q$, respectively, where the mode index $q$ encompasses both the wave vector ${\bf q}$ and the transverse polarization
$\pol_q$. The positive frequency component in terms of the quantization volume $V$ then reads~\cite{WallsMilburn}
\beq
\Dvhat{}^+(\rv)= \sum_q \zeta_q \pol_q \hat a_q e^{i\qv\cdot\rv},\quad
\zeta_q=\sqrt{\frac{\hbar \epsilon_0\omega_q}{2 V}}.
\label{eq:Dquantisation}
\eeq
The Hamiltonian for the free electromagnetic field energy is
\beq
\hat H_F = \sum_{q}\hbar\omega_{q}\hat{a}_q^{\dagger}\hat{a}_q.
\eeq

For notational simplicity,  we focus on the specific case of a $J=0\rightarrow J'=1$ atomic transition.  A more comprehensive treatment of the atomic level structure can be found in Ref.~\cite{Lee16}.
The positive frequency component of the atomic polarization density operator $\hat{\textbf{P}}{}^+(\textbf{r})$  for a set of atomic positions $\{\textbf{r}_1,\textbf{r}_2,\ldots, \textbf{r}_N\}$ then reads
 \begin{equation}\label{eq:polariz}
\hat{\textbf{P}}{}^+(\textbf{r})=\sum_{j\nu}\delta(\textbf{r}-\textbf{r}_j)\textbf{d}_{ge}^{(\nu)} \hat{\sigma}_{j\nu}^-,
\end{equation}
where $\hat{\sigma}_{j\nu}^{+}=(\hat{\sigma}_{j\nu}^{-})^\dagger=|e_{j\nu}\>\< g_j|$ is the raising operator to the excited state $\nu=\pm1,0$ for atom $j$, with the dipole matrix element $\textbf{d}_{ge}^{(\nu)}= {\cal D} \pol_\nu$,  $[\textbf{d}_{ge}^{(\nu)}]^*=\textbf{d}_{eg}^{(\nu)}$, and
the circular polarization unit vectors $\pol_\pm=\mp\frac{1}{\sqrt{2}} (\pol_x\pm i\pol_y)$, $\pol_0=\pol_z$.
Throughout the paper,  we refer to slowly varying amplitudes  for both atom and light field quantities, where the rapid oscillations at the laser frequency have been factored out, such as $\exp(-i \omega t)$ from $\hat{\sigma}_{j\nu}^{-}$.
For stationary atoms in the rotating wave approximation, the atom-light interaction part of the Hamiltonian reads
\beq\label{Eq:Hamiltonian}
\hat{H}_{\rm{a-l}} =-\sum_{j\nu}\hbar\Delta_{j\nu}\hat{\sigma}^{ee}_{j\nu}
-\frac{1}{\epsilon_0}\int d^3\textbf{r} \( \hat{\textbf{D}}{}^+(\textbf{r})\cdot \hat{\textbf{P}}{}^-(\textbf{r}) + {\rm H.c.}\).
\eeq
The first term corresponds to the laser frequency detuning  $\Delta_{j\nu}=\omega-\omega_{j\nu}$ from the level $\nu$ transition frequency $\omega_{j\nu}$ of atom $j$,
and $\hat{\sigma}^{ee}_{j\nu}=\hat{\sigma}_{j\nu}^{+}\hat{\sigma}_{j\nu}^{-}$ represents the excited level $\nu$ population.  
The second term describes the interaction of light with the atoms. Additionally, the Hamiltonian contains the self-polarization term,
$(2\epsilon_0)^{-1}\int\hat{\textbf{P}}(\textbf{r})\cdot \hat{\textbf{P}}(\textbf{r})d^3\textbf{r}$
that is inconsequential in our system of nonoverlapping point atoms. 

Typically, atoms are illuminated by a coherent light with the positive frequency component $\boldsymbol{\mathcal{E}}{}^{+}(\textbf{r})$, generating scattered 
electric field amplitude $\hat{\textbf{E}}{}^+_s$. This can be formally integrated~\cite{Ruostekoski1997a} using the Hamiltonian and is given by the contributions from all the atoms
\begin{align}\label{eq:scattlight}
\epsilon_0 \hat{\textbf{E}}{}^+_s(\textbf{r}) &=\int d^3 r'\, \mathsf{G}(\textbf{r}-\textbf{r}')\hat{\textbf{P}}{}^+(\textbf{r}')=\sum_{j\nu} \mathsf{G}(\mathbf{r}-\mathbf{r}_j)\textbf{d}_{ge}^{(\nu)} \hat{\sigma}_{j\nu}^-,
\end{align}
where $\mathsf{G}_{\nu\mu}({\bf r})=\unitvec{e}^\ast_{\nu} \cdot
  \radKernel(\rv) \unitvec{e}_{\mu}$ is the dipole radiation kernel 
\beq
{\sf G}_{\nu\mu}({\bf r}) =
\left[ {\partial\over\partial r_\nu}{\partial\over\partial r_\mu} -
\delta_{\nu\mu} \boldsymbol{\nabla}^2\right] {e^{ikr}\over4\pi r}
-\delta_{\nu\mu}\delta({\bf r})\,.
\label{eq:GDF}
\eeq
When acting on a dipole $\mathbf{d}$ at the origin, we obtain~\cite{Jackson,BOR99} 
\begin{align}\label{Gdef}
\mathsf{G}(\mathbf{r})\mathbf{d}&=-\frac{\mathbf{d}\delta(\mathbf{r})}{3}+\frac{k^3}{4\pi}\Bigg\{\left(\hat{\mathbf{r}}\times\mathbf{d}\right)\times\hat{\mathbf{r}}\frac{e^{ikr}}{kr}\nonumber\\
&\phantom{==}-\left[3\hat{\mathbf{r}}\left(\hat{\mathbf{r}}\cdot\mathbf{d}\right)-\mathbf{d}\right]\left[\frac{i}{(kr)^2}-\frac{1}{(kr)^3}\right]e^{ikr}\Bigg\},
\end{align}
where $\hat{\mathbf{r}}=\mathbf{r}/|\mathbf{r}|$ and we interpret the expression in such a way that the integral of the term inside the curly brackets over an infinitesimal volume enclosing the origin vanishes~\cite{Ruostekoski1997a}.
The total electric field amplitude operator $\hat{\textbf{E}}{}^{+}(\textbf{r})$  is the sum of the laser field and the fields scattered from all atoms
\beq\label{eq:totallight}
\hat{\textbf{E}}{}^{+}(\textbf{r})=\boldsymbol{\mathcal{E}}{}^{+}(\textbf{r})+\hat{\textbf{E}}{}_s^{+}(\textbf{r}).
\eeq
The system of Eqs.~\eqref{eq:scattlight} and~\eqref{eq:totallight} presents an integral form of Maxwell's wave equations, where ${\sf G}_{\nu\mu}$ is also known as a dyadic Green's function for the Helmholtz equation and the source term is provided by the electric dipole transitions of the atoms through the operator $\hat{\textbf{P}}{}^+$. Typically, multipole transitions are much weaker in comparison.

Equation~\eqref{eq:scattlight} provides the solution for the scattered light given a specific atomic polarization density. The main challenge arises in calculating the expectation values involving $\hat{\textbf{P}}$, as the interaction between the light and atoms can be strong.  The dipole amplitude of each atom, which represents the polarization density at its position, depends on the light scattered by all the other atoms in the sample. For closely spaced cold atoms, the scattering can be dominated by recurrent processes~\cite{Ishimaru1978,Lagendijk,vantiggelen90,Morice1995a,Ruostekoski1997a,Sokolov2011,Javanainen2014a,Kwong19} where the scattered photon is exchanged multiple times between the same atoms. 

To describe the effects of position fluctuations, it is useful to introduce second-quantized atomic field operators for the ground $\psihat{g}(\mathbf{r})$ and excited $\psihat{e\nu}(\mathbf{r})$ states. Thus, $\hat{\mathbf{P}}^+(\mathbf{r})= \sum_{\nu}  \mathop{\psihatdag{g}(\mathbf{r})}\textbf{d}_{ge}^{(\nu)}\mathop{\psihat{e\nu}(\mathbf{r})}$ becomes the field-theoretical version of Eq.~\eqref{eq:polariz}~\cite{Ruostekoski1997a}. The relationship between the two is obtained by considering the atomic correlation functions for atoms fixed at positions $\{\textbf{r}_1,\textbf{r}_2,\ldots, \textbf{r}_N\}$~\cite{Lee16}
\beq
\label{eq:pabExp} \sum_{\nu} \textbf{d}_{ge}^{(\nu)}
\left\langle \mathop{\psihatdag{g}(\mathbf{r})}\mathop{\psihat{e\nu}(\mathbf{r})}\right\rangle_{\{\mathbf{r}_1,\dots,\mathbf{r}_N\}}
 = 
\sum_{j\nu} \textbf{d}_{ge}^{(\nu)} \langle \hat{\sigma}^-_{j\nu}\rangle \mathop{\delta(\mathbf{r}-\mathbf{r}_j)} , 
\eeq
where the summation of the index $j$ runs over the set  $\{\textbf{r}_1,\textbf{r}_2,\ldots, \textbf{r}_N\}$. Similarly, 
the atomic level density operators in second quantization are $ \mathop{\psihatdag{g}(\mathbf{r})}\mathop{\psihat{g}(\mathbf{r})}$ and $\mathop{\psihatdag{e\nu}(\mathbf{r})}\mathop{\psihat{e\nu}(\mathbf{r})}$.

\subsection{Quantum master equation}
\label{sec:qme}

For stationary atoms the full dynamics is represented by the  quantum  many-body master equation (QME) for the reduced density matrix~\cite{Lehmberg1970,agarwal1970}, which for a configuration of positions $\{\textbf{r}_1,\textbf{r}_2,\ldots, \textbf{r}_N\}$, $\hat \rho=\hat \rho_{\{\textbf{r}_1,\textbf{r}_2,\ldots, \textbf{r}_N\} }$, reads
\begin{align}
\label{eq:rhoeom}
\dot{\hat \rho} &=- \frac{i}{\hbar} \big[ \sum_{j\nu} \hat H^{(0)}_{j\nu} -\sum_{j\ell\nu\mu (\ell\neq j)}\hbar\Omega^{(j\ell)}_{\nu\mu}\hat{\sigma}_{j\nu}^{+}\hat{\sigma}_{\ell\mu}^{-},\hat \rho\big] \nonumber\\
&+\sum_{j\nu}\gamma \left(
2\hat{\sigma}^{-}_{j\nu}\hat \rho\hat{\sigma}^{+}_{j\nu}-\hat{\sigma}_{j\nu}^{+}\hat{\sigma}_{j\nu}^{-}\hat \rho -\hat\rho\hat{\sigma}_{j\nu}^{+}\hat{\sigma}_{j\nu}^{-}\right)\nonumber\\
&+\sum_{j\ell\nu\mu (\ell\neq j)}\gamma^{(j\ell)}_{\nu\mu}\left(
2\hat{\sigma}^{-}_{\ell\mu}\hat \rho\hat{\sigma}^{+}_{j\nu}-\hat{\sigma}_{j\nu}^{+}\hat{\sigma}_{\ell\mu}^{-}\hat \rho -\hat\rho\hat{\sigma}_{j\nu}^{+}\hat{\sigma}_{\ell\mu}^{-}\right),
\end{align}
where the bracket denotes a commutator and the Hamiltonian for the incident field and the atoms [the coherent part of the light-matter coupling of \EQREF{Eq:Hamiltonian}] is given by
\beq \label{eq:Hsys}
\hat H^{(0)}_{j\nu} = -\hbar  \Delta_{j\nu}\hat{\sigma}_{j\nu}^{ee} - \big(\mathbf{d}^{(\nu)}_{eg} \cdot \mathop{{\cbE}^+(\mathbf{r}_j)} \hat{\sigma}_{j\nu}^{+} + {\rm H.c.}\big).
\eeq
The coupling between different atoms $j$ and $\ell$ in Eq.~\eqref{eq:rhoeom} results from the real and imaginary parts of 
\begin{equation}\label{eq:omga}
\Omega^{(j\ell)}_{\nu\mu}+i\gamma^{(j\ell)}_{\nu\mu}=
\xi \radKernel^{(j\ell)}_{\nu\mu}, \quad (j\neq\ell),  \quad \xi=\frac{6\pi\gamma}{k^3} \, ,
\end{equation}
corresponding to the dissipation and interaction terms, respectively, of  the DD interaction $ \radKernel^{(j\ell)}_{\nu\mu}= \radKernel_{\nu\mu}(\rv_j-\rv_\ell)$
between the atoms at  $\rv_j$ and $\rv_\ell$, with dipolar orientations $\unitvec{e}_{\nu}$ and $\unitvec{e}_{\mu}$.
The nonlocal terms in Eq.~\eqref{eq:omga} account for spatially correlated scattering between different atoms.
The single-atom (half-width at half-maximum) resonance linewidth~\cite{WallsMilburn} 
\beq\label{eq:WW}
\gamma=\frac{\mathcal{D}^2k^3}{6\pi\hbar\epsilon_0},
\eeq 
is given by the Wigner-Weisskopf expression.
The single-atom resonance shifts are absorbed in the detuning term $\Delta_{j\nu}$.

\subsection{Semiclassical approximation}
\label{sec:semi}

In many-atom cavity systems, it is common to factorize two-body and higher order correlations to obtain equations for the expectation values of one-body operators, see, e.g., Ref.~\cite{CarmichaelVol2}.
The equations of motion for one-body expectation values may be derived from the QME~\eqref{eq:rhoeom}. For example, for the coherences of the $j$th atom between the electronic ground level $| g_j\>$  and excited level $| e_{j\nu}\>$,
\beq
\rho_{ge\nu}^{(j)}={\rm Tr} [\hat{\sigma}_{j\nu}^{-} \hat \rho_{\{\textbf{r}_1,\textbf{r}_2,\ldots, \textbf{r}_N\} }],
\eeq
where $\hat \rho_{\{\textbf{r}_1,\textbf{r}_2,\ldots, \textbf{r}_N\} }$ is the solution to Eq.~\eqref{eq:rhoeom} and Tr denotes the trace.
Semiclassical dynamics that neglects all quantum fluctuations may then be obtained by factorizing the internal level correlations between the different atoms
\beq \label{eq:semiappr}
\langle \hat{\sigma}_{j\nu}^{\pm} \hat{\sigma}_{\ell\mu}^{\pm}\rangle\simeq \langle \hat{\sigma}_{j\nu}^{\pm}\rangle \langle \hat{\sigma}_{\ell\mu}^{\pm}\rangle, \quad j\neq \ell. 
\eeq
This results in coupled optical Bloch equations (OBEs) that form a compact set of equations for two-level atoms~\cite{Kramer2015a}. The full set of equations for an arbitrary level configuration can be found in Ref.~\cite{Lee16}.  For the $J=0\rightarrow J'=1$ transition, we have $6N$ equations for the coherences and $4N$ equations for the level populations, one of which may be eliminated due to the population conservation. We obtain~\cite{Lee16}
\begin{subequations}
	\begin{align}
	\dot{\rho}^{(j)}_{g e\eta}=&(i{\Delta}_{j\eta}-\gamma)\rho_{g e\eta}^{(j)}+ i\bar{\mathcal{R}}_{\eta}^{(j)}\rho_{gg}^{(j)}- i \bar{\mathcal{R}}^{(j)}_{\tau}\rho_{e\tau e\eta}^{(j)},\label{Pfullstochastic}\\
	\dot\rho_{e\nu e\eta}^{(j)} =&(i\bar{\Delta}^{(j)}_{\nu \eta}
-2\gamma)\rho_{e\nu e\eta}^{(j)}+ i \bar{\mathcal{R}}_{\eta}^{(j)}\rho_{e\nu g}^{(j)} -i[\bar{\mathcal{R}}_{\nu}^{(j)}\rho_{e\eta g}^{(j)}]^*,
\label{popfullstochastic}
	\end{align}
\label{eq:multilevelhighintensitystochastic}
\end{subequations}
where $\bar{\Delta}^{(j)}_{\nu \eta} = \Delta_{j\eta}-\Delta_{j\nu}$. Here the effective Rabi frequencies~\cite{Parmee2020} $\bar{\mathcal{R}}_{\nu}^{(j)} $ describe the driving of the atom $j$ at ${\bf r}_j$ by the Rabi frequency ${\cal R}^{(j)}_{\nu}=[\textbf{d}_{ge}^{(\nu)}]^\ast\cdot \boldsymbol{\mathcal{E}}{}^{+}(\textbf{r}_j)/\hbar$ of the incident field and the scattered field from all the other atoms  at positions ${\bf r}_\ell$
\begin{equation}\label{Eq:EffectiveRabi}
\bar{\mathcal{R}}_{\nu}^{(j)} = \mathcal{R}_{\nu}^{(j)} + \xi\sum_{\ell \neq j} \radKernel^{(j\ell)}_{\nu\mu} \rho_{ge\mu}^{(\ell)}.
\end{equation}  
The second term is responsible for light-mediated DD interactions via the dipole radiation. The incident field intensities on the atom $j$ are $I^{(j)}/I_{\rm sat}=2\sum_{\mu}|{\cal R}_{\mu}^{(j)}/\gamma|^2$, where 
\beq \label{eq:saturationint}
I_{\rm sat}= \frac{4\pi^2\hbar c \gamma}{3\lambda^3}
\eeq
is the saturation intensity. 
In  Eqs.~\eqref{eq:multilevelhighintensitystochastic} and~\eqref{Eq:EffectiveRabi}, $ \rho_{g g}^{(j)}$ is eliminated according to the substitution $  \rho_{g g}^{(j)} =1-\sum_\eta \rho_{e\eta e\eta}^{(j)}$ and the repeated indices $\tau,\mu$ are implicitly summed over. 

The system of equations \eqref{eq:multilevelhighintensitystochastic} describes the coupled dynamics of the one-body density matrix elements for each atom. 
In the absence of the radiative DD coupling terms $\radKernel^{(j\ell)}_{\nu\mu}$  between the different atoms, which
depend on the atomic positions, the equations reduce to the standard OBEs (when $\bar{\mathcal{R}}_{\nu}^{(j)}$ is replaced by ${\mathcal{R}}_{\nu}^{(j)}$).
Various forms of the semiclassical equations have been employed in the studies of atomic arrays~\cite{Kramer2015a,Parmee2018}.

It is important to note that in typical experimental scenarios with fluctuating atomic positions, Eqs.~\eqref{eq:multilevelhighintensitystochastic} do not represent the mean-field approximation because the light-induced
spatial correlations do not factorize. Despite the factorization of quantum correlations between atoms according to Eq.~\eqref{eq:semiappr}, significant spatial correlations between atoms can still be present due to the DD interactions. The correlations induced by position fluctuations can be accounted for by combining the dynamics of Eqs.~\eqref{eq:multilevelhighintensitystochastic} with stochastic sampling of atomic positions~\cite{Lee16,Bettles2020}, as explained in Sec.~\ref{sec:fluc}.

\subsection{Limit of low light intensity}
\label{sec:LLI}

In the limit of low light intensity (LLI), to first order in the incident field amplitude,  the atoms respond to light as classical linear oscillators, resulting in the so-called coupled-dipole model. Keeping the terms that include at most one of
the amplitudes $\rho_{g e\eta}$ or $ \mathop{\cbE}^\pm$, and no $\rho_{e\nu e\eta}$, Eqs.~\eqref{eq:multilevelhighintensitystochastic} reduce to a linear system. The only dynamical variable then is
$\rho_{ge\nu}^{(j)}= \langle \hat{\sigma}_{j\nu}^{-}\rangle$ that satisfies 
\begin{align}\label{eq:Peoms}
\frac{d}{dt}& \rho_{ge\nu}^{(j)}
  =  \left( i \Delta_{j\nu}- \gamma \right)
\rho_{ge\nu}^{(j)} + i\bar{\mathcal{R}}_{\nu}^{(j)} \nonumber\\
&=  \left( i \Delta_{j\nu}- \gamma \right)
\rho_{ge\nu}^{(j)} +   i {\mathcal{R}}_{\nu}^{(j)}+i\xi\sum_{\ell\eta(\ell\neq j)}\radKernel^{(j\ell)}_{\nu\eta}\rho_{ge\eta}^{(\ell)} ,
\end{align}
with the definitions of Sec.~\ref{sec:qme} and $\bar{\mathcal{R}}_{\nu}^{(j)}$ given by \EQREF{Eq:EffectiveRabi}.
The linear equations of motion can be cast in matrix form
\beq
\dot{{\bf b}} = i (\mathcal{H}+\delta\mathcal{H}){\bf b} + {\bf f} ,
\label{eq:eigensystem}
\eeq
where (with $\nu,\mu=\pm1,0$)  ${\bf b}_{3j-1+\nu}=\rho_{ge\nu}^{(j)}$, the driving term $ {\bf f}_{3j-1+\nu}= i {\mathcal{R}}_{\nu}^{(j)}$,  and the non-Hermitian matrix 
\begin{subequations}
\label{eq:hmatrix}
\begin{align}
\mathcal{H}_{3j-1+\nu,3j-1+\nu} &= i\gamma,  \\
\mathcal{H}_{3j-1+\nu,3\ell-1+\mu} &=\Omega^{(j\ell)}_{\mu\nu}+i\gamma^{(j\ell)}_{\mu\nu}, \quad j\neq \ell.
\end{align}
\end{subequations}
The diagonal matrix $\delta\mathcal{H}$ contains the laser detuning and level shifts of the atoms, with elements $\Delta^{(j)}-\mu\delta_{\mu}^{(j)}$.

The dynamics in the limit of LLI can be described in terms of collective radiative excitation eigenmodes~\cite{Rusek96, Jenkins_long16} of the non-Hermitian matrix $\mathcal{H}$, while $\delta\mathcal{H}$ introduces the resonance conditions and couplings between the eigenmodes. The eigenmodes are biorthogonal ${\bf w}_j^\dagger{\bf v}_\ell=\delta_{j\ell}$ between the left ${\bf w}_j$ and right ${\bf v}_\ell$ eigenvectors ($\mathcal{H}{\bf v}_j=\lambda_j{\bf v}_j$; ${\bf w}_j^\dagger \mathcal{H}=\lambda_j{\bf w}_j^\dagger$).
However, they are generally not orthogonal ${\bf v}_j^\dagger{\bf v}_\ell\neq \delta_{j\ell}$.
The eigenvalues $\lambda_j=\delta_j+i \upsilon_j$ have real and imaginary parts, given by the collective line shift $\delta_j=\omega-\omega_j$ from the single-atom resonance $\omega$ and the collective resonance linewidth $\upsilon_j$.
When eigenvalues are degenerate,  it is possible to have an exceptional point where two or more eigenvectors coalesce and become linearly dependent, such that ${\bf v}_j$ no longer form a basis, $\mathcal{H}$ cannot be diagonalized,
and the radiatiation no longer shows exponential decay of independent modes~\cite{Ballantine21PT}. 

Here we mostly consider the isotropic $J=0\rightarrow J'=1$ transition and assume that symmetry-breaking level shifts are encapsulated in $\delta\mathcal{H}$.
This makes $\mathcal{H}$ a symmetric matrix, resulting in a simple relation ${\bf w}_j^\dagger={\bf v}^T_j$, and so ${\bf v}^T_j {\bf v}_\ell=\delta_{j\ell}$, except some possible cases of ${\bf v}^T_j {\bf v}_j=0$. Due to the non-orthogonality of the eigenvectors, the definition
\beq
L_j= {|{\bf v}_j^T {\bf b}|^2\over \sum_\ell | {\bf v}_\ell^T  {\bf b} |^2}
\label{eq:measure}
\eeq
can be used as a measure of the occupation of an eigenmode ${\bf v}_j$ in the state ${\bf b}$. This describes accurately the contribution of the dominant collective mode occupation in the excitation decay~\cite{Facchinetti16}.  The measure \eqref{eq:measure} can also be used to determine a finite-array eigenmode that most closely matches a desired infinite lattice mode. When the eigenmodes form a basis, the excitation amplitudes can be expressed as
${\bf b} (t) = \sum_{n} c_n(t) {\bf v}_n$
for the amplitudes $c_n$ that satisfy
$c_n(t) = \exp{\left[t (i\delta_n-\upsilon_n)\right]} c_n(0)$ when acted by ${\cal H}$ alone.

The ``weak'' intensity requirement of the driving light  strongly depends on the collective eigenmodes coupled to the drive~\cite{Williamson2020,Cipris21}. 
Even at very low intensities, subradiant eigenmodes with narrow linewidths exhibit nonlinear responses. Numerical studies have shown that the intensity threshold for the validity of the LLI description scales with the resonance linewidth of the excited eigenmode as $\upsilon_j^{2.5}$~\cite{Williamson2020}. However, the absence of saturation effects can always be reached with single photon sources.
If we are not interested in quantum correlations between the atomic dipoles or the photon statistics, but rather in the dynamics of a single electronic excitation amplitude (induced by a single photon), the system is formally equivalent to the classical LLI model of coherently-driven linearly coupled dipoles~\cite{SVI10}.
However, it should be noted that higher-order correlations and entanglement cannot be obtained in this classical description~\cite{Ballantine21quantum}.
To formulate a single-photon model,
we restrict the Hilbert space to the sector with only one excitation and expand the density matrix $\rho(t) = \ket{\Psi(t)}\bra{\Psi(t)} +P\ket{G}\bra{G}$ in terms of pure states, following the approach outlined in Ref.~\cite{Ballantine20ant}. Here
$\ket{\Psi(t)} = \sum_{j,\nu} \Pc^{(j)}_{\nu}(t)\,\hat{\sigma}^{+}_{j\nu} \ket{G}$ represents a state with precisely one excitation, characterized by the amplitudes $ \Pc^{(j)}_{\nu}$. The state $\ket{G}$ corresponds to all the atoms being in the electronic ground level, and
the probability of the photon emission is $P$. The full QME for $\bar{\rho}^{(j k)}_{\nu\mu}=\bra{G}\hat{\sigma}^{-}_{j\nu}\rho\hat{\sigma}^{+}_{k\mu}\ket{G}$ simplifies to 
\begin{equation}
\dot{\bar{\rho}}^{(j k)}_{\nu\mu} = i\sum_{\ell\tau}(\bar{\mathcal{H}}_{\nu\tau}^{(j\ell)}\bar{\rho}^{(\ell k)}_{\tau\mu}-i\bar{\rho}^{(j\ell)}_{\nu\tau}[\bar{\mathcal{H}}_{\tau\mu}^{(\ell k})]^\ast),
\end{equation} 
where  $\bar{\mathcal{H}}^{(jk)}_{\nu\mu}=\mathcal{H}_{3j-1+\nu,3k-1+\mu}$. The amplitudes $\Pc^{(j)}_{\nu}(t)$ then satisfy the same dynamics as the coherences of the LLI system, given by $\mathcal{H}$ in Eq.~\eqref{eq:eigensystem}. An incident single-photon 
pulse may be approximated by a time-dependent ${\bf f}(t)$ in Eq.~\eqref{eq:eigensystem}.

\subsection{Truncated correlations and two-excitation sectors}

Full quantum simulations based on the QME~\eqref{eq:rhoeom} are inherently limited to small atom numbers due to the
exponential growth of the system size as the number of atoms increases. Considerably larger system sizes can be achieved with the semiclassical approach of Sec.~\ref{sec:semi}, which includes the nonlinear interaction but neglects quantum fluctuations. However, the semiclassical approach can be improved by incorporating the lowest-order quantum contributions of internal atomic level correlations. 
The complete quantum treatment can be formulated as a hierarchy of equations for atomic correlation functions. This hierarchy can be truncated by assuming that correlations beyond a certain order become less significant and can be disregarded.
The truncation is accomplished by reducing the expectation values of higher-order operator products to products of 
lower-order expectation values via a cumulant~\cite{Kubo62}, or closely related, expansion. This approach has been successfully applied to atomic arrays in Refs.~\cite{Kramer2015a,Robicheaux23,Rubies-Bigorda23} (see also Ref.~\cite{Robicheaux21}) and an automated quantum optics toolbox~\cite{Plankensteiner22} 
has been developed to facilitate these calculations by incorporating the lowest-order quantum fluctuations with internal level atomic correlation functions.
The truncation
\beq
\< \hat A \hat B \hat C \> \rightarrow \<  \hat A \hat B \>\<\hat C\> + \<\hat A\>  \<  \hat B \hat C \> + \<\hat B\>  \<  \hat A \hat C \>-2\<\hat A\>
\<\hat B\>\<\hat C\>,
\eeq
of three-operator correlations in Ref.~\cite{Robicheaux23} resulted in a closed set of equations for light transmission, which qualitatively agreed with the comparisons between the full quantum dynamics and the semiclassical equations in Ref.~\cite{Bettles2020}.

Another approach to limit the size of the Hilbert space in the full quantum system is truncation. By including only up to two electronic excitations, some nonlinear interactions can be incorporated while maintaining numerical tractability in larger arrays. Two-excitation systems exhibit intriguing phenomena such as fermionization of the excitations~\cite{Zhang2018,Henriet2018}, multiple subradiant excitations forming a superposition of singly excited subradiant states~\cite{Zhang2018}, entanglement~\cite{Ritsch_subr},  and bound dimer states~\cite{Zhang20,Parmee19}. Software packages are available to numerically solve these systems as well~\cite{Plankensteiner22}.

\subsection{Scattered light properties}
\label{sec:scatteredlight}

The scattered field consists of a mean field $\left<\hat{\mathbf{E}}_s^+\right>$ and fluctuations $\delta\hat{\mathbf{E}}_s^+=\hat{\mathbf{E}}_s^+-\left<\hat{\mathbf{E}}_s^+\right>$. 
We analyze different contributions to the scattered light by expanding the correlations that yield the total light intensity matrix
\begin{align}
I(\mathbf{r})=2\epsilon_0 c\left<\hat{\mathbf{E}}^-(\mathbf{r})\hat{\mathbf{E}}^+(\mathbf{r})\right>.
\end{align}
We obtain from Eq.~\eqref{eq:totallight}
\begin{align}
\label{eq:introducingFluctuations}
\left\langle 
\mathop{\hat{\mathbf{E}}^-(\mathbf{r})} 
\mathop{\hat{\mathbf{E}}^+(\mathbf{r})} \right\rangle
&=
\mathop{\cbEL^-(\mathbf{r})} 
\mathop{\cbEL^+(\mathbf{r})} +
\cbEL^-(\mathbf{r}) \left\langle\vecHat{E}{s}^+(\mathbf{r})\right\rangle 
+
\left\langle\vecHat{E}{s}^-(\mathbf{r})\right\rangle \cbEL^+(\mathbf{r}) \nonumber\\
&+
\left\langle \vecHat{E}{s}^-(\mathbf{r}) \right\rangle
\left\langle \vecHat{E}{s}^+(\mathbf{r}) \right\rangle 
+
\left\langle \mathop{\delta\vecHat{E}{s}^-(\mathbf{r})} 
\mathop{\delta\vecHat{E}{s}^+(\mathbf{r})}\right\rangle.
\end{align}
Here $\hat{\mathbf{E}}^{-}\hat{\mathbf{E}}^{+}$ is a dyadic product of elements $\hat{E}^-_{\alpha} \hat{E}^+_{\beta}$, with $\alpha,\beta \in\{1,2,3\}$ cycling over the polarization components, where the intensity is proportional to its diagonal elements.
The first term in Eq.~\eqref{eq:introducingFluctuations} corresponds to the incident field intensity alone. The next two terms give the interference between the incident field and coherently scattered field, which is revealed in homodyne measurements~\cite{WisemanMilburn}. The fourth term is proportional to the coherently scattered light intensity and the final term represents incoherent scattering
\begin{equation} \label{eq:incoherentScatteringFieldExpectation}
\left\langle \delta\vecHat{E}{s}^-(\mathbf{r}) \,\delta\vecHat{E}{s}^+(\mathbf{r}) \right\rangle = 
\left\langle\hat{\mathbf{E}}_s^-(\mathbf{r}) \,\hat{\mathbf{E}}_s^+(\mathbf{r}) \right\rangle - 
\left\langle\hat{\mathbf{E}}_s^-(\mathbf{r})\right\rangle
\left\langle\hat{\mathbf{E}}_s^+(\mathbf{r})\right\rangle.
\end{equation}
If only the scattered light is detected, e.g., by blocking the incident light by a thin wire, as in the dark-ground imaging~\cite{andrews1996}, incoherent scattering is obtained  by subtracting the coherently scattered intensity from the total intensity. 

We first consider the atoms at the fixed set of positions $\{\textbf{r}_1,\textbf{r}_2,\ldots, \textbf{r}_N\}$. The analysis of light scattering involving many-atom systems with fluctuating positions is more intricate and will be addressed in Sec.~\ref{sec:fluc}.  According to Eq.~\eqref{eq:scattlight}, the coherently scattered light reads
\begin{equation} \label{eq:cohscattlight}
\epsilon_0 \langle \hat{\textbf{E}}{}^+_s(\textbf{r}) \rangle =\sum_{j\nu}\mathsf{G}(\textbf{r}-\textbf{r}_j)\textbf{d}_{ge}^{(\nu)} \langle \hat{\sigma}_{j\nu}^-\rangle .
\end{equation}
For a detector sufficiently far away from the atoms $r\gg \lambda$,  the scattered light amplitudes can be evaluated in the far-field radiation zone~\cite{Jackson,BOR99},
\begin{align}\label{eq:rzone}
\mathsf{G}(\mathbf{r}-\mathbf{r}_j) \textbf{d}_{ge}^{(\nu)} \sim \frac{k^2}{4\pi r}e^{ikr-ik\hat{\mathbf{r}}\cdot\mathbf{r}_j}(\hat{\mathbf{r}}\times \textbf{d}_{ge}^{(\nu)} )\times \hat{\mathbf{r}},
\end{align}
where the unit vector $\hat{\mathbf{r}}$ joins a representative point in the sample to the observation point.
In Eq.~\eqref{eq:cohscattlight}, there is no distinction between quantum and semiclassical coherent scattering for a single atom, since in the semiclassical case $\hat{\sigma}_{j\nu}^{-}$ is replaced by $\langle\hat{\sigma}_{j\nu}^{-}\rangle $~\cite{meystre1998}. 
Therefore, any disparity between quantum and semiclassical coherent scattering in a many-atom ensemble is solely attributable to many-body quantum effects.
The intensity of the scattered light is given by
\begin{align}\label{intenscatt}
2\epsilon_0c &\left\langle \mathop{\hat{\mathbf{E}}_s^-(\mathbf{r}) }  \mathop{\hat{\mathbf{E}}_s^+(\mathbf{r}) }\right\rangle \nonumber\\
&=  \frac{2c}{\eo}\sum_{j\ell\nu\mu } \left[\mathsf{G}(\mathbf{r}-\mathbf{r}_j) \textbf{d}_{ge}^{(\nu)} \right]^*\mathsf{G}(\mathbf{r}-\mathbf{r}_\ell) \textbf{d}_{ge}^{(\mu)}\< \hat{\sigma}_{j\nu}^{+}\hat{\sigma}_{\ell\mu}^{-} \>\nonumber\\
& =  \frac{{\cal D}^2 k^4 c}{8\pi^2 \eo r^2}   \sum_{j\ell\nu\mu} [\delta_{\nu,\mu}-(\hat{\mathbf{r}}\cdot \pol^*_\nu ) (\hat{\mathbf{r}}\cdot \pol_\mu ) ] e^{ik\hat{\mathbf{r}}\cdot\mathbf{r}_{j\ell}} \< \hat{\sigma}_{j\nu}^{+}\hat{\sigma}_{\ell\mu}^{-} \>,
\end{align}
where the $j=\ell$ term is due to single-atom contributions. 
The $j\neq\ell$ terms originate from many-atom couplings and interferences. In the second equality, we employed the far-field limit [Eq.~\eqref{eq:rzone}] that simplifies the double summation over the atomic positions to  a sum of different phase factors and correlations,
where ${\mathbf{r}}_{j\ell}=\mathbf{r}_{\ell}-\mathbf{r}_{j}$ (see also Appendix~\ref{appen2}). In the semiclassical scattering version $\hat{\sigma}_{j\nu}^{+}$ is again replaced by $\langle\hat{\sigma}_{j\nu}^{+}\rangle $.

We can derive a simple formula for the total intensity of scattered light that illustrates the collective effects. 
The rate of photon scattering is calculated by integrating the scattered intensity per the energy of the photon over a closed surface $S$ enclosing the atoms. In Appendix~\ref{appen2}, we show
how this is expressed in terms of  a sum over the imaginary parts of the DD radiation coupling tensor between the different atoms 
\begin{align}\label{eq:scatrateformula} 
n_s &= \frac{1}{\hbar\omega} \int_{S} dS I_s = \frac{ 2\eo c}{\hbar\omega} \int_{S} dS \<  \hat{\textbf{E}}{}^-_s(\textbf{r}) \cdot \hat{\textbf{E}}{}^+_s(\textbf{r}) \rangle\nonumber\\
&=  2\gamma  \sum_{j\nu} \langle \hat{\sigma}_{j\nu}^+ \hat{\sigma}_{j\nu}^-\rangle  + 2  \sum_{j\ell\nu\mu (j\neq\ell)} \gamma_{\nu\mu}^{(j\ell)}  \langle \hat{\sigma}_{j\nu}^+ \hat{\sigma}_{\ell\mu}^-\rangle.
\end{align}
The first term represents the single-atom decay [Eq.~\eqref{eq:WW}] summed over all the atoms.
This is given by the excited-level population, according to $\hat{\sigma}_{j\nu}^{+}\hat{\sigma}_{j\nu}^{-} = \hat{\sigma}^{ee}_{j\nu}$. [Note that the terms proportional to $\hat{\sigma}_{j\nu}^{+}\hat{\sigma}_{j\mu}^{-} $, with $\nu\neq\mu$, in Eq.~\eqref{intenscatt} do not contribute to Eq.~\eqref{eq:scatrateformula}.]
The second term represents collective decay $\xi {\rm Im} [{\radKernel}^{(j\ell)}_{\nu\mu} ]$, introduced  in Eq.~\eqref{eq:omga}. The emission rates obtained here confirm the correct choice of the decay rates in the QME~\eqref{eq:rhoeom}. The simple rate formula~\eqref{eq:scatrateformula} applies to other observables of scattered light, such as coherent scattering, in which case the correlation functions are changed to
\beq\label{eq:cohscatrateformula} 
n_c =  2\gamma  \sum_{j\nu} \langle \hat{\sigma}_{j\nu}^+\>\< \hat{\sigma}_{j\nu}^-\rangle  + 2  \sum_{j\ell\nu\mu (j\neq\ell)} \gamma_{\nu\mu}^{(j\ell)} \langle \hat{\sigma}_{j\nu}^+\>\< \hat{\sigma}_{\ell\mu}^-\rangle.
\eeq

According to Eq.~\eqref{eq:incoherentScatteringFieldExpectation}, the incoherently scattered light contribution is obtained from
\begin{align}\label{eq:incoherentScattering}
\epsilon_0^2 \left\langle \delta\mathop{\hat{\mathbf{E}}_s^-(\mathbf{r}) }  \delta\mathop{\hat{\mathbf{E}}_s^+(\mathbf{r}) }\right\rangle 
& = \sum_{j\ell\nu\mu} \left[\mathsf{G}(\mathbf{r}-\mathbf{r}_j) \textbf{d}_{ge}^{(\nu)} \right]^*\mathsf{G}(\mathbf{r}-\mathbf{r}_\ell) \textbf{d}_{ge}^{(\mu)} \nonumber\\
&\times\( \< \hat{\sigma}_{j\nu}^{+}\hat{\sigma}_{\ell\mu}^{-} \> -  \< \hat{\sigma}_{j\nu}^{+}\> \<\hat{\sigma}_{\ell\mu}^{-} \> \).
\end{align}
In the far-field radiation zone, this can be simplified analogously to Eq.~\eqref{intenscatt}. In the case of a simple two-level atom, the single-atom contributions ($j=\ell$) are then proportional to the excited level population minus the the absolute square of the atomic coherences $ \langle  \hat{\sigma}^{ee}_{j}\rangle - | \langle \hat{\sigma}_{j}^-\rangle |^2 $.
In the semiclassical scattering approximation~\cite{meystre1998}, where the correlation functions for internal atomic operators factorize, the incoherently scattered light intensity in Eq.~\eqref{eq:incoherentScattering} vanishes. 
The semiclassical approximation to the scattered light is consistent with a systematic way of disregarding all quantum fluctuations when the atomic response is first calculated from the semiclassical atom dynamics of Eq.~\eqref{eq:multilevelhighintensitystochastic}. Therefore, any disparity between this approach and evaluating scattered light of Eq.~\eqref{eq:incoherentScattering} using the full solution of QME~\eqref{eq:rhoeom} provides a signature of \emph{quantum} effects in the collective atomic response~\cite{Bettles2020} -- the result of which is also valid in the case of fluctuating atomic positions when Eq.~\eqref{eq:incoherentScattering} is nonvanishing even semiclassically (see Sec.~\ref{sec:opticsfluctuations}).

For the full quantum analysis, the sum over single-atom contributions ($j=\ell$) in Eq.~\eqref{eq:incoherentScattering} yields a nonzero result.
Even for a single atom,  the incoherent scattering therefore differs depending on whether we treat it quantum-mechanically or semiclassically. 
This distinction contrasts with coherent scattering, where the single-atom contributions are identical in both quantum and semiclassical analyses.

 In the case of fixed atomic positions, many-body contributions to incoherent scattering arise from
nonnegligible many-body correlations $ \< \hat{\sigma}_{j\nu}^{+}\hat{\sigma}_{\ell\mu}^{-} \> \neq \< \hat{\sigma}_{j\nu}^{+}\> \<\hat{\sigma}_{\ell\mu}^{-} \>$, for $j\ne \ell$, which can be generated through light-mediated interactions.  These contributions represent quantum many-body effects in the scattered light that are absent in the semiclassical analysis. Without these quantum many-body effects and position fluctuations, all the incoherent scattering in  Eq.~\eqref{eq:incoherentScattering} originates solely from the single-atom quantum effects ($j=\ell$).

There exist straightforward methods to improve the semiclassical model of incoherent scattering without increasing computational complexity~\cite{Bettles2020}. These improvements provide greater accuracy  and offer a systematic approach to identify light-induced \emph{many-body} quantum effects in the optical response~\cite{Bettles2020}. The dynamical response of matter is once again calculated within the semiclassical approximation of Eq.~\eqref{eq:multilevelhighintensitystochastic} by neglecting all quantum fluctuations,
but we add to the semiclassical scattering description the sum of independent single-atom quantum contributions ($j=\ell$) in the incoherent scattering expression [Eq.~\eqref{eq:incoherentScattering}] and obtain
\begin{align}\label{eq:SAQincoherentScattering}
I_{\rm inc}^{\rm SAQ}=&\frac{2c}{\eo}
\sum_{j\nu\mu} \left[\mathsf{G}(\mathbf{r}-\mathbf{r}_j) \textbf{d}_{ge}^{(\nu)} \right]^*\mathsf{G}(\mathbf{r}-\mathbf{r}_j) \textbf{d}_{ge}^{(\mu)} \nonumber\\
&\times\( \< \hat{\sigma}_{j\nu}^{+}\hat{\sigma}_{j\mu}^{-} \> -  \< \hat{\sigma}_{j\nu}^{+}\> \<\hat{\sigma}_{j\mu}^{-} \> \).
\end{align}
For the case of only one excited level participating, there is no summation over $\mu,\nu$ and the first term provides the excited level population of the atom $j$. We obtain analogously to Eq.~\eqref{eq:scatrateformula} over a closed surface
\beq
n_{\rm inc}^{\rm SAQ} = \frac{1}{\hbar\omega} \int_{S} dS I_{\rm inc}^{\rm SAQ} = 2\gamma\sum_j \( \<  \hat{\sigma}_{j}^{ee} \> -|\<  \hat{\sigma}_{j}^{+} \>|^2 \).
\label{eq:saqscatrate} 
\eeq
The difference between Eq.~\eqref{eq:SAQincoherentScattering} when the dynamics is solved semiclassically and \EQREF{eq:incoherentScattering} when it is solved using the QME \eqref{eq:rhoeom} represents a signature of quantum many-body effects in the scattered light. $I_{\rm inc}^{\rm SAQ}$ in the presence of position fluctuations is discussed in Sec.~\ref{sec:opticsfluctuations}.

\subsection{Quantum trajectories}
\label{sec:traj}

The QME~\eqref{eq:rhoeom} 
can be solved using exact diagonalization methods. Alternatively, a numerically more efficient approach for large systems can be achieved by employing quantum trajectories of state vectors~\cite{dalibard1992,Tian92,Dum92,molmer1993}, where the dynamics is  unraveled into stochastic realizations of a wavefunction $|\psi(t)\>$.
In the quantum trajectory approach, the evolution of a density matrix of the master equation is replaced by the evolution of a smaller sized stochastic state vector that is run multiple times.
For the master equation
\beq
\dot{\hat\rho} = -\frac{i}{\hbar} [\hat H_s,\hat\rho]+\sum_j (2 \Jop_j\hat \rho \Jop_j^\dagger-\Jop_j^\dagger \Jop_j\hat\rho-\hat\rho \Jop_j^\dagger \Jop_j),
 \label{eq:modelmaster}
\eeq
the procedure involves evolving $|\psi(t)\>$ according to a non-unitary Hamiltonian $\hat H_s - {i\hbar}\sum_j\Jop_j^\dagger\Jop_j$, incorporating randomly determined quantum ``jumps'', followed by wave-function normalization. During the time interval $[t,t+dt]$, a quantum jump $\sqrt{2} \Jop_j|\psi(t)\>$ occurs with a probability $P_j=2\<\psi(t)| \Jop_j^\dagger \Jop_j|\psi(t)\>$. The additional advantage is that if the quantum jumps correspond to photon counts, the stochastic wavefunction is conditioned on photon detection records, describing quantum measurement-induced back-action.

In the context of many-atom light scattering with the QME~\eqref{eq:rhoeom}, this approach is efficiently formulated in terms of the source-mode jump operators~\cite{clemens2003a,carmichael2000}. The unitary part $\hat H_s$ of the stochastic wavefunction evolution is straightforwardly obtained by comparing Eqs.~\eqref{eq:rhoeom} and~\eqref{eq:modelmaster}.
The jump operators are expressed 
in the LLI excitation eigenmode basis ${\bf v}_j$, which diagonalizes $\Omega^{(j\ell)}_{\mu\nu}+i\gamma^{(j\ell)}_{\mu\nu}$ (Sec.~\ref{sec:LLI}).
The jump operators are then defined as $ \Jop_j = \sqrt{\upsilon_j} {\bf v}_j^T \boldvec{\h\Sigma} $ and $ \Jop_j^\dagger = \sqrt{\upsilon_j}\boldvec{\h\Sigma}^\dagger {\bf v}_j$, where $\boldvec{\h\Sigma}$ is a column vector composed of $\hat{\sigma}_{\ell\nu}^{-}$, and $\upsilon_j$ is the collective linewidth.  It is important to note that since the source-mode jump operators are expressed in terms of the LLI eigenmodes, they are generally not identified with photon detection events in individual trajectories~\cite{clemens2003a,carmichael2000}. 

The unraveling of the QME~\eqref{eq:rhoeom}  into quantum trajectories, where the quantum jumps represent photon counts, is achieved by resolving the emission time and direction~\cite{carmichael2000}. Consider the photon emission rate $n_s$ integrated over all directions in Eqs.~\eqref{eq:scatrateformula} and~\eqref{eq:emissionrateapp}. This rate can be divided into a discrete set of scattering directions $(\theta,\phi)$, each associated with a solid angle $d\Omega$, such that $n_s\simeq \sum_{j} n_s(\theta_j,\phi_j) d\Omega$. The unraveling then follows similar techniques to the integration of the total rate $n_s$ outlined in Appendix~\ref{appen2}. In the case of two-level atoms in the far field radiation zone [\EQREF{eq:rzone}], the jump operators have a simple analytic form [compare with \EQREF{Imnint}]
\beq
\Jop(\theta,\phi) = \left\{\frac{3\gamma}{4\pi}  [1-  (\hat{\mathbf{r}}\cdot \pol )^2 ]  d\Omega\right\}^{1/2}\sum_\ell e^{-ik\hat{\mathbf{r}}\cdot\mathbf{r}_{\ell}}  \hat{\sigma}_{\ell}^{-} ,
\eeq
where the unit vector $\hat{\mathbf{r}} = \hat{\mathbf{x}}\cos\phi\sin\theta+\hat{\mathbf{y}}\sin\phi\sin\theta+\hat{\mathbf{z}}\cos\theta$, and $\pol$ denotes the orientation of the atomic dipole.

\section{Fluctuations of atomic positions}
\label{sec:fluc}

\subsection{Stochastic simulations}
\label{sec:simufluc}

The atoms in a periodic array experience fluctuations of their positions~\cite{Morsch06}. The fluctuations can be thermal or zero-point quantum fluctuations at the ground state of the trapping potential at each lattice site. Such fluctuations only become negligible with very tight confinement. To describe the effect of position fluctuations of atoms in a periodic array, we follow the formalism introduced in Ref.~\cite{Jenkins2012a}.  

The position fluctuations of atoms in an array introduce characteristics of light propagation typically observed in disordered media. This considerably complicates the optical response even in the limit of LLI when the response of an atom to coherent incident field is that of an entirely classical linear harmonic oscillator (Sec.~\ref{sec:LLI}). The study of light propagation in disordered media has been actively pursued in mesoscopic physics for a considerable duration~\cite{Ishimaru1978,Lagendijk,vanRossum,Kupriyanov2017}. Recurrent scattering of light between closely spaced particles can induce strong position correlations within a classical framework. This phenomenon plays a crucial role, e.g., in localization of light, which is akin to the Anderson localization of electrons in solids. Despite numerous efforts to observe light localization in 3D media, it remains a subject of considerable controversy and debate~\cite{Sperling2016,Skipetrov14,Segev2013}.  Various proposals have been put forth to achieve 3D light localization using atomic ensembles~\cite{Skipetrov18}, offering the advantage of potentially incorporating quantum effects as well.

The impact of position correlations, resulting from position fluctuations, is most effectively described by employing atomic field operators for the ground and excited states $\psihat{g,e}(\mathbf{r})$~\cite{Ruostekoski1997a}, introduced in Sec.~\ref{sec:basic}.  To solve the scattered light of Eq.~\eqref{eq:scattlight}, 
it is necessary to calculate the expectation value for the atomic polarization density $\<\hat{\textbf{P}}(\mathbf{r})\>$.  

The analysis, specifically focusing on the limit of LLI, has revealed how multiply scattered light can establish correlations between atoms at fluctuating positions. These correlations give rise to a hierarchy of equations of motion for the correlation functions of atomic density and polarization~\cite{Morice1995a,Ruostekoski1997a}. Within this hierarchy, the atomic polarization density becomes coupled to a two-atom correlation function, such as $\langle \mathop{\psihatdag{g}(\mathbf{r})} \mathop{\psihatdag{g}(\mathbf{r}')} \mathop{\psihat{e}(\mathbf{r}')} \mathop{\psihat{g}(\mathbf{r})}\rangle$, which represents the correlations in the optical response of a ground-state atom at position ${\bf r}$ in the presence of an atomic dipole in the second atom at position ${\bf r}'$. These correlations originate from the resonant DD interaction between the atoms, which depends on their relative positions, as the strength of the DD interaction is sensitive to the atomic separation. In turn, the two-body correlation relies on the three-atom correlation function, and so on. Finding general solutions to this hierarchy of equations
is challenging when the atom density limit $\rho/k^3\ll 1$  in terms of the resonant wave number of light $k$ is not satisfied.
On the other hand, in the limit $\rho/k^3\ll 1$, perturbative solutions may be derived~\cite{Morice1995a,Ruostekoski1999a}. The hierarchy of equations for correlation functions can also be truncated for the case of inhomogeneously broadened samples~\cite{Javanainen2014a}. 

An alternative approach, applicable in the limit of LLI, involves utilizing coupled-dipole model equations~\eqref{eq:Peoms} for a fixed set of atomic positions $\{\rv_1,\rv_2, \ldots, \rv_N\}$. 
In this approach, the position coordinates are treated as stochastic variables and are sampled 
according to the probability distribution for the atomic positions in the absence of light.
This Monte Carlo sampling enables stochastic classical-electrodynamics simulations: In each realization, we sample the atomic positions
and subsequently solve the optical response for the given set of coordinates using Eqs.~\eqref{eq:Peoms} and~\eqref{eq:scattlight}. 
The required probability distribution is obtained as the absolute square of the many-body wave function $P(\rv_1,\rv_2, \ldots, \rv_N) = |\Psi(\rv_1,\rv_2, \ldots, \rv_N) |^2$~\cite{Javanainen1999a}. 
In the case of classically distributed atoms, it is typically sufficient to sample uncorrelated independent positions. However, for atoms obeying the Fermi-Dirac statistics, more sophisticated sampling methods, such as the Metropolis algorithm, have been implemented to solve the optical responses~\cite{Ruostekoski_waveguide}.

Ensemble-averaging over many realizations provides the expectation values of physical observables. In the limit of LLI, this can formally be shown to converge to an exact solution for stationary atoms at arbitrary densities by reproducing the correct hierarchy of correlation functions for both 1D scalar theory~\cite{Javanainen1999a} and 3D vector-electrodynamics with a single electronic ground level~\cite{Lee16}. 
Since each stochastic realization of atomic positions can be interpreted as an outcome of a quantum measurement process on scattered light that localizes the positions of the atoms, each stochastic trajectory also represents a possible outcome of a single experimental run. Numerically ensemble-averaging over many realizations then corresponds to an experimental ensemble-averaging over many measurement runs.
Within this interpretation, 
stochastic electrodynamics simulations can be extended beyond the LLI regime~\cite{Lee16,Bettles2020}. The nonlinear coupled-dipole equations~\eqref{eq:multilevelhighintensitystochastic}, which correspond to the semiclassical approximation, or the full QME~\eqref{eq:rhoeom}, are solved for a fixed set of atom positions in each stochastic realization. By ensemble-averaging over many realizations, the effects of position fluctuations on the physical observables are captured.

The semiclassical approximation, Eqs.~\eqref{eq:multilevelhighintensitystochastic}, does not represent a mean-field approximation when the atomic position fluctuate, since the light-induced correlations depend on the higher-order atomic correlation functions. In second quantization, the correlations are given as ensemble averages of individual realizations, sampled according to the probability distribution of atomic positions $P(\mathbf{r}_1,\dots,\mathbf{r}_N)$. Each realization corresponds to some fixed $N$-atom configuration of positions $\{\rv_1,\rv_2, \ldots, \rv_N\}$. For example,
\begin{widetext}
\begin{align} 
\label{eq:sampling2}
&\left\langle 
\mathop{\hat{\psi}^{\dag}_{e\nu}(\mathbf{r},t)} 
\mathop{\hat{\psi}^{\dag}_g(\mathbf{r}',t)} 
\mathop{\hat{\psi}_{e\mu}(\mathbf{r}',t)} 
\mathop{\hat{\psi}_g(\mathbf{r},t)} \right\rangle 
= 
  \int \mathop{\mathrm{d}^3 r_1} \ldots \mathop{\mathrm{d}^3 r_N} 
\left\langle 
\mathop{\hat{\psi}^{\dag}_{e\nu}(\mathbf{r},t)} 
\mathop{\hat{\psi}^{\dag}_g(\mathbf{r}',t)} 
\mathop{\hat{\psi}_{e\mu}(\mathbf{r}',t)} 
\mathop{\hat{\psi}_g(\mathbf{r},t)} \right\rangle_{\{\mathbf{r}_1,\dots,\mathbf{r}_N\}}  \mathop{P(\mathbf{r}_1,\dots,\mathbf{r}_N)},
\end{align}
\end{widetext}
where $\left\langle 
\mathop{\hat{\psi}^{\dag}_{e\nu}(\mathbf{r},t)} 
\mathop{\hat{\psi}^{\dag}_g(\mathbf{r}',t)} 
\mathop{\hat{\psi}_{e\mu}(\mathbf{r}',t)} 
\mathop{\hat{\psi}_g(\mathbf{r},t)} \right\rangle_{\{\mathbf{r}_1,\dots,\mathbf{r}_N\}}$ is calculated in a single realization of fixed positions $\{\rv_1,\rv_2, \ldots, \rv_N\}$,
and is given in terms of the  
operators $\hat\sigma^\pm_{j\nu}$ by
\begin{align} 
\label{eq:padbcExp}
&\left\langle \mathop{\hat{\psi}^{\dag}_{e\nu}(\mathbf{r},t)} \mathop{\hat{\psi}^{\dag}_g(\mathbf{r}',t)} \mathop{\hat{\psi}_{e\mu}(\mathbf{r}',t)} \mathop{\hat{\psi}_g(\mathbf{r},t)} \right\rangle_{\{\mathbf{r}_1,\dots,\mathbf{r}_N\}}\nonumber\\
& \;\;\;\;\;  =  \sum_{j\ell (j\neq\ell)} \< \hat{\sigma}_{j\nu}^{+}(t)\hat{\sigma}_{\ell\mu}^{-}(t) \>  \mathop{\delta(\mathbf{r} - \mathbf{r}_j)} \mathop{\delta(\mathbf{r}'-\mathbf{r}_{\ell})} , 
\end{align}
where the summations run over all the atoms, excluding the cases where the both operators refer to the same atom.

Ensemble-averaged solutions to semiclassical dynamics [Eqs.~\eqref{eq:multilevelhighintensitystochastic}] correspond to field-theoretical expectation values.  Even though we factorize quantum correlations between internal levels of the atoms $\< \hat{\sigma}_{j\nu}^{+}(t)\hat{\sigma}_{\ell\mu}^{-}(t) \>=\< \hat{\sigma}_{j\nu}^{+}(t)\>\<\hat{\sigma}_{\ell\mu}^{-}(t) \>$ $(j\neq\ell)$ [Eq.~\eqref{eq:semiappr}],  nonvanishing spatial correlations of Eq.~\eqref{eq:sampling2} induced by light do not factorize.  In general ($a,b,c,d \in \{g,e\nu\}$), 
\beq
\label{eq:2bodycorrelations}
\langle \hat{\psi}^{\dagger}_a({\bf r})\hat{\psi}^{\dagger}_b({\bf r}')\hat{\psi}_c({\bf r}')\hat{\psi}_d({\bf r}) \rangle \neq \langle\hat{\psi}^{\dagger}_a({\bf r})\hat{\psi}_d({\bf r})\rangle\langle\hat{\psi}^{\dagger}_b({\bf r}')\hat{\psi}_c({\bf r}')\rangle.
\eeq

\subsection{Optical response with position disorder}
\label{sec:opticsfluctuations}

Nonvanishing spatial correlations of Eq.~\eqref{eq:2bodycorrelations} due to position fluctuations act as additional sources of  incoherent scattering that are absent when the atoms are at fixed positions [Eq.~\eqref{eq:incoherentScattering}]. The scattered light intensity of \EQREF{intenscatt} is generalized in Ref.~\cite{Bettles2020} to include disorder in positions
\begin{align} \label{eq:EdEdExpectation}
2c\eo \left\langle \mathop{\hat{\mathbf{E}}_s^-(\mathbf{r}) }\mathop{\hat{\mathbf{E}}_s^+(\mathbf{r}) }\right\rangle 
= 
\frac{2c}{\epsilon_0} \int \mathop{\mathrm{d}^3 \rAtom}
\mathop{\mathrm{d}^3 \rAtom'} &
\left[ \mathsf{G}(\mathbf{r}-\rAtomBold) \right]^*
\left[ \mathsf{G}(\mathbf{r}-\rAtomBold') \right] \nonumber\\
&\times\left\langle \mathop{\hat{\mathbf{P}}^-(\rAtomBold)} \mathop{\hat{\mathbf{P}}^+(\rAtomBold')}  \right\rangle,
\end{align}
where  $[\mathsf{G}(\mathbf{r}-\rAtomBold)]^*$ acts on $\hat{\mathbf{P}}^-(\rAtomBold)$ and likewise $\mathsf{G}(\mathbf{r}-\rAtomBold')$ on $\hat{\mathbf{P}}^+(\rAtomBold')$.
Here $\left\langle \mathop{\hat{\mathbf{P}}^-(\rAtomBold)} \mathop{\hat{\mathbf{P}}^+(\rAtomBold')}  \right\rangle$ depends on the correlations in Eq.~\eqref{eq:sampling2} indicating how it is determined by the atom positions. As shown in Appendix~\ref{appen3}, the intensity in Eq.~\eqref{eq:EdEdExpectation} depends on the correlations $ \< \hat{\sigma}_{j\nu}^{+}\hat{\sigma}_{\ell\mu}^{-} \>$ for each realization of fixed positions  $\{\rv_1,\rv_2, \ldots, \rv_N\}$.
Solving the full coupled quantum dynamics from the QME~\eqref{eq:rhoeom} to obtain these correlations and ensemble-averaging over many such realizations provides the quantum solution of scattered light intensity.
In the semiclassical approximation of scattering~\cite{meystre1998}, we factorize the polarization correlations in each realization 
\begin{align} \label{semiresponse}
&\left\langle \hat{\mathbf{P}}^-(\rAtomBold) \hat{\mathbf{P}}^+(\rAtomBold')  \right\rangle \simeq 
 \int \mathop{\mathrm{d}^3 r_1} \ldots \mathop{\mathrm{d}^3 r_N} \nonumber\\  &\times
 \left\langle \hat{\mathbf{P}}^-(\rAtomBold) \right\rangle_{\{\mathbf{r}_1,\dots,\mathbf{r}_N\}}   \left\langle \hat{\mathbf{P}}^+(\rAtomBold') \right\rangle_{\{\mathbf{r}_1,\dots,\mathbf{r}_N\}} 
 \mathop{P(\mathbf{r}_1,\dots,\mathbf{r}_N)}.
\end{align}
When the atom-light dynamics is then also solved using the semiclassical equations, the quantum effects are then systematically neglected. Similarly to Sec.~\ref{sec:scatteredlight}, the difference between the semiclassical and full quantum solution reveals the signature of quantum effects in the scattered light~\cite{Bettles2020}. Analogous, albeit more involved, analysis is performed to improve the scattered light approximation to include the sum of single-atom quantum effects to the semiclassical scattering model to
generalize the fixed atomic position results of \EQREF{eq:SAQincoherentScattering}~\cite{Bettles2020}.

\subsection{Implementation in an optical lattice}
\label{sec:latticepot}

To characterize the influence of position fluctuations in an atomic array on the optical response, we consider the experimental setup~\cite{Rui2020,Srakaew22} of a sinusoidal optical lattice potential in a 
Mott-insulator state with one atom per site, following the simulation scheme introduced in Ref.~\cite{Jenkins2012a}. Experimentally, single-occupancy Mott-insulator states have been studied and manipulated in individual sites already for quite some time~\cite{BakrEtAlNature2009,ShersonEtAlNature2010,Weitenberg2011}. Achieving single-atom site occupancy can be accomplished by cooling atoms to the typical ``wedding-cake'' Mott-insulator ground state of an optical lattice superposed on a weak harmonic trap, and then manipulating the sites with excess occupancy~\cite{Weitenberg2011}.
For the purpose of classical stochastic-electrodynamics simulations to calculate the optical response with unit occupancy, the position coordinates of atoms within each site are treated as independent stochastic variables. Another interesting scenario would be to have two fermionic atoms per site~\cite{Orioli19,Orioli20}.
In the case of a 2D square lattice with a periodicity $a$ in the $yz$ plane, 
we express the potential in the units of the lattice recoil energy $E_R = \pi^2\hbar^2/(2ma^2)$ \cite{Morsch06}
\begin{equation}
  \label{eq:V_lat}
  V = sE_R\left[\sin^2\left(\pi {y}/{a}\right) +
    \sin^2\left(\pi {z}/{a}\right) \right],
\end{equation}
where  the lattice strength is denoted by $s$. The confinement of atoms in the $x=0$ plane is achieved through an additional potential, which can also be generated by an optical lattice.

In the optical imaging experiments, the atoms are confined in deep lattice potentials~\cite{Rui2020,Srakaew22}. In the vicinity of the lattice site minimum,  the potential $V$ is approximately harmonic, with
trapping frequencies $\omega_y = \omega_z = 2\sqrt{s}E_R/\hbar$ within the lattice plane and $\omega_x$ perpendicular to it.
The atoms occupy the vibrational ground states of the lattice sites at $\spvec{r}_j$, and their positions exhibit quantum fluctuations, resulting in Gaussian density distributions $|\phi_j(\rv)|^2$ of the Wannier wavefunctions $\phi_j(\rv) \equiv \phi(\rv - \spvec{r}_j)$. For the site $\spvec{r}_j=0$, we express it as:
\begin{equation}
  \label{eq:phi}
  |\phi(\rv)|^2 = \frac{1}{(\pi^3\ell^4\ell_z^2)^{1/2}}
  \exp\left(-\frac{y^2+z^2}{\ell^2} - \frac{x^2}{\ell_x^2}\right),
\end{equation}
with the width $\ell = as^{-1/4} / \pi$ in the $yz$ plane and $\ell_x = [\hbar/(m\omega_x)]^{1/2}$ in the $x$ direction. The width is determined by the lattice spacing and narrows
with increased  $s$.

\section{Optical response of a planar array}

\begin{figure}[htbp]
  \centering
   \includegraphics[width=0.9\columnwidth]{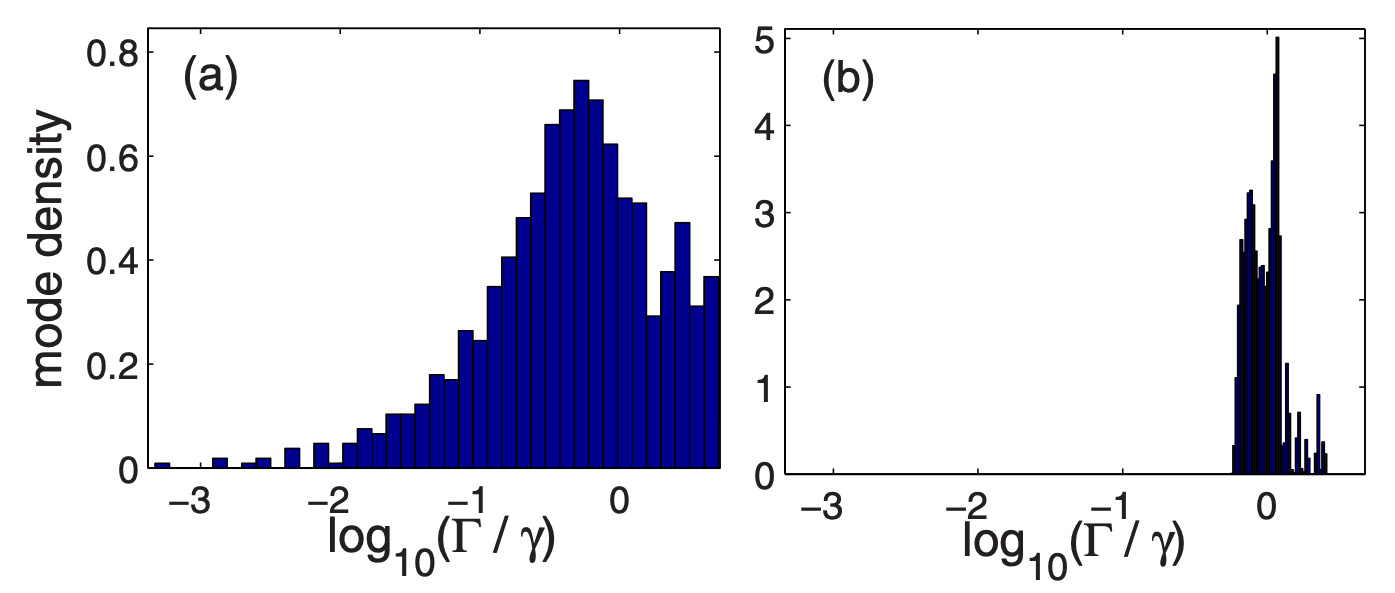}
      \vspace{-0.2cm}
 \caption{(Reproduced from Ref.~\cite{Lee16}). The probability distribution of collective linewidths $\Gamma$ of LLI eigenmodes for two-level atoms in a 32$\times$32 square array, with dipoles slightly tilted from the normal  to the array along $\pol_\perp+0.1\pol_\parallel$, where $\gamma$ is the single atom linewidth~\cite{Jenkins2012a,Lee16}. The long tail of subradiant modes with the lattice spacing $a=0.55\lambda$ and fixed atom positions in (a) becomes absent in (b) when $a=1.65\lambda$ and position fluctuate with $\ell\simeq0.12a$.
   }
  \label{fig:subradiant}
  \end{figure}

We now turn our attention to cooperatively responding planar arrays of atoms that can exhibit a range of LLI subradiant collective excitation eigenmodes (Fig.~\ref{fig:subradiant}). The array acts as a diffraction grating with wavefunctions of the individual lattice sites determining the analogy of the Debye-Waller factor. For normal incidence, only the zeroth-order Bragg scattering peak exists in a lattice with subwavelength spacing, resulting in coherent
scattering at the exact forward and backward directions. At other incident angles, the maximum spacing allowed for the existence of only a single Bragg peak varies between half a wavelength and a full wavelength.
The following analysis relies on this highly collimated scattering that leads to light propagation similar to that in 1D electrodynamics.

\subsection{Uniformly excited array}
\label{sec:uniform}

The coupling of a coherent incident light beam to a planar array is greatly simplified by the phase-matching of the light profile to only the most dominant collective excitation eigenmodes.
This allows for the introduction of superatom models where the collective many-atom optical response of the array  can be described qualitatively by one (Secs.~\ref{sec:normalinci} and~\ref{sec:trans}), two (Sec.~\ref{sec:twomode}), three (Sec.~\ref{sec:arbitrary}) or four (Sec.~\ref{sec:storage}) eigenmodes. To simplify the analysis, we begin by considering a driving light with a uniform phase and intensity profile perpendicular to the propagation direction of light and a sufficiently large square array, 
neglecting edge effects and treating the system as translationally invariant. Additionally, we assume degenerate electronically excited atomic levels with equal resonance frequencies for each atom. At the normal incidence, the light then couples to a collective (``coherent in-plane'') mode in which all the dipoles oscillate coherently in phase along the direction of the light polarization~\cite{CAIT}. By applying a uniform Zeeman shift to the atoms, they can also couple to a uniform phase-coherent out-of-plane (``coherent perpendicular'') eigenmode where all the atomic dipoles oscillate in phase normal to the plane~\cite{Facchinetti16,Facchinetti18}. This analysis will be discussed in Sec.~\ref{sec:twomode}.

\subsubsection{Normal incidence: effective 1D propagation}
\label{sec:normalinci}

For a normally incident uniform plane wave, the response of a sufficiently large array is dominated by the  LLI eigenmode of the uniform phase profile (vanishing wavevector $\qv={\bf 0}$), due to the phase-matching of the incident light (although this can be violated at sufficiently high intensities and small lattice spacings). 
For finite arrays, the matching is further improved by a Gaussian beam with the width comparable with the lattice size~\cite{Facchinetti16,Manzoni18}.
The resonance of the $\qv={\bf 0}$ LLI eigenmode becomes the key parameter in the optical response and the system can be analyzed as a \emph{single mode} model of collective dipole oscillations~\cite{CAIT}. In essence, a planar array of atoms behaves like a single  \emph{superatom}, exhibiting a Lorentzian-shaped response to the incident light, similar to that of an individual atom,
but with modified resonance frequency and linewidth.

To analyze the response of a planar array to normally incident light, we examine the steady-state solutions of a square lattice within the semiclassical dynamics of Eqs.~\eqref{eq:multilevelhighintensitystochastic} in which case the LLI solutions are obtained as straightforward limiting cases. The derivation closely follows Ref.~\cite{Parmee2021}. 
The uniform stationary solution to Eqs.~\eqref{eq:multilevelhighintensitystochastic} for the atomic coherence and excited level population in terms of $\bar{\mathcal{R}}_\nu^{(\ell)}=\bar{\mathcal{R}}$ along the direction of the excited atomic dipole $\textbf{d}_{ge}^{(\nu)}={\cal D}\pol_\nu$ 
is then given by $\Delta_{\ell\nu} = \Delta$, $\rho^{(\ell)}_{ee}=\rho_{ee}$, and $\rho_{ge}^{(\ell)}=\rho_{ge}$~\cite{Parmee2020,Parmee2021}
\begin{subequations}\label{Eq:EOMSolns}
	\begin{align}
	\rho_{ge} &=\bar{\mathcal{R}}\frac{-\Delta+\text{i}\gamma}{\Delta^2+\gamma^2+2|\bar{\mathcal{R}}|^2},\label{Eq:Coherence}\\
	\rho_{ee} &=\frac{|\bar{\mathcal{R}}|^2}{\Delta^2+\gamma^2+2|\bar{\mathcal{R}}|^2} \label{Eq:Excitations}.
	\end{align}
\end{subequations}
These solutions share a similar form with the solutions of single-atom OBEs,
 with the distinction that the Rabi frequency, $\mathcal{R}$, which drives each atom, is substituted by the effective Rabi frequency $\bar{\mathcal{R}}$ [Eq.~\eqref{Eq:EffectiveRabi}]. $\bar{\mathcal{R}}$ includes both the incident field and the dipole radiation emitted by all other atoms in the array. For the uniform solution, we can express Eq.~\eqref{Eq:EffectiveRabi} as
\begin{equation}\label{Eq:EDef}
\bar{\mathcal{R}}= {\cal R}+(\tilde{\Omega}+\text{i}\tilde {\gamma}) \rho_{ge},
\end{equation}
where 
\begin{equation}\label{FourierTransformsq0}
\tilde\Omega=\sum_{\ell\neq j}\Omega^{(j\ell)}_{\nu\nu},\quad\tilde{\gamma}=\sum_{\ell\neq j}\gamma^{(j\ell)}_{\nu\nu},
\end{equation}
are obtained from the real and imaginary parts of the dipole kernel, Eq.~\eqref{eq:omga}, respectively, and
$\pol_\nu$ defines the direction of the excited atomic dipole.
Since the edge effects are neglected, the lattice is translationally invariant and the site $j$ is arbitrary in Eq.~\eqref{FourierTransformsq0}.
The eigenvalue of the phase-uniform LLI collective radiative excitation eigenmode of ${\cal H}$ in \EQREF{eq:eigensystem} reads $\tilde{\Omega}+i(\tilde{\gamma}+\gamma)$. Here
$\tilde{\gamma}$ represents the change of the resonance linewidth due to collective effects for the LLI eigenmode $\qv=0$, and $\tilde{\Omega}$ is the corresponding collective line shift. 

We obtain from Eq.~\eqref{Eq:Coherence} an equation for collective radiation field alone by eliminating $\rho_{ge}$ with Eq.~\eqref{Eq:EDef}
\begin{equation}\label{Eq:yequation}
\frac{\mathcal{R}}{\bar{\mathcal{R}}}=1+\frac{2C(\Delta^2+\gamma^2)}{\Delta^2+\gamma^2+2|\bar{\mathcal{R}}|^2},
\end{equation}
where we have introduced a notation similar to that used in cavity systems~\cite{Bonifacio1978}, and defined the \emph{cooperativity parameter}~\cite{Parmee2020,Parmee2021},
\begin{equation}\label{Eq:GsumDefinition}
C = \frac{1}{2}\frac{\tilde{\Omega}+i\tilde{\gamma}}{\Delta+i\gamma},
\end{equation} 
which is a measure of the collective behavior in the array and plays an important role in describing bistability in Sec.~\ref{sec:bistable}.

The subradiant resonance narrowing in the transmitted light through a planar lattice of about 200 atoms was experimentally measured in Ref.~\cite{Rui2020} (see Sec.~\ref{sec:trans}). 
The observed resonance belongs to the $\qv={\bf 0}$ LLI eigenmode in which all the atomic dipoles coherently oscillate in phase, parallel to the laser polarization. 
The collective linewidth of the $\qv={\bf 0}$ LLI eigenmode for an infinite dipole array at fixed positions can be analytically calculated, e.g., using diffraction theory and taking the limit of an infinite lattice, which was performed in Ref.~\cite{CAIT}  (for the spacing $a<\lambda$)
\begin{equation}\label{Analytic}
\gamma+\tilde{\gamma}=\frac{3\pi\gamma}{(ka)^2}.
\end{equation}
In Appendix~\ref{appen1}, we present a physically intuitive derivation of the linewidth directly integrating the scattered light in the spatial representation.
The observed transmission resonance narrowing to $0.68\gamma$~\cite{Rui2020} is consistent with the collective linewidth value $\gamma+\tilde{\gamma} \simeq 0.52\gamma$ of Eq.~\eqref{Analytic} for $a\simeq0.68\lambda$, considering the finite array size and position fluctuations that broaden the resonance.
 It is worth noting that according to Eq.~\eqref{Analytic}, reducing the lattice spacing leads to significant linewidth broadening and the occurrence of super-radiance in the same system.

Due to the existence of only the zeroth-order Bragg peak in propagating light, 
the highly collimated coherent scattering from a subwavelength atomic array leads to light propagation that corresponds to the effective 1D electrodynamics~\cite{Shahmoon,Facchinetti16,Facchinetti18,Guimond2019,Javanainen19}. 
This behavior arises from the array's single superatom characteristics, which emerge at sufficient distance from a uniformly excited array, 
causing the atomic dipoles to appear as if they are continuously spread across a plane. 
This principle can be extended to stacked planar arrays, where each individual array behaves similarly to a ``1D atom'' coupled to the other arrays through
scattered light~\cite{Facchinetti16,Facchinetti18,Javanainen19,Yoo20}.  A stack of infinite 2D atom arrays therefore emulates regularly spaced atoms in a lossless 1D waveguide. Finite-size effects and defects due to missing atoms, that lead to losses, can be incorporated by introducing phenomenological scattering rates in other radiation modes~\cite{Javanainen19,Ballantine21PT}, similar to losses in waveguides and fibers. 
In Appendix~\ref{appen1}, we formally derive the emerging 1D electrodynamics from the spatial distribution of the scattered light, employing coarse-graining integration approximations introduced in Ref.~\cite{Sargent_laserphys}. 
The derivation relies on two critical assumptions:  the uniform excitation of atoms within each individual planar array and the separation between layers, which allows us to disregard the discreteness of the atoms in the light propagation. The latter assumption holds true for separations $d$ that satisfy 
$\lambda \lesssim d \ll \sqrt{\mathcal{A}} $ (or $d\agt 0.5\lambda$, for $a\alt 0.7\lambda$), where $\mathcal{A}$ is the array area~\cite{Javanainen19}.  The scattered light from a planar atom array, with  the effective 1D atomic polarization amplitude given by the uniform excitation $\rho_{ge} $, then satisfy
\beq
\label{eq:1dfield}
\epsilon_0 \langle \hat{E}{}^+_s(x) \rangle= G(x)
\bar{\cal D} \rho_{ge} ,\quad G(x)= \frac{  ik  }{2} e^{ik|x|},\quad \bar{\cal D}=\frac{\cal D}{{\cal A}'}
\eeq
where $G(x)$ denotes the 1D dipole radiation kernel (the Green's function of the 1D Helmholtz equation~\cite{BOR99}) and $\bar{\cal D}$ is the density of the atomic dipoles in the array (${\cal A}'={\cal A}/N$ is the area of the unit cell); see Appendix~\ref{appen1}. 
In the limit of LLI, the atomic polarization amplitudes of the stacked uniformly excited atom arrays obey the dynamics
\beq
\frac{d}{dt}\varrho_{ge}^{(j)}  =  (i \Delta_{\rm 1D}-\gamma_{\rm 1D})\varrho_{ge}^{(j)} +i{\cal R}(x_j)
-\gamma_{\rm 1D}  \sum_{\ell\neq j} e^{ik |x_j-x_\ell |}\,\varrho_{ge}^{(\ell)},
\label{TIMEDEPEQ1}
\eeq
where $\varrho_{ge}^{(j)} $ denotes the uniform polarization amplitude $\rho_{ge} $ of the $j$th array in the 1D electrodynamics. The 1D electrodynamics linewidth $\gamma_{\rm 1D}=\gamma+\tilde\gamma$ equals the linewidth of the uniform mode. The detuning $\Delta_{\rm 1D}$ represents the single-atom resonance  shifted by the collective line shift $\tilde\Omega$. 

Neglecting the discreteness of the atoms in the coarse-graining approximation 
in effective 1D propagation 
represents a mean-field type of approximation. The effects of reduced filling factors (defects)~\cite{Javanainen19}, scattering to other modes, and incoherent scattering due to position fluctuations~\cite{Ballantine21PT} can similarly be incorporated as a mean-field approximation by multiplying the collective dipole excitation amplitude by the filling factor~\cite{Javanainen19} or the ratio of the linewidths between the coherent scattering to the target mode and total scattering (Purcell factor)~\cite{Ballantine21PT}. The latter allows to use Eq.~\eqref{TIMEDEPEQ1} also for the case of modes with nonuniform phase profiles. Numerical studies~\cite{Javanainen19} for intralayer spacings close to half a wavelength confirm the accuracy of the mean-field approximation. This is interpreted as a result of two beneficial factors. Firstly, although deviations from continuous media electrodynamics can be detected at such densities, the density is too low to lead to qualitative failure of the mean-field model~\cite{Javanainen2014a,JavanainenMFT}. Secondly, unlike in disordered media, in the array atoms are trapped at specific locations which prevents small interatomic separations where the $1/r^3$ DD terms rapidly increase.

\subsubsection{Transmission and reflection}
\label{sec:trans}

The efficacy of the single-mode model is effectively demonstrated through the examination of transmission and reflection properties in an array.
We begin by analyzing the coherent transmission and reflection of normally incident light through a single array. We express the reflectance $r$ and the transmittance $t$ amplitudes in terms of the incident $\boldsymbol{\mathcal{E}}{}^+(\textbf{r})$, scattered $\<\hat{\textbf{E}}{}^+_{s}(\textbf{r})\>$, and total $\<\hat{\textbf{E}}{}^+(\textbf{r})\>= \boldsymbol{\mathcal{E}}{}^+_s(\textbf{r})+\<\hat{\textbf{E}}{}^+_s(\textbf{r})\>$ field amplitudes
\begin{equation}\label{Eq:TransmissionAmplitude}
t=\frac{\int_{x>0}\hat{\textbf{e}}\cdot\<\hat{\textbf{E}}{}^+(\textbf{r})\>d\Omega}{\int_{x>0}\hat{\textbf{e}}\cdot\boldsymbol{\mathcal{E}}{}^+(\textbf{r})d\Omega},\quad r=\frac{\int_{x<0}\hat{\textbf{e}}\cdot\<\hat{\textbf{E}}{}^+_s(\textbf{r})\>d\Omega}{\int_{x>0}\hat{\textbf{e}}\cdot\boldsymbol{\mathcal{E}}{}^+(\textbf{r})d\Omega},
\end{equation}
where the subscript $x\gtrless 0$ represents the integration over a solid angle enclosing the light component $\pol$ propagating to the region $x\gtrless0$.
Due to the symmetry, the amplitudes satisfy
\begin{equation}\label{Eq:tr}
t = 1+r.
\end{equation}

We consider the more general scenario of 
incident laser powers that are not confined to the LLI limit but employ the semiclassical approximation (Sec.~\ref{sec:semi}) to neglect quantum fluctuations. Using Eqs.~\eqref{eq:1dfield} and~\eqref{Analytic}, we obtain 
for the mode with a uniform phase profile ($\textbf{q}=\textbf{0}$)~\cite{Parmee2021},
\begin{equation}\label{Eq:Transmission2}
r = i (\gamma+\tilde{\gamma})\frac{\rho_{ge}}{\mathcal{R}}.
\end{equation}
Solving the semiclassical dynamics for the coherence gives the reflection amplitude as a function of the excited level population. 
In Eq.~\eqref{Eq:Coherence}, the effective field ${\cal R}_{\rm eff}$ depends on $\rho_{ge}$ that can be solved for the $\textbf{q}=\textbf{0}$ mode~\cite{Parmee2021}, with
\begin{equation}\label{SScoherence}
\rho_{ge}=\frac{i{\cal R} Z}{i (\Delta-Z\tilde\Omega)-(\gamma-Z\tilde{\gamma})},
\end{equation} 
where the excited level population is encapsulated in $Z$, 
\begin{equation}
Z=2\rho_{ee}-1.
\end{equation}

In the LLI limit, $\rho_{ee}=0$ in Eq.~\eqref{SScoherence}, and
\beq \label{eq:singlemodeamplitude}
\rho_{ge}= \frac{-{\cal R}}{\Delta+\tilde\Omega+i(\gamma+\tilde\gamma)},
\eeq
resulting in the reflection and transmission amplitudes
\begin{equation}\label{LLItrans}
r=\frac{-i(\gamma+\tilde\gamma)}{\Delta+\tilde\Omega+i (\gamma+\tilde\gamma)}, \quad
t=\frac{\Delta+\tilde\Omega}{\Delta+\tilde\Omega+i(\gamma+\tilde\gamma)}.
\end{equation} 
At the resonance $\Delta=-\tilde\Omega$ of the LLI $\textbf{q}=\textbf{0}$ eigenmode, Eq.~\eqref{LLItrans} indicates
the total reflection $r=-1$, analogously to the total reflection of a resonant atom in 1D electrodynamics~\cite{Javanainen1999a}. 
For a square array, there are two values of the lattice constant for $a<\lambda$,  $a/\lambda\simeq0.2$ and 0.8, when $\tilde\Omega\simeq0$ and the atomic resonance equals the array resonance~\cite{Bettles2016,Shahmoon}. 
The total resonance reflection from a planar array of linear dipole scatterers has been well-established in classical electrodynamics~\cite{Tretyakov}. More recently, it has gained attention in nanophotonics~\cite{Laroche2006,Abajo07,CAIT}.
Experimental studies have demonstrated near-perfect reflection in small-scale~\cite{Moitra2014}  and large-scale~\cite{Moitra2015} dielectric planar arrays of dipoles, based on electron beam lithography and silicon cylinder resonators, respectively. Dipolar Mie resonances of the nanocylinders achieved an average reflection of 99.7\%~\cite{Moitra2015}.
The same dipolar reflection has been highlighted in atomic arrays~\cite{Bettles2016,Facchinetti16,Shahmoon}. Increasing disorder in atomic positions has the effect of decreasing extinction, both in the LLI limit~\cite{Bettles2016} and at higher intensities~\cite{Bettles2020}, eventually reducing the cross-section below the independent-atom value. The reason is easy to grasp, since in an infinite regular array the resonance shifts of each atom caused by the DD interactions with other atoms are equal, while in disordered ensembles the interaction shifts within any close atom pair fluctuate and rapidly grow at small atom separations, resulting in inhomogenously-broadened resonance frequencies~\cite{Jenkins_long16}. 
The degradation of reflectivity due to fluctuations is estimated to scale as $\ell^2/d^2$~\cite{Shahmoon}, where $\ell$ is defined in \EQREF{eq:phi}. 

Another thin planar array system closely related to fabricated metasurfaces can be created using transition metal dichalcogenides, such as MoSe$_2$. These materials are direct band-gap semiconductors without free charge carriers. In this context, optical excitations occur as excitons with high binding energy, leading to high reflectivity~\cite{Scuri18,Back18}, rapid switching~\cite{Andersen22}, and the potential for nonlinear responses~\cite{Atac18,Walther22}.

\begin{figure}[htbp]
  \centering
   \includegraphics[width=0.9\columnwidth]{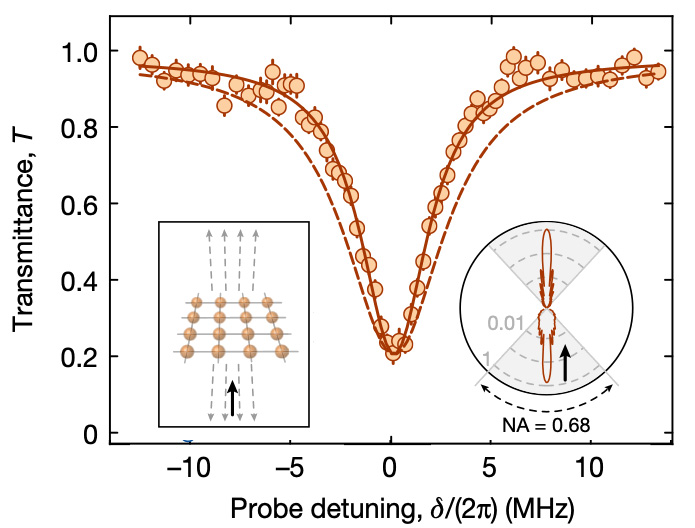}
      \vspace{-0.2cm}
 \caption{(Reproduced from Ref.~\cite{Rui2020}). Cooperative narrowing of a resonance linewidth below the single-atom quantum limit for the transmission of light through a near-unity-filled optical lattice of Rb atoms. The dashed line corresponds to the reference spectrum with the natural linewidth of a single atom. The measured Lorentzian linewidth and resonance transmission for the array were $2\gamma/(2\pi)=4.09(11)$MHz (for a single atom $2\gamma/(2\pi)\simeq6.06$MHz) and $T=0.23(1)$, respectively. The inset shows numerically simulated differential cross-section of the scattered field.
   }
  \label{fig:linewidth}
  \end{figure}

In the atomic case Fig.~\ref{fig:linewidth} shows the observed cooperative spectral narrowing in the LLI transmission through a square array~\cite{Rui2020}, surpassing the quantum-limited Wigner-Weisskopf linewidth of an isolated atom. Ultracold $^{87}$Rb atoms were trapped in an optical lattice, undergoing a quantum phase transition from a Bose-Einstein condensate to a Mott-insulator state. This allowed for precise control over the occupation numbers per lattice site (see also Sec.~\ref{sec:latticepot}). Approximately 200 atoms, with a near-unit filling fraction of 0.92, were probed using an isolated two-level cycling transition, and the forward or backward scattering was detected. The resonance (power) transmission $T=|t|^2\simeq0.23$ and reflection  $R=|r|^2\simeq 0.58$ ($R+T<1$ because of the finite collection angle) were not $T\simeq0$, $R\simeq1$ due to the position fluctuations of the atoms, although the corresponding resonance linewidths, $0.68\gamma$ and $0.66\gamma$, qualitatively approached the ideal value expected for an infinite array of fixed atomic positions [Eq.~\eqref{Analytic} for $a=0.68\lambda$]. Increasing the lattice height beyond $300E_r$ deteriorated subradiance due to motional spreading of atoms in the antitrapped electronically excited state $5^2$P$_{3/2}$ in the lattice potential. This spreading caused heating and inhomogeneous broadening from position-dependent atomic resonance shifts  inside each lattice site.

To assess the impact of a regular atomic pattern, the optical response was compared under two conditions: for randomized atomic positions along the propagation direction of light, and for reduced atomic filling factors~\cite{Rui2020}. Increased randomization of the DD interactions between the atoms due to disorder rapidly reduced reflection~\cite{Bettles2016} and increased the transmission linewidth beyond the linewidth of an isolated atom. At low fillings, the linewidth was similar to that of a single atom, while becoming increasingly subradiant as the site occupancy approached unity. Additionally, changing the occupancy also notably shifted the transmission resonance.

While the experiment~\cite{Rui2020} used a weak light probe, it is important to consider the optical response beyond the LLI limit. From Eqs.~\eqref{Eq:Coherence}, \eqref{Eq:EDef}, and~\eqref{SScoherence}, we can find the leading collective contribution (for $|\tilde\gamma|/\gamma\ll1$) to the power-broadened resonance linewidth due to the incident light intensity $I$ as $\gamma_{\rm PB}\simeq \gamma[1+I/I_{\rm sat}(1-2b\tilde\gamma) ]^{1/2}$, where $b\equiv-Z/(\gamma-Z\tilde\gamma)>0$ and $I_{\rm sat}$ is defined in \EQREF{eq:saturationint}. The power broadening of independent atoms~\cite{meystre1998} (corresponding to $\tilde\gamma=0$) is enhanced for subradiant states ($\tilde\gamma<0$) and reduced for super-radiant states ($\tilde\gamma>0$). However, for subradiant (super-radiant) states it remains narrower (broader) than the power-broadened single-atom linewidth.
Using Eqs.~\eqref{SScoherence}, \eqref{Eq:Transmission2}, and~\eqref{Eq:tr}, we obtain reflectivity $R$ and extinction $1-T$ beyond the LLI limit~\cite{Parmee2021}
\begin{subequations}\label{Eq:Reflectivity-2DUniformArray-Z}
\begin{align}
\begin{split}
1-T &= -\frac{Z(\gamma+\tilde{\gamma})\left[2(\gamma-Z\tilde{\gamma})+Z(\gamma+\tilde{\gamma})\right]}{(\Delta-Z\tilde{\Omega})^2+(\gamma-Z\tilde{\gamma})^2},
\end{split}\label{T(Z)}\\
\begin{split}
R &= \frac{Z^2(\gamma+\tilde{\gamma})^2}{(\Delta-Z\tilde{\Omega})^2+(\gamma-Z\tilde{\gamma})^2}.
\end{split}\label{R(Z)}
\end{align}
\end{subequations}
The presence of incoherent scattering, however, implies $R+T<1$. We can formulate a self-consistent theoretical model that conserves the energy by incorporating the flux
of incoherently scattered light in the semiclassical approximation, amended by the single-atom quantum contributions in \EQREF{eq:SAQincoherentScattering}. From the expression of the scattered light in Eq.~\eqref{eq:introducingFluctuations}, we consider the last term of incoherent scattering [Eq.~\eqref{eq:incoherentScatteringFieldExpectation}] that is generally due to position and quantum fluctuations. Since in this example we consider atoms at fixed positions, the incoherent contribution is solely due to quantum fluctuations. In the semiclassical approximation, amended by the single-atom quantum contributions [\EQREF{eq:SAQincoherentScattering}], the normalized flux for the incoherent light integrated over a closed surface is given by
\beq
\label{IncFluc1}
F_{\rm{inc}} =\frac{ \int\langle\delta\hat{\textbf{E}}{}_{s}^-(\textbf{r})\cdot\delta\hat{\textbf{E}}{}_{s}^+(\textbf{r})\rangle d\Omega}{\int |\boldsymbol{\mathcal{E}}{}^+(\textbf{r})|^2 d\Omega}.
\eeq
The flux contribution in the denominator for an incident plane wave yields ${\cal A}\,|\boldsymbol{\mathcal{E}}{}^+|^2$ and the flux in the numerator is evaluated straightforwardly~\cite{Parmee2021} using Eq.~\eqref{eq:saqscatrate}
\begin{align}
F_{\rm{inc}} &=2\gamma(\gamma+\tilde{\gamma})\left(\frac{\rho_{ee}}{|\mathcal{R}|^2}-\left|\frac{\rho_{ge}}{\mathcal{R}}\right|^2\right)\nonumber\\
&=2(\gamma+\tilde{\gamma})\text{Im}\left[\frac{\rho_{ge}}{\mathcal{R}}\right]-2(\gamma+\tilde{\gamma})^2\left|\frac{\rho_{ge}}{\mathcal{R}}\right|^2,
\end{align}
where in the last line, we have used the stationary solutions to the coupled many-atom OBEs \eqref{Eq:EOMSolns} to eliminate the excited level population. 
Summing up the incident, coherent and incoherent normalized fluxes equals to one, implying that the model conserves energy~\cite{Parmee2021}
\begin{equation}\label{normalised flux2}
R+T + F_{\rm{inc}} = 1.
\end{equation} 
The experimental result $R+T <1$~\cite{Rui2020} can imply considerable incoherent scattering that was not collected by the lens. This is not surprising, given the zero-point position fluctuations of the atoms in the lattice sites. However, another potential source is $F_{\rm{inc}}\neq0$, indicating that there were scattering effects beyond the limit of LLI
that are expected to become more dominant the narrower the subradiant resonance~\cite{Williamson2020,Cipris21}.

In Ref.~\cite{Bettles2020}, light transmission beyond the LLI limit was analyzed and many-body quantum fluctuations in the coherent transmission (Secs.~\ref{sec:scatteredlight} and~\ref{sec:opticsfluctuations}) were identified 
that increased with increasing DD interaction, reaching their maximum at intermediate intensities $I\simeq I_{\rm sat}$. Outside this regime and at larger spacings ($a\agt0.4\lambda$), the semiclassical approximation accurately described coherently transmitted light, except when exciting subradiant modes. When augmented with the single-atom quantum contributions, the semiclassical model also qualitatively reproduced incoherent transmission in the same regime. Beyond the LLI limit, nonlinearities can also result in nonuniform excitations including antiferromagnetic and oscillatory phases, despite a uniform drive~\cite{Parmee2020}.

\subsubsection{Bistable transmission}
\label{sec:bistable}

In the semiclassical approximation, where the quantum fluctuations are neglected, the transmitted light can exhibit optical bistability and hysteresis~\cite{Parmee2021,Parmee2020}. Equation~\eqref{Eq:yequation} provides the effective field on the atoms (the incident field plus the scattered light from all the other atoms). Taking the absolute square of both sides of Eq.~\eqref{Eq:yequation} yields a cubic polynomial equation for $|\mathcal{R}_{\rm eff}|^2$, which can have either one or two dynamically stable solutions. Interestingly, this formulation closely resembles the one used to describe bistability in cavity systems~\cite{Lugiato1984,bonifacio1976,Bonifacio1978,Carmichael1977,Agrawal79,Carmichael1986a,CarmichaelVol2}. 
In both cavity systems and atomic arrays, the number of stable solutions depends on the cooperativity parameter $C$~\cite{Parmee2021}. For the atomic array, it is
given by 
Eq.~\eqref{Eq:GsumDefinition} that depends on the collective linewidth and line shift.
In the optical cavity, 
$C=Ng^2/2\gamma\kappa$~\cite{Bonifacio1978,CarmichaelVol2}, where $\kappa$ denotes the cavity linewidth and $g$ the atom-cavity coupling coefficient. The condition $C\gg1$ corresponds to the strong coupling regime of optical cavities.  In optical cavities,
a large value of $C$ indicates multiple recurrent scattering events of an atom with light bouncing back and forth between the cavity mirrors. In arrays in free space,
large $|C|\agt4$ represents recurrent scattering events between neighboring atoms at high densities, with $ka \sim 1$ when $\tilde\gamma$ assumes the role of the atom cloud coupling to the cavity given by $Ng$~\cite{Parmee2021}. 

\begin{figure}[htbp]
  \centering
   \includegraphics[width=0.95\columnwidth]{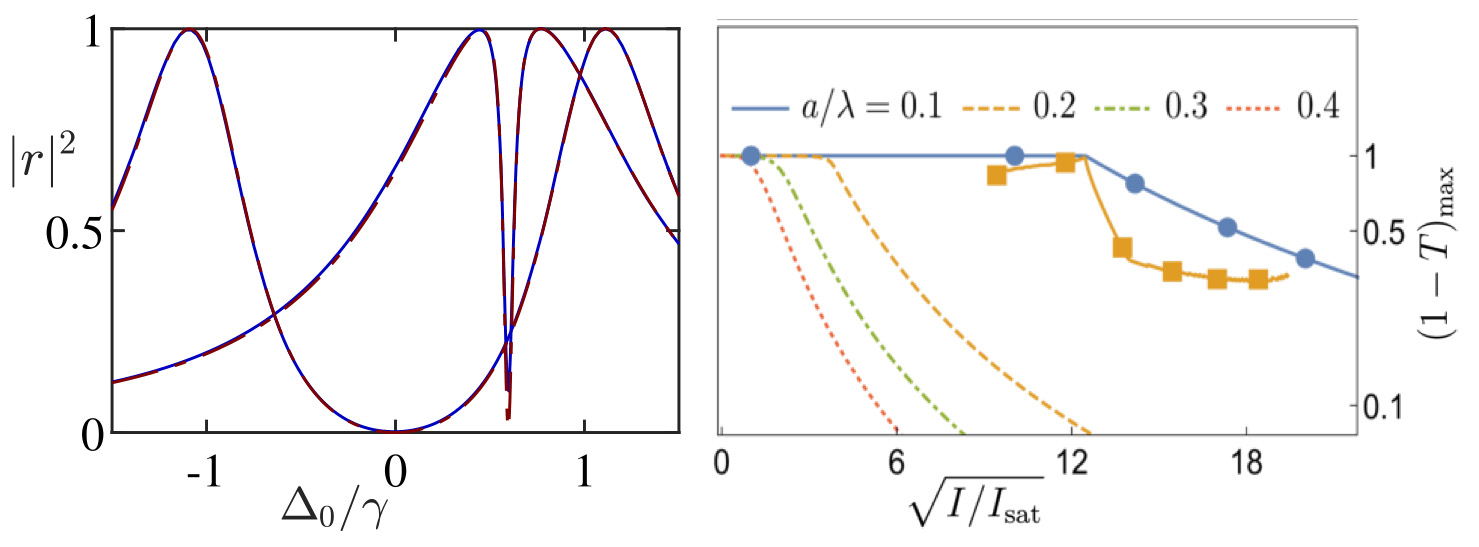}
      \vspace{-0.4cm}
 \caption{[Reproduced from Ref.~\cite{Facchinetti16} (left) from Ref.~\cite{Parmee2021} (right)]. 
 Left: Reflectivity from a 20$\times$20 square array as a function of the detuning, with $a=0.55\lambda$, nonzero level shifts, and subradiant linewidth $0.0031\gamma$ (Sec.~\ref{sec:twomode}).
 The reflection is near-perfect at two values of the detuning on each side of the subradient resonance with near-perfect transmission due to
the dipoles being excited in the light propagation direction. The two-mode model of Sec.~\ref{sec:twomode}  [\EQREF{Eq:ZeemanShift}] (red, dashed curves) agrees well with the full LLI numerics [\EQREF{eq:Peoms}] (blue, solid curves). The narrow (broad) resonance corresponds to small (large) level shift $\bar\delta/\gamma=0.15$  ($1.1$). Right: Bistable light transmission through a square array at different lattice spacings, showing maximum extinction at any detuning as a function of incident light intensity (Sec.~\ref{sec:bistable}). The extinction remains close to one at the higher intensities the smaller the spacing. At $a=0.1\lambda$, there are two stable solutions: ``single-atom'' (orange squares) and ``cooperative'' (blue circles) over a range of intensities.
  }
  \label{fig:twomode}
  \end{figure}

For sufficiently large lattice spacings,  there is only a single solution for the transmitted light, resulting in no bistability for any intensity or detuning; Fig.~\ref{fig:twomode}. The maximum lattice spacing at which the bistable transmission through an infinite square array can be observed is $a \simeq 0.165\lambda$~\cite{Parmee2021,Parmee2020}. This value is very close to the analytically derived condition $k a < (\pi/3)^{1/2}$ that can be obtained from Eq.~\eqref{Eq:yequation} at specific parameter values. The maximum spacing $ka\sim 1$ 
to observe bistable behavior corresponds to the atom separation at which the induced level shift due to the DD interaction exceeds 
the single-atom linewidth.  

Analogous to the optical cavity bistability~\cite{Bonifacio1978}, the two stable solutions for the transmitted light can be referred to as the ``cooperative'' and ``single-atom'' solutions
due to their distinct optical responses; Fig.~\ref{fig:twomode}. Analytic estimates for these solutions have been derived in limiting cases from the cubic equation for $|\mathcal{R}_{\rm eff}|^2$~\cite{Parmee2021}.
 In the cooperative solution, the atoms exhibit collective behavior, generating a field that counteracts the incident light and leading to strong absorption, particularly at higher atom densities~\cite{Parmee2021}.
Cooperative behavior is most evident in the LLI limit, where the effective field $\mathcal{R}_{\rm{eff}}\approx\mathcal{R}/(2C+1)$ inversely scales with $C$,  resulting in a small $\mathcal{R}_{\rm{eff}}$ for strong collective behavior.
In the single-atom solution, the atoms saturate, causing weak absorption and rendering the medium transparent as the atoms respond to the incident light independently.
At high-intensities, the effective field scales linearly with the incident field, $\mathcal{R}_{\rm{eff}}\approx\mathcal{R}$, and there is no dependence on the cooperativity parameter $C$ as collective behavior among the atoms is lost.

\subsubsection{Transfer matrices in 1D propagation}
\label{sec:transfer}

For stacked (subwavelength) planar arrays of atoms with sufficiently large spacing between them, the light propagation through the layers obeys effective 1D electrodynamics of Eqs.~\eqref{eq:1dfield} and~\eqref{TIMEDEPEQ1} (Sec.~\ref{sec:normalinci}). This holds approximately true at normal incidence when the nearest-neighbor layer separation $d$ satisfies
$\lambda \lesssim d \ll \sqrt{\mathcal{A}}$ (or $d\agt 0.5\lambda$, for $a\alt 0.7\lambda$), where $\mathcal{A}$ is the array area~\cite{Javanainen19}. The transmission and reflection through the system of stacked arrays may then be solved in the limit of LLI using transfer matrices. We consider one polarization component within the array plane and treat light as a scalar field. 
When incident light with an amplitude $E^+_i$ arrives from the left, it results in a reflected amplitude  $E^+_r$ and a transmitted amplitude  $E^+_t$, which are related by the transfer matrix ${\cal T}_{\rm sys}$
of the entire system of stacked arrays
\beq
\left[
\begin{array}{c}
E^+_{t} \\
0
\end{array}
\right]
={\cal T}_{\rm sys}  \left[
\begin{array}{c}
E^+_{i} \\
E^+_{r}
\end{array}
\right],
\label{eq:systrans}
\eeq
where ${\cal T}_{\rm sys}$ consists of the transfer matrices for individual arrays and free propagation between the arrays. 
The elements of the inverse of ${\cal T}_{\rm sys}$ then give the analytic solutions to
the transmission and reflection amplitudes~\cite{Ruostekoski17}
\beq \label{eq:gentr}
t_{\rm sys} = \frac{ E^+_{t}}{E^+_{i}}=\frac{1}{\big[{\cal T}_{\rm sys}^{-1} \big]_{11}},\quad
r_{\rm sys} = \frac{E^+_{r}}{E^+_{i}}=\frac{ \big[{\cal T}_{\rm sys}^{-1} \big]_{21} }{\big[{\cal T}_{\rm sys}^{-1} \big]_{11}}.
\eeq
For instance, a transfer matrix for two arrays at $x_j$ and $x_\ell$ is ${\cal T}_{\rm sys} = {\cal T} \Phi(x_\ell,x_j) {\cal T}$, where ${\cal T}$ describes a single array and the light propagation phases for both 
the right- and left-propagating waves between $x_j$ to $x_\ell$ are governed by 
\beq
\Phi(x_\ell,x_j)  = \left[\begin{array}{cc}
e^{ik(x_\ell-x_j)} & 0\\
0&e^{-ik(x_\ell-x_j)}
\end{array}\right].
\eeq

A transfer matrix ${\cal T}$ for an individual array is derived  by separating the outgoing field amplitudes $E^+_{l}$ on the left side of the array from those on the right side
$E^+_{r}$, and, similarly, the incoming (from the left/right) field amplitudes ${E}^+_{{\rm ext},l/r}$,
\beq
\left[
\begin{array}{c}
E_l^+ \\
E_{{\rm ext},l}^+
\end{array}
\right]
=
{\cal T}
\left[
\begin{array}{c}
E_{{\rm ext},r}^+\\
E^+_r
\end{array}
\right].
\eeq
We then obtain 
\beq
{\cal T} =  \frac{1}{{r +1} }\left[
\begin{array}{cc}
 {2 r+1} & { r }\\
 -{r } & {1} \label{eq:transfer_singleatom}\\
\end{array}
\right],
\eeq
where \EQREF{LLItrans} gives the reflection amplitude $r$ and the transmission amplitude $1+r$.

\subsection{Nonuniform excitations}
\label{sec:nonuniform}

\subsubsection{Collective eigenmodes and band structure}
\label{sec:band}

The study of excitations and band structure in regular 3D emitter arrays has been a topic of long-standing theoretical interest~\cite{Hopfield58,Coevorden96,devries98,Tip00,Klugkist06,Antezza2009,Antezza09b,Castin13}. In this section, we explore
collective excitation eigenmodes with nonuniform phase profiles in the LLI limit beyond the uniform ones of Sec.~\ref{sec:normalinci}. Driving nonuniform modes requires some nonuniform coupling mechanism of light to atoms, except at higher intensities when this can occur solely due to nonlinearities, resulting in an excitation phase diagram~\cite{Parmee2020}.
The LLI collective excitation eigenmodes, collective linewidths, and collective line shifts are obtained from the matrix ${\cal H}$ [Eq.~\eqref{eq:hmatrix}], as explained in Sec.~\ref{sec:LLI}. 
According to Bloch's theorem, in an infinite square lattice, the eigenmodes are plane waves with the wavevector $\qv$, such that $\rho_{ge\nu}^{(\ell)} = \rho_{ge\nu} e^{i\textbf{q}\cdot\textbf{r}_{\ell}}$.
$\Omega^{(j\ell)}_{\nu\mu}+i\gamma^{(j\ell)}_{\nu\mu}$ represents the real and imaginary parts of the dipole radiation kernel [Eq.~\eqref{eq:omga}], and in an infinite array, they depend on $\qv$,
\begin{equation}\label{FourierTransforms}
\tilde\Omega_{\nu\mu}(\textbf{q})=\sum_{\ell\neq j}\Omega^{(j\ell)}_{\nu\mu}e^{i\textbf{q}\cdot\textbf{r}_{j\ell}},\quad\tilde{\gamma}_{\nu\mu}(\textbf{q})=\sum_{\ell\neq j}\gamma^{(j\ell)}_{\nu\mu} e^{i\textbf{q}\cdot\textbf{r}_{j\ell}},
\end{equation}
where $\textbf{r}_{j\ell}=\rv_\ell-\rv_j$.
Due to the translational invariance, the reference atom $j$  is arbitrary. The collective line shifts and linewidths relevant for normally incident light in a square array can be given as $\tilde\Omega_{\mu}(\textbf{q})$ and $\gamma+\tilde{\gamma}_{\mu}(\textbf{q})$ along the atomic polarization $\mu$, with the uniform phase profile modes corresponding to $\qv={\bf 0}$. In other symmetries, such as triangular and triangular-like~\cite{Parmee22b,Bettles_lattice}, or Kagome~\cite{Bettles_lattice,Yoo2016}, atomic dipoles can form more complex nonuniform profiles, involving in the latter case also Fano transmission resonances even for normal incidence.

The dipole radiation kernel [Eqs.~\eqref{eq:GDF} and~\eqref{Gdef}] for the scattered light can be expressed in momentum space as~\cite{devries98}
\beq \label{eq:dipoleradmom}
\tilde{\sf G}_{\nu\mu}(\pv) = \frac{1}{k^2} \frac{k^2\delta_{\nu\mu}-p_\nu p_\mu}{k^2-p^2+i\epsilon}, \quad {\sf G}_{\nu\mu}(\rv) = \int \frac{d^3\pv}{(2\pi)^3}e^{i\pv\cdot\rv} \tilde{\sf G}_{\nu\mu}(\pv),
\eeq
with infinitesimal $\epsilon>0$. The derivation of nonrelativistic electrodynamics implicitly assumes a high-frequency cut-off.  
For example, to ensure that the system satisfies Maxwell's wave equation, which requires the contact term in Eq.~\eqref{Gdef}, it is necessary in the derivation to remove high-frequency modes $|\pv|\gg 1/\eta$, where $\eta$ represents a characteristic length scale~\cite{Ruostekoski1997a}. This is achieved by introducing a cut-off term $\exp(-p^2 \eta^2/4)$. As a result, the delta function in Eq.~\eqref{Gdef} is replaced by a Gaussian of $1/e$ width $\eta$ and height $\pi^{-3/2}\eta^{-3}$~\cite{Ruostekoski1997a}. The same cut-off expression regularizes the dipole radiation kernel in an infinite lattice for the calculation of band structures and excitations~\cite{Antezza2009,Castin13}. By employing this cut-off, momentum space summations converge rapidly and this resolves the ambiguity in the summation of infinite series expressions that are not absolutely convergent. From a physical standpoint, the Gaussian smearing of the lattice site positions can be understood as a consequence of a finite-sized harmonic oscillator potential with a size of $\eta=[\hbar/ (m\omega_{\rm ho})]^{1/2}$, where quantum or thermal position fluctuations of the emitter are determined by the strength of the trap frequency $\omega_{\rm ho}$~\cite{Antezza2009}.

In a translationally-invariant 3D lattice, the scattered field on atom $j$ from all the other atoms is transformed to the momentum space by 
\beq \label{eq:transform}
\sum_{\ell\neq j} {\sf G}_{\nu\mu}(\rv_{j\ell})  e^{i\textbf{q}\cdot\textbf{r}_{j\ell}} =  \frac{1}{{\cal V}}\sum_j \tilde{\sf G}_{\nu\mu}(\qv+{\bf g}_j) -   {\sf G}_{\nu\mu}(0) ,
\eeq
where we have added and subtracted the $\ell=j$ term on the right-hand-side to complete the Poisson summation formula. The reciprocal-lattice vectors ${\bf g}_j$ are related to the lattice vectors ${\bf g}_j\cdot {\bf r}_{\ell}=2\pi n$, for integer $n$, and ${\cal V}$ is the unit cell volume.

The momentum representation of the dipole radiation kernel with the high-momentum cut-off is denoted by $\tilde{\sf G}^*_{\nu\mu}(\pv)= \tilde{\sf G}_{\nu\mu}(\pv)  \exp(-p^2 \eta^2/4) $. Then its Fourier transform ${\sf G}^*_{\nu\mu}(\rv)$ is a convolution of ${\sf G}_{\nu\mu}(\rv)$ and the smeared-out Gaussian for the density profile of the lattice site [the Fourier transform of $ \exp(-p^2 \eta^2/4)$]. When the terms on the right-hand-side of \EQREF{eq:transform} are replaced by their regularized versions $\tilde{\sf G}^*_{\nu\mu}(\pv)$ and ${\sf G}^*_{\nu\mu}(\rv)$ [${\sf G}^*_{\nu\mu}(0)$ becomes a Gaussian multiplied by $-\delta_{\nu\mu}/3$ from \EQREF{Gdef}], the integrals are well defined and can be calculated~\cite{Antezza2009,Perczel2017a}. Taking the limit $\eta\rightarrow0$ of 
\beq \label{eq:eigensum}
\sum_{\ell\neq j} {\sf G}_{\nu\mu}(\rv_{j\ell})  e^{i\textbf{q}\cdot\textbf{r}_{j\ell}} \simeq \frac{ e^{p^2 \eta^2/4} }{{\cal V}}\sum_j \tilde{\sf G}^*_{\nu\mu}(\qv+{\bf g}_j) -   {\sf G}^*_{\nu\mu}(0) 
\eeq
yields the solutions for fixed atomic positions ($\eta\ll 1/k$), while the effect of position fluctuations due to finite confinement can be estimated by a finite value of $\eta$. 

Near planar surfaces it is customary to expand the spherical wave of the dipole radiation kernel, \EQREF{Gdef}, in a plane wave basis, see Appendix~\ref{appen4}. In a 2D planar array, this provides Eqs.~\eqref{eq:transform} and~\eqref{eq:eigensum} in a more suitable form when we replace $ \tilde{\sf G}_{\nu\mu}(\pv)/{\cal V}$ and $ \tilde{\sf G}^*_{\nu\mu}(\pv)/{\cal V}$ by $ \tilde{\sf G}^{\parallel}_{\nu\mu}(\pv)/{\cal A}'$ and $ \tilde{\sf G}^{\parallel *}_{\nu\mu}(\pv)/{\cal A}'$, respectively. Here ${\cal A}'$  is the unit cell area, $\tilde{\sf G}^{\parallel *}_{\nu\mu}(\pv)= \tilde{\sf G}^\parallel_{\nu\mu}(\pv)  \exp(-p^2 \eta^2/4) $, and $\tilde{\sf G}^{\parallel}_{\nu\mu}(\pv)$ is the Fourier transform of the radiation kernel on the array plane
\beq
\label{eq:2dfourier}
{\sf G}_{\nu\mu}(\rv) = \int \frac{d^2\pv_\parallel}{(2\pi)^2}e^{i\pv_\parallel\cdot\rv} \tilde{\sf G}^{\parallel}_{\nu\mu}(\pv),
\eeq 
where $\pv_\parallel=(p_y,p_z)$ denotes a wavevector on the array plane. Appendix~\ref{appen4} presents the expressions for collective linewidths and line shifts in Eqs.~\eqref{eq:ananonuni}.
The technique is adapted to the 2D planar arrays in Ref.~\cite{Perczel2017a} where the appropriate regularization cut-off integrals of the radiation kernel were evaluated. The method offers a rapidly converging sum for the collective radiative linewidths and line shifts. It can also be easily extended to non-Bravais lattices~\cite{Perczel2017a}. 

\subsubsection{Topological bands and edge states}
\label{sec:topo}

Planar arrays of atoms interacting with light can exhibit topologically protected quantum optical behavior, similar to topologically protected photonic systems found in photonic crystals and metamaterials composed of resonator arrays~\cite{Ozawa_review,Khanikaev17}. 
However, unlike topological phases of matter observed in solid-state crystals~\cite{Haldane17} or cold-atom systems~\cite{Cooper19}, the presence of light-matter coupling introduces dissipation due to light scattering and $1/r$ long-range DD interactions.

Topologically nontrivial band structures in the limit of LLI have been explored in honeycomb lattices~\cite{Bettles_17topo,Perczel2017b}, non-Bravais square lattices~\cite{Bettles_17topo,Perczel2017a}, and triangular lattices~\cite{Perczel2017a} by breaking the time-reversal symmetries with atomic level splittings of the excited electronic levels of the $J=0\rightarrow J'=1$ transition. Two of these transitions occur within the array plane, leading to Bloch bands with nontrivial Chern numbers,
\beq
C= \frac{1}{2\pi i}\int d^2 \qv\, [\partial_{q_x} A_y(\qv)- \partial_{q_y} A_x(\qv)],
\eeq
where $A_\mu(\qv) = \< n(\qv)| \partial_{q_\mu} | n(\qv)\>$ is the Berry connection, $ | n(\qv)\>$ is an eigenstate, and the integration is over the Brillouin zone. Generally, the bands are not continuous due to the Bragg resonances and the calculation of the Chern numbers must avoid the divergences. The band structure can be evaluated using the techniques explained in Sec.~\ref{sec:band}.

\begin{figure}[htbp]
  \centering
   \includegraphics[width=1\columnwidth]{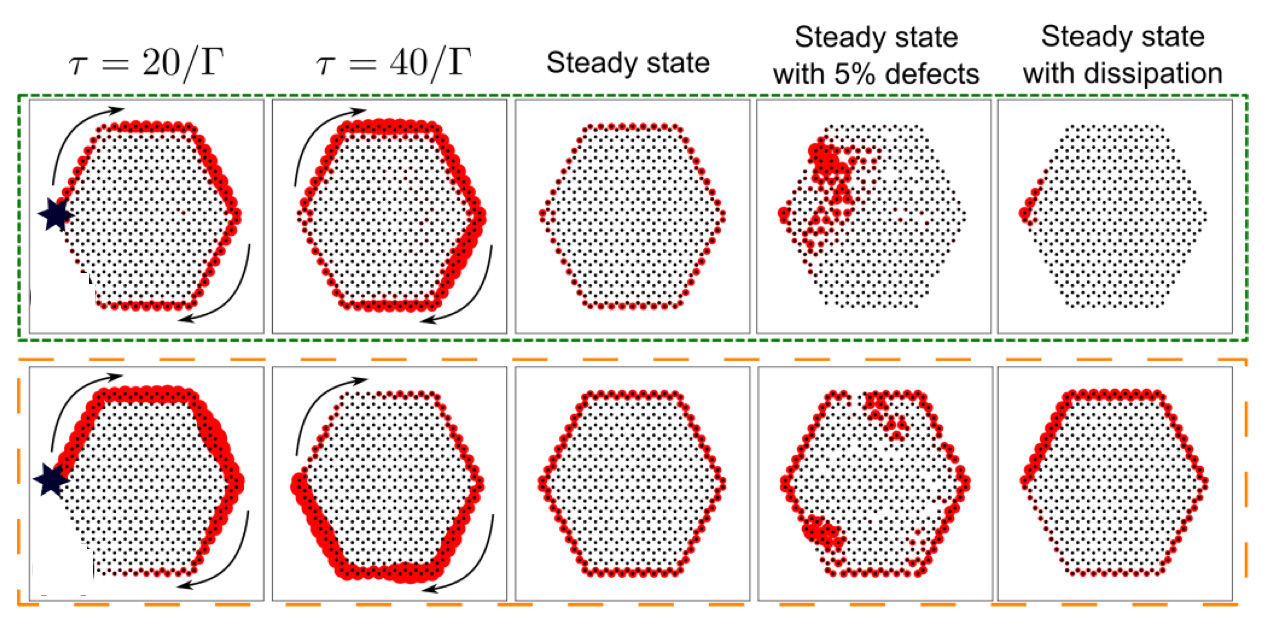}
      \vspace{-0.4cm}
  \caption{(Reproduced from Ref.~\cite{Bettles_17topo}). The dynamics of atomic polarization magnitude on a hexagonal array with the lattice spacing $a/\lambda=0.1$ when two different collective eigenmodes (in different rows) located at the edge of the array are resonantly targeted by driving a single edge atom  (indicated by the star in the leftmost panels).  Both cases  demonstrate chiral clockwise propagation, as indicated by snapshot images, and the steady state profile at the edge. However, only one of the edge states (the bottom row) exhibits topological robustness against defects and dissipation, despite the sum of the Chern numbers of the lower bands being equal in the both cases.   }
  \label{fig:topo}
  \end{figure}

Simulations conducted on arrays with small spacings of $a/\lambda =0.05$ and 0.1 have revealed the presence of topologically robust edge mode propagation, associated with strongly subradiant eigenmodes outside the light cone $|\qv|>|\kv|$~\cite{Bettles_17topo,Perczel2017b}. These edge modes offer appealing features such as reflection-free unidirectionality and resilience against defects (Fig.~\ref{fig:topo}). Interestingly, the presence of the long-range interactions were found to violate the standard bulk-boundary relation and only some of the edge states showed topological robustness, despite the band Chern numbers~\cite{Bettles_17topo}.

\subsection{Atomic level shifts and cooperative transparency}
\label{sec:twomode}

In Sec.~\ref{sec:trans}, we examined the case of light with spatially uniform phase profile normally incident on the array when the excited atomic levels are degenerate. 
This was analyzed in the \emph{single-mode} model using a collective ``coherent in-plane'' mode that  describes transmission and reflection. By introducing nondegenerate atomic level energies through a uniform level shift, the light can also couple to a collective ``coherent perpendicular'' eigenmode, where all the atomic dipoles oscillate in phase perpendicular to the plane~\cite{Facchinetti16,Facchinetti18}. 
In the limit of LLI, this system is accurately described by an effective \emph{two-mode} model, where each mode represents one of the perpendicular uniform collective eigenmodes. The two-mode superatom model extends beyond the conventional Lorentzian profile of the full reflection of the single-mode superatom, introducing also a narrow transparency window (Fig.~\ref{fig:twomode}).

In short 1D atom chains, subradiant eigenmodes can be excited by strong illumination~\cite{Williamson2020,Holzinger21}, and this excitation is further enhanced by a spatially-varying level shift along the chain~\cite{Plankensteiner2015,Jen2016,Ferioli21}. In planar arrays, achieving phase-matching between the incident field and deeply subradiant eigenmodes is more challenging due to the variation of field amplitude across the entire 2D plane.
However, by coupling the two modes in a planar array through a uniform level shift, it becomes possible to selectively excite a subradiant mode whose resonance linewidth approaches zero in the infinite lattice limit.
In numerical simulations up to 98-99\% of the total excitation at the resonance occupies a \emph{single} many-atom subradiant eigenmode~\cite{Facchinetti16,Facchinetti18}.
This behavior resembles giant spatially extended subradiance observed in strongly coupled planar scatterer arrays~\cite{Jenkins17}.

We consider a square lattice 
in the $yz$ plane and the $y$-polarized light propagating along the $x$ direction. 
In the case of degenerate excited states in the $J=0\rightarrow J'=1$ transition, the system exhibits isotropy, and any orientation of the orthogonal basis forms an eigenbasis. However, introducing linear Zeeman shift (along the $z$ axis) breaks this symmetry, and the dipoles pointing along the $y$ or $x$ direction
no longer form an eigenbasis or evolve independently.
The dynamics can be understood as follows: In the degenerate case, the $m=\pm1$ atomic polarization components represent equal and opposite circulations, resulting in a linear net polarization along the $y$ axis. The level shift tunes the $m=\pm1$ components out-of-phase, leading to elliptical polarization composed of both the $y$ and $x$ linear polarization components~\footnote{Classical analogy is an electron that starts rotating on a circular orbit under the Lorentz force.}. In the limiting case, the $y$ component completely disappears.

In this context, $\rho_{gey}$ ($\rho_{gex}$) denotes the collective in-plane (perpendicular) mode where the atomic dipoles oscillate along $y$ ($x$) axis. For an infinite array, these are exact collective eigenmodes of ${\cal H}$ [\EQREF{eq:hmatrix}] (the eigenmodes in the absence of  Zeeman shifts);  see Sec.~\ref{sec:LLI}.
However, due to the level shifts,  the light driving $\rho_{gey}$ becomes coupled to $\rho_{gex}$.
The dynamics in the LLI limit, given by \EQREF{eq:eigensystem}, in the two-mode model approximation  reads~\cite{Facchinetti16,Facchinetti18}
\begin{subequations}
\begin{align}
&\dot{\rho}_{gex}= (i\Delta_{P}-i\tilde{\delta}-\upsilon_P) \rho_{ge x}-\bar{\delta}\rho_{gey},\\
&\dot{\rho}_{gey} = (i\Delta_{I}-i\tilde{\delta}-\upsilon_I) \rho_{ge y}+\bar{\delta}\rho_{gex} +i{\cal R},
\end{align}
\label{Eq:ZeemanShift}
\end{subequations}
where  $\upsilon_{P}$ and $\upsilon_{I}$ are the collective linewidths of $\rho_{gex}$ and $\rho_{gey}$,
respectively, 
$
\Delta_{P/I}=\Delta_0+\delta_{P/I}
$
are the laser detunings from the mode resonances, and $\delta_{P/I}$ are the collective line shifts. 
The excited level shifts $\tilde{\delta}=(\delta_+-\delta_-)/2$, $\bar{\delta}=(\delta_-+\delta_+)/2$ in \EQREF{Eq:ZeemanShift}, which are encapsulated in $\delta{\cal H}$ in  \EQREF{eq:eigensystem}, break the isotropy of the transition via $\bar{\delta}$, and allow dipoles to be excited parallel to the light propagation by coupling  $\rho_{gex}$ to $\rho_{gey}$.

For an infinite subwavelength lattice, a uniform mode can only emit perpendicular to the array. Since no emission occurs along the dipole axis, the perpendicular mode $\rho_{gex}$ cannot scatter in that direction, resulting in strong subradiance, with $\lim_{N\rightarrow \infty} \upsilon_P = 0$ in the limit of infinite atom number $N$, despite the eigenmode residing inside the light cone. Numerically,  $\upsilon_P/\gamma \simeq N^{-0.9}$~\cite{Facchinetti16}, which is in contrast to the behavior of the \emph{most} subradiant eigenmode of the system, whose linewidth in typical situations narrows as $\propto N^{-3}$~\cite{Asenjo-Garcia2017a,Zhang20b}. Even in a shallow lattice of 50$E_r$ (see Sec.~\ref{sec:latticepot}), $\upsilon_P\simeq 0.15\gamma$ for $N\agt 100$~\cite{Facchinetti16}.

The steady-state solutions of Eqs.~\eqref{Eq:ZeemanShift} read
\beq  \label{eq:ssI}
 	\rho_{ge y}= \frac{Z_P(\Delta_0){\cal R}}{\bar{\delta}^{2} - Z_P(\Delta_0)Z_I(\Delta_0)} ,\quad
 	\rho_{gex} = -i\frac{\bar{\delta} \rho_{ge y}}{Z_P(\Delta_0)} \textrm{,}
\eeq
where $  Z_{P/I}(\Delta_0)  \equiv  \Delta_0 + \delta_{P/I} -\tilde{\delta} + i \upsilon_{P/I}$. Without level shifts ($\bar\delta=\tilde\delta=0$), $\rho_{ge y}$ trivially reduces to the single-mode model of \EQREF{eq:singlemodeamplitude}, with a Lorentzian profile and total resonance reflection ($\upsilon_I=\gamma+\tilde\gamma$ of Sec.~\ref{sec:trans}). 
Forward and backward scattering is solely produced by the in-plane amplitude $\rho_{ge y}$.
The steady-state reflection amplitude is shown in Fig.~\ref{fig:twomode} and is analytically obtained from Eqs.~\eqref{Eq:Transmission2} and~\eqref{eq:ssI} 
  \begin{equation}
  \label{reflect1}
    r =\frac{ i\upsilon_I Z_P(\Delta_0)}{\bar{\delta}^{2} - Z_P(\Delta_0)Z_I(\Delta_0)}\,.
\end{equation}
Since $\lim_{N\rightarrow\infty}\upsilon_P=0$, the reflection is always perfect $|r|^2=1$ at two values of the detuning $ \Delta_0 + \delta_{P} -\tilde{\delta}=\delta_d\pm(\bar\delta^2+\delta_d^2)^{1/2}$, where $\delta_d=(\delta_P-\delta_I)/2$. This happens on each side of the detuning that gives the complete transmission $r=0$ at the resonance $\Delta_0+\delta_P-\tilde{\delta}=0$ of the perpendicular mode~\cite{Facchinetti16,Facchinetti18}. 
 The shifts induced by ac Stark effect of lasers or microwaves~\cite{gerbier_pra_2006,Ballantine22str} can be rapidly turned off once the system has reached a steady-state with the population in the subradiant mode $\rho_{gex}$. In the absence of the level shifts, $\rho_{gex}$ is no longer coupled to $\rho_{gey}$ in \EQREF{Eq:ZeemanShift} and the decay is determined by the subradiant perpendicular mode with the narrow linewidth $\upsilon_P$.

The array exhibits dynamics similar to that of the electromagnetically-induced transparency (EIT) of independently scattering atoms~\cite{FleischhauerEtAlRMP2005}.
The two-mode model accurately predicts the behavior of the arrays, see Fig.~\ref{fig:twomode}, for given values of  $\upsilon_{P/I}$ and $\delta_{P/I}$.
The coherent in-plane mode corresponds to a `bright' mode of the EIT and the coherent perpendicular mode represents a deeply subradiant `dark' mode. 
In the independent-atom EIT, the single-atom linewidth defines a spectral transparency window where the absorption is suppressed in an otherwise opaque medium.
In the many-atom array, the transparency window is determined by the ultrasharp collective resonance $\upsilon_{P}$ and the level shift $\bar{\delta}$ serves to control the population of the dark (subradiant) mode, similarly to the coupling field control in the EIT~\cite{LiuEtAlNature2001}. 
The emergence of two full reflection maxima in \EQREF{reflect1} can be understood as a resonance splitting due to the effective ``coupling field'' $\bar\delta$, similarly to the EIT doublet splitting. 
At $a\simeq0.54\lambda$, we have $\delta_{P}\simeq\delta_I$ and the resonances are symmetrically split at $\pm\bar\delta$.

For finite-size arrays, for which $\upsilon_P\neq0$ but small, the reflectivity has a simple expression for $\delta_{P}\simeq\delta_I$.
At the resonance of the perpendicular mode~\cite{Facchinetti16}
\begin{equation}
  \label{eq:R_delta_M}
  r \approx - \frac{\upsilon_I\upsilon_P}{\bar{\delta}^2 + \upsilon_I\upsilon_P}.
\end{equation}
The transmission is close to one for $
\bar{\delta}^{2} \gg \upsilon_P\upsilon_I$. As shown in Fig.~\ref{fig:twomode}, while for large $\bar\delta$ the resonance width is $[(\upsilon_I^2+4\bar\delta^2)^{1/2}-\upsilon_I]/2$, for small  $\bar\delta$ the resonance narrows and eventually only depends on the very narrow  linewidth $\upsilon_P$, being a direct manifestation of subradiance~\cite{Facchinetti18}.

The presence of a narrow resonance results in significant group delays for a transmitted pulse due to its sharp dispersion~\cite{Facchinetti16,Facchinetti18}. 
The EIT magnetometry~\cite{Fleisch_magneto} relies on the significant dispersion that is sensitive to the magnetic field, while simultaneously suppressing absorption. 
The analogy of the array with the EIT suggests potential applications in cooperative magnetometry~\cite{Facchinetti18}. The accuracy of the cooperative magnetometer, however, is not limited by the single-atom resonance linewidth $\gamma$, but rather by the much narrower collective linewidth $\upsilon_P$. Sharp transmission variation in stacked layers has also been proposed as a band-stop filter~\cite{Javanainen20}.

\subsubsection{Parity-time symmetry and coherent perfect absorption}
\label{sec:pt}

The two-mode model in \EQREF{Eq:ZeemanShift} immediately allows to identify exceptional points when the decay becomes non-exponential by neglecting all the other modes in ${\cal H}+\delta{\cal H}$~\cite{Ballantine21PT}. Assuming $\delta_{P}\simeq\delta_I$  (at $a\simeq0.54\lambda$),  the two eigenmodes coalesce at $|\bar\delta|=|\upsilon_I-\upsilon_P|/2$ at which point the two-mode ${\cal H}+\delta{\cal H}$ possesses only one right eigenmode. 
The two-mode model also demonstrates spontaneous breaking of parity-time (${\cal PT}$) symmetry when scattering from the collective bright in-plane mode balances loss from the collective dark perpendicular mode~\cite{Ballantine21PT}. 
In this scenario, the array is symmetrically illuminated from both directions, and the total coherently scattered light exactly vanishes when the ${\cal PT}$ symmetry is broken, resulting in coherent perfect absorption with all the scattered light becoming incoherent. 
The incoherent scattering, e.g.\ due to position fluctuations, can be included in the mean-field approximation of 1D light propagation (Sec.~\ref{sec:normalinci}).

\subsection{Arbitrary angle of incidence}
\label{sec:arbitrary}

So far, we have focused on the case of normally incident light onto the array.  However, the analysis becomes more intricate when considering arbitrary angles of incidence, as discussed in Refs.~\cite{Shahmoon,Javanainen19}.
A simplified scenario arises when the dipoles are excited solely in the plane of a square array by a tilted incident field projected onto the lattice plane,
leading to the excitation of higher $\qv$-wavevector modes~\cite{Parmee2020,Parmee2021}.  In this case, the Rabi frequency ${\cal R}^{(\ell)}_\nu={\cal R} e^{i \qv\cdot \rv_\ell}$ 
and the atomic polarization density $\rho_{ge}^{(\ell)}=\rho_{ge}e^{i\qv\cdot \rv_\ell}$ in \EQREF{Eq:EOMSolns} only acquire phase variations. 
This situation can arise, e.g., with an isolated two-level transition within the lattice plane. 
The excitation of different $\qv$-modes is associated with changes in the bistability conditions (Sec.~\ref{sec:bistable}) and phase transitions~\cite{Parmee2020}.

We begin by examining a single array in the LLI limit and separate the wavevectors of the incident plane wave, ${\bf k}={\bf k}_\perp+{\bf k}_\parallel$, into those parallel to the $yz$ lattice plane, $\kv_\parallel$,  and those perpendicular to it, ${\bf k}_\perp=|{\bf k}_\perp| \pol_x$. The excitation eigenmodes for which $|\qv|> |{\bf k}|$ are outside the light cone and cannot be directly coupled to light in an infinite lattice at any incident angle, due to the rapid phase variation required. Equations~\eqref{Eq:EOMSolns} are valid for finite wavevector excitations $\qv=\kv_\parallel$ when we substitute ${\cal R}^{(\ell)}_\nu={\cal R}_\nu e^{i \qv\cdot \rv_\ell}$, $\bar{\mathcal{R}}^{(\ell)}_\nu=\bar{\mathcal{R}}_\nu e^{i \qv\cdot \rv_\ell}$, 
and  $\rho_{ge\nu}^{(\ell)}=\rho_{ge\nu}e^{i\qv\cdot \rv_\ell}$. However, \EQREF{Eq:EDef} is replaced by the matrix equation~\cite{Javanainen19}
\begin{align}\label{Eq:EDefmat}
\bar{\mathcal{R}}_\nu &= \mathcal{R}_\nu + [\tilde{\Omega}_{\nu\mu}(\qv)+i\tilde {\gamma}_{\nu\mu}(\qv)]\rho_{ge\mu} \nonumber\\
&= \mathcal{R}_\nu +  \frac{\alpha}{\xi} [\tilde{\Omega}_{\nu\mu}(\qv)+i\tilde {\gamma}_{\nu\mu}(\qv)] \bar{\mathcal{R}}_\mu ,
\end{align}
where $\tilde{\Omega}_{\nu\mu}(\qv)$ and $\tilde {\gamma}_{\nu\mu}(\qv) $ are defined in \EQREF{FourierTransforms}.
In \EQREF{Eq:EDefmat}, we have introduced the atomic polarizability 
\beq \label{eq:polarizability}
\alpha = -\frac{{\cal D}^2}{\hbar\eo (\Delta+i\gamma)},
\eeq
representing in the limit of LLI the ratio of the induced atomic dipole to the effective field of the incident light plus the dipole radiation from all the other atoms in the array.

For the linearly responding atoms, it is then straightforward to formally solve \EQREF{Eq:EDefmat} in the matrix form~\cite{Javanainen19},
\begin{subequations}
\label{eq:matrixsinglearray}
\begin{align}
\bar{ \boldsymbol{\mathcal{R}}} &= \big\{ 1-\frac{\alpha}{\xi} [\tilde{\boldsymbol{\Omega}}(\qv)+i\tilde {\boldsymbol{\gamma}}(\qv)] \big\}^{-1}  \boldsymbol{\mathcal{R}},\\
\rho_{ge\nu} &=  \pol_\nu^*\cdot\frac{\alpha}{\xi} \big\{ 1-\frac{\alpha}{\xi} [\tilde{\boldsymbol{\Omega}}(\qv)+i\tilde {\boldsymbol{\gamma}}(\qv)] \big\}^{-1}  \boldsymbol{\mathcal{R}}.
\end{align}
\end{subequations}
The excitation wavevector, $\qv=(k_\perp,\kv_\parallel)$ (assuming only one Bragg peak is excited, for a more general case, see Appendix~\ref{appen4}), is determined by the incident angle.  In the case of non-normal incidence and for given $\qv$, the matrix $ [\tilde{\boldsymbol{\Omega}}(\qv)+i\tilde {\boldsymbol{\gamma}}(\qv)]$ generally exhibits three distinct resonances, each with its own collective linewidth~\cite{Javanainen19}. The atomic polarization acquires an $x$-component 
that is decoupled from the resonances in the $yz$ lattice plane. When considering a general incident angle, two elliptically polarized eigenmodes are excited in the $yz$ plane (even for a square lattice) with distinct resonances.

\begin{figure}[htbp]
  \centering
   \includegraphics[width=0.8\columnwidth]{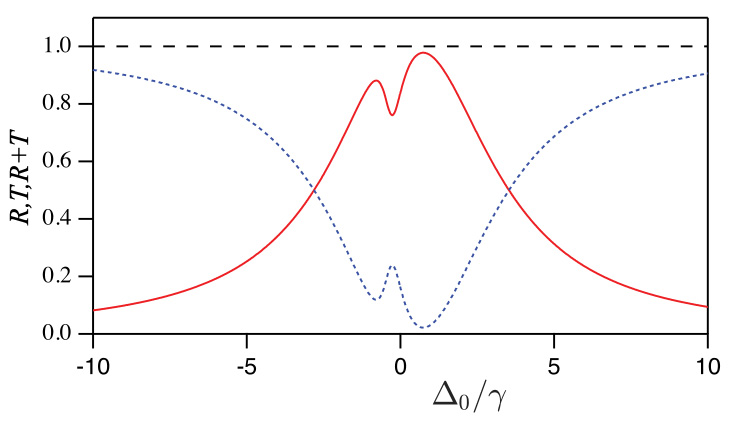}
      \vspace{-0.3cm}
  \caption{(Reproduced from Ref.~\cite{Javanainen19}).  Reflection $R=|r|^2$ (solid red line), transmission $T=|t|^2$ (dotted blue line), and $R+T$ (dashed black line) as a function of detuning for non-normal incidence in an infinite square lattice. The  lattice spacing $a=0.5\lambda$, the incident angle with the normal $\theta=0.4\pi$, and the angle of $\kv_\parallel$ with the lattice axis $\phi=\pi/8$. Due to the non-normal incidence and $\phi\neq n\pi/4$ ($n$ integer), $R$ never reaches one and there are two distinct resonances representing the eigenvalues $(-0.325+0.389i)\gamma$ and $(0.399+ 3.00i)\gamma$, but no third one since the incident light is polarized along the array plane.
  }
  \label{fig:nonnormal}
  \end{figure}

The lattice plane component, $\kv_\parallel$,  also has an impact on the light propagation between stacked atomic layers. The coupling of 1D electrodynamics  between the planar arrays is altered by the phase variation within each plane.  In principle, the derivation could be conducted similarly to the case without phase variation in Appendix~\ref{appen1}, with the exception that \EQREF{eq:scattlightplanar} is replaced by a matrix equation that also incorporates the momentum contribution from $\kv_\parallel$~\cite{Javanainen19},
\begin{equation} \label{eq:scattlightplanarB}
\epsilon_0 \pol_\nu^* \cdot \langle \hat{\textbf{E}}{}^+_s(\textbf{r}) \rangle  = {\cal D} \sum_{j} e^{i \kv_\parallel\cdot \textbf{R}_j}  \pol_\nu^* \cdot \mathsf{G}(\textbf{R}_j)\pol_\mu \,\rho_{ge\mu} = {\sf T}_{\nu\mu}  {\cal D} \rho_{ge\mu},
\end{equation}
where $\textbf{R}_j = x\,\hat{\bf e}_x\! -y_j \,\hat{\bf e}_y\! -z_j\, \hat{\bf e}_z$ defines the vector joining the $j$th atom and the observation point $(x,0,0)$. From \EQREF{eq:matrixsinglearray}, we obtain forward and backward scattered fields for sufficiently large $|x|$, with, e.g., the forward scattered Rabi frequency $\boldsymbol{\mathcal{R}}_{s}$ 
\beq \label{eq:scatterednonnormal}
\mathcal{R}_{s,\nu} =   \pol_\nu^*\cdot {\sf T}  {\alpha} \big\{ 1-\frac{\alpha}{\xi} [\tilde{\boldsymbol{\Omega}}(\qv)+i\tilde {\boldsymbol{\gamma}}(\qv)] \big\}^{-1}  \boldsymbol{\mathcal{R}}.
\eeq
By comparing with the numerical results, it was possible in Ref.~\cite{Javanainen19} to establish  an analytic expression for ${\sf T}$ ($x>0$),
\beq \label{eq:T}
{\sf T} = \frac{ik}{2{\cal A}'} {\sf R}(-\phi) {\sf M}(\theta)  {\sf R}(\phi) e^{i k_\perp x},
\eeq 
where $ {\sf R}(\phi) $ denotes the rotation matrix around the $x$ axis with an angle $\phi$ and 
 \begin{equation}\label{eq:M}
{\sf M} (\theta) = \left(
\begin{array}{ccc}
 \sin^2\theta/\cos\theta   & -\sin\theta & 0 \\
-\sin\theta &   \cos\theta &  0 \\
 0 &  0 & 1/\cos\theta  \\
\end{array}
\right),
\end{equation}
where the coordinates $(\theta,\phi)$ are determined by the incident light wavevector $\kv = \cos\theta\, \pol_x+ \sin\theta\cos\phi\, \pol_y+\sin\theta\sin\phi\,\pol_z$. Equation~\eqref{eq:T} represents
projection to the subspace orthogonal to the light propagation $\kv$, forcing the radiated field to be transverse, with the density of the atomic dipoles adjusted by the viewing angle $\theta$. This yields
a compact expression that is also derived in Appendix~\ref{appen4}
\beq
\label{eq:projectionpro}
{\sf T} = \frac{ik}{2{\cal A}'\cos\theta} {\sf P}_\perp(\kv) e^{i k_\perp |x|},
\eeq
where ${\sf P}_\perp(\kv)$ denotes the projection operator~\cite{Javanainen19}. An example case is shown in Fig.~\ref{fig:nonnormal}.

By carefully selecting the propagation direction and polarization of light, it is possible to find applications in canceling density-dependent collective resonance shifts caused by DD interactions, which can have a significant detrimental effect in precision spectroscopy~\cite{Javanainen19}.
Transfer matrix analysis of Sec.~\ref{sec:transfer} can be used to estimate the resonance shift of the peak transmission through stacked arrays for normal incidence
$\delta\simeq \cot(2kd)\gamma_{\rm 1D}/2$~\cite{Ruostekoski17}.
When the layer spacing is close, but not exactly equal, to an integer multiple of half of the wavelength, the shift diverges.
This can be understood as a  resonance shift due to cavity-like effects, where each pair of successive layers defines a cavity. However, the shift is affected by the angle dependence of the collective mode coupling, resulting in specific incident angles where it vanishes.

\subsubsection{Arbitrary angle of incidence and level splitting}
\label{sec:arbitrarysplit}

In the previous section, we assumed that the $J=0\rightarrow J'=1$ transition is isotropic due to the degeneracy of the excited levels. However, in Sec.~\ref{sec:twomode}, it was demonstrated that the atomic polarization can acquire a component parallel to light propagation even for normally incident light beam when the degeneracy is broken. The findings presented in Sec.~\ref{sec:arbitrary}, which pertained to arbitrary angles of incidence,  can be extended to the case of anisotropic polarizability by replacing the scalar polarizability $\alpha$ in Eqs.~\eqref{eq:matrixsinglearray} and~\eqref{eq:scatterednonnormal} with a $3\times3$ tensor polarizability $\boldsymbol{\alpha}$~\cite{Javanainen19}.

\subsection{Wavefront engineering}
\label{sec:wavefront}

Thin layers of artificially fabricated metamaterials, known as metasurfaces, have been developed to manipulate and control classical light propagation~\cite{Yu14,Chen2016_review,Luo18,Qiu21_review}. Unlike in bulky 3D metamaterials, absorption is suppressed and microscopic patterned surface coating can introduce abrupt phase modulation on transmitted and reflected light. Metasurfaces function as optical elements, enabling various beam-shaping capabilities without altering the material's geometric shape. 
Nevertheless, metasurfaces face certain constraints in their functionality. Fabrication irregularities result in inhomogeneous resonance broadening, and preventing absorptive losses at optical frequencies poses a significant challenge.  Furthermore, their operation is primarily confined to the classical domain, with limited quantum capabilities~\cite{Solntsev21}.

Atomic planar arrays are more amenable to quantum-optical control and offer other key advantages. 
Here we consider atomic single and bilayer arrays as nanophotonic surfaces for wavefront control. A lattice excitation can be engineered to target some superposition of  transverse modes $\varphi_n(\rho,\phi)$, which form some suitable basis in the polar coordinates, subsequently affecting coherent transmission.
In the paraxial approximation  [see \EQREF{eq:1dfield}] normal to the array
\beq
\label{eq:1dfield2}
\epsilon_0 \langle \hat{E}{}^+_s(x) \rangle= \frac{  ik  }{2} \sum_n c_n \varphi_n(\rho,\phi) e^{ik|x|} \<\hat P^+(\rv)\>.
\eeq
Moreover, two independent light polarization components in transmission realizes an optical Jones matrix.
In Ref.~\cite{Bassler23}, a single layer with excited-level Zeeman shifts (see Sec.~\ref{sec:twomode}) was proposed as polarizer, allowing the transmission of only one of the polarization components. However, implementing an optical wave-plate with a single layer results in unavoidable losses in transmission~\cite{Bassler23}.

Cooperatively interacting planar arrays of atoms can be engineered to exhibit collective excitation eigenmodes that show magnetic responses at optical frequencies, despite individual atoms having negligible coupling to the magnetic component of light~\cite{Ballantine20Huygens,Alaee20}. Coupling light, in particular, to bilayer arrays that can have different unit cell orientations allows great engineering flexibility.  It becomes possible to precisely control the phase, polarization, and direction of transmitted light, and to realize Huygens' surfaces~\cite{Ballantine20Huygens,Ballantine21wavefront,Ballantine22str}. Huygens' surfaces are based on Huygens' principle that every point in a propagating wave acts as an independent source of forward-propagating waves~\cite{Huygens,Love1901}. These surfaces can be physically realized by implementing fictitious sources of wavefront based on Huygens' concept, using crossed electric and magnetic dipoles~\cite{Pfeiffer13,Decker15,Yu15}. As any wave can be represented by Huygens' principle, achieving complete control over the forward-propagating wavefront is theoretically possible through a Huygens' surface.

A collective LLI eigenmode consisting of an effective magnetic dipole can be formed, e.g., by arranging four atoms at the corners of a square unit cell, with electric dipoles oriented in a circular fashion resembling a circular loop of a continuous azimuthal electric polarization density~\cite{Ballantine20Huygens}. Such a unit cell results in a net zero electric dipole and nonvanishing perpendicular magnetic dipole. 
The overall array eigenmode can approximate a phase-uniform repetition of coherent magnetic dipoles in each unit cell. The LLI eigenmode with a uniform phase profile, considered in Sec.~\ref{sec:uniform}, exhibits even symmetry between forward and backward scattered light $\<\hat{\textbf{E}}{}^+_{s,f}(\textbf{r})\>=\<\hat{\textbf{E}}{}^+_{s,b}(\textbf{r})\>$. However, for the mode composed of magnetic dipoles pointing along the bilayer plane, the field is antisymmetric around the lattice plane, leading to  $-\<\hat{\textbf{E}}{}^+_{s,f}(\textbf{r})\>=\<\hat{\textbf{E}}{}^+_{s,b}(\textbf{r})\>$. 
For a uniform distribution of effective magnetic dipoles ${\bf m}={\cal M} \pol_z$ we obtain the reflection and transmission amplitudes [analogously to those in \EQREF{LLItrans}]
\begin{equation}\label{LLItransmag}
r=\frac{i \gamma_{\rm 1D}^M}{\Delta_{\rm 1D}^M+i \gamma_{\rm 1D}^M}, \quad
t=\frac{\Delta_{\rm 1D}^M}{\Delta_{\rm 1D}^M+i \gamma_{\rm 1D}^M},\quad \gamma_{\rm 1D}^M= \frac{\mu_0 k{\cal M}^2}{2{\cal A}'\hbar },
\end{equation} 
where $\gamma_{\rm 1D}^M$ is the collective eigenmode linewidth with the area density ${\cal M}/{\cal A}'$ of magnetic dipoles [the magnetic analog of \EQREF{eq:planarwidth1}] and $\Delta_{\rm 1D}^M$ is the laser detuning from the collective mode resonance. The transmission retains the same form as in Eq.~\eqref{LLItrans}, while
the resonance reflection $r=-1$ in Eq.~\eqref{LLItrans} changes to $r=1$ in Eq.~\eqref{LLItransmag}. This represents a magnetic mirror, observed in metal~\cite{Sievenpiper99,Schwanecke06} and dielectric~\cite{Liu14,Lin16} structures. Due to the sign change, the conventional half-wave loss of fields near the array interface is completely eliminated.

Despite having different resonance frequencies, the collective electric and magnetic dipole modes, responsible for reflection and transmission of Eqs.~\eqref{LLItrans} and~\eqref{LLItransmag}, can be simultaneously excited through level shifts incorporated in $\delta{\cal H}$ in \EQREF{eq:eigensystem}, enabling the realization of a Huygens' surface~\cite{Ballantine20Huygens,Ballantine21wavefront,Ballantine22str}. Due to the different phase shifts for the reflected waves, the scattered light from the two modes can destructively interfere in the backward direction, while constructively interfering in the forward direction. This constructive interference approximately compensates for any reduction in transmission near the resonance, resulting in $|t|\simeq1 $ at any frequency, while allowing the phase of the total light amplitude to vary over the full $2\pi$ range. The simultaneous excitation of these two collective modes requires synthesizing spatially nonuniform ac Stark shifts for the excited electronic levels, which can potentially be achieved, e.g., in Sr atoms with off-resonant $5s4d\,{^3D}_1\rightarrow 5s6p\,{^3P}_J $ transitions, while the probe is tuned to the $5s5p\, {^3}P_0 \rightarrow 5s4d\, ^3D_1$ transition~\cite{Ballantine22str}. However, the procedure can be simplified by the observation that crossed electric and magnetic dipoles are not strictly  necessarily for a Huygens' surface;  what matters is the simultaneous excitation of collective uniform modes with even and odd parities 
that may be formed by any combinations of electromagnetic multipoles, and can be achieved even without externally induced level shifts~\cite{Ballantine23}.

A functional Huygens' surface serves as a versatile tool for wavefront engineering. It can be used as an effective variable wave-plate by introducing a delay in the incoming $y$-polarized light, which couples to the Huygens' surface resonance, relative to the $z$-polarized light, which does not~\cite{Ballantine21quantum}. Applying linear level shift gradients can generate beam-steering, focusing, or can transform a Gaussian beam to a Poincar\'e beam~\cite{Ballantine22str}.

\subsection{Photon storage}
\label{sec:storage}

Planar atomic arrays offer great potential as a platform for quantum networks~\cite{Grankin18,Kimble08,Ritter12} due to
the presence of strongly subradiant states and highly collimated coherent light emission. In Sec.~\ref{sec:twomode}, we discussed how,  in the steady state, nearly the entire optical excitation can be driven to a deeply subradiant collective excitation eigenmode~\cite{Facchinetti16,Facchinetti18}. To achieve reversible quantum memory of light and high-fidelity storage of a single photon in a planar array, efficient absorption of a time-dependent single-photon pulse and subsequent storage of the absorbed pulse need to be considered. However, in a single-layer 2D lattice, efficient absorption or emission of light  occurs symmetrically in both forward and backward directions. The maximum absorption efficiency for a normally incident unidirectional light pulse is 0.5. This can be understood by considering a plane wave
$\exp(ikx)=\cos(kx)+i \sin(kx)$. In this case, at the lattice position $x=0$, only the symmetric component $\cos(kx)$ couples to the atoms, resulting in a loss of 50\% of the intensity~\cite{Ballantine22bilayer}. 

In a general scheme, the limits of storage error in a planar array for a photon arriving by some nonspecific mechanism were analyzed in Ref.~\cite{Manzoni18}.
The analysis considered a subwavelength square array composed of three-level atoms with the electronic ground state $|g\>$, excited state $|e\>$, and ``storage'' state $|s\>$. The incoming photon couples  $|g\>$ to $|e\>$, while a coherent field drives the transition between $|e\>$ and $|s\>$. The minimum retrieval/storage error was calculated by varying the waist of a Gaussian beam-shaped incident photon.  The main sources of
error were identified as the fraction of the energy carried by the beam beyond the array boundaries and the range of wavevector components that reach the target mode~\cite{Manzoni18}. For the beam waist values $w_0$ less than the array size, the latter scales as $\propto (\lambda/w_0)^4$.
After optimizing the beam waist, the leading term for the error for the atoms at fixed positions scales with the atom number as $\propto (\log N)^2/(4N^2)$. For a $4\times4$ array, this corresponds to an efficiency above 99\%. In a dilute disordered ensemble of three-level atoms, the storage efficiency depends on the optical depth $D\sim \sigma_0 N/w_0^2$, where $\sigma_0=3\lambda^2/(2\pi)$ denotes the resonance cross-section of a single atom~\cite{Gorshkov07}. Only values close to $D\sim 600$ (or $N\sim10^6$-$10^7$) provide similarly small errors~\cite{Manzoni18}. The storage efficiency was found to deteriorate as a function of the position fluctuations [see \EQREF{eq:phi}] as $\propto\ell^2/d^2$.

Highly excited Rydberg states offer the possibility of generating a single excitation in an atomic array~\cite{Dudin12,Srakaew22,Bekenstein2020,Moreno2021,Zhang2022}, see Sec.~\ref{sec:blockade}. Within a sufficiently small spatial region determined by the blockade radius a single excitation could be stored and released as a photon, as DD interactions between the atoms can suppress more than one Rydberg excitation.
Alternatively, the approach proposed in Refs.~\cite{Petrosyan18,Grankin18} involves a single isolated atom in a Rydberg state. This excitation then is transferred to an array through a resonant exchange interaction with a collective Rydberg state of the array atoms, employing techniques such as adiabatic passage. A complete Rydberg blockade among the array atoms is not necessary. The photon could subsequently be stored in the array or the array could serve as a collimated single photon source, or a quantum antenna~\cite{Grankin18}. 
Here a localized source atom can create an excitation with a nonuniform phase profile, which could be compensated by introducing spatially nonuniform detunings in the array atoms~\cite{Grankin18}. If only small localized region of the array is excited by a single photon, the excitation can be transferred to a perpendicular mode by rotating the dipoles, as described in Sec.~\ref{sec:twomode} and the following paragraph. The localized excitation then propagates across the array and can probabilistically reach the deeply subradiant perpendicular eigenmode with a uniform phase profile across the entire array~\cite{Ballantine20ant}.

The storage of a photon pulse in a deeply subradiant  eigenmode, where dipoles oscillate in phase perpendicular to the array plane (Sec.~\ref{sec:twomode}), has been examined in Ref.~\cite{Ballantine21quantum}.
To address the challenge of simultaneously suppressing both forward and backward scattering through interference with the incident beam, the pulse was directed through a beam splitter to symmetrically illuminate the lattice from both sides.
This configuration enables the lattice to couple to a standing wave $\cos(kx)$ at $x=0$, minimizing losses. 
The storage process follows a similar principle to driving the subradiant perpendicular mode described in Sec.~\ref{sec:twomode}. The incident light directly excites the in-plane 
mode that is polarized in the same direction. However, since the degeneracy of the excited electronic levels is lifted through optical- or microwave-induced ac Stark shifts,
the in-plane mode is coupled to the perpendicular mode. This facilitates the excitation transfer to the deeply subradiant phase-uniform perpendicular mode. 
Once the photon is absorbed,  the level shifts are rapidly removed, and the degeneracy is restored, resulting in the storage of the photon in a subradiant state. 
A closely related storage scheme was examined in Ref.~\cite{Rubies22} (see also Ref.~\cite{Fayard2023}), where the excitation was coupled to a checkerboard pattern of atomic level shifts. Such a nonuniform profile of atomic resonance frequencies couples the incident light to a collective eigenmode with antiferromagentic spin pattern~\cite{Parmee2020}. In the absence of atomic level shifts, the subradiant antiferromagnetic eigenmode can lie outside the light cone. 

To establish a quantum network using atomic planar arrays and enable coherent quantum links between them, it is important to achieve directional control over efficient absorption, storage, and release of photons. The need to illuminate from both sides the array~\cite{Ballantine21quantum,Rubies22} can be overcome by considering bilayer arrays~\cite{Ballantine22bilayer}. 
In a bilayer array, the collective eigenmodes of interest are characterized by atoms within each layer oscillating in phase, while
the top and bottom layers exhibit either symmetric (in-phase) or antisymmetric ($\pi$ out-of-phase) modes.
The control of emission and absorption directionality follows from the different parity of these modes: the symmetric mode, with even parity, scatters equally in both directions, while for the antisymmetric mode, with odd parity, the scattering in the forward and backward direction is $\pi$ out-of-phase. 
As a result, scattering losses are suppressed during photon absorption~\cite{Ballantine22bilayer}: in the forward direction, the symmetric and antisymmetric modes are in-phase, and destructively interfere with the incident photon [see Eqs.~\eqref{eq:1dfield} and~\eqref{eq:singlemodeamplitude}], while in the backward direction the scattered light from these modes destructively interferes. 
The storage process after absorption follows again the same principle of transferring  the photon to the subradiant perpendicular mode as described in Sec.~\ref{sec:twomode} and above. 
Numerical simulations on a 20$\times$20$\times$2 array for $a\simeq0.91\lambda$ and a layer separation $d=0.25\lambda$ demonstrated a storage efficiency of 0.93.
For $d=0.9\lambda$, an efficiency of 0.85 was achieved. 
The perpendicular mode, uniform in each layer, consists of symmetric and antisymmetric modes.
By adjusting the relative phase between the symmetric and antisymmetric amplitudes using an induced atomic level shift between the two layers,  the excitation can be transferred back to the in-plane modes in such a way that it is emitted in the forward direction, backward direction, or an arbitrary combination of the two, independent of the incident beam direction. The process is qualitatively described by four eigenmodes.

\subsection{Dipole blockade and nonclassical light emission}
\label{sec:blockade}

Quantum solutions for stationary atoms (Secs.~\ref{sec:qme} and~\ref{sec:traj}) demonstrate nonlinearities at the few-photon level.
Strong light-mediated interactions between the atoms in a planar array lead to correlated responses, suppressed joint photon detection events, and dipole blockade~\cite{Cidrim20,Williamson2020b}. These nonlinearities can be further enhanced by involving
Rydberg excitations coupled to optical transitions~\cite{Bekenstein2020,Moreno2021,Zhang2022}, as demonstrated in recent experiments~\cite{Srakaew22}.  

A well-known result of quantum optics for a two-level atom~\cite{Carmichael_1976,kimble1977,kimble1978,dagenais1978,walls1979} reveals quantum correlations $g_2(0)<g_2(\tau)$  in the joint probability of two photon detection events occurring $\tau$ apart
\begin{equation} \label{eq:g2}
g_2(\tau)\equiv\lim_{t\rightarrow\infty}\frac{\langle\,:\!\hat{n}(t+\tau)\hat{n}(t)\!:\,\rangle}{\langle\hat{n}(t)\rangle^2},
\end{equation}
where $:\,:$ denotes normal ordering and $\hat{n}(t)$ is the photon number operator. The emission times of two back-to-back photons are modified: two photons cannot closely follow each other, demonstrating the particlelike behavior of light. This is a direct consequence of the two-level nature of the atom, as the electron, after each photon emission, returns to the ground state and cannot re-emit until excited by laser light again.

Strong light-mediated interactions between atoms are required for these correlations to persist in a many-atom system because, in a noninteracting ensemble, atoms emit photons independently and emission events from different atoms quickly wash out any nonclassical correlations. At sufficiently small lattice spacings ($a\alt 0.15$),  multiple scattering events give rise to a correlated response, enhancing the nonclassical nature of light emission~\cite{Cidrim20,Williamson2020b}. Photon emission events become synchronized, and the atomic ensemble behaves as a single superatom. The dipole blockade, which inhibits transitions into all but singly-excited states, can then survive over a collective correlated state with a size of about $\lambda$~\cite{Williamson2020b}. The dipole blockade is analogous to the microwave dipole blockade in Rydberg atoms~\cite{Jaksch00,Lukin01,Urban09,Grangier09,Saffman10,Schauss12,ripka2018}. 

Due to the syncronized emission events in the atomic array, the functional form of $g_2(\tau)$~\cite{Carmichael_1976,carmichael1976} for a single isolated atom remains valid for the entire array in the
strongly correlated regime of interest~\cite{Williamson2020b}
\begin{equation}\label{eq:g2single-atom}
g_2^{(\gamma,\kappa)}(\tau)\equiv 1-e^{-3\gamma \tau/2}\left(\cosh \kappa \gamma \tau+\frac{3}{2}\frac{\sinh \kappa \gamma \tau}{\kappa}\right),
\end{equation}
where $\kappa \equiv \frac{1}{2}[1-8I_\text{in}/I_s(\gamma)]^{1/2}$, and $I_\text{in}$ and $I_s(\gamma)$ are the incident light and saturation~[\EQREF{eq:saturationint}] intensities, respectively. 
In the many-atom case, the single-atom result \eqref{eq:g2single-atom} is modified by multiplying $I_{\rm in}$ by the projection of the incident field onto the excited LLI eigenmode, by an overall normalization of light emission at zero delay, 
 and by the replacement of the single-atom linewidth $\gamma$ by the linewidth $\upsilon$  of the underlying LLI collective excitation eigenmode in \EQREF{eq:g2single-atom} and in $I_s(\gamma)$ of \EQREF{eq:saturationint}~\cite{Williamson2020b}. 
As a result, the nonclassical nature of emitted light, conveyed by photon antibunching, is significantly enhanced by driving a subradiant collective excitation with a narrow linewidth, surpassing the performance of a single isolated atom. Additional nonlinearities in a bilayer array can further suppress joint photon detection events~\cite{Pedersen23}.

The statistics of emitted photons from a planar array were calculated using quantum trajectories in Ref.~\cite{Cano21}, considering the scenario where the incident light is indistinguishable from the scattered photons. It was observed that the transmitted light exhibited photon bunching, while the reflected light at low intensities displayed photon antibunching. Interestingly, this work also demonstrated quantum measurement-induced back-action on the state of the array:
when a photon is detected in the forward direction, it projects the atoms into a collective state with reduced reflectivity and increased excited level population. 
This, in turn, enhances the probability of subsequent photon emissions and leads to photon bunching.

\begin{figure}[htbp]
  \centering
   \includegraphics[width=0.9\columnwidth]{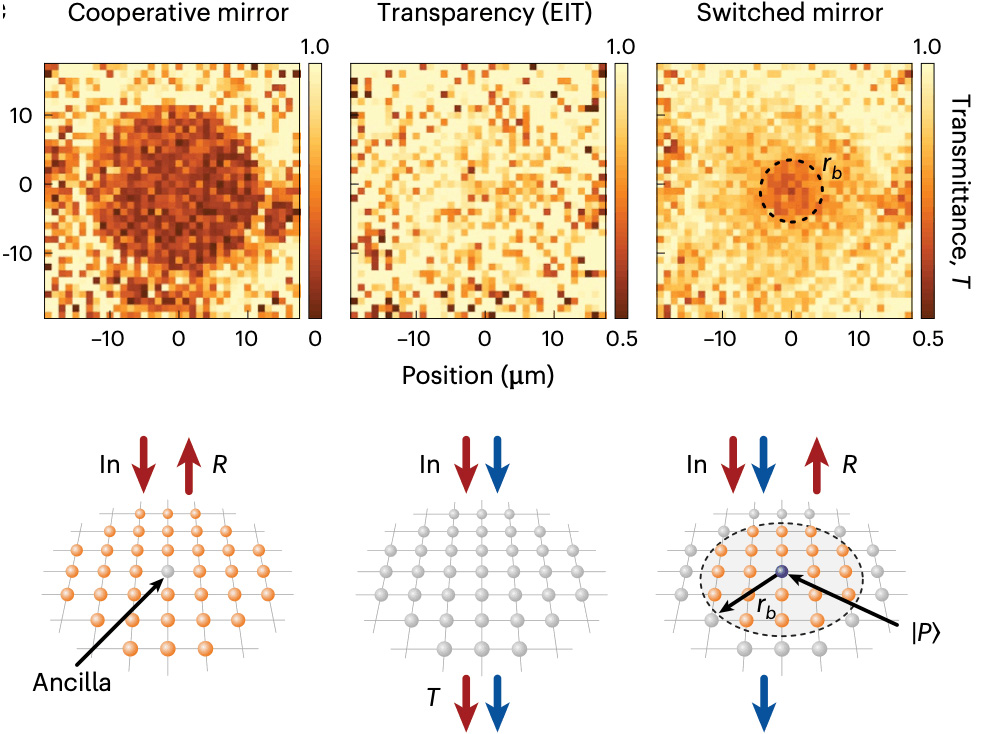}
      \vspace{-0.4cm}
  \caption{ (Reproduced from Ref.~\cite{Srakaew22}). An experiment in an analogous system to Fig.~\ref{fig:linewidth},
 when the transmission 
is controlled by a single ancilla atom at the center of the array. The top row shows the spatially resolved transmitted light. Left: strong resonant reflection when only the probe beam is on.
Middle: the additional control field coupled to the Rydberg state establishes an EIT resonance, rendering the array transparent. Right: the ancilla atom restores the reflectivity within a finite radius around the ancilla through the dipolar Rydberg interaction that shifts the control field out of resonance.}
  \label{fig:blockade}
  \end{figure}

Strong nonlinear interactions were successfully achieved by utilizing Rydberg states, which enabled coherent control through dipole blockade, leading  to the switching of the planar atomic array between transmission and reflection of light~\cite{Srakaew22}. This experimental breakthrough was preceded by theoretical investigations~\cite{Bekenstein2020,Moreno2021,Zhang2022}. In the experiment, 
a subwavelength planar array consisting of up to 1500 $^{87}$Rb atoms in a Mott-insulator state of an optical square lattice with a single atom per lattice site was controlled by using a single ancilla atom excited to a Rydberg state. The ancilla atom within the array was prepared in the electronic ground state $|g'\>$ and coupled to a Rydberg $P$-state $|r'\>$ by an ultraviolet beam, while the remaining atoms in the lattice occupied a different hyperfine electronic ground state $|g\>$.
An incident light pulse with the Rabi frequency ${\cal R}$ coupled $|g\>$ to an excited state $|e\>$.  By employing a control field  with the Rabi frequency ${\cal R}_c$ that coupled $|e\>$ to a highly excited Rydberg $S$-state $|r\>$, Rydberg-EIT was induced. In the weak driving limit, the populations of $|e\>$  and $|r\>$ are negligible and the dynamics can be described by the EIT response in terms of the atomic coherences $ \rho_{gr}$ and $ \rho_{ge}$:
\begin{subequations}
\begin{align}
&\dot{\rho}_{gr}= (i\Delta_{r}+iU-\gamma_r) \rho_{gr}+ i {\cal R}^*_c \rho_{ge},\\
&\dot{\rho}_{ge} = (i\Delta_{I}-\upsilon_I) \rho_{ge}+i {\cal R}_c \rho_{gr} +i{\cal R},
\end{align}
\label{Eq:Rydberg}
\end{subequations}
where  $\upsilon_{I}$ is the collective linewidth of the targeted in-plane eigenmode,  $\Delta_{I}=\Delta+\delta_{I}$ is the laser detuning from this mode resonance, $\delta_{I}$ is the collective line shift, and $\gamma_r$ and $\Delta_r$ are the linewidth and the detuning of the Rydberg state. The Rydberg states $|r'\>$ and $|r\>$ experience a strong long-wavelength dipolar interaction that establishes a distance-dependent level shift that is denoted by $U(r)$. By defining  $  Z_{I}(\Delta)  \equiv  \Delta + \delta_{I} + i \upsilon_{I}$ and $  Z_{r}(\Delta_r)  \equiv  \Delta_r + U+ i \gamma_r$, the steady-state reflection amplitude reads [compare with the level shift generated EIT-like responses of Eqs.~\eqref{Eq:ZeemanShift}, \eqref{eq:ssI}, and~\eqref{reflect1}]
  \begin{equation}
  \label{rydreflect1}
    r =\frac{ i\upsilon_I Z_r(\Delta_r)}{|{\cal R}_c|^2 - Z_r(\Delta_r)Z_I(\Delta)}\,,
\end{equation}
which displays an EIT resonance doublet and transparency window. However, this also represents an optical switch, prepared in Ref.~\cite{Srakaew22}, as
the control field creates an admixture of $|e\>$ and $|r\>$, allowing $|e\>$ to inherit the characteristics of the Rydberg long-range interactions through controlled level shift $U$ facilitated  by the ancilla atom. The dipole blockade effect induced by the ancilla atom then enables  the array to switch between subradiant transmission and reflection, as shown in Fig.~\ref{fig:blockade}. Numerical simulations involving an ancilla atom~\cite{Zhang2022} and Rydberg-dressed interactions~\cite{Moreno2021} demonstrated antibunching in the second-order correlation function~\EQREF{eq:g2}.

\subsubsection{Entanglement}
\label{sec:entangle}

The experiment~\cite{Srakaew22} on the optical switch using the Rydberg-atom blockage was conducted in a classical regime. However, by considering quantum superpositions of the ancilla atom $(|g'\>+|r'\>)/2^{1/2}$, it is possible to entangle the array and the Rydberg atom to achieve quantum control over the optical response~\cite{Bekenstein2020}. By using non-overlapping Gaussian beams focused on different locations of the array, single photons in the transmitted and reflected modes can represent qubit states. Photons in different locations could, in principle, be entangled, or projective measurements could be used to prepare entangled Greenberger-Horne-Zeilinger (GHZ) or cluster photonic states~\cite{Bekenstein2020}. The properties of a coherent photon-photon gate were analyzed in Ref.~\cite{Moreno2021}.

A photon-mediated entanglement of two spatially distant planar arrays was studied in Ref.~\cite{Guimond2019}. As each layer can behave as a superatom governed by 1D electrodynamics, the collective state of the arrays supports a nonlocal subradiant  Bell superposition state. This excitation can be understood as an analog of an antisymmetric pair of atoms in a waveguide, coupled through 1D electrodynamics. The arrays thus serve as a resource of nonlocal entanglement,  allowing for coherent and deterministic exchange of quantum information between them,  mediated by the subradiant state.

In Ref.~\cite{Rusconi2021}, $\Lambda$-three-level atoms in a bilayer lattice were studied when one of the two electronic ground levels was off-resonantly coupled to the excited level. With different resonance linewidths between the two transitions, the atom can effectively behave as a two-level system between the two electronic ground levels. It can be driven to a subradiant eigenmode outside the light cone by two noncopropagating beams, with each beam coupled to a different ground state. The nonlinear DD coupling between the layers can be utilized to prepare an entangled state shared by the two arrays, realizing a $\sqrt{{\rm iSWAP}}$ gate where the occupation states of the two arrays approximately undergo transformations:  $|00\>\rightarrow |00\>$, $|10\>\rightarrow (|10\>-i|01\>)/2^{1/2}$, $|01\>\rightarrow (|01\>-i|10\>)/2^{1/2}$, $|11\>\rightarrow |11\>$.

The wavefront engineering of transmitted light through a planar array, as discussed in Sec.~\ref{sec:wavefront}, can also be controlled by a coupled qubit, such as a Rydberg atom. An alternative example of an optical cavity was considered in Ref.~\cite{Ballantine21quantum}. In this scenario, a photon inside a cavity induces an ac Stark shift, which in turn controls the resonance of the transmitted light through the array or facilitates a photon storage process.
By shifting the resonance of an atomic Huygens' surface, it becomes possible to entangle the state of the cavity with different properties of the transmitted photon in various Huygens' surface realizations, including its polarization in birefringence or its angle of transmission.

\subsection{Atomic nanorings}
\label{sec:rings}

A different type of periodic array with a planar geometry is formed by a regular polygon (nanoring), or by multiple polygons within a plane~\cite{Asenjo-Garcia2017a,Jen2018,Moreno-Cardoner19,Needham19,Holzinger20,Moreno-Cardoner22}. The LLI excitation eigenmodes of a single nanoring exhibit distinct characteristics. For the atomic dipoles oriented perpendicular to the plane of the ring or tangentially along the ring, the eigenmodes are translationally invariant with well-defined angular momentum. The amplitudes of these modes satisfy $\rho_{ge}^{(\ell)}=\rho_{ge}\exp(i m\phi_\ell)$, where $\phi_\ell$ is the angular coordinate of the atom $\ell$ and $m$ represents integer-valued angular momentum~\cite{Moreno-Cardoner19}. For a nanoring with a fixed radius, the most super-radiant mode exhibits a linewidth that is linearly proportional to the atom number. However, there exist subradiant modes with exponentially decreasing linewidths as a function of the atom number~\cite{Asenjo-Garcia2017a}.

Energy transfer between two rings on a plane can exhibit similarities to the light-harvesting mechanism in biological systems~\cite{Moreno-Cardoner19}. Moreover, by placing an atom at the center of the ring, a steady-state coherent light source  can be achieved with a spectral linewidth narrower than that of a single isolated atom, resembling a miniature laser~\cite{Holzinger20}. The atoms within the ring behave analogously to a cavity resonator and exhibit a narrow collective linewidth with efficient coupling to the gain atom located at the center.

\section{Discussion and outlook}
\label{sec:outlook}

\subsection{Optical manipulation and quantum networks}

The similarity between atomic planar arrays and nanofabricated metasurfaces paves the way for a wide range of applications of ultraflat optics with atomic layers in the generation, manipulation, and detection of light. Metasurfaces are thin nanostructured films, composed of metallic and dielectric nanoresonators that are arranged in a regular 2D pattern with subwavelength spacing~\cite{Yu14,Chen2016_review,Luo18,Qiu21_review}. They have gained popularity as alternatives to traditional bulky classical optical elements. Metasurfaces are used for designing light-matter interactions where  they offer versatile control over the amplitude, phase, and polarization of light, functioning as antenna arrays that can tailor near-field responses and redirect light in ways that would not be possible with conventional refractive optics. 

Although the manipulation of light using atomic planar arrays is still in its early stages and presents significant technical challenges, these arrays offer several potential advantages over nanofabricated metasurfaces. Atoms are free from manufacturing imperfections and provide precise control over internal transitions. They have well-defined resonance frequencies and long coherence times. Additionally, all absorbed photons are eventually re-emitted, instead of being captured by the material and turned into heat. Importantly, reaching the quantum regime with nanofabricated metasurfaces is challenging~\cite{Solntsev21}, while quantum interfaces between atoms and light are already well-established. Atomic metasurfaces can simultaneously also act as diverse physical systems with multiple processes that have functionalities different from controlling light. It was recently proposed that atomic planar arrays coupled by light could serve as a quantum network~\cite{Grankin18}. The network would consists of many nodes, formed by individual arrays or array systems, and communication channels established by light to transfer quantum states between the nodes. Each node would be capable of performing quantum operations.

To establish a quantum network, one requirement is for quantum information to be processed and stored locally in quantum memories of quantum nodes~\cite{Kimble08,Ritter12}. Localized, individual nodes are linked by high-speed photonic quantum channels, enabling the distribution of entanglement and the transfer of quantum states with high fidelity across the network. Crucial elements are efficient quantum interfaces between light and matter that facilitate reversible mapping of quantum states between the two, coherent control over the light-matter interactions at the single-photon level, and long-lived controllable quantum memories~\cite{Kimble08}. The system should also be robust in the present of imperfections.

Atomic planar arrays offer ultrathin surfaces with significantly enhanced optical cross sections compared with a single atom in free space. In free space, the extinction of light by a single atom  is limited by the ratio of the resonant cross section to the minimum beam waist area, which is typically well below one. Strong coupling between an atom and light can be achieved in optical cavities with a small mode volume or by engineering confined modes in waveguides, fibers, or close to dielectric surfaces. However, creating optical free-space links in a network offers many advantages by avoiding challenging surface effects. Moreover, e.g., trapping atoms near optical nanofibers or in waveguides can be plagued by losses due to scattering into undesired modes. 

Atomic arrays offer highly collimated coherent light emission and light propagation that effectively follows 1D electrodynamics, efficiently linking different nodes in a network and eliminating a significant loss channel of spontaneous emission in undesired directions 
in disordered ensembles. Free-space schemes for quantum information processing in disordered atomic ensembles~\cite{HAM10}, such as  the Duan-Lukin-Cirac-Zoller (DLCZ) protocol~\cite{Duan01}, require a high optical depth $D$. As discussed in Sec.~\ref{sec:storage}, only values close to $D\sim 600$, or $10^6$-$10^7$ atoms, provide photon storage efficiencies comparable with small $4\times4$ atomic arrays~\cite{Manzoni18}.

Schemes employing subradiance provide reversible quantum memories that can be rapidly accessed (Sec.~\ref{sec:storage}). Wavefront control in atomic planar arrays (Sec.~\ref{sec:wavefront}) offers a platform for operating flat surface optics for manipulating light that resembles nanofabricated metasurfaces where the amplitude, phase, and polarization can be controlled. The operations of the arrays can be controlled by qubits and gate operations can be performed within the arrays (Sec.~\ref{sec:entangle}). 
Recent experiments combining optical excitations in arrays with Rydberg excitations~\cite{Srakaew22} show promising potential, although Rydberg states are sensitive to stray electric and magnetic fields. 

Atomic planar arrays share common features with cavity QED and nanofabricated metasurfaces, and studies in these areas may serve as a guide for potential future developments. Exciting examples related to quantum network operations in cavity QED include teleportation of atomic states~\cite{Enk98}, entanglement generation between atomic internal states and light polarization~\cite{Duan03}, and photon-photon interactions~\cite{Duan04}. 
Although artificial solid-state metasurfaces are susceptible to significant decoherence, there is considerable interest in quantum applications, e.g., in quantum sensing, ghost imaging, multiphoton and angular momentum states of light~\cite{Solntsev21}.

\subsection{Many-body phenomena}

Despite the differences in physics, atomic planar arrays in free space exhibit surprisingly similar behaviors to atoms in optical cavities, extending beyond the light-atom interfaces discussed in the previous section. In Sec.~\ref{sec:bistable}, we highlighted the similarities between the cooperativity parameters in both systems, which are determined by the recurrent scattering of light by an atom either from the neighboring atoms (in arrays) or facilitated by mirrors (in optical cavities). 
In both systems, the value of the cooperativity parameter defines a ``strongly-coupled'' regime and the emerging bistability. In planar arrays, this occurs when the atom separation is small enough for the level shift due to the DD interaction to exceed the single-atom linewidth. Dipole blockade (Sec.~\ref{sec:blockade}) by optical transitions relies on similarly strong DD coupling. Moreover, the optical response of a planar array beyond the LLI limit can exhibit many-body analogs of vacuum Rabi splitting of transmission resonances~\cite{Bettles2020}, a well-known phenomenon in cavities~\cite{Thompson92,Khitrova2006}. Additional analogies to cavities arise in lasing, where neighboring atoms act as resonators (Sec.~\ref{sec:rings}, see also Ref.~\cite{Mkhitaryan18}), and high-precision measurements and optomechanics~\cite{Shahmoon2019_opto}.
The control of light transmission by a Rydberg excitation (Sec.~\ref{sec:blockade}) can also be achieved in cavities~\cite{Vaneecloo22}.

Ultracold atoms have been actively investigated as simulators of strongly interacting quantum systems~\cite{Gross2017},  and it is anticipated that more research will explore analogous topics for interacting many-body systems of atoms and photons in arrays in the future.  The presence of long-range interactions and dissipation mediated by light, as well as the potential for designing spin-dependent couplings between the atoms, opens up avenues for studying rich many-body phenomena that are challenging to achieve in other ultracold atom systems. Phase transitions are already actively investigated for single- and multimode cavities~\cite{Domokos2002,Baumann2010,Gopalakrishnan09,Strack11}. Other examples include the study of topological phases (Sec.~\ref{sec:topo}), excitation statistics~\cite{Zhang2018}, thermalization, and magnetism and frustration~\cite{Parmee22b}.

\subsection{Experimental challenges}

Experiments on light transmission through planar atomic arrays in Refs.~\cite{Rui2020,Srakaew22} employed $^{87}$Rb atoms. 
Alkali-metal atoms, including Rb, typically possess multiple electronic ground states.
Even in the limit of LLI, the involvement of multiple ground levels can significantly complicate the dynamics, introducing, e.g., quantum entanglement~\cite{Lee16,Ritsch_subr}. This can be avoided by employing cycling transitions of maximally polarized hyperfine levels while tuning off-resonance the other transitions, e.g., by magnetic fields.

Although relatively long optical lattice spacing $a\simeq 0.68\lambda$ was needed to observe the subradiant transmission resonance narrowing~\cite{Rui2020,Srakaew22}  [see \EQREF{Analytic}], several intriguing strongly interacting phenomena necessitate significantly smaller spacing. For example, the dipole blockade (Sec.~\ref{sec:blockade}), bistability (Sec.~\ref{sec:bistable}), phase transitions~\cite{Parmee2020}, and strongly correlated effects arising from optical transitions rely on the spacing below 0.2$\lambda$. 
Alkaline-earth-metal and rare-earth-metal atoms are promising constituent elements for arrays due to a rich variety of optical transitions. They offer  great flexibility in experimental control~\cite{Daley08,Fukuhara09,Ye08} and Mott-insulator states have been observed in such systems~\cite{Fukuhara09, Stellmer12}. Atomic arrays may find applications in atomic clocks~\cite{Kramer2016,Henriet2018,Qu19} that employ narrow optical transitions in Sr~\cite{Bothwell19}.
In bosonic isotopes of Sr and Yb, the nuclear spin vanishes, making them prototype models for spatially isotropic $J=0\rightarrow J^\prime=1$ transitions that 
enables simple 
classical simulation models for light propagation in the limit of LLI. However, the utilization of extremely narrow transitions may present challenges due to the influence of atomic recoil or longer time scales in which dipolar forces can take effect, unless the atoms are tightly confined. Additionally, the recoil effects in a lattice can be collective, further complicating the situation~\cite{Robicheaux19}. 

In Ref.~\cite{Olmos13} (see also Ref.~\cite{Ballantine22str}), a method was proposed to create an optical lattice using Sr with especially short lattice spacing $a\simeq0.08\lambda$, resulting in strong DD interactions.  The  low-lying metastable triplet state $5s5p\,^3P_0$ is coupled to $5s4d\,^3D_1$ via resonance wavelength $\lambda\simeq 2.6\,\mu$m, providing a $J=0\rightarrow J^\prime=1$ transition with the linewidth $\gamma\simeq 1.45\times 10^{5} {\rm s}^{-1}$~\cite{Zhou10,Werij92}. Additional resonances~\cite{Olmos13} are utilized to trap the atoms at a magic wavelength $\lambda\sim 415\,$nm.
Furthermore, $5s4d\,^3D_1$ state can be off-resonantly coupled to $5s6p\,^3P_j$ states, enabling  a wide range of light-induced level shifts through the ac Stark effect, e.g., around $\lambda\simeq636$nm, which facilitates engineering the collective responses of the array~\cite{Ballantine22str}.

Yb atoms exhibit a telecom resonance  transition $\lambda\simeq 1.4\,\mu$m from the metastable $6s6p\,^3P_0$ state to $5d6s\,^3D_1$, with $\gamma\simeq 1.0\times 10^{6} {\rm s}^{-1}$~\cite{Beloy12} and a large branching ratio. This transition was analyzed in Ref.~\cite{Covey19} for a fermionic $^{171}$Yb isotope with the nuclear spin $I=1/2$, as part of a spin-photon entanglement scheme. Two such fermionic spin states confined in each lattice site can also support dark states due to the Fermi blocking~\cite{Orioli19}. The system has magic wavelengths of 473nm and 532nm~\cite{Beloy12,Yamamoto2016,Covey19}, providing short lattice spacings $a/\lambda\simeq0.17$ and 0.19, respectively. 

In atomic arrays created using optical tweezers~\cite{Kim16,Endres16,Barredo18,Cooper18,Saskin19,Schymik22}, achieving spacing even below $\lambda$ presents a challenge. The proposal of Ref.~\cite{Covey19} considers Yb with the $\lambda\simeq 1.4\,\mu$m  transition and an optical tweeezer wavelength of 470nm, resulting in subwavelength array spacing and tight atom confinement.

In the experiments~\cite{Rui2020,Srakaew22}, the atoms were confined in an optical lattice with the depths of 300$E_r$ and 100$E_r$, respectively (Sec.~\ref{sec:latticepot}). More than an order-of-magnitude deeper potentials would be achievable;
however, increasing the lattice height beyond 300$E_r$ led to a deterioration of subradiant narrowing in transmission due to motional spreading of atoms in the antitrapped electronically excited state  in the lattice potential. Position fluctuations of the atoms can be reduced in deeper lattices that operate at a magic wavelength to  limit the spread of the atomic wavefunction.
However, improving the confinement significantly in the presence of a sinusoidal optical lattice potential generated by a standing-wave laser is not straightforward, as the ratio of the
$1/e$ width of the confinement to the lattice spacing scales very slowly with the lattice potential height $\ell/a = s^{-1/4} / \pi$~\cite{Morsch06} (see Sec.~\ref{sec:latticepot}). 
There has been a recent surge of interest in developing alternative periodic trapping schemes aimed at altering the potential shape and reducing the effective lattice spacing~\cite{Wang18,Bienias20,Tsui20,Anderson20,Kubala21}. In the optical response experiments, small spacing and tightening the atom confinement for the electronic ground and excited levels are crucial parameters, while the trap lifetime can often be much shorter than in typical ultracold atom experiments.

\begin{acknowledgments}
We acknowledge financial support from the EPSRC (Grant No.\ EP/S002952/1) and discussions with J.\ Javanainen, K.\ Ballantine, C.\ Parmee, L.\ Williamson, L.\ Ruks,  D.\ Wilkowski, M.\ Borgh, R.\ Holzinger, H.\ Carmichael, C.\ Gross, I.\ Bloch, J.\ Zeiher, S.\ Yelin, and O.\ Rubies-Bigorda.
\end{acknowledgments}

\appendix

\section{Derivation of 1D propagation from a planar array}\label{appen1}

A large planar array of atoms with a phase-uniform excitation exhibits behavior akin to 1D electrodynamics for light.  When the atoms are spaced at a subwavelength distance, the array functions as a diffraction grating with only the zeroth-order Bragg peak present, representing coherent forward and back scattering.
Here we demonstrate a formal derivation of effective 1D electrodynamics for such systems in a physically intuitive way by directly analyzing the spatial distribution of the scattered light.  The techniques introduced in Ref.~\cite{Sargent_laserphys} are employed
to integrate the response of the array of $N$ atoms in the continuum limit. Related spatial integrals for coherent light transmission, without formal derivations, can also be found in
Refs.~\cite{dalibardexp,Javanainen17,Facchinetti18,Javanainen19}.
Let us consider a dipole located in the $yz$ plane when the plane is centered at the origin, and an observation point at $y=0$, $z=0$, $|x|>0$ (where $x<0$ corresponds to reflection and $x>0$ transmission).  In the argument, we consider each dipole to be polarized in the $y$ direction, $\textbf{d}_{ge}^{(\nu)} = {\cal D}\, \hat{\bf e}_y$. 
We assume an approximately uniform excitation over the entire lattice and set $\langle \hat{\sigma}_{j\nu}^-\rangle =\langle \hat{\sigma}^-\rangle $, so that
the scattered light amplitude from Eq.~\eqref{eq:scattlight} simplifies to
\begin{equation} \label{eq:scattlightplanar}
\epsilon_0 \langle \hat{\textbf{E}}{}^+_s(\textbf{r}) \rangle =\sum_{j\nu}\mathsf{G}(\textbf{r}-\textbf{r}_j)\textbf{d}_{ge}^{(\nu)} \langle \hat{\sigma}_{j\nu}^-\rangle = {\cal D}\langle \hat{\sigma}^-\rangle\sum_{j}\mathsf{G}(\textbf{R}_j)\hat{\bf e}_y  ,
\end{equation}
where $\textbf{R}_j = x\,\hat{\bf e}_x\! -y_j \,\hat{\bf e}_y\! -z_j\, \hat{\bf e}_z$ defines the vector joining the $j$th atom and the observation point. Next, we make the assumption that the observation point is sufficiently far from the atomic array and that the lattice is much larger than the distance to the observation point $\lambda\alt |x|\ll \sqrt{{\cal A}}$  (numerically it can be shown that it is sufficient to have $|x|\agt 0.5\lambda$ when the spacing $a\alt 0.7\lambda$~\cite{Javanainen19}), where ${\cal A}$ is the total area of the array. This allows us to neglect the discreteness of the atoms and we
replace the summation by integration over the lattice plane defining coordinates $y$ and $z$. When working out the explicit expression for $\mathsf{G}(\textbf{R}_j)\hat{\bf e}_y $ using the expanded form of the radiation kernel [\EQREF{Gdef}], we can immediately discard terms that are odd in $y$ since they cancel out in the integration. This simplifies the expression to $\mathsf{G}_{y,{\rm even}}(\textbf{R}_j)\hat{\bf e}_y $. 
Transforming the atomic coordinates $(y,z)$ in the lattice to the polar coordinates $(\rho \cos\phi, \rho\sin\phi)$ then yields
\begin{align}
\epsilon_0 \langle \hat{\bE}{}^+_s(x\,\hat{\bf e}_x) \rangle&=
 \frac{{\cal D} \rho_{ge} }{{\cal A}'}\!\!\int_0^{\rho_0}\!\!\! \rho\, d\rho\int_{-\pi}^{\pi}\!\!\!d\phi\,{\sf G}_{y,{\rm even}}(\textbf{R} ) \,\hat{\bf e}_y,
\end{align}
where we have written $ \rho_{ge}=\langle \hat{\sigma}^-\rangle$, ${\cal A}'={\cal A}/N$ the area per atomic dipole, $\textbf{R} = \textbf{R} (\rho,\phi)= x\,\hat{\bf e}_x\! -\rho \cos\phi \,\hat{\bf e}_y\! - \rho\sin\phi\, \hat{\bf e}_z$, and where we take the limit $\rho_0\rightarrow\infty$ of an infinite lattice size.
We find the explicit expression 
\begin{align}
\int_{-\pi}^{\pi} &\!\!\!d\phi\,{\sf G}_{y,{\rm even}}(\textbf{R} ) \,\hat{\bf e}_y \nonumber\\
&= \frac{\hat{\bf e}_y k^2}{4 R} e^{ik R} \left[ \frac{\rho^2+2x^2}{R^2} + \(\frac{1}{k^2R^2}-\frac{i}{kR}\)\(\frac{3\rho^2}{R^2}-2\)\right].
\end{align}
Next, the integral over $\rho$ can be performed using the techniques introduced in Ref.~\cite{Sargent_laserphys}. We make the substitution $R=\sqrt{x^2+\rho^2}$, with $\rho d\rho= R dR$. Then
\begin{align}\label{eq:recurrel}
\epsilon_0 \langle \hat{\bE}{}^+_s(x\,\hat{\bf e}_x) \rangle=
 \frac{\hat{\bf e}_y  {\cal D} \rho_{ge} }{{4\cal A}'} &(F_2-ikF_1-3x^2F_4+3ikx^2F_3\nonumber\\
 &+k^2F_0+k^2x^2 F_2),
\end{align}
where we have introduced the functions $F_n$, for integer $n$,
\beq
F_n= \int_{|x|}^\infty dR \frac{e^{ikR}}{R^n}.
\eeq
With the help of a convergence factor that ensures that the fields vanish for $R\rightarrow\infty$, the first term can be integrated straightforwardly and yields $F_0=i \exp{(ik|x|)} /k$. The remaining terms are obtained
by deriving a recursion formula
\beq \label{eq:recur}
F_n= \[ \frac{e^{ikR}}{ik R^n}\]_{|x|}^\infty +\frac{n}{ik} \int_{|x|}^\infty dR \frac{e^{ikR}}{R^{n+1}} = \frac{1}{ik} \big(n F_{n+1} - \frac{e^{ik|x|}}{|x|^n}\big).
\eeq
Substituting first $F_1$ and $F_2$ in Eq.~\eqref{eq:recurrel}, using Eq.~\eqref{eq:recur}, gives
\beq \label{eq:recurrel2}
\epsilon_0 \langle \hat{\bE}{}^+_s(x\,\hat{\bf e}_x) \rangle=
 \frac{\hat{\bf e}_y  {\cal D} \rho_{ge} }{{4\cal A}'} 
\big( \frac{e^{ik|x|}}{|x|} -3 x^2 F_4 +ikx^2F_3+2ik e^{ik|x|} \big).
\eeq
Expressing then all the integrals in Eq.~\eqref{eq:recurrel2} in terms of $F_4$ cancels out all the unknown terms and we are left with
\begin{align}\label{eq:1dfirst}
\epsilon_0 \langle \hat{\bE}{}^+_s(x\,\hat{\bf e}_x) \rangle=
\hat{\bf e}_y {\cal D}\rho_{ge} \frac{  ik  }{2{\cal A}'} e^{ik|x|}
\end{align}
The derivation works identically for a dipole oriented along the $z$ axis, so for any dipoles in the $yz$ plane of the array we obtain an effective 1D propagation of light along the $x$ axis. 

We can then write the field component along the $y$ axis as a scalar field and express Eq.~\eqref{eq:1dfirst} as 
\begin{align}
\epsilon_0 \langle \hat{E}{}^+_s(x) \rangle= G(x)
 \bar{\cal D} \rho_{ge},\quad G(x)= \frac{  ik  }{2} e^{ik|x|}, \quad  \bar{\cal D} = \frac{{\cal D}}{{\cal A}'},
\end{align}
where $G(x)$ is a 1D scalar dipole radiation kernel~\cite{BOR99}, $\bar{\cal D}$ is the density of atomic dipoles in the plane, and ${\cal A}'$ is the previously introduced area of a single lattice unit cell.

We can now provide a straightforward derivation of the collective resonance linewidth for the excitation eigenmode that exhibits a uniform phase profile. Since
Eq.~\eqref{eq:1dfirst} captures all the emitted radiation by the eigenmode, we can directly apply Eq.~\eqref{eq:omga} for Eq.~\eqref{eq:1dfirst} 
\begin{align}
\label{eq:planarwidth1}
\gamma_{\rm 1D} &=\gamma  +  \sum_{j\ell (j\neq\ell)} \gamma^{(j\ell)}_{\nu\nu}  =  \xi \sum_{j\ell} {\rm Im}  \[ \radKernel^{(j\ell)}_{\nu\nu} \] \nonumber\\
& = \xi\lim_{x\rightarrow0}  {\rm Im}  \[ G(x) \]  = \frac{k{\cal D}^2}{2{\cal A}' \hbar \eo}.
\end{align}
In the square lattice of lattice constant $a$, we have ${\cal A}'=a^2$. We then have in terms of the single-atom linewidth $\gamma$ [Eq.~\eqref{eq:WW}]
\beq
\label{eq:planarwidth2}
\gamma_{\rm 1D} = \gamma +\tilde\gamma  =  \frac{3\pi\gamma}{k^2 a^2}.
\eeq

Similarly, we consider the equations of motion for the atomic polarization density amplitudes from a uniformly excited atomic array in the limit of LLI [see Eq.~\eqref{eq:Peoms}]: 
\beq\label{eq:1dPeoms}
\dot \rho_{ge}
  =  \left( i \Delta_{\rm 1D}- \gamma_{\rm 1D} \right)
 \rho_{ge} + i {\cal R}_{{\rm ext},y}.
\eeq
 Here $\rho_{ge}$ represents the uniform excitation of an array and ${\cal R}_{{\rm ext},y}$ is the $y$-component of the Rabi frequency of the external light impinging on the array. For a single isolated array, 
 ${\cal R}_{{\rm ext},y}$ represents the
 incident light. However, for a number of stacked, parallel planar arrays, we also include the scattering from all other arrays. 
 This can be calculated similarly to the previous example of the scattered light, with the exception that now the coordinate $\textbf{R}^{(\ell)}_j$ refers to the $j$th atom
 of the $\ell$th array. We write the uniform excitation of the polarization amplitude $\rho_{ge}$ of the $\ell$th array as $\varrho^{(\ell)}_{ge}$, such that
 \begin{align}
 i{\cal R}_{{\rm ext},y}&=  i{\cal R}_{{\rm in},y} +  i \xi  \sum_{j,\ell} \hat{\bf e}_y \cdot\mathsf{G}(\textbf{R}^{(\ell)}_j) \hat{\bf e}_y \varrho^{(\ell)}_{ge} 
\nonumber\\
&=i{\cal R}_{{\rm in},y}  -\gamma_{\rm 1D} \sum_{\ell (x\neq0)} e^{ik|x_\ell|}  \varrho^{(\ell)}_{ge} ,
\end{align}
where we have explicitly highlighted in the subscript that ${\cal R}_{{\rm in}}$ denotes the Rabi frequency of the incident light.
Dropping the subscript $y$, we can then express Eq.~\eqref{eq:1dPeoms} as a compact, coupled set of equations for stacked planar arrays
\beq
\frac{d}{dt} \varrho_{ge}^{(j)}  =  (i \Delta_{\rm 1D}-\gamma_{\rm 1D}) \varrho_{ge}^{(j)} + i{\cal R}_{{\rm in}}(x_j) -\gamma_{\rm 1D}  \sum_{\ell\neq j} e^{ik |x_j-x_\ell |}\, \varrho_{ge}^{(\ell)}.
\label{TIMEDEPEQ}
\eeq

\section{Calculation of photon scattering rate}\label{appen2}

We calculate the total photon scattering rate, introduced in Eq.~\eqref{eq:scatrateformula}. The following treatment extends the two-level analysis of Ref.~\cite{carmichael2000}  for the $J=0\rightarrow J'=1$ transition. For different atom operator expectation
values, the result also provides the rates for coherently and incoherently scattered light.

The photon scattering rate over a surface $S$ can be obtained by integrating the scattered intensity per the photon energy
\begin{align} \label{eq:emissionrateapp}
n_s &= \frac{1}{\hbar\omega} \int_{S} dS I_s = \frac{ 2\eo c}{\hbar\omega} \int_{S} dS \<  \hat{\textbf{E}}{}^-_s(\textbf{r}) \cdot\hat{\textbf{E}}{}^+_s(\textbf{r}) \rangle\nonumber\\
&=  \sum_{j\nu} \Gamma^{(jj)}_{\nu\nu}  \langle \hat{\sigma}_{j\nu}^+ \hat{\sigma}_{j\nu}^-\rangle + \sum_{j\ell\nu\mu (j\neq\ell)} \Gamma^{(j\ell)}_{\nu\mu}  \langle \hat{\sigma}_{j\nu}^+ \hat{\sigma}_{\ell\mu}^-\rangle, 
\end{align}
where the result depends on the integrals $\Gamma^{(j\ell)}_{\nu\mu}$. Substituting  Eq.~\eqref{eq:scattlight} yields
\begin{align}\label{intG}
\Gamma^{(j\ell)}_{\nu\mu}  =\frac{2c}{\hbar\epsilon_0\omega}\int_S dS \left[\mathsf{G}(\mathbf{r}-\mathbf{r}_j) \textbf{d}_{ge}^{(\nu)} \right]^*\mathsf{G}(\mathbf{r}-\mathbf{r}_\ell) \textbf{d}_{ge}^{(\mu)}.
\end{align}
We evaluate the integral sufficiently far away from the atoms $r\gg \lambda$ in the far-field radiation zone using Eq.~\eqref{eq:rzone}.
To calculate the total scattering rate, we assume that the integrated surface completely encloses the atoms. 
In the spherical coordinates $(\theta,\phi)$ (we consider a spherical surface with the solid angle of $4\pi$), this gives
\begin{align}\label{Imnint}
\Gamma^{(j\ell)}_{\nu\mu} =\frac{3\gamma}{4\pi}\int_{-1}^{1} d(\cos\theta) \int_0^{2\pi}d\phi \left[\delta_{\nu,\mu}-(\hat{\mathbf{r}}\cdot \pol^*_\nu ) (\hat{\mathbf{r}}\cdot \pol_\mu ) \right]e^{ik\hat{\mathbf{r}}\cdot\mathbf{r}_{j\ell}} .
\end{align}
We choose the unit vector  $\hat{\mathbf{r}}_{j\ell}={\bf r}_{j\ell}/r_{j\ell}=(\mathbf{r}_{\ell}-\mathbf{r}_{j})/|\mathbf{r}_{\ell}-\mathbf{r}_{j}| $ to be along the $z$ axis ($\hat{\mathbf{r}}_{j\ell} = \hat{\mathbf{z}}$) and substitute
$\hat{\mathbf{r}} = \hat{\mathbf{x}}\cos\phi\sin\theta+\hat{\mathbf{y}}\sin\phi\sin\theta+\hat{\mathbf{z}}\cos\theta$. The integration over $\phi$ is then straightforward. For $j=\ell$, we obtain (the terms  $\Gamma^{(j j)}_{\nu\mu} $, with $\nu\neq\mu$, vanish)
\beq
\Gamma^{(j j)}_{\nu\nu} =2\gamma,
\eeq
and for  $j\neq \ell$,
\begin{align}
\Gamma^{(j\ell)}_{\nu\mu} &=\frac{3\gamma}{8}  \int_{-1}^1 d(\cos\theta) \left\{3\delta_{\nu,\mu}-(\hat{\mathbf{r}}_{j\ell}\cdot\pol^*_\nu)(\hat{\mathbf{r}}_{j\ell}\cdot\pol_\mu)\right. 
 \nonumber\\
& \left.  -[\delta_{\nu,\mu} -3 (\hat{\mathbf{r}}_{j\ell}\cdot\pol^*_\nu)(\hat{\mathbf{r}}_{j\ell}\cdot\pol_\mu)]\cos2\theta\right\} e^{ikr_{j\ell}\cos\theta} \nonumber\\
&=3\gamma[\delta_{\nu,\mu}-(\hat{\mathbf{r}}_{j\ell}\cdot\pol^*_\nu)(\hat{\mathbf{r}}_{j\ell}\cdot\pol_\mu)] \frac{\sin kr_{j\ell}}{kr_{j\ell}}
 \nonumber\\
&  +3\gamma[\delta_{\nu,\mu}-3 (\hat{\mathbf{r}}_{j\ell}\cdot\pol^*_\nu)(\hat{\mathbf{r}}_{j\ell}\cdot\pol_\mu)]\left(\frac{\cos kr_{j\ell}}{k^2r_{j\ell}^2}-\frac{\sin kr_{j\ell}}{k^3r_{j\ell}^3}\right)\nonumber\\
&=2\gamma^{(j\ell)}_{\nu\mu}= 2 \xi {\rm Im} \[\radKernel^{(j\ell)}_{\nu\mu} \],
\end{align}
where we have used Eq.~\eqref{eq:omga}.
The total photon scattering rate is therefore given by
\beq
n_s =   2\gamma  \sum_{j\nu} \langle \hat{\sigma}_{j\nu}^+ \hat{\sigma}_{j\nu}^-\rangle  + 2  \sum_{j\ell\nu\mu (j\neq\ell)} \gamma^{(j\ell)}_{\nu\mu} \langle \hat{\sigma}_{j\nu}^+ \hat{\sigma}_{\ell\mu}^-\rangle .
\eeq
The result is consistent with the single-atom and collective decay terms in the QME~\eqref{eq:rhoeom}.

\section{Incoherent scattering with position disorder}\label{appen3}

For fluctuating atomic positions, Eq.~\eqref{eq:EdEdExpectation} represents an ensemble average calculated over many realizations. We would like to express Eq.~\eqref{eq:EdEdExpectation} in terms of the solutions of the correlation functions in the coupled dynamics between the light and atoms for each stochastic run, Eq.~\eqref{eq:padbcExp}, and
\begin{align} 
\label{eq:pabExp}
\left\langle \mathop{\hat{\psi}^{\dag}_{e\nu}(\mathbf{r},t)} \mathop{\hat{\psi}_{e\mu}(\mathbf{r},t)} \right\rangle_{\{\mathbf{r}_1,\dots,\mathbf{r}_N\}}
&  = 
\sum_j \< \hat{\sigma}_{j\nu}^{+}(t)\hat{\sigma}_{j\mu}^{-} (t)\>  \mathop{\delta(\mathbf{r}-\mathbf{r}_j)}.
\end{align}
We rearrange the terms by placing the atomic operators in the normal order. This yields for the correlation function on the right hand side of Eq.~\eqref{eq:EdEdExpectation} (for both fermionic and bosonic atoms)
\begin{align} 
\label{eq:normalOrder}
\left\langle  
\mathop{\hat{\psi}_{e\nu}^{\dag}(\rAtomBold)}
\mathop{\hat{\psi}_g(\rAtomBold)}
\mathop{\hat{\psi}_g^{\dag}(\rAtomBold')} 
\mathop{\hat{\psi}_{e\mu}(\rAtomBold')} \right\rangle
&= 
\left\langle  \mathop{\hat{\psi}_{e\nu}^{\dag}(\rAtomBold)}
\mathop{\hat{\psi}_{e\mu}(\rAtomBold')}\right\rangle 
\mathop{\delta(\rAtomBold-\rAtomBold')} \nonumber\\
& \!\!\!  \!\!\!  \!\!\!  \!\!\!  + 
\left\langle  
\mathop{\hat{\psi}_{e\nu}^{\dag}(\rAtomBold)} 
\mathop{\hat{\psi}_g^{\dag}(\rAtomBold')} 
\mathop{\hat{\psi}_{e\mu}(\rAtomBold')} 
\mathop{\hat{\psi}_g(\rAtomBold)} \right\rangle.
\end{align}
Substituting this into Eq.~\eqref{eq:EdEdExpectation} and expressing the correlation functions in terms of single stochastic run values according to Eq.~\eqref{eq:sampling2}, integrated over all trajectories that represents the ensemble-averaging procedure, we obtain~\cite{Bettles2020}
\begin{widetext}
\begin{align} 
\label{eq:scatteredFieldflucpos}
&2c\eo \left\langle \mathop{\hat{\mathbf{E}}_s^-(\mathbf{r}) }\mathop{\hat{\mathbf{E}}_s^+(\mathbf{r}) }\right\rangle
= 
\frac{2c}{\epsilon_0} \int \mathop{\mathrm{d}^3 \rAtom}  \sum_{\nu\mu}\left\{
\left[ \mathop{\mathsf{G}(\mathbf{r}-\rAtomBold)} \mathbf{d}^{(\nu)}_{ge}\right]^{*}
\left[ \mathop{\mathsf{G}(\mathbf{r}-\rAtomBold)} \mathbf{d}^{(\mu)}_{ge} \right]  \int \mathop{\mathrm{d}^3 r_1} \ldots \mathop{\mathrm{d}^3 r_N}   
\left\langle  \mathop{\hat{\psi}_{e\nu}^{\dag}(\rAtomBold)}  
\mathop{\hat{\psi}_{e\mu}(\rAtomBold)}\right\rangle_{\{\mathbf{r}_1,\dots,\mathbf{r}_N\}}  
 \mathop{P(\mathbf{r}_1,\dots,\mathbf{r}_N)} \right\}
\nonumber\\
&+  \frac{2c}{\epsilon_0} \int' \mathop{\mathrm{d}^3 \rAtom} 
\mathop{\mathrm{d}^3 \rAtom'} \sum_{\nu\mu}\left\{
\left[ \mathop{\mathsf{G}(\mathbf{r}-\rAtomBold)} \mathbf{d}^{(\nu)}_{ge}\right]^{*}
\left[ \mathop{\mathsf{G}(\mathbf{r}-\rAtomBold')} \mathbf{d}^{(\mu)}_{ge} \right]
 \int \mathop{\mathrm{d}^3 r_1} \ldots \mathop{\mathrm{d}^3 r_N}   
 \left\langle  
\mathop{\hat{\psi}_{e\nu}^{\dag}(\rAtomBold)} 
\mathop{\hat{\psi}_g^{\dag}(\rAtomBold')} 
\mathop{\hat{\psi}_{e\mu}(\rAtomBold')} 
\mathop{\hat{\psi}_g(\rAtomBold)} \right\rangle_{\{\mathbf{r}_1,\dots,\mathbf{r}_N\}}  
 \mathop{P(\mathbf{r}_1,\dots,\mathbf{r}_N)}  \right\},
\end{align}
\end{widetext}
where $\int'$ denotes a double integral over all $( \rAtom, \rAtom')$ excluding $ \rAtom= \rAtom'$.  The single-run expectation value of the atomic solution
$\left\langle 
\mathop{\hat{\psi}^{\dag}_{e\nu}(\rAtomBold)} 
\mathop{\hat{\psi}^{\dag}_g(\rAtomBold')} 
\mathop{\hat{\psi}_{e\mu}(\rAtomBold')} 
\mathop{\hat{\psi}_g(\rAtomBold)} \right\rangle_{\{\mathbf{r}_1,\dots,\mathbf{r}_N\}}$ is defined in terms of the atomic operator expectation values $ \< \hat{\sigma}_{j\nu}^{+}\hat{\sigma}_{\ell\mu}^{-} \>$  $(j\neq\ell)$ in Eq.~\eqref{eq:padbcExp} and 
$\left\langle \mathop{\hat{\psi}^{\dag}_{e\nu}(\mathbf{r},t)} \mathop{\hat{\psi}_{e\mu}(\mathbf{r},t)} \right\rangle_{\{\mathbf{r}_1,\dots,\mathbf{r}_N\}}$ in terms of 
$\< \hat{\sigma}_{j\nu}^{+}\hat{\sigma}_{j\mu}^{-} \>$ in \EQREF{eq:pabExp}.
The scattered intensity in Eq.~\eqref{eq:scatteredFieldflucpos} then depends on the correlations $\< \hat{\sigma}_{j\nu}^{+}\hat{\sigma}_{j\mu}^{-} \>$ and $ \< \hat{\sigma}_{j\nu}^{+}\hat{\sigma}_{\ell\mu}^{-} \>$ of each stochastic realization of fixed atomic positions  $\{\rv_1,\rv_2, \ldots, \rv_N\}$ through  Eqs.~\eqref{eq:padbcExp} and~\eqref{eq:pabExp}, and 
ensemble-averaging over stochastic realizations of atomic positions. The quantum solutions of the correlations can be obtained from  the QME~\eqref{eq:rhoeom}.

The effect of position fluctuations on the incoherently scattered light becomes obvious in Eq.~\eqref{eq:scatteredFieldflucpos} when we consider nonvanishing spatial correlations
in Eq.~\eqref{eq:2bodycorrelations}. Independently, whether we have the full quantum solution or use the semiclassical approximation, the second term on the right-hand side of 
Eq.~\eqref{eq:scatteredFieldflucpos} can substantially differ from the corresponding coherent contribution
\begin{widetext}
\beq
\frac{2c}{\epsilon_0} 
 \int \mathop{\mathrm{d}^3 r_1} \ldots \mathop{\mathrm{d}^3 r_N}   
\left| \int \mathop{\mathrm{d}^3 \rAtom} \sum_{\nu}
\left[ \mathop{\mathsf{G}(\mathbf{r}-\rAtomBold)} \mathbf{d}^{(\nu)}_{ge}\right]
 \left\langle  
\mathop{\hat{\psi}_{e\nu}^{\dag}(\rAtomBold)} 
\mathop{\hat{\psi}_g(\rAtomBold)} \right\rangle_{\{\mathbf{r}_1,\dots,\mathbf{r}_N\}}  \right|^2
 \mathop{P(\mathbf{r}_1,\dots,\mathbf{r}_N)},
\eeq
with the difference between the two indicating incoherently scattered light.
\end{widetext}

\section{Expansion of spherical waves in a plane wave basis}\label{appen4}

Here we employ the standard optics techniques of expressing a spherical wave from the dipole radiation near a planar surface as a plane wave expansion, see, e.g. Refs.~\cite{Belov05,Novotny_nanooptics,Greffet_nanophotonics,Shahmoon,Asenjo-Garcia2017a,Bassler23}. The results are used in calculating nonuniform excitations in the array Sec.~\ref{sec:nonuniform} and for nonnormal light incidence in Sec.~\ref{sec:arbitrary}.
The procedure is analogous to connecting Huygens' principle to Kirchhoff-Sommerfeld formulation of diffraction.
By also keeping track of the contact interaction terms, we can express the dipole radiation kernel in \EQREF{eq:GDF} as
\begin{align}
{\sf G}_{\nu\mu}({\bf r}) &=
\left[ {\partial\over\partial r_\nu}{\partial\over\partial r_\mu} +
\delta_{\nu\mu} k^2\right] {e^{ikr}\over4\pi r}\nonumber\\
&= \frac{i}{8\pi^2}  \int {d^2\qv_\parallel}\left[ {\partial\over\partial r_\nu}{\partial\over\partial r_\mu} +
\delta_{\nu\mu} k^2\right]\frac{1}{k_\perp} e^{i \qv_\parallel\cdot\rv} e^{i k_\perp |x|}
\label{eq:GDF2}
\end{align}
where $k_\perp=(k^2-q_\parallel^2)^{1/2}$, $\qv_\parallel = (q_y,q_z)$ defines the wavevector on the array plane, and in the second line we have used the Weyl identity~\cite{Greffet_nanophotonics}. A comparison with \EQREF{eq:2dfourier} shows that the 2D Fourier transform of the dipole radiation kernel reads
\beq
\label{eq:2dimefourier}
\tilde{\sf G}^{\parallel}_{\nu\mu}(\qv) =  \frac{i}{2} ( -q_\nu q_\mu +
\delta_{\nu\mu} k^2 ) \frac{1}{k_\perp}  e^{i k_\perp |x|},
\eeq
where $\qv=[{\rm sgn}(x) k_\perp,\qv_\parallel]$. Substituting this to the Poisson summation formula in Sec.~\ref{sec:band} [2D analogy of \EQREF{eq:transform}] results in replacing the summation over  the lattice sites by the summation over the reciprocal-lattice vectors ${\bf g}_j$
\begin{align} \label{eq:transformx}
&\sum_{\ell\neq j} {\sf G}_{\nu\mu}(\rv_{j\ell})  e^{i\textbf{q}\cdot\textbf{r}_{j\ell}} =  \frac{1}{{\cal A}'}\sum_j \tilde{\sf G}^{\parallel}_{\nu\mu}(\qv+{\bf g}_j) -   {\sf G}_{\nu\mu}(0) \nonumber\\
& =  \frac{i}{2{\cal A}'} \sum_j [- (q_\nu+g_{j\nu}) (q_\mu+g_{j\mu})  +
\delta_{\nu\mu} k^2 ]   \frac{1}{k_\perp({\bf g}_j)} e^{i  k_\perp({\bf g}_j) |x|}\nonumber\\
&\;\;\;\; -   {\sf G}_{\nu\mu}(0) ,
\end{align}
where we have now incorporated ${\bf g}_j$ in $k_\perp$
\beq
k_\perp({\bf g}_j) = { \sqrt{k^2-(q_{y}+g_{jy})^2-(q_z+g_{jz})^2}}.
\eeq
The sums can be performed by using the momentum-regularized $\tilde{\sf G}^{\parallel *}_{\nu\mu}(\pv)= \tilde{\sf G}^\parallel_{\nu\mu}(\pv)  \exp(-p^2 \eta^2/4) $ in the place of $\tilde{\sf G}^{\parallel }_{\nu\mu}(\pv)$ in \EQREF{eq:transformx}, as discussed in Sec.~\ref{sec:band}.

The collective linewidths and line shifts can be evaluated by setting $x=0$ in \EQREF{eq:transformx}
\begin{subequations} \label{eq:ananonuni}
\begin{align}
\Omega_{\nu\mu}^{(j\ell)} (\qv)& = {\rm Re}\big[ \frac{{\cal D}^2}{\hbar\eo{\cal A}'}\sum_j \tilde{\sf G}^{\parallel}_{\nu\mu}(\qv+{\bf g}_j) -    \frac{{\cal D}^2}{\hbar\eo}{\sf G}_{\nu\mu}(0) \big], \label{eq:ananonuniwidth} \\
\gamma_{\nu\mu}^{(j\ell)}(\qv)  & = {\rm Im} \big[  \frac{{\cal D}^2}{\hbar\eo{\cal A}'}\sum_j \tilde{\sf G}^{\parallel}_{\nu\mu}(\qv+{\bf g}_j) -   \frac{{\cal D}^2}{\hbar\eo} {\sf G}_{\nu\mu}(0) \big]. \label{eq:ananonunishift}
\end{align}
\end{subequations}

For radiation normal to the lattice plane, there only exists the zeroth-order Bragg peak with ${\bf g}=0$ when $a<\lambda$, because for nonzero reciprocal-lattice vectors $k_\perp({\bf g}_j)$ is imaginary representing evanescent fields. For light radiation along the lattice plane $\kv=\kv_\parallel$, we have $\qv_\parallel=\kv$ and again the only contribution to propagating waves is ${\bf g}=0$ whenever $a<\lambda/2$. 

The light scattered from the planar array from the excitation with the wavevector $\qv_\parallel$ and the dipole ${\cal D}\pol_\mu$  is given by
\begin{equation} \label{eq:scattexit}
\epsilon_0  \langle \hat{\textbf{E}}{}^+_s(\textbf{r}) \rangle  = {\cal D} \sum_{\ell} e^{i \qv_\parallel\cdot \textbf{r}_\ell}   \mathsf{G}(\textbf{r}-\textbf{r}_\ell)\pol_\mu \,\rho_{ge\mu} .
\end{equation}
Analogously to the earlier derivation we obtain 
\begin{align} \label{eq:transformx2}
 \sum_{\ell} e^{i \qv_\parallel\cdot \textbf{r}_\ell}   \mathsf{G}(\textbf{r}-\textbf{r}_\ell) =  \frac{i}{2{\cal A}'} \sum_j & [- (q_\nu+g_{j\nu}) (q_\mu+g_{j\mu})  +
\delta_{\nu\mu} k^2 ]  \nonumber\\
&  \times  \frac{1}{k_\perp({\bf g}_j)} e^{i \qv_\parallel\cdot\rv} e^{i  k_\perp({\bf g}_j) |x|}.
\end{align}
For ${\bf g}=0$, an intuitive description can be formulated in terms of the light beam propagation direction when we express $k_\perp=(k^2-q_\parallel^2)^{1/2}=k\cos\theta$, where $\kv = \cos\theta\, \pol_x+ \sin\theta\cos\phi\, \pol_y+\sin\theta\sin\phi\,\pol_z$. The term
$  -q_\mu q_\nu/k^2 +\delta_{\mu\nu} = - \qv\times(\qv\times \mathds{1})/k^2$  in Eqs.~\eqref{eq:2dimefourier} and~\eqref{eq:transformx} describes projection of light polarization to the direction perpendicular to the light propagation and we obtain [compare with \EQREF{eq:projectionpro}, evaluated at $y=z=0$]
\beq
 \sum_{\ell} e^{i \qv_\parallel\cdot \textbf{r}_\ell}   \mathsf{G}(\textbf{r}-\textbf{r}_\ell) =  \frac{ik}{2{\cal A}'\cos\theta} {\sf P}_\perp(\kv) e^{i \qv_\parallel\cdot\rv}  e^{i k_\perp |x|}  .
\eeq

For ${\bf g}=\qv=0$,
we obtain again for the normal incidence the effective 1D electrodynamics [compare this with the spatially-integrated \EQREF{eq:1dfirst}] with
\beq
\sum_{\ell } {\sf G}_{\nu\mu}(\rv_{j\ell})  e^{i\textbf{q}\cdot\textbf{r}_{j\ell}} = \frac{i k}{2{\cal A}'} 
\delta_{\mu\nu}  e^{i  k_\perp |x|}.
\eeq


\begin{thebibliography}{241}%
\makeatletter
\providecommand \@ifxundefined [1]{%
 \@ifx{#1\undefined}
}%
\providecommand \@ifnum [1]{%
 \ifnum #1\expandafter \@firstoftwo
 \else \expandafter \@secondoftwo
 \fi
}%
\providecommand \@ifx [1]{%
 \ifx #1\expandafter \@firstoftwo
 \else \expandafter \@secondoftwo
 \fi
}%
\providecommand \natexlab [1]{#1}%
\providecommand \enquote  [1]{``#1''}%
\providecommand \bibnamefont  [1]{#1}%
\providecommand \bibfnamefont [1]{#1}%
\providecommand \citenamefont [1]{#1}%
\providecommand \href@noop [0]{\@secondoftwo}%
\providecommand \href [0]{\begingroup \@sanitize@url \@href}%
\providecommand \@href[1]{\@@startlink{#1}\@@href}%
\providecommand \@@href[1]{\endgroup#1\@@endlink}%
\providecommand \@sanitize@url [0]{\catcode `\\12\catcode `\$12\catcode
  `\&12\catcode `\#12\catcode `\^12\catcode `\_12\catcode `\%12\relax}%
\providecommand \@@startlink[1]{}%
\providecommand \@@endlink[0]{}%
\providecommand \url  [0]{\begingroup\@sanitize@url \@url }%
\providecommand \@url [1]{\endgroup\@href {#1}{\urlprefix }}%
\providecommand \urlprefix  [0]{URL }%
\providecommand \Eprint [0]{\href }%
\providecommand \doibase [0]{https://doi.org/}%
\providecommand \selectlanguage [0]{\@gobble}%
\providecommand \bibinfo  [0]{\@secondoftwo}%
\providecommand \bibfield  [0]{\@secondoftwo}%
\providecommand \translation [1]{[#1]}%
\providecommand \BibitemOpen [0]{}%
\providecommand \bibitemStop [0]{}%
\providecommand \bibitemNoStop [0]{.\EOS\space}%
\providecommand \EOS [0]{\spacefactor3000\relax}%
\providecommand \BibitemShut  [1]{\csname bibitem#1\endcsname}%
\let\auto@bib@innerbib\@empty
\bibitem [{\citenamefont {Bothwell}\ \emph {et~al.}(2019)\citenamefont
  {Bothwell}, \citenamefont {Kedar}, \citenamefont {Oelker}, \citenamefont
  {Robinson}, \citenamefont {Bromley}, \citenamefont {Tew}, \citenamefont
  {Ye},\ and\ \citenamefont {Kennedy}}]{Bothwell19}%
  \BibitemOpen
  \bibfield  {author} {\bibinfo {author} {\bibfnamefont {T.}~\bibnamefont
  {Bothwell}}, \bibinfo {author} {\bibfnamefont {D.}~\bibnamefont {Kedar}},
  \bibinfo {author} {\bibfnamefont {E.}~\bibnamefont {Oelker}}, \bibinfo
  {author} {\bibfnamefont {J.~M.}\ \bibnamefont {Robinson}}, \bibinfo {author}
  {\bibfnamefont {S.~L.}\ \bibnamefont {Bromley}}, \bibinfo {author}
  {\bibfnamefont {W.~L.}\ \bibnamefont {Tew}}, \bibinfo {author} {\bibfnamefont
  {J.}~\bibnamefont {Ye}},\ and\ \bibinfo {author} {\bibfnamefont {C.~J.}\
  \bibnamefont {Kennedy}},\ }\bibfield  {title} {\bibinfo {title} {{JILA} {SrI}
  optical lattice clock with uncertainty of $2.0 \times 10^{-18}$},\ }\href
  {https://doi.org/10.1088/1681-7575/ab4089} {\bibfield  {journal} {\bibinfo
  {journal} {Metrologia}\ }\textbf {\bibinfo {volume} {56}},\ \bibinfo {pages}
  {065004} (\bibinfo {year} {2019})}\BibitemShut {NoStop}%
\bibitem [{\citenamefont {Hammerer}\ \emph {et~al.}(2010)\citenamefont
  {Hammerer}, \citenamefont {S\o{}rensen},\ and\ \citenamefont
  {Polzik}}]{HAM10}%
  \BibitemOpen
  \bibfield  {author} {\bibinfo {author} {\bibfnamefont {K.}~\bibnamefont
  {Hammerer}}, \bibinfo {author} {\bibfnamefont {A.~S.}\ \bibnamefont
  {S\o{}rensen}},\ and\ \bibinfo {author} {\bibfnamefont {E.~S.}\ \bibnamefont
  {Polzik}},\ }\bibfield  {title} {\bibinfo {title} {Quantum interface between
  light and atomic ensembles},\ }\href
  {https://doi.org/10.1103/RevModPhys.82.1041} {\bibfield  {journal} {\bibinfo
  {journal} {Rev. Mod. Phys.}\ }\textbf {\bibinfo {volume} {82}},\ \bibinfo
  {pages} {1041} (\bibinfo {year} {2010})}\BibitemShut {NoStop}%
\bibitem [{\citenamefont {Dicke}(1954)}]{Dicke54}%
  \BibitemOpen
  \bibfield  {author} {\bibinfo {author} {\bibfnamefont {R.~H.}\ \bibnamefont
  {Dicke}},\ }\bibfield  {title} {\bibinfo {title} {Coherence in spontaneous
  radiation processes},\ }\href {https://doi.org/10.1103/PhysRev.93.99}
  {\bibfield  {journal} {\bibinfo  {journal} {Phys. Rev.}\ }\textbf {\bibinfo
  {volume} {93}},\ \bibinfo {pages} {99} (\bibinfo {year} {1954})}\BibitemShut
  {NoStop}%
\bibitem [{\citenamefont {Gross}\ and\ \citenamefont
  {Haroche}(1982)}]{GrossHarochePhysRep1982}%
  \BibitemOpen
  \bibfield  {author} {\bibinfo {author} {\bibfnamefont {M.}~\bibnamefont
  {Gross}}\ and\ \bibinfo {author} {\bibfnamefont {S.}~\bibnamefont
  {Haroche}},\ }\bibfield  {title} {\bibinfo {title} {Superradiance: An essay
  on the theory of collective spontaneous emission},\ }\href@noop {} {\bibfield
   {journal} {\bibinfo  {journal} {Phys. Rep.}\ }\textbf {\bibinfo {volume}
  {93}},\ \bibinfo {pages} {301} (\bibinfo {year} {1982})}\BibitemShut
  {NoStop}%
\bibitem [{\citenamefont {DeVoe}\ and\ \citenamefont {Brewer}(1996)}]{DeVoe}%
  \BibitemOpen
  \bibfield  {author} {\bibinfo {author} {\bibfnamefont {R.~G.}\ \bibnamefont
  {DeVoe}}\ and\ \bibinfo {author} {\bibfnamefont {R.~G.}\ \bibnamefont
  {Brewer}},\ }\bibfield  {title} {\bibinfo {title} {Observation of
  superradiant and subradiant spontaneous emission of two trapped ions},\
  }\href {https://doi.org/10.1103/PhysRevLett.76.2049} {\bibfield  {journal}
  {\bibinfo  {journal} {Phys. Rev. Lett.}\ }\textbf {\bibinfo {volume} {76}},\
  \bibinfo {pages} {2049} (\bibinfo {year} {1996})}\BibitemShut {NoStop}%
\bibitem [{\citenamefont {Hettich}\ \emph {et~al.}(2002)\citenamefont
  {Hettich}, \citenamefont {Schmitt}, \citenamefont {Zitzmann}, \citenamefont
  {K\"ohn}, \citenamefont {Gerhardt},\ and\ \citenamefont
  {Sandoghdar}}]{Hettich}%
  \BibitemOpen
  \bibfield  {author} {\bibinfo {author} {\bibfnamefont {C.}~\bibnamefont
  {Hettich}}, \bibinfo {author} {\bibfnamefont {C.}~\bibnamefont {Schmitt}},
  \bibinfo {author} {\bibfnamefont {J.}~\bibnamefont {Zitzmann}}, \bibinfo
  {author} {\bibfnamefont {S.}~\bibnamefont {K\"ohn}}, \bibinfo {author}
  {\bibfnamefont {I.}~\bibnamefont {Gerhardt}},\ and\ \bibinfo {author}
  {\bibfnamefont {V.}~\bibnamefont {Sandoghdar}},\ }\bibfield  {title}
  {\bibinfo {title} {Nanometer resolution and coherent optical dipole coupling
  of two individual molecules},\ }\href
  {https://doi.org/10.1126/science.1075606} {\bibfield  {journal} {\bibinfo
  {journal} {Science}\ }\textbf {\bibinfo {volume} {298}},\ \bibinfo {pages}
  {385} (\bibinfo {year} {2002})}\BibitemShut {NoStop}%
\bibitem [{\citenamefont {McGuyer}\ \emph {et~al.}(2015)\citenamefont
  {McGuyer}, \citenamefont {McDonald}, \citenamefont {Iwata}, \citenamefont
  {Tarallo}, \citenamefont {Skomorowski}, \citenamefont {Moszynski},\ and\
  \citenamefont {Zelevinsky}}]{McGuyer}%
  \BibitemOpen
  \bibfield  {author} {\bibinfo {author} {\bibfnamefont {B.~H.}\ \bibnamefont
  {McGuyer}}, \bibinfo {author} {\bibfnamefont {M.}~\bibnamefont {McDonald}},
  \bibinfo {author} {\bibfnamefont {G.~Z.}\ \bibnamefont {Iwata}}, \bibinfo
  {author} {\bibfnamefont {M.~G.}\ \bibnamefont {Tarallo}}, \bibinfo {author}
  {\bibfnamefont {W.}~\bibnamefont {Skomorowski}}, \bibinfo {author}
  {\bibfnamefont {R.}~\bibnamefont {Moszynski}},\ and\ \bibinfo {author}
  {\bibfnamefont {T.}~\bibnamefont {Zelevinsky}},\ }\bibfield  {title}
  {\bibinfo {title} {Precise study of asymptotic physics with subradiant
  ultracold molecules},\ }\href {https://doi.org/10.1038/NPHYS3182} {\bibfield
  {journal} {\bibinfo  {journal} {Nat. Phys.}\ }\textbf {\bibinfo {volume}
  {11}},\ \bibinfo {pages} {32} (\bibinfo {year} {2015})}\BibitemShut {NoStop}%
\bibitem [{\citenamefont {Takasu}\ \emph {et~al.}(2012)\citenamefont {Takasu},
  \citenamefont {Saito}, \citenamefont {Takahashi}, \citenamefont {Borkowski},
  \citenamefont {Ciury\l{}o},\ and\ \citenamefont {Julienne}}]{Takasu}%
  \BibitemOpen
  \bibfield  {author} {\bibinfo {author} {\bibfnamefont {Y.}~\bibnamefont
  {Takasu}}, \bibinfo {author} {\bibfnamefont {Y.}~\bibnamefont {Saito}},
  \bibinfo {author} {\bibfnamefont {Y.}~\bibnamefont {Takahashi}}, \bibinfo
  {author} {\bibfnamefont {M.}~\bibnamefont {Borkowski}}, \bibinfo {author}
  {\bibfnamefont {R.}~\bibnamefont {Ciury\l{}o}},\ and\ \bibinfo {author}
  {\bibfnamefont {P.~S.}\ \bibnamefont {Julienne}},\ }\bibfield  {title}
  {\bibinfo {title} {Controlled production of subradiant states of a diatomic
  molecule in an optical lattice},\ }\href
  {https://doi.org/10.1103/PhysRevLett.108.173002} {\bibfield  {journal}
  {\bibinfo  {journal} {Phys. Rev. Lett.}\ }\textbf {\bibinfo {volume} {108}},\
  \bibinfo {pages} {173002} (\bibinfo {year} {2012})}\BibitemShut {NoStop}%
\bibitem [{\citenamefont {Lovera}\ \emph {et~al.}(2013)\citenamefont {Lovera},
  \citenamefont {Gallinet}, \citenamefont {Nordlander},\ and\ \citenamefont
  {Martin}}]{Lovera}%
  \BibitemOpen
  \bibfield  {author} {\bibinfo {author} {\bibfnamefont {A.}~\bibnamefont
  {Lovera}}, \bibinfo {author} {\bibfnamefont {B.}~\bibnamefont {Gallinet}},
  \bibinfo {author} {\bibfnamefont {P.}~\bibnamefont {Nordlander}},\ and\
  \bibinfo {author} {\bibfnamefont {O.~J.}\ \bibnamefont {Martin}},\ }\bibfield
   {title} {\bibinfo {title} {Mechanisms of fano resonances in coupled
  plasmonic systems},\ }\href {https://doi.org/10.1021/nn401175j} {\bibfield
  {journal} {\bibinfo  {journal} {ACS Nano}\ }\textbf {\bibinfo {volume} {7}},\
  \bibinfo {pages} {4527} (\bibinfo {year} {2013})}\BibitemShut {NoStop}%
\bibitem [{\citenamefont {Frimmer}\ \emph {et~al.}(2012)\citenamefont
  {Frimmer}, \citenamefont {Coenen},\ and\ \citenamefont
  {Koenderink}}]{Frimmer}%
  \BibitemOpen
  \bibfield  {author} {\bibinfo {author} {\bibfnamefont {M.}~\bibnamefont
  {Frimmer}}, \bibinfo {author} {\bibfnamefont {T.}~\bibnamefont {Coenen}},\
  and\ \bibinfo {author} {\bibfnamefont {A.~F.}\ \bibnamefont {Koenderink}},\
  }\bibfield  {title} {\bibinfo {title} {{Signature of a Fano Resonance in a
  Plasmonic Metamolecule's Local Density of Optical States}},\ }\href
  {https://doi.org/{10.1103/PhysRevLett.108.077404}} {\bibfield  {journal}
  {\bibinfo  {journal} {{Phys. Rev. Lett.}}\ }\textbf {\bibinfo {volume}
  {{108}}},\ \bibinfo {pages} {{077404}} (\bibinfo {year}
  {{2012}})}\BibitemShut {NoStop}%
\bibitem [{\citenamefont {Guerin}\ \emph {et~al.}(2016)\citenamefont {Guerin},
  \citenamefont {Ara\'ujo},\ and\ \citenamefont {Kaiser}}]{Guerin_subr16}%
  \BibitemOpen
  \bibfield  {author} {\bibinfo {author} {\bibfnamefont {W.}~\bibnamefont
  {Guerin}}, \bibinfo {author} {\bibfnamefont {M.~O.}\ \bibnamefont
  {Ara\'ujo}},\ and\ \bibinfo {author} {\bibfnamefont {R.}~\bibnamefont
  {Kaiser}},\ }\bibfield  {title} {\bibinfo {title} {Subradiance in a large
  cloud of cold atoms},\ }\href
  {https://doi.org/10.1103/PhysRevLett.116.083601} {\bibfield  {journal}
  {\bibinfo  {journal} {Phys. Rev. Lett.}\ }\textbf {\bibinfo {volume} {116}},\
  \bibinfo {pages} {083601} (\bibinfo {year} {2016})}\BibitemShut {NoStop}%
\bibitem [{\citenamefont {Guerin}(2023)}]{Guerin2023}%
  \BibitemOpen
  \bibfield  {author} {\bibinfo {author} {\bibfnamefont {W.}~\bibnamefont
  {Guerin}},\ }\bibfield  {title} {\bibinfo {title} {Chapter four - super- and
  subradiance in dilute disordered cold atomic samples: observations and
  interpretations},\ }in\ \href
  {https://doi.org/https://doi.org/10.1016/bs.aamop.2023.04.002} {\emph
  {\bibinfo {booktitle} {Advances in Atomic, Molecular, and Optical
  Physics}}},\ Vol.~\bibinfo {volume} {72},\ \bibinfo {editor} {edited by\
  \bibinfo {editor} {\bibfnamefont {L.~F.}\ \bibnamefont {DiMauro}}, \bibinfo
  {editor} {\bibfnamefont {H.}~\bibnamefont {Perrin}},\ and\ \bibinfo {editor}
  {\bibfnamefont {S.~F.}\ \bibnamefont {Yelin}}}\ (\bibinfo  {publisher}
  {Academic Press},\ \bibinfo {year} {2023})\ pp.\ \bibinfo {pages}
  {253--296}\BibitemShut {NoStop}%
\bibitem [{\citenamefont {Jenkins}\ \emph {et~al.}(2017)\citenamefont
  {Jenkins}, \citenamefont {Ruostekoski}, \citenamefont {Papasimakis},
  \citenamefont {Savo},\ and\ \citenamefont {Zheludev}}]{Jenkins17}%
  \BibitemOpen
  \bibfield  {author} {\bibinfo {author} {\bibfnamefont {S.~D.}\ \bibnamefont
  {Jenkins}}, \bibinfo {author} {\bibfnamefont {J.}~\bibnamefont
  {Ruostekoski}}, \bibinfo {author} {\bibfnamefont {N.}~\bibnamefont
  {Papasimakis}}, \bibinfo {author} {\bibfnamefont {S.}~\bibnamefont {Savo}},\
  and\ \bibinfo {author} {\bibfnamefont {N.~I.}\ \bibnamefont {Zheludev}},\
  }\bibfield  {title} {\bibinfo {title} {Many-body subradiant excitations in
  metamaterial arrays: Experiment and theory},\ }\href
  {https://doi.org/10.1103/PhysRevLett.119.053901} {\bibfield  {journal}
  {\bibinfo  {journal} {Phys. Rev. Lett.}\ }\textbf {\bibinfo {volume} {119}},\
  \bibinfo {pages} {053901} (\bibinfo {year} {2017})}\BibitemShut {NoStop}%
\bibitem [{\citenamefont {Rui}\ \emph {et~al.}(2020)\citenamefont {Rui},
  \citenamefont {Wei}, \citenamefont {Rubio-Abadal}, \citenamefont {Hollerith},
  \citenamefont {Zeiher}, \citenamefont {Stamper-Kurn}, \citenamefont {Gross},\
  and\ \citenamefont {Bloch}}]{Rui2020}%
  \BibitemOpen
  \bibfield  {author} {\bibinfo {author} {\bibfnamefont {J.}~\bibnamefont
  {Rui}}, \bibinfo {author} {\bibfnamefont {D.}~\bibnamefont {Wei}}, \bibinfo
  {author} {\bibfnamefont {A.}~\bibnamefont {Rubio-Abadal}}, \bibinfo {author}
  {\bibfnamefont {S.}~\bibnamefont {Hollerith}}, \bibinfo {author}
  {\bibfnamefont {J.}~\bibnamefont {Zeiher}}, \bibinfo {author} {\bibfnamefont
  {D.~M.}\ \bibnamefont {Stamper-Kurn}}, \bibinfo {author} {\bibfnamefont
  {C.}~\bibnamefont {Gross}},\ and\ \bibinfo {author} {\bibfnamefont
  {I.}~\bibnamefont {Bloch}},\ }\bibfield  {title} {\bibinfo {title} {{A
  subradiant optical mirror formed by a single structured atomic layer}},\
  }\href {https://doi.org/10.1038/s41586-020-2463-x} {\bibfield  {journal}
  {\bibinfo  {journal} {Nature}\ }\textbf {\bibinfo {volume} {583}},\ \bibinfo
  {pages} {369} (\bibinfo {year} {2020})}\BibitemShut {NoStop}%
\bibitem [{Note1()}]{Note1}%
  \BibitemOpen
  \bibinfo {note} {Subradiant decay in lattices was also observed in Ref.~\cite
  {Ferioli21}}\BibitemShut {NoStop}%
\bibitem [{\citenamefont {Shahmoon}\ \emph {et~al.}(2017)\citenamefont
  {Shahmoon}, \citenamefont {Wild}, \citenamefont {Lukin},\ and\ \citenamefont
  {Yelin}}]{Shahmoon}%
  \BibitemOpen
  \bibfield  {author} {\bibinfo {author} {\bibfnamefont {E.}~\bibnamefont
  {Shahmoon}}, \bibinfo {author} {\bibfnamefont {D.~S.}\ \bibnamefont {Wild}},
  \bibinfo {author} {\bibfnamefont {M.~D.}\ \bibnamefont {Lukin}},\ and\
  \bibinfo {author} {\bibfnamefont {S.~F.}\ \bibnamefont {Yelin}},\ }\bibfield
  {title} {\bibinfo {title} {Cooperative resonances in light scattering from
  two-dimensional atomic arrays},\ }\href
  {https://doi.org/10.1103/PhysRevLett.118.113601} {\bibfield  {journal}
  {\bibinfo  {journal} {Phys. Rev. Lett.}\ }\textbf {\bibinfo {volume} {118}},\
  \bibinfo {pages} {113601} (\bibinfo {year} {2017})}\BibitemShut {NoStop}%
\bibitem [{\citenamefont {Facchinetti}\ \emph {et~al.}(2016)\citenamefont
  {Facchinetti}, \citenamefont {Jenkins},\ and\ \citenamefont
  {Ruostekoski}}]{Facchinetti16}%
  \BibitemOpen
  \bibfield  {author} {\bibinfo {author} {\bibfnamefont {G.}~\bibnamefont
  {Facchinetti}}, \bibinfo {author} {\bibfnamefont {S.~D.}\ \bibnamefont
  {Jenkins}},\ and\ \bibinfo {author} {\bibfnamefont {J.}~\bibnamefont
  {Ruostekoski}},\ }\bibfield  {title} {\bibinfo {title} {Storing light with
  subradiant correlations in arrays of atoms},\ }\href
  {https://doi.org/10.1103/PhysRevLett.117.243601} {\bibfield  {journal}
  {\bibinfo  {journal} {Phys. Rev. Lett.}\ }\textbf {\bibinfo {volume} {117}},\
  \bibinfo {pages} {243601} (\bibinfo {year} {2016})}\BibitemShut {NoStop}%
\bibitem [{\citenamefont {Facchinetti}\ and\ \citenamefont
  {Ruostekoski}(2018)}]{Facchinetti18}%
  \BibitemOpen
  \bibfield  {author} {\bibinfo {author} {\bibfnamefont {G.}~\bibnamefont
  {Facchinetti}}\ and\ \bibinfo {author} {\bibfnamefont {J.}~\bibnamefont
  {Ruostekoski}},\ }\bibfield  {title} {\bibinfo {title} {Interaction of light
  with planar lattices of atoms: Reflection, transmission, and cooperative
  magnetometry},\ }\href {https://doi.org/10.1103/PhysRevA.97.023833}
  {\bibfield  {journal} {\bibinfo  {journal} {Phys. Rev. A}\ }\textbf {\bibinfo
  {volume} {97}},\ \bibinfo {pages} {023833} (\bibinfo {year}
  {2018})}\BibitemShut {NoStop}%
\bibitem [{\citenamefont {Guimond}\ \emph {et~al.}(2019)\citenamefont
  {Guimond}, \citenamefont {Grankin}, \citenamefont {Vasilyev}, \citenamefont
  {Vermersch},\ and\ \citenamefont {Zoller}}]{Guimond2019}%
  \BibitemOpen
  \bibfield  {author} {\bibinfo {author} {\bibfnamefont {P.-O.}\ \bibnamefont
  {Guimond}}, \bibinfo {author} {\bibfnamefont {A.}~\bibnamefont {Grankin}},
  \bibinfo {author} {\bibfnamefont {D.~V.}\ \bibnamefont {Vasilyev}}, \bibinfo
  {author} {\bibfnamefont {B.}~\bibnamefont {Vermersch}},\ and\ \bibinfo
  {author} {\bibfnamefont {P.}~\bibnamefont {Zoller}},\ }\bibfield  {title}
  {\bibinfo {title} {Subradiant bell states in distant atomic arrays},\ }\href
  {https://doi.org/10.1103/PhysRevLett.122.093601} {\bibfield  {journal}
  {\bibinfo  {journal} {Phys. Rev. Lett.}\ }\textbf {\bibinfo {volume} {122}},\
  \bibinfo {pages} {093601} (\bibinfo {year} {2019})}\BibitemShut {NoStop}%
\bibitem [{\citenamefont {Javanainen}\ and\ \citenamefont
  {Rajapakse}(2019)}]{Javanainen19}%
  \BibitemOpen
  \bibfield  {author} {\bibinfo {author} {\bibfnamefont {J.}~\bibnamefont
  {Javanainen}}\ and\ \bibinfo {author} {\bibfnamefont {R.}~\bibnamefont
  {Rajapakse}},\ }\bibfield  {title} {\bibinfo {title} {Light propagation in
  systems involving two-dimensional atomic lattices},\ }\href
  {https://doi.org/10.1103/PhysRevA.100.013616} {\bibfield  {journal} {\bibinfo
   {journal} {Phys. Rev. A}\ }\textbf {\bibinfo {volume} {100}},\ \bibinfo
  {pages} {013616} (\bibinfo {year} {2019})}\BibitemShut {NoStop}%
\bibitem [{\citenamefont {Yu}\ and\ \citenamefont {Capasso}(2014)}]{Yu14}%
  \BibitemOpen
  \bibfield  {author} {\bibinfo {author} {\bibfnamefont {N.}~\bibnamefont
  {Yu}}\ and\ \bibinfo {author} {\bibfnamefont {F.}~\bibnamefont {Capasso}},\
  }\bibfield  {title} {\bibinfo {title} {Flat optics with designer
  metasurfaces},\ }\href {https://doi.org/10.1038/nmat3839} {\bibfield
  {journal} {\bibinfo  {journal} {Nature Materials}\ }\textbf {\bibinfo
  {volume} {13}},\ \bibinfo {pages} {139} (\bibinfo {year} {2014})}\BibitemShut
  {NoStop}%
\bibitem [{\citenamefont {Chen}\ \emph {et~al.}(2016)\citenamefont {Chen},
  \citenamefont {Taylor},\ and\ \citenamefont {Yu}}]{Chen2016_review}%
  \BibitemOpen
  \bibfield  {author} {\bibinfo {author} {\bibfnamefont {H.-T.}\ \bibnamefont
  {Chen}}, \bibinfo {author} {\bibfnamefont {A.~J.}\ \bibnamefont {Taylor}},\
  and\ \bibinfo {author} {\bibfnamefont {N.}~\bibnamefont {Yu}},\ }\bibfield
  {title} {\bibinfo {title} {A review of metasurfaces: physics and
  applications},\ }\href {https://doi.org/10.1088/0034-4885/79/7/076401}
  {\bibfield  {journal} {\bibinfo  {journal} {Reports on Progress in Physics}\
  }\textbf {\bibinfo {volume} {79}},\ \bibinfo {pages} {076401} (\bibinfo
  {year} {2016})}\BibitemShut {NoStop}%
\bibitem [{\citenamefont {Luo}(2018)}]{Luo18}%
  \BibitemOpen
  \bibfield  {author} {\bibinfo {author} {\bibfnamefont {X.}~\bibnamefont
  {Luo}},\ }\bibfield  {title} {\bibinfo {title} {Subwavelength optical
  engineering with metasurface waves},\ }\href
  {https://doi.org/https://doi.org/10.1002/adom.201701201} {\bibfield
  {journal} {\bibinfo  {journal} {Advanced Optical Materials}\ }\textbf
  {\bibinfo {volume} {6}},\ \bibinfo {pages} {1701201} (\bibinfo {year}
  {2018})}\BibitemShut {NoStop}%
\bibitem [{\citenamefont {Qiu}\ \emph {et~al.}(2021)\citenamefont {Qiu},
  \citenamefont {Zhang}, \citenamefont {Hu},\ and\ \citenamefont
  {Kivshar}}]{Qiu21_review}%
  \BibitemOpen
  \bibfield  {author} {\bibinfo {author} {\bibfnamefont {C.-W.}\ \bibnamefont
  {Qiu}}, \bibinfo {author} {\bibfnamefont {T.}~\bibnamefont {Zhang}}, \bibinfo
  {author} {\bibfnamefont {G.}~\bibnamefont {Hu}},\ and\ \bibinfo {author}
  {\bibfnamefont {Y.}~\bibnamefont {Kivshar}},\ }\bibfield  {title} {\bibinfo
  {title} {Quo vadis, metasurfaces?},\ }\href
  {https://doi.org/10.1021/acs.nanolett.1c00828} {\bibfield  {journal}
  {\bibinfo  {journal} {Nano Letters}\ }\textbf {\bibinfo {volume} {21}},\
  \bibinfo {pages} {5461} (\bibinfo {year} {2021})}\BibitemShut {NoStop}%
\bibitem [{\citenamefont {Solntsev}\ \emph {et~al.}(2021)\citenamefont
  {Solntsev}, \citenamefont {Agarwal},\ and\ \citenamefont
  {Kivshar}}]{Solntsev21}%
  \BibitemOpen
  \bibfield  {author} {\bibinfo {author} {\bibfnamefont {A.~S.}\ \bibnamefont
  {Solntsev}}, \bibinfo {author} {\bibfnamefont {G.~S.}\ \bibnamefont
  {Agarwal}},\ and\ \bibinfo {author} {\bibfnamefont {Y.~S.}\ \bibnamefont
  {Kivshar}},\ }\bibfield  {title} {\bibinfo {title} {Metasurfaces for quantum
  photonics},\ }\href {https://doi.org/10.1038/s41566-021-00793-z} {\bibfield
  {journal} {\bibinfo  {journal} {Nature Photonics}\ }\textbf {\bibinfo
  {volume} {15}},\ \bibinfo {pages} {327} (\bibinfo {year} {2021})}\BibitemShut
  {NoStop}%
\bibitem [{\citenamefont {Walls}\ and\ \citenamefont
  {Milburn}(1994)}]{WallsMilburn}%
  \BibitemOpen
  \bibfield  {author} {\bibinfo {author} {\bibfnamefont {D.~F.}\ \bibnamefont
  {Walls}}\ and\ \bibinfo {author} {\bibfnamefont {G.~J.}\ \bibnamefont
  {Milburn}},\ }\href@noop {} {\emph {\bibinfo {title} {Quantum Optics}}}\
  (\bibinfo  {publisher} {Springer-Verlag},\ \bibinfo {address} {Berlin},\
  \bibinfo {year} {1994})\BibitemShut {NoStop}%
\bibitem [{\citenamefont {Sheremet}\ \emph {et~al.}(2023)\citenamefont
  {Sheremet}, \citenamefont {Petrov}, \citenamefont {Iorsh}, \citenamefont
  {Poshakinskiy},\ and\ \citenamefont {Poddubny}}]{Sheremet23}%
  \BibitemOpen
  \bibfield  {author} {\bibinfo {author} {\bibfnamefont {A.~S.}\ \bibnamefont
  {Sheremet}}, \bibinfo {author} {\bibfnamefont {M.~I.}\ \bibnamefont
  {Petrov}}, \bibinfo {author} {\bibfnamefont {I.~V.}\ \bibnamefont {Iorsh}},
  \bibinfo {author} {\bibfnamefont {A.~V.}\ \bibnamefont {Poshakinskiy}},\ and\
  \bibinfo {author} {\bibfnamefont {A.~N.}\ \bibnamefont {Poddubny}},\
  }\bibfield  {title} {\bibinfo {title} {Waveguide quantum electrodynamics:
  Collective radiance and photon-photon correlations},\ }\href
  {https://doi.org/10.1103/RevModPhys.95.015002} {\bibfield  {journal}
  {\bibinfo  {journal} {Rev. Mod. Phys.}\ }\textbf {\bibinfo {volume} {95}},\
  \bibinfo {pages} {015002} (\bibinfo {year} {2023})}\BibitemShut {NoStop}%
\bibitem [{\citenamefont {Chang}\ \emph {et~al.}(2018)\citenamefont {Chang},
  \citenamefont {Douglas}, \citenamefont {Gonz\'alez-Tudela}, \citenamefont
  {Hung},\ and\ \citenamefont {Kimble}}]{Chang18}%
  \BibitemOpen
  \bibfield  {author} {\bibinfo {author} {\bibfnamefont {D.~E.}\ \bibnamefont
  {Chang}}, \bibinfo {author} {\bibfnamefont {J.~S.}\ \bibnamefont {Douglas}},
  \bibinfo {author} {\bibfnamefont {A.}~\bibnamefont {Gonz\'alez-Tudela}},
  \bibinfo {author} {\bibfnamefont {C.-L.}\ \bibnamefont {Hung}},\ and\
  \bibinfo {author} {\bibfnamefont {H.~J.}\ \bibnamefont {Kimble}},\ }\bibfield
   {title} {\bibinfo {title} {Colloquium: Quantum matter built from nanoscopic
  lattices of atoms and photons},\ }\href
  {https://doi.org/10.1103/RevModPhys.90.031002} {\bibfield  {journal}
  {\bibinfo  {journal} {Rev. Mod. Phys.}\ }\textbf {\bibinfo {volume} {90}},\
  \bibinfo {pages} {031002} (\bibinfo {year} {2018})}\BibitemShut {NoStop}%
\bibitem [{\citenamefont {Reitz}\ \emph {et~al.}(2022)\citenamefont {Reitz},
  \citenamefont {Sommer},\ and\ \citenamefont {Genes}}]{Reitz22}%
  \BibitemOpen
  \bibfield  {author} {\bibinfo {author} {\bibfnamefont {M.}~\bibnamefont
  {Reitz}}, \bibinfo {author} {\bibfnamefont {C.}~\bibnamefont {Sommer}},\ and\
  \bibinfo {author} {\bibfnamefont {C.}~\bibnamefont {Genes}},\ }\bibfield
  {title} {\bibinfo {title} {Cooperative quantum phenomena in light-matter
  platforms},\ }\href {https://doi.org/10.1103/PRXQuantum.3.010201} {\bibfield
  {journal} {\bibinfo  {journal} {PRX Quantum}\ }\textbf {\bibinfo {volume}
  {3}},\ \bibinfo {pages} {010201} (\bibinfo {year} {2022})}\BibitemShut
  {NoStop}%
\bibitem [{\citenamefont {Power}\ and\ \citenamefont
  {Zienau}(1959)}]{PowerZienauPTRS1959}%
  \BibitemOpen
  \bibfield  {author} {\bibinfo {author} {\bibfnamefont {E.~A.}\ \bibnamefont
  {Power}}\ and\ \bibinfo {author} {\bibfnamefont {S.}~\bibnamefont {Zienau}},\
  }\bibfield  {title} {\bibinfo {title} {Coulomb gauge in non-relativistic
  quantum electro-dynamics and the shape of spectral lines},\ }\href
  {https://doi.org/10.1098/rsta.1959.0008} {\bibfield  {journal} {\bibinfo
  {journal} {Philos. Trans. R. Soc.}\ }\textbf {\bibinfo {volume} {251}},\
  \bibinfo {pages} {427} (\bibinfo {year} {1959})}\BibitemShut {NoStop}%
\bibitem [{\citenamefont {Woolley}(1971)}]{Woolley1971a}%
  \BibitemOpen
  \bibfield  {author} {\bibinfo {author} {\bibfnamefont {R.~G.}\ \bibnamefont
  {Woolley}},\ }\bibfield  {title} {\bibinfo {title} {Molecular quantum
  electrodynamics},\ }\href {https://doi.org/10.1098/rspa.1971.0049} {\bibfield
   {journal} {\bibinfo  {journal} {Proc. R. Soc. Lond. A}\ }\textbf {\bibinfo
  {volume} {321}},\ \bibinfo {pages} {557} (\bibinfo {year}
  {1971})}\BibitemShut {NoStop}%
\bibitem [{\citenamefont {{Cohen-Tannaudji}}\ \emph {et~al.}(1989)\citenamefont
  {{Cohen-Tannaudji}}, \citenamefont {{Dupont-Roc}},\ and\ \citenamefont
  {Grynberg}}]{CohenT}%
  \BibitemOpen
  \bibfield  {author} {\bibinfo {author} {\bibfnamefont {C.}~\bibnamefont
  {{Cohen-Tannaudji}}}, \bibinfo {author} {\bibfnamefont {J.}~\bibnamefont
  {{Dupont-Roc}}},\ and\ \bibinfo {author} {\bibfnamefont {G.}~\bibnamefont
  {Grynberg}},\ }\href@noop {} {\emph {\bibinfo {title} {Photons and Atoms:
  Introduction to Quantum Electrodynamics}}}\ (\bibinfo  {publisher} {John
  Wiley \& Sons},\ \bibinfo {address} {New York},\ \bibinfo {year}
  {1989})\BibitemShut {NoStop}%
\bibitem [{\citenamefont {Lee}\ \emph {et~al.}(2016)\citenamefont {Lee},
  \citenamefont {Jenkins},\ and\ \citenamefont {Ruostekoski}}]{Lee16}%
  \BibitemOpen
  \bibfield  {author} {\bibinfo {author} {\bibfnamefont {M.~D.}\ \bibnamefont
  {Lee}}, \bibinfo {author} {\bibfnamefont {S.~D.}\ \bibnamefont {Jenkins}},\
  and\ \bibinfo {author} {\bibfnamefont {J.}~\bibnamefont {Ruostekoski}},\
  }\bibfield  {title} {\bibinfo {title} {Stochastic methods for light
  propagation and recurrent scattering in saturated and nonsaturated atomic
  ensembles},\ }\href {https://doi.org/10.1103/PhysRevA.93.063803} {\bibfield
  {journal} {\bibinfo  {journal} {Phys. Rev. A}\ }\textbf {\bibinfo {volume}
  {93}},\ \bibinfo {pages} {063803} (\bibinfo {year} {2016})}\BibitemShut
  {NoStop}%
\bibitem [{\citenamefont {Ruostekoski}\ and\ \citenamefont
  {Javanainen}(1997)}]{Ruostekoski1997a}%
  \BibitemOpen
  \bibfield  {author} {\bibinfo {author} {\bibfnamefont {J.}~\bibnamefont
  {Ruostekoski}}\ and\ \bibinfo {author} {\bibfnamefont {J.}~\bibnamefont
  {Javanainen}},\ }\bibfield  {title} {\bibinfo {title} {Quantum field theory
  of cooperative atom response: Low light intensity},\ }\href
  {https://doi.org/10.1103/PhysRevA.55.513} {\bibfield  {journal} {\bibinfo
  {journal} {Phys. Rev. A}\ }\textbf {\bibinfo {volume} {55}},\ \bibinfo
  {pages} {513} (\bibinfo {year} {1997})}\BibitemShut {NoStop}%
\bibitem [{\citenamefont {Jackson}(1999)}]{Jackson}%
  \BibitemOpen
  \bibfield  {author} {\bibinfo {author} {\bibfnamefont {J.~D.}\ \bibnamefont
  {Jackson}},\ }\href@noop {} {\emph {\bibinfo {title} {Classical
  Electrodynamics}}},\ \bibinfo {edition} {3rd}\ ed.\ (\bibinfo  {publisher}
  {Wiley, New York},\ \bibinfo {year} {1999})\BibitemShut {NoStop}%
\bibitem [{\citenamefont {Born}\ and\ \citenamefont {Wolf}(1999)}]{BOR99}%
  \BibitemOpen
  \bibfield  {author} {\bibinfo {author} {\bibfnamefont {M.}~\bibnamefont
  {Born}}\ and\ \bibinfo {author} {\bibfnamefont {E.}~\bibnamefont {Wolf}},\
  }\href@noop {} {\emph {\bibinfo {title} {Principles of Optics}}},\ \bibinfo
  {edition} {7th}\ ed.\ (\bibinfo  {publisher} {Cambridge University Press,
  Cambridge, UK},\ \bibinfo {year} {1999})\BibitemShut {NoStop}%
\bibitem [{\citenamefont {Ishimaru}(1978)}]{Ishimaru1978}%
  \BibitemOpen
  \bibfield  {author} {\bibinfo {author} {\bibfnamefont {A.}~\bibnamefont
  {Ishimaru}},\ }\href@noop {} {\emph {\bibinfo {title} {Wave Propagation and
  Scattering in Random Media: Multiple Scattering, Turbulence, Rough Surfaces,
  and Remote-Sensing}}},\ Vol.~\bibinfo {volume} {2}\ (\bibinfo  {publisher}
  {Academic Press},\ \bibinfo {address} {St. Louis, Missouri},\ \bibinfo {year}
  {1978})\BibitemShut {NoStop}%
\bibitem [{\citenamefont {Lagendijk}\ and\ \citenamefont {van
  Tiggelen}(1996)}]{Lagendijk}%
  \BibitemOpen
  \bibfield  {author} {\bibinfo {author} {\bibfnamefont {A.}~\bibnamefont
  {Lagendijk}}\ and\ \bibinfo {author} {\bibfnamefont {B.~A.}\ \bibnamefont
  {van Tiggelen}},\ }\bibfield  {title} {\bibinfo {title} {Resonant multiple
  scattering of light},\ }\href {https://doi.org/DOI:
  10.1016/0370-1573(95)00065-8} {\bibfield  {journal} {\bibinfo  {journal}
  {Phys. Rep.}\ }\textbf {\bibinfo {volume} {270}},\ \bibinfo {pages} {143 }
  (\bibinfo {year} {1996})}\BibitemShut {NoStop}%
\bibitem [{\citenamefont {van Tiggelen}\ \emph {et~al.}(1990)\citenamefont {van
  Tiggelen}, \citenamefont {Lagendijk},\ and\ \citenamefont
  {Tip}}]{vantiggelen90}%
  \BibitemOpen
  \bibfield  {author} {\bibinfo {author} {\bibfnamefont {B.~A.}\ \bibnamefont
  {van Tiggelen}}, \bibinfo {author} {\bibfnamefont {A.}~\bibnamefont
  {Lagendijk}},\ and\ \bibinfo {author} {\bibfnamefont {A.}~\bibnamefont
  {Tip}},\ }\bibfield  {title} {\bibinfo {title} {Multiple-scattering effects
  for the propagation of light in 3d slabs},\ }\href@noop {} {\bibfield
  {journal} {\bibinfo  {journal} {J. Phys. Cond. Mat.}\ }\textbf {\bibinfo
  {volume} {2}},\ \bibinfo {pages} {7653} (\bibinfo {year} {1990})}\BibitemShut
  {NoStop}%
\bibitem [{\citenamefont {Morice}\ \emph {et~al.}(1995)\citenamefont {Morice},
  \citenamefont {Castin},\ and\ \citenamefont {Dalibard}}]{Morice1995a}%
  \BibitemOpen
  \bibfield  {author} {\bibinfo {author} {\bibfnamefont {O.}~\bibnamefont
  {Morice}}, \bibinfo {author} {\bibfnamefont {Y.}~\bibnamefont {Castin}},\
  and\ \bibinfo {author} {\bibfnamefont {J.}~\bibnamefont {Dalibard}},\
  }\bibfield  {title} {\bibinfo {title} {Refractive index of a dilute {Bose}
  gas},\ }\href {https://doi.org/10.1103/PhysRevA.51.3896} {\bibfield
  {journal} {\bibinfo  {journal} {Phys. Rev. A}\ }\textbf {\bibinfo {volume}
  {51}},\ \bibinfo {pages} {3896} (\bibinfo {year} {1995})}\BibitemShut
  {NoStop}%
\bibitem [{\citenamefont {Sokolov}\ \emph {et~al.}(2011)\citenamefont
  {Sokolov}, \citenamefont {Kupriyanov},\ and\ \citenamefont
  {Havey}}]{Sokolov2011}%
  \BibitemOpen
  \bibfield  {author} {\bibinfo {author} {\bibfnamefont {I.~M.}\ \bibnamefont
  {Sokolov}}, \bibinfo {author} {\bibfnamefont {D.~V.}\ \bibnamefont
  {Kupriyanov}},\ and\ \bibinfo {author} {\bibfnamefont {M.~D.}\ \bibnamefont
  {Havey}},\ }\bibfield  {title} {\bibinfo {title} {Microscopic theory of
  scattering of weak electromagnetic radiation by a dense ensemble of ultracold
  atoms},\ }\href {https://doi.org/10.1134/S106377611101016X} {\bibfield
  {journal} {\bibinfo  {journal} {Journal of Experimental and Theoretical
  Physics}\ }\textbf {\bibinfo {volume} {112}},\ \bibinfo {pages} {246}
  (\bibinfo {year} {2011})}\BibitemShut {NoStop}%
\bibitem [{\citenamefont {Javanainen}\ \emph {et~al.}(2014)\citenamefont
  {Javanainen}, \citenamefont {Ruostekoski}, \citenamefont {Li},\ and\
  \citenamefont {Yoo}}]{Javanainen2014a}%
  \BibitemOpen
  \bibfield  {author} {\bibinfo {author} {\bibfnamefont {J.}~\bibnamefont
  {Javanainen}}, \bibinfo {author} {\bibfnamefont {J.}~\bibnamefont
  {Ruostekoski}}, \bibinfo {author} {\bibfnamefont {Y.}~\bibnamefont {Li}},\
  and\ \bibinfo {author} {\bibfnamefont {S.-M.}\ \bibnamefont {Yoo}},\
  }\bibfield  {title} {\bibinfo {title} {Shifts of a resonance line in a dense
  atomic sample},\ }\href {https://doi.org/10.1103/PhysRevLett.112.113603}
  {\bibfield  {journal} {\bibinfo  {journal} {Phys. Rev. Lett.}\ }\textbf
  {\bibinfo {volume} {112}},\ \bibinfo {pages} {113603} (\bibinfo {year}
  {2014})}\BibitemShut {NoStop}%
\bibitem [{\citenamefont {Kwong}\ \emph {et~al.}(2019)\citenamefont {Kwong},
  \citenamefont {Wilkowski}, \citenamefont {Delande},\ and\ \citenamefont
  {Pierrat}}]{Kwong19}%
  \BibitemOpen
  \bibfield  {author} {\bibinfo {author} {\bibfnamefont {C.~C.}\ \bibnamefont
  {Kwong}}, \bibinfo {author} {\bibfnamefont {D.}~\bibnamefont {Wilkowski}},
  \bibinfo {author} {\bibfnamefont {D.}~\bibnamefont {Delande}},\ and\ \bibinfo
  {author} {\bibfnamefont {R.}~\bibnamefont {Pierrat}},\ }\bibfield  {title}
  {\bibinfo {title} {Coherent light propagation through cold atomic clouds
  beyond the independent scattering approximation},\ }\href
  {https://doi.org/10.1103/PhysRevA.99.043806} {\bibfield  {journal} {\bibinfo
  {journal} {Phys. Rev. A}\ }\textbf {\bibinfo {volume} {99}},\ \bibinfo
  {pages} {043806} (\bibinfo {year} {2019})}\BibitemShut {NoStop}%
\bibitem [{\citenamefont {Lehmberg}(1970)}]{Lehmberg1970}%
  \BibitemOpen
  \bibfield  {author} {\bibinfo {author} {\bibfnamefont {R.~H.}\ \bibnamefont
  {Lehmberg}},\ }\bibfield  {title} {\bibinfo {title} {Radiation from an
  {$N$-Atom System. I. General Formalism}},\ }\href
  {https://doi.org/10.1103/PhysRevA.2.883} {\bibfield  {journal} {\bibinfo
  {journal} {Phys. Rev. A}\ }\textbf {\bibinfo {volume} {2}},\ \bibinfo {pages}
  {883} (\bibinfo {year} {1970})}\BibitemShut {NoStop}%
\bibitem [{\citenamefont {Agarwal}(1970)}]{agarwal1970}%
  \BibitemOpen
  \bibfield  {author} {\bibinfo {author} {\bibfnamefont {G.~S.}\ \bibnamefont
  {Agarwal}},\ }\bibfield  {title} {\bibinfo {title} {Master-equation approach
  to spontaneous emission},\ }\href {https://doi.org/10.1103/PhysRevA.2.2038}
  {\bibfield  {journal} {\bibinfo  {journal} {Phys. Rev. A}\ }\textbf {\bibinfo
  {volume} {2}},\ \bibinfo {pages} {2038} (\bibinfo {year} {1970})}\BibitemShut
  {NoStop}%
\bibitem [{\citenamefont {Carmichael}(2007)}]{CarmichaelVol2}%
  \BibitemOpen
  \bibfield  {author} {\bibinfo {author} {\bibfnamefont {H.}~\bibnamefont
  {Carmichael}},\ }\href@noop {} {\emph {\bibinfo {title} {Statistical Methods
  in Quantum Optics}}},\ Vol.~\bibinfo {volume} {2}\ (\bibinfo  {publisher}
  {Springer},\ \bibinfo {address} {Berlin},\ \bibinfo {year}
  {2007})\BibitemShut {NoStop}%
\bibitem [{\citenamefont {Kr{\"{a}}mer}\ and\ \citenamefont
  {Ritsch}(2015)}]{Kramer2015a}%
  \BibitemOpen
  \bibfield  {author} {\bibinfo {author} {\bibfnamefont {S.}~\bibnamefont
  {Kr{\"{a}}mer}}\ and\ \bibinfo {author} {\bibfnamefont {H.}~\bibnamefont
  {Ritsch}},\ }\bibfield  {title} {\bibinfo {title} {{Generalized mean-field
  approach to simulate the dynamics of large open spin ensembles with long
  range interactions}},\ }\href {https://doi.org/10.1140/epjd/e2015-60266-5}
  {\bibfield  {journal} {\bibinfo  {journal} {Eur. Phys. J. D}\ }\textbf
  {\bibinfo {volume} {69}},\ \bibinfo {pages} {282} (\bibinfo {year}
  {2015})}\BibitemShut {NoStop}%
\bibitem [{\citenamefont {Parmee}\ and\ \citenamefont
  {Ruostekoski}(2020)}]{Parmee2020}%
  \BibitemOpen
  \bibfield  {author} {\bibinfo {author} {\bibfnamefont {C.~D.}\ \bibnamefont
  {Parmee}}\ and\ \bibinfo {author} {\bibfnamefont {J.}~\bibnamefont
  {Ruostekoski}},\ }\bibfield  {title} {\bibinfo {title} {{Signatures of
  optical phase transitions in superradiant and subradiant atomic arrays}},\
  }\href {https://doi.org/10.1038/s42005-020-00476-1} {\bibfield  {journal}
  {\bibinfo  {journal} {Commun. Phys.}\ }\textbf {\bibinfo {volume} {3}},\
  \bibinfo {pages} {205} (\bibinfo {year} {2020})}\BibitemShut {NoStop}%
\bibitem [{\citenamefont {Parmee}\ and\ \citenamefont
  {Cooper}(2018)}]{Parmee2018}%
  \BibitemOpen
  \bibfield  {author} {\bibinfo {author} {\bibfnamefont {C.~D.}\ \bibnamefont
  {Parmee}}\ and\ \bibinfo {author} {\bibfnamefont {N.~R.}\ \bibnamefont
  {Cooper}},\ }\bibfield  {title} {\bibinfo {title} {Phases of driven two-level
  systems with nonlocal dissipation},\ }\href
  {https://doi.org/10.1103/PhysRevA.97.053616} {\bibfield  {journal} {\bibinfo
  {journal} {Phys. Rev. A}\ }\textbf {\bibinfo {volume} {97}},\ \bibinfo
  {pages} {053616} (\bibinfo {year} {2018})}\BibitemShut {NoStop}%
\bibitem [{\citenamefont {Bettles}\ \emph {et~al.}(2020)\citenamefont
  {Bettles}, \citenamefont {Lee}, \citenamefont {Gardiner},\ and\ \citenamefont
  {Ruostekoski}}]{Bettles2020}%
  \BibitemOpen
  \bibfield  {author} {\bibinfo {author} {\bibfnamefont {R.~J.}\ \bibnamefont
  {Bettles}}, \bibinfo {author} {\bibfnamefont {M.~D.}\ \bibnamefont {Lee}},
  \bibinfo {author} {\bibfnamefont {S.~A.}\ \bibnamefont {Gardiner}},\ and\
  \bibinfo {author} {\bibfnamefont {J.}~\bibnamefont {Ruostekoski}},\
  }\bibfield  {title} {\bibinfo {title} {{Quantum and nonlinear effects in
  light transmitted through planar atomic arrays}},\ }\href
  {https://doi.org/10.1038/s42005-020-00404-3} {\bibfield  {journal} {\bibinfo
  {journal} {Commun. Phys.}\ }\textbf {\bibinfo {volume} {3}},\ \bibinfo
  {pages} {141} (\bibinfo {year} {2020})}\BibitemShut {NoStop}%
\bibitem [{\citenamefont {Rusek}\ \emph {et~al.}(1996)\citenamefont {Rusek},
  \citenamefont {Or\l{}owski},\ and\ \citenamefont {Mostowski}}]{Rusek96}%
  \BibitemOpen
  \bibfield  {author} {\bibinfo {author} {\bibfnamefont {M.}~\bibnamefont
  {Rusek}}, \bibinfo {author} {\bibfnamefont {A.}~\bibnamefont {Or\l{}owski}},\
  and\ \bibinfo {author} {\bibfnamefont {J.}~\bibnamefont {Mostowski}},\
  }\bibfield  {title} {\bibinfo {title} {Localization of light in
  three-dimensional random dielectric media},\ }\href
  {https://doi.org/10.1103/PhysRevE.53.4122} {\bibfield  {journal} {\bibinfo
  {journal} {Phys. Rev. E}\ }\textbf {\bibinfo {volume} {53}},\ \bibinfo
  {pages} {4122} (\bibinfo {year} {1996})}\BibitemShut {NoStop}%
\bibitem [{\citenamefont {Jenkins}\ \emph {et~al.}(2016)\citenamefont
  {Jenkins}, \citenamefont {Ruostekoski}, \citenamefont {Javanainen},
  \citenamefont {Jennewein}, \citenamefont {Bourgain}, \citenamefont
  {Pellegrino}, \citenamefont {Sortais},\ and\ \citenamefont
  {Browaeys}}]{Jenkins_long16}%
  \BibitemOpen
  \bibfield  {author} {\bibinfo {author} {\bibfnamefont {S.~D.}\ \bibnamefont
  {Jenkins}}, \bibinfo {author} {\bibfnamefont {J.}~\bibnamefont
  {Ruostekoski}}, \bibinfo {author} {\bibfnamefont {J.}~\bibnamefont
  {Javanainen}}, \bibinfo {author} {\bibfnamefont {S.}~\bibnamefont
  {Jennewein}}, \bibinfo {author} {\bibfnamefont {R.}~\bibnamefont {Bourgain}},
  \bibinfo {author} {\bibfnamefont {J.}~\bibnamefont {Pellegrino}}, \bibinfo
  {author} {\bibfnamefont {Y.~R.~P.}\ \bibnamefont {Sortais}},\ and\ \bibinfo
  {author} {\bibfnamefont {A.}~\bibnamefont {Browaeys}},\ }\bibfield  {title}
  {\bibinfo {title} {Collective resonance fluorescence in small and dense atom
  clouds: Comparison between theory and experiment},\ }\href
  {https://doi.org/10.1103/PhysRevA.94.023842} {\bibfield  {journal} {\bibinfo
  {journal} {Phys. Rev. A}\ }\textbf {\bibinfo {volume} {94}},\ \bibinfo
  {pages} {023842} (\bibinfo {year} {2016})}\BibitemShut {NoStop}%
\bibitem [{\citenamefont {Ballantine}\ and\ \citenamefont
  {Ruostekoski}(2021{\natexlab{a}})}]{Ballantine21PT}%
  \BibitemOpen
  \bibfield  {author} {\bibinfo {author} {\bibfnamefont {K.~E.}\ \bibnamefont
  {Ballantine}}\ and\ \bibinfo {author} {\bibfnamefont {J.}~\bibnamefont
  {Ruostekoski}},\ }\bibfield  {title} {\bibinfo {title} {Parity-time symmetry
  and coherent perfect absorption in a cooperative atom response},\ }\href
  {https://doi.org/doi:10.1515/nanoph-2020-0635} {\bibfield  {journal}
  {\bibinfo  {journal} {Nanophotonics}\ }\textbf {\bibinfo {volume} {10}},\
  \bibinfo {pages} {1357} (\bibinfo {year} {2021}{\natexlab{a}})}\BibitemShut
  {NoStop}%
\bibitem [{\citenamefont {Williamson}\ and\ \citenamefont
  {Ruostekoski}(2020)}]{Williamson2020}%
  \BibitemOpen
  \bibfield  {author} {\bibinfo {author} {\bibfnamefont {L.~A.}\ \bibnamefont
  {Williamson}}\ and\ \bibinfo {author} {\bibfnamefont {J.}~\bibnamefont
  {Ruostekoski}},\ }\bibfield  {title} {\bibinfo {title} {Optical response of
  atom chains beyond the limit of low light intensity: The validity of the
  linear classical oscillator model},\ }\href
  {https://doi.org/10.1103/PhysRevResearch.2.023273} {\bibfield  {journal}
  {\bibinfo  {journal} {Phys. Rev. Research}\ }\textbf {\bibinfo {volume}
  {2}},\ \bibinfo {pages} {023273} (\bibinfo {year} {2020})}\BibitemShut
  {NoStop}%
\bibitem [{\citenamefont {Cipris}\ \emph {et~al.}(2021)\citenamefont {Cipris},
  \citenamefont {Moreira}, \citenamefont {do~Espirito~Santo}, \citenamefont
  {Weiss}, \citenamefont {Villas-Boas}, \citenamefont {Kaiser}, \citenamefont
  {Guerin},\ and\ \citenamefont {Bachelard}}]{Cipris21}%
  \BibitemOpen
  \bibfield  {author} {\bibinfo {author} {\bibfnamefont {A.}~\bibnamefont
  {Cipris}}, \bibinfo {author} {\bibfnamefont {N.~A.}\ \bibnamefont {Moreira}},
  \bibinfo {author} {\bibfnamefont {T.~S.}\ \bibnamefont {do~Espirito~Santo}},
  \bibinfo {author} {\bibfnamefont {P.}~\bibnamefont {Weiss}}, \bibinfo
  {author} {\bibfnamefont {C.~J.}\ \bibnamefont {Villas-Boas}}, \bibinfo
  {author} {\bibfnamefont {R.}~\bibnamefont {Kaiser}}, \bibinfo {author}
  {\bibfnamefont {W.}~\bibnamefont {Guerin}},\ and\ \bibinfo {author}
  {\bibfnamefont {R.}~\bibnamefont {Bachelard}},\ }\bibfield  {title} {\bibinfo
  {title} {Subradiance with saturated atoms: Population enhancement of the
  long-lived states},\ }\href {https://doi.org/10.1103/PhysRevLett.126.103604}
  {\bibfield  {journal} {\bibinfo  {journal} {Phys. Rev. Lett.}\ }\textbf
  {\bibinfo {volume} {126}},\ \bibinfo {pages} {103604} (\bibinfo {year}
  {2021})}\BibitemShut {NoStop}%
\bibitem [{\citenamefont {Svidzinsky}\ \emph {et~al.}(2010)\citenamefont
  {Svidzinsky}, \citenamefont {Chang},\ and\ \citenamefont {Scully}}]{SVI10}%
  \BibitemOpen
  \bibfield  {author} {\bibinfo {author} {\bibfnamefont {A.~A.}\ \bibnamefont
  {Svidzinsky}}, \bibinfo {author} {\bibfnamefont {J.-T.}\ \bibnamefont
  {Chang}},\ and\ \bibinfo {author} {\bibfnamefont {M.~O.}\ \bibnamefont
  {Scully}},\ }\bibfield  {title} {\bibinfo {title} {Cooperative spontaneous
  emission of $n$ atoms: Many-body eigenstates, the effect of virtual {L}amb
  shift processes, and analogy with radiation of $n$ classical oscillators},\
  }\href {https://doi.org/10.1103/PhysRevA.81.053821} {\bibfield  {journal}
  {\bibinfo  {journal} {Phys. Rev. A}\ }\textbf {\bibinfo {volume} {81}},\
  \bibinfo {pages} {053821} (\bibinfo {year} {2010})}\BibitemShut {NoStop}%
\bibitem [{\citenamefont {Ballantine}\ and\ \citenamefont
  {Ruostekoski}(2021{\natexlab{b}})}]{Ballantine21quantum}%
  \BibitemOpen
  \bibfield  {author} {\bibinfo {author} {\bibfnamefont {K.~E.}\ \bibnamefont
  {Ballantine}}\ and\ \bibinfo {author} {\bibfnamefont {J.}~\bibnamefont
  {Ruostekoski}},\ }\bibfield  {title} {\bibinfo {title} {Quantum single-photon
  control, storage, and entanglement generation with planar atomic arrays},\
  }\href {https://doi.org/10.1103/PRXQuantum.2.040362} {\bibfield  {journal}
  {\bibinfo  {journal} {PRX Quantum}\ }\textbf {\bibinfo {volume} {2}},\
  \bibinfo {pages} {040362} (\bibinfo {year} {2021}{\natexlab{b}})}\BibitemShut
  {NoStop}%
\bibitem [{\citenamefont {Ballantine}\ and\ \citenamefont
  {Ruostekoski}(2020{\natexlab{a}})}]{Ballantine20ant}%
  \BibitemOpen
  \bibfield  {author} {\bibinfo {author} {\bibfnamefont {K.~E.}\ \bibnamefont
  {Ballantine}}\ and\ \bibinfo {author} {\bibfnamefont {J.}~\bibnamefont
  {Ruostekoski}},\ }\bibfield  {title} {\bibinfo {title} {Subradiance-protected
  excitation spreading in the generation of collimated photon emission from an
  atomic array},\ }\href {https://doi.org/10.1103/PhysRevResearch.2.023086}
  {\bibfield  {journal} {\bibinfo  {journal} {Phys. Rev. Research}\ }\textbf
  {\bibinfo {volume} {2}},\ \bibinfo {pages} {023086} (\bibinfo {year}
  {2020}{\natexlab{a}})}\BibitemShut {NoStop}%
\bibitem [{\citenamefont {Kubo}(1962)}]{Kubo62}%
  \BibitemOpen
  \bibfield  {author} {\bibinfo {author} {\bibfnamefont {R.}~\bibnamefont
  {Kubo}},\ }\bibfield  {title} {\bibinfo {title} {Generalized cumulant
  expansion method},\ }\href {https://doi.org/10.1143/JPSJ.17.1100} {\bibfield
  {journal} {\bibinfo  {journal} {Journal of the Physical Society of Japan}\
  }\textbf {\bibinfo {volume} {17}},\ \bibinfo {pages} {1100} (\bibinfo {year}
  {1962})}\BibitemShut {NoStop}%
\bibitem [{\citenamefont {Robicheaux}\ and\ \citenamefont
  {Suresh}(2023)}]{Robicheaux23}%
  \BibitemOpen
  \bibfield  {author} {\bibinfo {author} {\bibfnamefont {F.}~\bibnamefont
  {Robicheaux}}\ and\ \bibinfo {author} {\bibfnamefont {D.~A.}\ \bibnamefont
  {Suresh}},\ }\href@noop {} {\bibinfo {title} {Intensity effects of light
  coupling to one- or two-atom arrays of infinite extent}} (\bibinfo {year}
  {2023}),\ \Eprint {https://arxiv.org/abs/2304.09740} {arXiv:2304.09740
  [quant-ph]} \BibitemShut {NoStop}%
\bibitem [{\citenamefont {Rubies-Bigorda}\ \emph {et~al.}(2023)\citenamefont
  {Rubies-Bigorda}, \citenamefont {Ostermann},\ and\ \citenamefont
  {Yelin}}]{Rubies-Bigorda23}%
  \BibitemOpen
  \bibfield  {author} {\bibinfo {author} {\bibfnamefont {O.}~\bibnamefont
  {Rubies-Bigorda}}, \bibinfo {author} {\bibfnamefont {S.}~\bibnamefont
  {Ostermann}},\ and\ \bibinfo {author} {\bibfnamefont {S.~F.}\ \bibnamefont
  {Yelin}},\ }\bibfield  {title} {\bibinfo {title} {Characterizing superradiant
  dynamics in atomic arrays via a cumulant expansion approach},\ }\href
  {https://doi.org/10.1103/PhysRevResearch.5.013091} {\bibfield  {journal}
  {\bibinfo  {journal} {Phys. Rev. Res.}\ }\textbf {\bibinfo {volume} {5}},\
  \bibinfo {pages} {013091} (\bibinfo {year} {2023})}\BibitemShut {NoStop}%
\bibitem [{\citenamefont {Robicheaux}\ and\ \citenamefont
  {Suresh}(2021)}]{Robicheaux21}%
  \BibitemOpen
  \bibfield  {author} {\bibinfo {author} {\bibfnamefont {F.}~\bibnamefont
  {Robicheaux}}\ and\ \bibinfo {author} {\bibfnamefont {D.~A.}\ \bibnamefont
  {Suresh}},\ }\bibfield  {title} {\bibinfo {title} {Beyond lowest order
  mean-field theory for light interacting with atom arrays},\ }\href
  {https://doi.org/10.1103/PhysRevA.104.023702} {\bibfield  {journal} {\bibinfo
   {journal} {Phys. Rev. A}\ }\textbf {\bibinfo {volume} {104}},\ \bibinfo
  {pages} {023702} (\bibinfo {year} {2021})}\BibitemShut {NoStop}%
\bibitem [{\citenamefont {Plankensteiner}\ \emph {et~al.}(2022)\citenamefont
  {Plankensteiner}, \citenamefont {Hotter},\ and\ \citenamefont
  {Ritsch}}]{Plankensteiner22}%
  \BibitemOpen
  \bibfield  {author} {\bibinfo {author} {\bibfnamefont {D.}~\bibnamefont
  {Plankensteiner}}, \bibinfo {author} {\bibfnamefont {C.}~\bibnamefont
  {Hotter}},\ and\ \bibinfo {author} {\bibfnamefont {H.}~\bibnamefont
  {Ritsch}},\ }\bibfield  {title} {\bibinfo {title} {Quantum{C}umulants.jl: {A}
  {J}ulia framework for generalized mean-field equations in open quantum
  systems},\ }\href {https://doi.org/10.22331/q-2022-01-04-617} {\bibfield
  {journal} {\bibinfo  {journal} {{Quantum}}\ }\textbf {\bibinfo {volume}
  {6}},\ \bibinfo {pages} {617} (\bibinfo {year} {2022})}\BibitemShut {NoStop}%
\bibitem [{\citenamefont {Zhang}\ and\ \citenamefont
  {M\o{}lmer}(2019)}]{Zhang2018}%
  \BibitemOpen
  \bibfield  {author} {\bibinfo {author} {\bibfnamefont {Y.-X.}\ \bibnamefont
  {Zhang}}\ and\ \bibinfo {author} {\bibfnamefont {K.}~\bibnamefont
  {M\o{}lmer}},\ }\bibfield  {title} {\bibinfo {title} {Theory of subradiant
  states of a one-dimensional two-level atom chain},\ }\href
  {https://doi.org/10.1103/PhysRevLett.122.203605} {\bibfield  {journal}
  {\bibinfo  {journal} {Phys. Rev. Lett.}\ }\textbf {\bibinfo {volume} {122}},\
  \bibinfo {pages} {203605} (\bibinfo {year} {2019})}\BibitemShut {NoStop}%
\bibitem [{\citenamefont {Henriet}\ \emph {et~al.}(2019)\citenamefont
  {Henriet}, \citenamefont {Douglas}, \citenamefont {Chang},\ and\
  \citenamefont {Albrecht}}]{Henriet2018}%
  \BibitemOpen
  \bibfield  {author} {\bibinfo {author} {\bibfnamefont {L.}~\bibnamefont
  {Henriet}}, \bibinfo {author} {\bibfnamefont {J.~S.}\ \bibnamefont
  {Douglas}}, \bibinfo {author} {\bibfnamefont {D.~E.}\ \bibnamefont {Chang}},\
  and\ \bibinfo {author} {\bibfnamefont {A.}~\bibnamefont {Albrecht}},\
  }\bibfield  {title} {\bibinfo {title} {Critical open-system dynamics in a
  one-dimensional optical-lattice clock},\ }\href
  {https://doi.org/10.1103/PhysRevA.99.023802} {\bibfield  {journal} {\bibinfo
  {journal} {Phys. Rev. A}\ }\textbf {\bibinfo {volume} {99}},\ \bibinfo
  {pages} {023802} (\bibinfo {year} {2019})}\BibitemShut {NoStop}%
\bibitem [{\citenamefont {Hebenstreit}\ \emph {et~al.}(2017)\citenamefont
  {Hebenstreit}, \citenamefont {Kraus}, \citenamefont {Ostermann},\ and\
  \citenamefont {Ritsch}}]{Ritsch_subr}%
  \BibitemOpen
  \bibfield  {author} {\bibinfo {author} {\bibfnamefont {M.}~\bibnamefont
  {Hebenstreit}}, \bibinfo {author} {\bibfnamefont {B.}~\bibnamefont {Kraus}},
  \bibinfo {author} {\bibfnamefont {L.}~\bibnamefont {Ostermann}},\ and\
  \bibinfo {author} {\bibfnamefont {H.}~\bibnamefont {Ritsch}},\ }\bibfield
  {title} {\bibinfo {title} {Subradiance via entanglement in atoms with several
  independent decay channels},\ }\href
  {https://doi.org/10.1103/PhysRevLett.118.143602} {\bibfield  {journal}
  {\bibinfo  {journal} {Phys. Rev. Lett.}\ }\textbf {\bibinfo {volume} {118}},\
  \bibinfo {pages} {143602} (\bibinfo {year} {2017})}\BibitemShut {NoStop}%
\bibitem [{\citenamefont {Zhang}\ \emph {et~al.}(2020)\citenamefont {Zhang},
  \citenamefont {Yu},\ and\ \citenamefont {M\o{}lmer}}]{Zhang20}%
  \BibitemOpen
  \bibfield  {author} {\bibinfo {author} {\bibfnamefont {Y.-X.}\ \bibnamefont
  {Zhang}}, \bibinfo {author} {\bibfnamefont {C.}~\bibnamefont {Yu}},\ and\
  \bibinfo {author} {\bibfnamefont {K.}~\bibnamefont {M\o{}lmer}},\ }\bibfield
  {title} {\bibinfo {title} {Subradiant bound dimer excited states of emitter
  chains coupled to a one dimensional waveguide},\ }\href
  {https://doi.org/10.1103/PhysRevResearch.2.013173} {\bibfield  {journal}
  {\bibinfo  {journal} {Phys. Rev. Research}\ }\textbf {\bibinfo {volume}
  {2}},\ \bibinfo {pages} {013173} (\bibinfo {year} {2020})}\BibitemShut
  {NoStop}%
\bibitem [{\citenamefont {Parmee}\ and\ \citenamefont
  {Cooper}(2019)}]{Parmee19}%
  \BibitemOpen
  \bibfield  {author} {\bibinfo {author} {\bibfnamefont {C.~D.}\ \bibnamefont
  {Parmee}}\ and\ \bibinfo {author} {\bibfnamefont {N.~R.}\ \bibnamefont
  {Cooper}},\ }\bibfield  {title} {\bibinfo {title} {Decay rates and energies
  of free magnons and bound states in dissipative $xxz$ chains},\ }\href
  {https://doi.org/10.1103/PhysRevA.99.063615} {\bibfield  {journal} {\bibinfo
  {journal} {Phys. Rev. A}\ }\textbf {\bibinfo {volume} {99}},\ \bibinfo
  {pages} {063615} (\bibinfo {year} {2019})}\BibitemShut {NoStop}%
\bibitem [{\citenamefont {Wiseman}\ and\ \citenamefont
  {Milburn}(2010)}]{WisemanMilburn}%
  \BibitemOpen
  \bibfield  {author} {\bibinfo {author} {\bibfnamefont {H.~M.}\ \bibnamefont
  {Wiseman}}\ and\ \bibinfo {author} {\bibfnamefont {G.~J.}\ \bibnamefont
  {Milburn}},\ }\href@noop {} {\emph {\bibinfo {title} {Quantum Measurement and
  Control}}},\ \bibinfo {edition} {1st}\ ed.\ (\bibinfo  {publisher} {Cambridge
  University Press, Cambridge},\ \bibinfo {year} {2010})\BibitemShut {NoStop}%
\bibitem [{\citenamefont {Andrews}\ \emph {et~al.}(1996)\citenamefont
  {Andrews}, \citenamefont {Mewes}, \citenamefont {Van~Druten}, \citenamefont
  {Durfee}, \citenamefont {Kurn},\ and\ \citenamefont
  {Ketterle}}]{andrews1996}%
  \BibitemOpen
  \bibfield  {author} {\bibinfo {author} {\bibfnamefont {M.}~\bibnamefont
  {Andrews}}, \bibinfo {author} {\bibfnamefont {M.-O.}\ \bibnamefont {Mewes}},
  \bibinfo {author} {\bibfnamefont {N.}~\bibnamefont {Van~Druten}}, \bibinfo
  {author} {\bibfnamefont {D.}~\bibnamefont {Durfee}}, \bibinfo {author}
  {\bibfnamefont {D.}~\bibnamefont {Kurn}},\ and\ \bibinfo {author}
  {\bibfnamefont {W.}~\bibnamefont {Ketterle}},\ }\bibfield  {title} {\bibinfo
  {title} {Direct, nondestructive observation of a {Bose} condensate},\ }\href
  {https://doi.org/10.1126/science.273.5271.84} {\bibfield  {journal} {\bibinfo
   {journal} {Science}\ }\textbf {\bibinfo {volume} {273}},\ \bibinfo {pages}
  {84} (\bibinfo {year} {1996})}\BibitemShut {NoStop}%
\bibitem [{\citenamefont {Meystre}\ and\ \citenamefont
  {Sargent}(1998)}]{meystre1998}%
  \BibitemOpen
  \bibfield  {author} {\bibinfo {author} {\bibfnamefont {P.}~\bibnamefont
  {Meystre}}\ and\ \bibinfo {author} {\bibfnamefont {M.}~\bibnamefont
  {Sargent}},\ }\href {https://books.google.co.uk/books?id=dWnIOHloxoEC} {\emph
  {\bibinfo {title} {Elements of Quantum Optics}}}\ (\bibinfo  {publisher}
  {Springer Berlin Heidelberg},\ \bibinfo {year} {1998})\BibitemShut {NoStop}%
\bibitem [{\citenamefont {Dalibard}\ \emph {et~al.}(1992)\citenamefont
  {Dalibard}, \citenamefont {Castin},\ and\ \citenamefont
  {M\o{}lmer}}]{dalibard1992}%
  \BibitemOpen
  \bibfield  {author} {\bibinfo {author} {\bibfnamefont {J.}~\bibnamefont
  {Dalibard}}, \bibinfo {author} {\bibfnamefont {Y.}~\bibnamefont {Castin}},\
  and\ \bibinfo {author} {\bibfnamefont {K.}~\bibnamefont {M\o{}lmer}},\
  }\bibfield  {title} {\bibinfo {title} {Wave-function approach to dissipative
  processes in quantum optics},\ }\href
  {https://doi.org/10.1103/PhysRevLett.68.580} {\bibfield  {journal} {\bibinfo
  {journal} {Phys. Rev. Lett.}\ }\textbf {\bibinfo {volume} {68}},\ \bibinfo
  {pages} {580} (\bibinfo {year} {1992})}\BibitemShut {NoStop}%
\bibitem [{\citenamefont {Tian}\ and\ \citenamefont
  {Carmichael}(1992)}]{Tian92}%
  \BibitemOpen
  \bibfield  {author} {\bibinfo {author} {\bibfnamefont {L.}~\bibnamefont
  {Tian}}\ and\ \bibinfo {author} {\bibfnamefont {H.~J.}\ \bibnamefont
  {Carmichael}},\ }\bibfield  {title} {\bibinfo {title} {Quantum trajectory
  simulations of two-state behavior in an optical cavity containing one atom},\
  }\href {https://doi.org/10.1103/PhysRevA.46.R6801} {\bibfield  {journal}
  {\bibinfo  {journal} {Phys. Rev. A}\ }\textbf {\bibinfo {volume} {46}},\
  \bibinfo {pages} {R6801} (\bibinfo {year} {1992})}\BibitemShut {NoStop}%
\bibitem [{\citenamefont {Dum}\ \emph {et~al.}(1992)\citenamefont {Dum},
  \citenamefont {Zoller},\ and\ \citenamefont {Ritsch}}]{Dum92}%
  \BibitemOpen
  \bibfield  {author} {\bibinfo {author} {\bibfnamefont {R.}~\bibnamefont
  {Dum}}, \bibinfo {author} {\bibfnamefont {P.}~\bibnamefont {Zoller}},\ and\
  \bibinfo {author} {\bibfnamefont {H.}~\bibnamefont {Ritsch}},\ }\bibfield
  {title} {\bibinfo {title} {Monte carlo simulation of the atomic master
  equation for spontaneous emission},\ }\href
  {https://doi.org/10.1103/PhysRevA.45.4879} {\bibfield  {journal} {\bibinfo
  {journal} {Phys. Rev. A}\ }\textbf {\bibinfo {volume} {45}},\ \bibinfo
  {pages} {4879} (\bibinfo {year} {1992})}\BibitemShut {NoStop}%
\bibitem [{\citenamefont {M\o{}lmer}\ \emph {et~al.}(1993)\citenamefont
  {M\o{}lmer}, \citenamefont {Castin},\ and\ \citenamefont
  {Dalibard}}]{molmer1993}%
  \BibitemOpen
  \bibfield  {author} {\bibinfo {author} {\bibfnamefont {K.}~\bibnamefont
  {M\o{}lmer}}, \bibinfo {author} {\bibfnamefont {Y.}~\bibnamefont {Castin}},\
  and\ \bibinfo {author} {\bibfnamefont {J.}~\bibnamefont {Dalibard}},\
  }\bibfield  {title} {\bibinfo {title} {{M}onte {C}arlo wave-function method
  in quantum optics},\ }\href {https://doi.org/10.1364/JOSAB.10.000524}
  {\bibfield  {journal} {\bibinfo  {journal} {J. Opt. Soc. Am. B}\ }\textbf
  {\bibinfo {volume} {10}},\ \bibinfo {pages} {524} (\bibinfo {year}
  {1993})}\BibitemShut {NoStop}%
\bibitem [{\citenamefont {Clemens}\ \emph {et~al.}(2003)\citenamefont
  {Clemens}, \citenamefont {Horvath}, \citenamefont {Sanders},\ and\
  \citenamefont {Carmichael}}]{clemens2003a}%
  \BibitemOpen
  \bibfield  {author} {\bibinfo {author} {\bibfnamefont {J.~P.}\ \bibnamefont
  {Clemens}}, \bibinfo {author} {\bibfnamefont {L.}~\bibnamefont {Horvath}},
  \bibinfo {author} {\bibfnamefont {B.~C.}\ \bibnamefont {Sanders}},\ and\
  \bibinfo {author} {\bibfnamefont {H.~J.}\ \bibnamefont {Carmichael}},\
  }\bibfield  {title} {\bibinfo {title} {Collective spontaneous emission from a
  line of atoms},\ }\href {https://doi.org/10.1103/PhysRevA.68.023809}
  {\bibfield  {journal} {\bibinfo  {journal} {Phys. Rev. A}\ }\textbf {\bibinfo
  {volume} {68}},\ \bibinfo {pages} {023809} (\bibinfo {year}
  {2003})}\BibitemShut {NoStop}%
\bibitem [{\citenamefont {Carmichael}\ and\ \citenamefont
  {Kim}(2000)}]{carmichael2000}%
  \BibitemOpen
  \bibfield  {author} {\bibinfo {author} {\bibfnamefont {H.}~\bibnamefont
  {Carmichael}}\ and\ \bibinfo {author} {\bibfnamefont {K.}~\bibnamefont
  {Kim}},\ }\bibfield  {title} {\bibinfo {title} {A quantum trajectory
  unraveling of the superradiance master equation},\ }\href
  {https://doi.org/10.1016/S0030-4018(99)00694-X} {\bibfield  {journal}
  {\bibinfo  {journal} {Opt. Commun.}\ }\textbf {\bibinfo {volume} {179}},\
  \bibinfo {pages} {417} (\bibinfo {year} {2000})}\BibitemShut {NoStop}%
\bibitem [{\citenamefont {Morsch}\ and\ \citenamefont
  {Oberthaler}(2006)}]{Morsch06}%
  \BibitemOpen
  \bibfield  {author} {\bibinfo {author} {\bibfnamefont {O.}~\bibnamefont
  {Morsch}}\ and\ \bibinfo {author} {\bibfnamefont {M.}~\bibnamefont
  {Oberthaler}},\ }\bibfield  {title} {\bibinfo {title} {Dynamics of
  bose-einstein condensates in optical lattices},\ }\href
  {https://doi.org/10.1103/RevModPhys.78.179} {\bibfield  {journal} {\bibinfo
  {journal} {Rev. Mod. Phys.}\ }\textbf {\bibinfo {volume} {78}},\ \bibinfo
  {pages} {179} (\bibinfo {year} {2006})}\BibitemShut {NoStop}%
\bibitem [{\citenamefont {Jenkins}\ and\ \citenamefont
  {Ruostekoski}(2012)}]{Jenkins2012a}%
  \BibitemOpen
  \bibfield  {author} {\bibinfo {author} {\bibfnamefont {S.~D.}\ \bibnamefont
  {Jenkins}}\ and\ \bibinfo {author} {\bibfnamefont {J.}~\bibnamefont
  {Ruostekoski}},\ }\bibfield  {title} {\bibinfo {title} {Controlled
  manipulation of light by cooperative response of atoms in an optical
  lattice},\ }\href {https://doi.org/10.1103/PhysRevA.86.031602} {\bibfield
  {journal} {\bibinfo  {journal} {Phys. Rev. A}\ }\textbf {\bibinfo {volume}
  {86}},\ \bibinfo {pages} {031602} (\bibinfo {year} {2012})}\BibitemShut
  {NoStop}%
\bibitem [{\citenamefont {van Rossum}\ and\ \citenamefont
  {Nieuwenhuizen}(1999)}]{vanRossum}%
  \BibitemOpen
  \bibfield  {author} {\bibinfo {author} {\bibfnamefont {M.~C.~W.}\
  \bibnamefont {van Rossum}}\ and\ \bibinfo {author} {\bibfnamefont {T.~M.}\
  \bibnamefont {Nieuwenhuizen}},\ }\bibfield  {title} {\bibinfo {title}
  {Multiple scattering of classical waves: microscopy, mesoscopy, and
  diffusion},\ }\href {https://doi.org/10.1103/RevModPhys.71.313} {\bibfield
  {journal} {\bibinfo  {journal} {Rev. Mod. Phys.}\ }\textbf {\bibinfo {volume}
  {71}},\ \bibinfo {pages} {313} (\bibinfo {year} {1999})}\BibitemShut
  {NoStop}%
\bibitem [{\citenamefont {Kupriyanov}\ \emph {et~al.}(2017)\citenamefont
  {Kupriyanov}, \citenamefont {Sokolov},\ and\ \citenamefont
  {Havey}}]{Kupriyanov2017}%
  \BibitemOpen
  \bibfield  {author} {\bibinfo {author} {\bibfnamefont {D.}~\bibnamefont
  {Kupriyanov}}, \bibinfo {author} {\bibfnamefont {I.}~\bibnamefont
  {Sokolov}},\ and\ \bibinfo {author} {\bibfnamefont {M.}~\bibnamefont
  {Havey}},\ }\bibfield  {title} {\bibinfo {title} {Mesoscopic coherence in
  light scattering from cold, optically dense and disordered atomic systems},\
  }\href {https://doi.org/https://doi.org/10.1016/j.physrep.2016.12.004}
  {\bibfield  {journal} {\bibinfo  {journal} {Physics Reports}\ }\textbf
  {\bibinfo {volume} {671}},\ \bibinfo {pages} {1} (\bibinfo {year}
  {2017})}\BibitemShut {NoStop}%
\bibitem [{\citenamefont {Sperling}\ \emph {et~al.}(2016)\citenamefont
  {Sperling}, \citenamefont {Schertel}, \citenamefont {Ackermann},
  \citenamefont {Aubry}, \citenamefont {Aegerter},\ and\ \citenamefont
  {Maret}}]{Sperling2016}%
  \BibitemOpen
  \bibfield  {author} {\bibinfo {author} {\bibfnamefont {T.}~\bibnamefont
  {Sperling}}, \bibinfo {author} {\bibfnamefont {L.}~\bibnamefont {Schertel}},
  \bibinfo {author} {\bibfnamefont {M.}~\bibnamefont {Ackermann}}, \bibinfo
  {author} {\bibfnamefont {G.~J.}\ \bibnamefont {Aubry}}, \bibinfo {author}
  {\bibfnamefont {C.~M.}\ \bibnamefont {Aegerter}},\ and\ \bibinfo {author}
  {\bibfnamefont {G.}~\bibnamefont {Maret}},\ }\bibfield  {title} {\bibinfo
  {title} {Can {3D} light localization be reached in `white paint'?},\ }\href
  {https://doi.org/10.1088/1367-2630/18/1/013039} {\bibfield  {journal}
  {\bibinfo  {journal} {New J. Phys.}\ }\textbf {\bibinfo {volume} {18}},\
  \bibinfo {pages} {013039} (\bibinfo {year} {2016})}\BibitemShut {NoStop}%
\bibitem [{\citenamefont {Skipetrov}\ and\ \citenamefont
  {Sokolov}(2014)}]{Skipetrov14}%
  \BibitemOpen
  \bibfield  {author} {\bibinfo {author} {\bibfnamefont {S.~E.}\ \bibnamefont
  {Skipetrov}}\ and\ \bibinfo {author} {\bibfnamefont {I.~M.}\ \bibnamefont
  {Sokolov}},\ }\bibfield  {title} {\bibinfo {title} {Absence of anderson
  localization of light in a random ensemble of point scatterers},\ }\href
  {https://doi.org/10.1103/PhysRevLett.112.023905} {\bibfield  {journal}
  {\bibinfo  {journal} {Phys. Rev. Lett.}\ }\textbf {\bibinfo {volume} {112}},\
  \bibinfo {pages} {023905} (\bibinfo {year} {2014})}\BibitemShut {NoStop}%
\bibitem [{\citenamefont {Segev}\ \emph {et~al.}(2013)\citenamefont {Segev},
  \citenamefont {Silberberg},\ and\ \citenamefont
  {Christodoulides}}]{Segev2013}%
  \BibitemOpen
  \bibfield  {author} {\bibinfo {author} {\bibfnamefont {M.}~\bibnamefont
  {Segev}}, \bibinfo {author} {\bibfnamefont {Y.}~\bibnamefont {Silberberg}},\
  and\ \bibinfo {author} {\bibfnamefont {D.~N.}\ \bibnamefont
  {Christodoulides}},\ }\bibfield  {title} {\bibinfo {title} {Anderson
  localization of light},\ }\href@noop {} {\bibfield  {journal} {\bibinfo
  {journal} {Nat. Phot.}\ }\textbf {\bibinfo {volume} {7}},\ \bibinfo {pages}
  {197} (\bibinfo {year} {2013})}\BibitemShut {NoStop}%
\bibitem [{\citenamefont {Skipetrov}(2018)}]{Skipetrov18}%
  \BibitemOpen
  \bibfield  {author} {\bibinfo {author} {\bibfnamefont {S.~E.}\ \bibnamefont
  {Skipetrov}},\ }\bibfield  {title} {\bibinfo {title} {Localization transition
  for light scattering by cold atoms in an external magnetic field},\ }\href
  {https://doi.org/10.1103/PhysRevLett.121.093601} {\bibfield  {journal}
  {\bibinfo  {journal} {Phys. Rev. Lett.}\ }\textbf {\bibinfo {volume} {121}},\
  \bibinfo {pages} {093601} (\bibinfo {year} {2018})}\BibitemShut {NoStop}%
\bibitem [{\citenamefont {Ruostekoski}\ and\ \citenamefont
  {Javanainen}(1999)}]{Ruostekoski1999a}%
  \BibitemOpen
  \bibfield  {author} {\bibinfo {author} {\bibfnamefont {J.}~\bibnamefont
  {Ruostekoski}}\ and\ \bibinfo {author} {\bibfnamefont {J.}~\bibnamefont
  {Javanainen}},\ }\bibfield  {title} {\bibinfo {title} {Optical linewidth of a
  low density fermi-dirac gas},\ }\href
  {https://doi.org/10.1103/PhysRevLett.82.4741} {\bibfield  {journal} {\bibinfo
   {journal} {Phys. Rev. Lett.}\ }\textbf {\bibinfo {volume} {82}},\ \bibinfo
  {pages} {4741} (\bibinfo {year} {1999})}\BibitemShut {NoStop}%
\bibitem [{\citenamefont {Javanainen}\ \emph {et~al.}(1999)\citenamefont
  {Javanainen}, \citenamefont {Ruostekoski}, \citenamefont {Vestergaard},\ and\
  \citenamefont {Francis}}]{Javanainen1999a}%
  \BibitemOpen
  \bibfield  {author} {\bibinfo {author} {\bibfnamefont {J.}~\bibnamefont
  {Javanainen}}, \bibinfo {author} {\bibfnamefont {J.}~\bibnamefont
  {Ruostekoski}}, \bibinfo {author} {\bibfnamefont {B.}~\bibnamefont
  {Vestergaard}},\ and\ \bibinfo {author} {\bibfnamefont {M.~R.}\ \bibnamefont
  {Francis}},\ }\bibfield  {title} {\bibinfo {title} {One-dimensional modeling
  of light propagation in dense and degenerate samples},\ }\href
  {https://doi.org/10.1103/PhysRevA.59.649} {\bibfield  {journal} {\bibinfo
  {journal} {Phys. Rev. A}\ }\textbf {\bibinfo {volume} {59}},\ \bibinfo
  {pages} {649} (\bibinfo {year} {1999})}\BibitemShut {NoStop}%
\bibitem [{\citenamefont {Ruostekoski}\ and\ \citenamefont
  {Javanainen}(2016)}]{Ruostekoski_waveguide}%
  \BibitemOpen
  \bibfield  {author} {\bibinfo {author} {\bibfnamefont {J.}~\bibnamefont
  {Ruostekoski}}\ and\ \bibinfo {author} {\bibfnamefont {J.}~\bibnamefont
  {Javanainen}},\ }\bibfield  {title} {\bibinfo {title} {Emergence of
  correlated optics in one-dimensional waveguides for classical and quantum
  atomic gases},\ }\href {https://doi.org/10.1103/PhysRevLett.117.143602}
  {\bibfield  {journal} {\bibinfo  {journal} {Phys. Rev. Lett.}\ }\textbf
  {\bibinfo {volume} {117}},\ \bibinfo {pages} {143602} (\bibinfo {year}
  {2016})}\BibitemShut {NoStop}%
\bibitem [{\citenamefont {Srakaew}\ \emph {et~al.}(2023)\citenamefont
  {Srakaew}, \citenamefont {Weckesser}, \citenamefont {Hollerith},
  \citenamefont {Wei}, \citenamefont {Adler}, \citenamefont {Bloch},\ and\
  \citenamefont {Zeiher}}]{Srakaew22}%
  \BibitemOpen
  \bibfield  {author} {\bibinfo {author} {\bibfnamefont {K.}~\bibnamefont
  {Srakaew}}, \bibinfo {author} {\bibfnamefont {P.}~\bibnamefont {Weckesser}},
  \bibinfo {author} {\bibfnamefont {S.}~\bibnamefont {Hollerith}}, \bibinfo
  {author} {\bibfnamefont {D.}~\bibnamefont {Wei}}, \bibinfo {author}
  {\bibfnamefont {D.}~\bibnamefont {Adler}}, \bibinfo {author} {\bibfnamefont
  {I.}~\bibnamefont {Bloch}},\ and\ \bibinfo {author} {\bibfnamefont
  {J.}~\bibnamefont {Zeiher}},\ }\bibfield  {title} {\bibinfo {title} {A
  subwavelength atomic array switched by a single rydberg atom},\ }\href
  {https://doi.org/10.1038/s41567-023-01959-y} {\bibfield  {journal} {\bibinfo
  {journal} {Nature Physics}\ }\textbf {\bibinfo {volume} {19}},\ \bibinfo
  {pages} {714} (\bibinfo {year} {2023})}\BibitemShut {NoStop}%
\bibitem [{\citenamefont {Bakr}\ \emph {et~al.}(2009)\citenamefont {Bakr},
  \citenamefont {Gillen}, \citenamefont {Peng}, \citenamefont {F{\"{o}}lling},\
  and\ \citenamefont {Greiner}}]{BakrEtAlNature2009}%
  \BibitemOpen
  \bibfield  {author} {\bibinfo {author} {\bibfnamefont {W.~S.}\ \bibnamefont
  {Bakr}}, \bibinfo {author} {\bibfnamefont {J.~I.}\ \bibnamefont {Gillen}},
  \bibinfo {author} {\bibfnamefont {A.}~\bibnamefont {Peng}}, \bibinfo {author}
  {\bibfnamefont {S.}~\bibnamefont {F{\"{o}}lling}},\ and\ \bibinfo {author}
  {\bibfnamefont {M.}~\bibnamefont {Greiner}},\ }\bibfield  {title} {\bibinfo
  {title} {A quantum gas microscope for detecting single atoms in a
  hubbard-regime optical lattice},\ }\href
  {https://doi.org/10.1038/nature08482} {\bibfield  {journal} {\bibinfo
  {journal} {Nature}\ }\textbf {\bibinfo {volume} {462}},\ \bibinfo {pages}
  {74} (\bibinfo {year} {2009})}\BibitemShut {NoStop}%
\bibitem [{\citenamefont {Sherson}\ \emph {et~al.}(2010)\citenamefont
  {Sherson}, \citenamefont {Weitenberg}, \citenamefont {Endres}, \citenamefont
  {Cheneau}, \citenamefont {Bloch},\ and\ \citenamefont
  {Kuhr}}]{ShersonEtAlNature2010}%
  \BibitemOpen
  \bibfield  {author} {\bibinfo {author} {\bibfnamefont {J.~F.}\ \bibnamefont
  {Sherson}}, \bibinfo {author} {\bibfnamefont {C.}~\bibnamefont {Weitenberg}},
  \bibinfo {author} {\bibfnamefont {M.}~\bibnamefont {Endres}}, \bibinfo
  {author} {\bibfnamefont {M.}~\bibnamefont {Cheneau}}, \bibinfo {author}
  {\bibfnamefont {I.}~\bibnamefont {Bloch}},\ and\ \bibinfo {author}
  {\bibfnamefont {S.}~\bibnamefont {Kuhr}},\ }\bibfield  {title} {\bibinfo
  {title} {Single-atom-resolved fluorescence imaging of an atomic mott
  insulator},\ }\href {https://doi.org/10.1038/nature09378} {\bibfield
  {journal} {\bibinfo  {journal} {Nature}\ }\textbf {\bibinfo {volume} {467}},\
  \bibinfo {pages} {68} (\bibinfo {year} {2010})}\BibitemShut {NoStop}%
\bibitem [{\citenamefont {Weitenberg}\ \emph {et~al.}(2011)\citenamefont
  {Weitenberg}, \citenamefont {Endres}, \citenamefont {Sherson}, \citenamefont
  {Cheneau}, \citenamefont {Schau{\ss}}, \citenamefont {Fukuhara},
  \citenamefont {Bloch},\ and\ \citenamefont {Kuhr}}]{Weitenberg2011}%
  \BibitemOpen
  \bibfield  {author} {\bibinfo {author} {\bibfnamefont {C.}~\bibnamefont
  {Weitenberg}}, \bibinfo {author} {\bibfnamefont {M.}~\bibnamefont {Endres}},
  \bibinfo {author} {\bibfnamefont {J.~F.}\ \bibnamefont {Sherson}}, \bibinfo
  {author} {\bibfnamefont {M.}~\bibnamefont {Cheneau}}, \bibinfo {author}
  {\bibfnamefont {P.}~\bibnamefont {Schau{\ss}}}, \bibinfo {author}
  {\bibfnamefont {T.}~\bibnamefont {Fukuhara}}, \bibinfo {author}
  {\bibfnamefont {I.}~\bibnamefont {Bloch}},\ and\ \bibinfo {author}
  {\bibfnamefont {S.}~\bibnamefont {Kuhr}},\ }\bibfield  {title} {\bibinfo
  {title} {{Single-spin addressing in an atomic Mott insulator.}},\ }\href
  {https://doi.org/10.1038/nature09827} {\bibfield  {journal} {\bibinfo
  {journal} {Nature}\ }\textbf {\bibinfo {volume} {471}},\ \bibinfo {pages}
  {319} (\bibinfo {year} {2011})}\BibitemShut {NoStop}%
\bibitem [{\citenamefont {Pi\~neiro Orioli}\ and\ \citenamefont
  {Rey}(2019)}]{Orioli19}%
  \BibitemOpen
  \bibfield  {author} {\bibinfo {author} {\bibfnamefont {A.}~\bibnamefont
  {Pi\~neiro Orioli}}\ and\ \bibinfo {author} {\bibfnamefont {A.~M.}\
  \bibnamefont {Rey}},\ }\bibfield  {title} {\bibinfo {title} {Dark states of
  multilevel fermionic atoms in doubly filled optical lattices},\ }\href
  {https://doi.org/10.1103/PhysRevLett.123.223601} {\bibfield  {journal}
  {\bibinfo  {journal} {Phys. Rev. Lett.}\ }\textbf {\bibinfo {volume} {123}},\
  \bibinfo {pages} {223601} (\bibinfo {year} {2019})}\BibitemShut {NoStop}%
\bibitem [{\citenamefont {Pi\~neiro Orioli}\ and\ \citenamefont
  {Rey}(2020)}]{Orioli20}%
  \BibitemOpen
  \bibfield  {author} {\bibinfo {author} {\bibfnamefont {A.}~\bibnamefont
  {Pi\~neiro Orioli}}\ and\ \bibinfo {author} {\bibfnamefont {A.~M.}\
  \bibnamefont {Rey}},\ }\bibfield  {title} {\bibinfo {title} {Subradiance of
  multilevel fermionic atoms in arrays with filling $n\ensuremath{\ge}2$},\
  }\href {https://doi.org/10.1103/PhysRevA.101.043816} {\bibfield  {journal}
  {\bibinfo  {journal} {Phys. Rev. A}\ }\textbf {\bibinfo {volume} {101}},\
  \bibinfo {pages} {043816} (\bibinfo {year} {2020})}\BibitemShut {NoStop}%
\bibitem [{\citenamefont {Jenkins}\ and\ \citenamefont
  {Ruostekoski}(2013)}]{CAIT}%
  \BibitemOpen
  \bibfield  {author} {\bibinfo {author} {\bibfnamefont {S.~D.}\ \bibnamefont
  {Jenkins}}\ and\ \bibinfo {author} {\bibfnamefont {J.}~\bibnamefont
  {Ruostekoski}},\ }\bibfield  {title} {\bibinfo {title} {Metamaterial
  transparency induced by cooperative electromagnetic interactions},\ }\href
  {https://doi.org/10.1103/PhysRevLett.111.147401} {\bibfield  {journal}
  {\bibinfo  {journal} {Phys. Rev. Lett.}\ }\textbf {\bibinfo {volume} {111}},\
  \bibinfo {pages} {147401} (\bibinfo {year} {2013})}\BibitemShut {NoStop}%
\bibitem [{\citenamefont {Manzoni}\ \emph {et~al.}(2018)\citenamefont
  {Manzoni}, \citenamefont {Moreno-Cardoner}, \citenamefont {Asenjo-Garcia},
  \citenamefont {Porto}, \citenamefont {Gorshkov},\ and\ \citenamefont
  {Chang}}]{Manzoni18}%
  \BibitemOpen
  \bibfield  {author} {\bibinfo {author} {\bibfnamefont {M.~T.}\ \bibnamefont
  {Manzoni}}, \bibinfo {author} {\bibfnamefont {M.}~\bibnamefont
  {Moreno-Cardoner}}, \bibinfo {author} {\bibfnamefont {A.}~\bibnamefont
  {Asenjo-Garcia}}, \bibinfo {author} {\bibfnamefont {J.~V.}\ \bibnamefont
  {Porto}}, \bibinfo {author} {\bibfnamefont {A.~V.}\ \bibnamefont
  {Gorshkov}},\ and\ \bibinfo {author} {\bibfnamefont {D.~E.}\ \bibnamefont
  {Chang}},\ }\bibfield  {title} {\bibinfo {title} {{Optimization of photon
  storage fidelity in ordered atomic arrays}},\ }\href
  {https://doi.org/10.1088/1367-2630/aadb74} {\bibfield  {journal} {\bibinfo
  {journal} {New J. Phys.}\ }\textbf {\bibinfo {volume} {20}},\ \bibinfo
  {pages} {083048} (\bibinfo {year} {2018})}\BibitemShut {NoStop}%
\bibitem [{\citenamefont {Parmee}\ and\ \citenamefont
  {Ruostekoski}(2021)}]{Parmee2021}%
  \BibitemOpen
  \bibfield  {author} {\bibinfo {author} {\bibfnamefont {C.~D.}\ \bibnamefont
  {Parmee}}\ and\ \bibinfo {author} {\bibfnamefont {J.}~\bibnamefont
  {Ruostekoski}},\ }\bibfield  {title} {\bibinfo {title} {Bistable optical
  transmission through arrays of atoms in free space},\ }\href
  {https://doi.org/10.1103/PhysRevA.103.033706} {\bibfield  {journal} {\bibinfo
   {journal} {Phys. Rev. A}\ }\textbf {\bibinfo {volume} {103}},\ \bibinfo
  {pages} {033706} (\bibinfo {year} {2021})}\BibitemShut {NoStop}%
\bibitem [{\citenamefont {Bonifacio}\ and\ \citenamefont
  {Lugiato}(1978)}]{Bonifacio1978}%
  \BibitemOpen
  \bibfield  {author} {\bibinfo {author} {\bibfnamefont {R.}~\bibnamefont
  {Bonifacio}}\ and\ \bibinfo {author} {\bibfnamefont {L.~A.}\ \bibnamefont
  {Lugiato}},\ }\bibfield  {title} {\bibinfo {title} {{Optical bistability and
  cooperative effects in resonance fluorescence}},\ }\href
  {https://doi.org/10.1103/PhysRevA.18.1129} {\bibfield  {journal} {\bibinfo
  {journal} {Phys. Rev. A}\ }\textbf {\bibinfo {volume} {18}},\ \bibinfo
  {pages} {1129} (\bibinfo {year} {1978})}\BibitemShut {NoStop}%
\bibitem [{\citenamefont {Yoo}\ and\ \citenamefont {Javanainen}(2020)}]{Yoo20}%
  \BibitemOpen
  \bibfield  {author} {\bibinfo {author} {\bibfnamefont {S.-M.}\ \bibnamefont
  {Yoo}}\ and\ \bibinfo {author} {\bibfnamefont {J.}~\bibnamefont
  {Javanainen}},\ }\bibfield  {title} {\bibinfo {title} {Light reflection and
  transmission in planar lattices of cold atoms},\ }\href
  {https://doi.org/10.1364/OE.389570} {\bibfield  {journal} {\bibinfo
  {journal} {Opt. Express}\ }\textbf {\bibinfo {volume} {28}},\ \bibinfo
  {pages} {9764} (\bibinfo {year} {2020})}\BibitemShut {NoStop}%
\bibitem [{\citenamefont {Sargent}\ \emph {et~al.}(1977)\citenamefont
  {Sargent}, \citenamefont {Scully},\ and\ \citenamefont
  {Lamb}}]{Sargent_laserphys}%
  \BibitemOpen
  \bibfield  {author} {\bibinfo {author} {\bibfnamefont {M.}~\bibnamefont
  {Sargent}}, \bibinfo {author} {\bibfnamefont {M.~O.}\ \bibnamefont
  {Scully}},\ and\ \bibinfo {author} {\bibfnamefont {W.~E.}\ \bibnamefont
  {Lamb}},\ }\href@noop {} {\emph {\bibinfo {title} {Laser Physics}}},\
  \bibinfo {edition} {1st}\ ed.\ (\bibinfo  {publisher} {Addison-Wesley,
  Reading, MA},\ \bibinfo {year} {1977})\BibitemShut {NoStop}%
\bibitem [{\citenamefont {Javanainen}\ and\ \citenamefont
  {Ruostekoski}(2016)}]{JavanainenMFT}%
  \BibitemOpen
  \bibfield  {author} {\bibinfo {author} {\bibfnamefont {J.}~\bibnamefont
  {Javanainen}}\ and\ \bibinfo {author} {\bibfnamefont {J.}~\bibnamefont
  {Ruostekoski}},\ }\bibfield  {title} {\bibinfo {title} {Light propagation
  beyond the mean-field theory of standard optics},\ }\href
  {https://doi.org/10.1364/OE.24.000993} {\bibfield  {journal} {\bibinfo
  {journal} {Opt. Express}\ }\textbf {\bibinfo {volume} {24}},\ \bibinfo
  {pages} {993} (\bibinfo {year} {2016})}\BibitemShut {NoStop}%
\bibitem [{\citenamefont {Bettles}\ \emph {et~al.}(2016)\citenamefont
  {Bettles}, \citenamefont {Gardiner},\ and\ \citenamefont
  {Adams}}]{Bettles2016}%
  \BibitemOpen
  \bibfield  {author} {\bibinfo {author} {\bibfnamefont {R.~J.}\ \bibnamefont
  {Bettles}}, \bibinfo {author} {\bibfnamefont {S.~A.}\ \bibnamefont
  {Gardiner}},\ and\ \bibinfo {author} {\bibfnamefont {C.~S.}\ \bibnamefont
  {Adams}},\ }\bibfield  {title} {\bibinfo {title} {Enhanced optical cross
  section via collective coupling of atomic dipoles in a {2D} array},\ }\href
  {https://doi.org/10.1103/PhysRevLett.116.103602} {\bibfield  {journal}
  {\bibinfo  {journal} {Phys. Rev. Lett.}\ }\textbf {\bibinfo {volume} {116}},\
  \bibinfo {pages} {103602} (\bibinfo {year} {2016})}\BibitemShut {NoStop}%
\bibitem [{\citenamefont {Tretyakov}(2003)}]{Tretyakov}%
  \BibitemOpen
  \bibfield  {author} {\bibinfo {author} {\bibfnamefont {S.}~\bibnamefont
  {Tretyakov}},\ }\href@noop {} {\emph {\bibinfo {title} {Analytical Modeling
  in Applied Electromagnetics}}},\ \bibinfo {edition} {1st}\ ed.\ (\bibinfo
  {publisher} {Norwood, MA: Artech House},\ \bibinfo {year} {2003})\BibitemShut
  {NoStop}%
\bibitem [{\citenamefont {Laroche}\ \emph {et~al.}(2006)\citenamefont
  {Laroche}, \citenamefont {Albaladejo}, \citenamefont {G\'omez-Medina},\ and\
  \citenamefont {S\'aenz}}]{Laroche2006}%
  \BibitemOpen
  \bibfield  {author} {\bibinfo {author} {\bibfnamefont {M.}~\bibnamefont
  {Laroche}}, \bibinfo {author} {\bibfnamefont {S.}~\bibnamefont {Albaladejo}},
  \bibinfo {author} {\bibfnamefont {R.}~\bibnamefont {G\'omez-Medina}},\ and\
  \bibinfo {author} {\bibfnamefont {J.~J.}\ \bibnamefont {S\'aenz}},\
  }\bibfield  {title} {\bibinfo {title} {Tuning the optical response of
  nanocylinder arrays: An analytical study},\ }\href
  {https://doi.org/10.1103/PhysRevB.74.245422} {\bibfield  {journal} {\bibinfo
  {journal} {Phys. Rev. B}\ }\textbf {\bibinfo {volume} {74}},\ \bibinfo
  {pages} {245422} (\bibinfo {year} {2006})}\BibitemShut {NoStop}%
\bibitem [{\citenamefont {Garc\'{\i}a~de Abajo}(2007)}]{Abajo07}%
  \BibitemOpen
  \bibfield  {author} {\bibinfo {author} {\bibfnamefont {F.~J.}\ \bibnamefont
  {Garc\'{\i}a~de Abajo}},\ }\bibfield  {title} {\bibinfo {title} {Colloquium:
  Light scattering by particle and hole arrays},\ }\href
  {https://doi.org/10.1103/RevModPhys.79.1267} {\bibfield  {journal} {\bibinfo
  {journal} {Rev. Mod. Phys.}\ }\textbf {\bibinfo {volume} {79}},\ \bibinfo
  {pages} {1267} (\bibinfo {year} {2007})}\BibitemShut {NoStop}%
\bibitem [{\citenamefont {Moitra}\ \emph {et~al.}(2014)\citenamefont {Moitra},
  \citenamefont {Slovick}, \citenamefont {{Gang Yu}}, \citenamefont
  {Krishnamurthy},\ and\ \citenamefont {Valentine}}]{Moitra2014}%
  \BibitemOpen
  \bibfield  {author} {\bibinfo {author} {\bibfnamefont {P.}~\bibnamefont
  {Moitra}}, \bibinfo {author} {\bibfnamefont {B.~A.}\ \bibnamefont {Slovick}},
  \bibinfo {author} {\bibfnamefont {Z.}~\bibnamefont {{Gang Yu}}}, \bibinfo
  {author} {\bibfnamefont {S.}~\bibnamefont {Krishnamurthy}},\ and\ \bibinfo
  {author} {\bibfnamefont {J.}~\bibnamefont {Valentine}},\ }\bibfield  {title}
  {\bibinfo {title} {{Experimental demonstration of a broadband all-dielectric
  metamaterial perfect reflector}},\ }\href {https://doi.org/10.1063/1.4873521}
  {\bibfield  {journal} {\bibinfo  {journal} {Appl. Phys. Lett.}\ }\textbf
  {\bibinfo {volume} {104}},\ \bibinfo {pages} {171102} (\bibinfo {year}
  {2014})}\BibitemShut {NoStop}%
\bibitem [{\citenamefont {Moitra}\ \emph {et~al.}(2015)\citenamefont {Moitra},
  \citenamefont {Slovick}, \citenamefont {Li}, \citenamefont {Kravchencko},
  \citenamefont {Briggs}, \citenamefont {Krishnamurthy},\ and\ \citenamefont
  {Valentine}}]{Moitra2015}%
  \BibitemOpen
  \bibfield  {author} {\bibinfo {author} {\bibfnamefont {P.}~\bibnamefont
  {Moitra}}, \bibinfo {author} {\bibfnamefont {B.~A.}\ \bibnamefont {Slovick}},
  \bibinfo {author} {\bibfnamefont {W.}~\bibnamefont {Li}}, \bibinfo {author}
  {\bibfnamefont {I.~I.}\ \bibnamefont {Kravchencko}}, \bibinfo {author}
  {\bibfnamefont {D.~P.}\ \bibnamefont {Briggs}}, \bibinfo {author}
  {\bibfnamefont {S.}~\bibnamefont {Krishnamurthy}},\ and\ \bibinfo {author}
  {\bibfnamefont {J.}~\bibnamefont {Valentine}},\ }\bibfield  {title} {\bibinfo
  {title} {{Large-Scale All-Dielectric Metamaterial Perfect Reflectors}},\
  }\href {https://doi.org/10.1021/acsphotonics.5b00148} {\bibfield  {journal}
  {\bibinfo  {journal} {ACS Photonics}\ }\textbf {\bibinfo {volume} {2}},\
  \bibinfo {pages} {692} (\bibinfo {year} {2015})}\BibitemShut {NoStop}%
\bibitem [{\citenamefont {Scuri}\ \emph {et~al.}(2018)\citenamefont {Scuri},
  \citenamefont {Zhou}, \citenamefont {High}, \citenamefont {Wild},
  \citenamefont {Shu}, \citenamefont {De~Greve}, \citenamefont {Jauregui},
  \citenamefont {Taniguchi}, \citenamefont {Watanabe}, \citenamefont {Kim},
  \citenamefont {Lukin},\ and\ \citenamefont {Park}}]{Scuri18}%
  \BibitemOpen
  \bibfield  {author} {\bibinfo {author} {\bibfnamefont {G.}~\bibnamefont
  {Scuri}}, \bibinfo {author} {\bibfnamefont {Y.}~\bibnamefont {Zhou}},
  \bibinfo {author} {\bibfnamefont {A.~A.}\ \bibnamefont {High}}, \bibinfo
  {author} {\bibfnamefont {D.~S.}\ \bibnamefont {Wild}}, \bibinfo {author}
  {\bibfnamefont {C.}~\bibnamefont {Shu}}, \bibinfo {author} {\bibfnamefont
  {K.}~\bibnamefont {De~Greve}}, \bibinfo {author} {\bibfnamefont {L.~A.}\
  \bibnamefont {Jauregui}}, \bibinfo {author} {\bibfnamefont {T.}~\bibnamefont
  {Taniguchi}}, \bibinfo {author} {\bibfnamefont {K.}~\bibnamefont {Watanabe}},
  \bibinfo {author} {\bibfnamefont {P.}~\bibnamefont {Kim}}, \bibinfo {author}
  {\bibfnamefont {M.~D.}\ \bibnamefont {Lukin}},\ and\ \bibinfo {author}
  {\bibfnamefont {H.}~\bibnamefont {Park}},\ }\bibfield  {title} {\bibinfo
  {title} {Large excitonic reflectivity of monolayer ${\mathrm{mose}}_{2}$
  encapsulated in hexagonal boron nitride},\ }\href
  {https://doi.org/10.1103/PhysRevLett.120.037402} {\bibfield  {journal}
  {\bibinfo  {journal} {Phys. Rev. Lett.}\ }\textbf {\bibinfo {volume} {120}},\
  \bibinfo {pages} {037402} (\bibinfo {year} {2018})}\BibitemShut {NoStop}%
\bibitem [{\citenamefont {Back}\ \emph {et~al.}(2018)\citenamefont {Back},
  \citenamefont {Zeytinoglu}, \citenamefont {Ijaz}, \citenamefont {Kroner},\
  and\ \citenamefont {Imamo\ifmmode~\breve{g}\else \u{g}\fi{}lu}}]{Back18}%
  \BibitemOpen
  \bibfield  {author} {\bibinfo {author} {\bibfnamefont {P.}~\bibnamefont
  {Back}}, \bibinfo {author} {\bibfnamefont {S.}~\bibnamefont {Zeytinoglu}},
  \bibinfo {author} {\bibfnamefont {A.}~\bibnamefont {Ijaz}}, \bibinfo {author}
  {\bibfnamefont {M.}~\bibnamefont {Kroner}},\ and\ \bibinfo {author}
  {\bibfnamefont {A.}~\bibnamefont {Imamo\ifmmode~\breve{g}\else
  \u{g}\fi{}lu}},\ }\bibfield  {title} {\bibinfo {title} {Realization of an
  electrically tunable narrow-bandwidth atomically thin mirror using monolayer
  ${\mathrm{mose}}_{2}$},\ }\href
  {https://doi.org/10.1103/PhysRevLett.120.037401} {\bibfield  {journal}
  {\bibinfo  {journal} {Phys. Rev. Lett.}\ }\textbf {\bibinfo {volume} {120}},\
  \bibinfo {pages} {037401} (\bibinfo {year} {2018})}\BibitemShut {NoStop}%
\bibitem [{\citenamefont {Andersen}\ \emph {et~al.}(2022)\citenamefont
  {Andersen}, \citenamefont {Gelly}, \citenamefont {Scuri}, \citenamefont
  {Dwyer}, \citenamefont {Wild}, \citenamefont {Bekenstein}, \citenamefont
  {Sushko}, \citenamefont {Sung}, \citenamefont {Zhou}, \citenamefont {Zibrov},
  \citenamefont {Liu}, \citenamefont {Joe}, \citenamefont {Watanabe},
  \citenamefont {Taniguchi}, \citenamefont {Yelin}, \citenamefont {Kim},
  \citenamefont {Park},\ and\ \citenamefont {Lukin}}]{Andersen22}%
  \BibitemOpen
  \bibfield  {author} {\bibinfo {author} {\bibfnamefont {T.~I.}\ \bibnamefont
  {Andersen}}, \bibinfo {author} {\bibfnamefont {R.~J.}\ \bibnamefont {Gelly}},
  \bibinfo {author} {\bibfnamefont {G.}~\bibnamefont {Scuri}}, \bibinfo
  {author} {\bibfnamefont {B.~L.}\ \bibnamefont {Dwyer}}, \bibinfo {author}
  {\bibfnamefont {D.~S.}\ \bibnamefont {Wild}}, \bibinfo {author}
  {\bibfnamefont {R.}~\bibnamefont {Bekenstein}}, \bibinfo {author}
  {\bibfnamefont {A.}~\bibnamefont {Sushko}}, \bibinfo {author} {\bibfnamefont
  {J.}~\bibnamefont {Sung}}, \bibinfo {author} {\bibfnamefont {Y.}~\bibnamefont
  {Zhou}}, \bibinfo {author} {\bibfnamefont {A.~A.}\ \bibnamefont {Zibrov}},
  \bibinfo {author} {\bibfnamefont {X.}~\bibnamefont {Liu}}, \bibinfo {author}
  {\bibfnamefont {A.~Y.}\ \bibnamefont {Joe}}, \bibinfo {author} {\bibfnamefont
  {K.}~\bibnamefont {Watanabe}}, \bibinfo {author} {\bibfnamefont
  {T.}~\bibnamefont {Taniguchi}}, \bibinfo {author} {\bibfnamefont {S.~F.}\
  \bibnamefont {Yelin}}, \bibinfo {author} {\bibfnamefont {P.}~\bibnamefont
  {Kim}}, \bibinfo {author} {\bibfnamefont {H.}~\bibnamefont {Park}},\ and\
  \bibinfo {author} {\bibfnamefont {M.~D.}\ \bibnamefont {Lukin}},\ }\bibfield
  {title} {\bibinfo {title} {Beam steering at the nanosecond time scale with an
  atomically thin reflector},\ }\href
  {https://doi.org/10.1038/s41467-022-29976-0} {\bibfield  {journal} {\bibinfo
  {journal} {Nature Communications}\ }\textbf {\bibinfo {volume} {13}},\
  \bibinfo {pages} {3431} (\bibinfo {year} {2022})}\BibitemShut {NoStop}%
\bibitem [{\citenamefont {Zeytino\ifmmode~\breve{g}\else \u{g}\fi{}lu}\ and\
  \citenamefont {\ifmmode \dot{I}\else \.{I}\fi{}mamo\ifmmode~\breve{g}\else
  \u{g}\fi{}lu}(2018)}]{Atac18}%
  \BibitemOpen
  \bibfield  {author} {\bibinfo {author} {\bibfnamefont {S.}~\bibnamefont
  {Zeytino\ifmmode~\breve{g}\else \u{g}\fi{}lu}}\ and\ \bibinfo {author}
  {\bibfnamefont {A.}~\bibnamefont {\ifmmode \dot{I}\else
  \.{I}\fi{}mamo\ifmmode~\breve{g}\else \u{g}\fi{}lu}},\ }\bibfield  {title}
  {\bibinfo {title} {Interaction-induced photon blockade using an atomically
  thin mirror embedded in a microcavity},\ }\href
  {https://doi.org/10.1103/PhysRevA.98.051801} {\bibfield  {journal} {\bibinfo
  {journal} {Phys. Rev. A}\ }\textbf {\bibinfo {volume} {98}},\ \bibinfo
  {pages} {051801} (\bibinfo {year} {2018})}\BibitemShut {NoStop}%
\bibitem [{\citenamefont {Walther}\ \emph {et~al.}(2022)\citenamefont
  {Walther}, \citenamefont {Zhang}, \citenamefont {Yelin},\ and\ \citenamefont
  {Pohl}}]{Walther22}%
  \BibitemOpen
  \bibfield  {author} {\bibinfo {author} {\bibfnamefont {V.}~\bibnamefont
  {Walther}}, \bibinfo {author} {\bibfnamefont {L.}~\bibnamefont {Zhang}},
  \bibinfo {author} {\bibfnamefont {S.~F.}\ \bibnamefont {Yelin}},\ and\
  \bibinfo {author} {\bibfnamefont {T.}~\bibnamefont {Pohl}},\ }\bibfield
  {title} {\bibinfo {title} {Nonclassical light from finite-range interactions
  in a two-dimensional quantum mirror},\ }\href
  {https://doi.org/10.1103/PhysRevB.105.075307} {\bibfield  {journal} {\bibinfo
   {journal} {Phys. Rev. B}\ }\textbf {\bibinfo {volume} {105}},\ \bibinfo
  {pages} {075307} (\bibinfo {year} {2022})}\BibitemShut {NoStop}%
\bibitem [{\citenamefont {Lugiato}(1984)}]{Lugiato1984}%
  \BibitemOpen
  \bibfield  {author} {\bibinfo {author} {\bibfnamefont {L.~A.}\ \bibnamefont
  {Lugiato}},\ }\bibfield  {title} {\bibinfo {title} {{II Theory of Optical
  Bistability}}\ }(\bibinfo  {publisher} {Elsevier},\ \bibinfo {year} {1984})\
  pp.\ \bibinfo {pages} {69 -- 216}\BibitemShut {NoStop}%
\bibitem [{\citenamefont {Bonifacio}\ and\ \citenamefont
  {Lugiato}(1976)}]{bonifacio1976}%
  \BibitemOpen
  \bibfield  {author} {\bibinfo {author} {\bibfnamefont {R.}~\bibnamefont
  {Bonifacio}}\ and\ \bibinfo {author} {\bibfnamefont {L.}~\bibnamefont
  {Lugiato}},\ }\bibfield  {title} {\bibinfo {title} {Cooperative effects and
  bistability for resonance fluorescence},\ }\href
  {https://doi.org/https://doi.org/10.1016/0030-4018(76)90335-7} {\bibfield
  {journal} {\bibinfo  {journal} {Optics Communications}\ }\textbf {\bibinfo
  {volume} {19}},\ \bibinfo {pages} {172 } (\bibinfo {year}
  {1976})}\BibitemShut {NoStop}%
\bibitem [{\citenamefont {Carmichael}\ and\ \citenamefont
  {Walls}(1977)}]{Carmichael1977}%
  \BibitemOpen
  \bibfield  {author} {\bibinfo {author} {\bibfnamefont {H.~J.}\ \bibnamefont
  {Carmichael}}\ and\ \bibinfo {author} {\bibfnamefont {D.~F.}\ \bibnamefont
  {Walls}},\ }\bibfield  {title} {\bibinfo {title} {Hysteresis in the spectrum
  for cooperative resonance fluorescence},\ }\href
  {https://doi.org/10.1088/0022-3700/10/18/002} {\bibfield  {journal} {\bibinfo
   {journal} {Journal of Physics B: Atomic and Molecular Physics}\ }\textbf
  {\bibinfo {volume} {10}},\ \bibinfo {pages} {L685} (\bibinfo {year}
  {1977})}\BibitemShut {NoStop}%
\bibitem [{\citenamefont {Agrawal}\ and\ \citenamefont
  {Carmichael}(1979)}]{Agrawal79}%
  \BibitemOpen
  \bibfield  {author} {\bibinfo {author} {\bibfnamefont {G.~P.}\ \bibnamefont
  {Agrawal}}\ and\ \bibinfo {author} {\bibfnamefont {H.~J.}\ \bibnamefont
  {Carmichael}},\ }\bibfield  {title} {\bibinfo {title} {{Optical bistability
  through nonlinear dispersion and absorption}},\ }\href
  {https://doi.org/10.1103/PhysRevA.19.2074} {\bibfield  {journal} {\bibinfo
  {journal} {Phys. Rev. A}\ }\textbf {\bibinfo {volume} {19}},\ \bibinfo
  {pages} {2074} (\bibinfo {year} {1979})}\BibitemShut {NoStop}%
\bibitem [{\citenamefont {Carmichael}(1986)}]{Carmichael1986a}%
  \BibitemOpen
  \bibfield  {author} {\bibinfo {author} {\bibfnamefont {H.}~\bibnamefont
  {Carmichael}},\ }\bibfield  {title} {\bibinfo {title} {{"Theory of Quantum
  Fluctuations in Optical Bistability"}},\ }in\ \href@noop {} {\emph {\bibinfo
  {booktitle} {Front. Quantum Opt.}}}\ (\bibinfo  {publisher} {Adam Hilger,
  Bristol},\ \bibinfo {year} {1986})\ pp.\ \bibinfo {pages}
  {120--203}\BibitemShut {NoStop}%
\bibitem [{\citenamefont {Ruostekoski}\ and\ \citenamefont
  {Javanainen}(2017)}]{Ruostekoski17}%
  \BibitemOpen
  \bibfield  {author} {\bibinfo {author} {\bibfnamefont {J.}~\bibnamefont
  {Ruostekoski}}\ and\ \bibinfo {author} {\bibfnamefont {J.}~\bibnamefont
  {Javanainen}},\ }\bibfield  {title} {\bibinfo {title} {Arrays of strongly
  coupled atoms in a one-dimensional waveguide},\ }\href
  {https://doi.org/10.1103/PhysRevA.96.033857} {\bibfield  {journal} {\bibinfo
  {journal} {Phys. Rev. A}\ }\textbf {\bibinfo {volume} {96}},\ \bibinfo
  {pages} {033857} (\bibinfo {year} {2017})}\BibitemShut {NoStop}%
\bibitem [{\citenamefont {Hopfield}(1958)}]{Hopfield58}%
  \BibitemOpen
  \bibfield  {author} {\bibinfo {author} {\bibfnamefont {J.~J.}\ \bibnamefont
  {Hopfield}},\ }\bibfield  {title} {\bibinfo {title} {Theory of the
  contribution of excitons to the complex dielectric constant of crystals},\
  }\href {https://doi.org/10.1103/PhysRev.112.1555} {\bibfield  {journal}
  {\bibinfo  {journal} {Phys. Rev.}\ }\textbf {\bibinfo {volume} {112}},\
  \bibinfo {pages} {1555} (\bibinfo {year} {1958})}\BibitemShut {NoStop}%
\bibitem [{\citenamefont {van Coevorden}\ \emph {et~al.}(1996)\citenamefont
  {van Coevorden}, \citenamefont {Sprik}, \citenamefont {Tip},\ and\
  \citenamefont {Lagendijk}}]{Coevorden96}%
  \BibitemOpen
  \bibfield  {author} {\bibinfo {author} {\bibfnamefont {D.~V.}\ \bibnamefont
  {van Coevorden}}, \bibinfo {author} {\bibfnamefont {R.}~\bibnamefont
  {Sprik}}, \bibinfo {author} {\bibfnamefont {A.}~\bibnamefont {Tip}},\ and\
  \bibinfo {author} {\bibfnamefont {A.}~\bibnamefont {Lagendijk}},\ }\bibfield
  {title} {\bibinfo {title} {Photonic band structure of atomic lattices},\
  }\href {https://doi.org/10.1103/PhysRevLett.77.2412} {\bibfield  {journal}
  {\bibinfo  {journal} {Phys. Rev. Lett.}\ }\textbf {\bibinfo {volume} {77}},\
  \bibinfo {pages} {2412} (\bibinfo {year} {1996})}\BibitemShut {NoStop}%
\bibitem [{\citenamefont {de~Vries}\ \emph {et~al.}(1998)\citenamefont
  {de~Vries}, \citenamefont {van Coevorden},\ and\ \citenamefont
  {Lagendijk}}]{devries98}%
  \BibitemOpen
  \bibfield  {author} {\bibinfo {author} {\bibfnamefont {P.}~\bibnamefont
  {de~Vries}}, \bibinfo {author} {\bibfnamefont {D.~V.}\ \bibnamefont {van
  Coevorden}},\ and\ \bibinfo {author} {\bibfnamefont {A.}~\bibnamefont
  {Lagendijk}},\ }\bibfield  {title} {\bibinfo {title} {Point scatterers for
  classical waves},\ }\href {https://doi.org/10.1103/RevModPhys.70.447}
  {\bibfield  {journal} {\bibinfo  {journal} {Rev. Mod. Phys.}\ }\textbf
  {\bibinfo {volume} {70}},\ \bibinfo {pages} {447} (\bibinfo {year}
  {1998})}\BibitemShut {NoStop}%
\bibitem [{\citenamefont {Tip}\ \emph {et~al.}(2000)\citenamefont {Tip},
  \citenamefont {Moroz},\ and\ \citenamefont {Combes}}]{Tip00}%
  \BibitemOpen
  \bibfield  {author} {\bibinfo {author} {\bibfnamefont {A.}~\bibnamefont
  {Tip}}, \bibinfo {author} {\bibfnamefont {A.}~\bibnamefont {Moroz}},\ and\
  \bibinfo {author} {\bibfnamefont {J.~M.}\ \bibnamefont {Combes}},\ }\bibfield
   {title} {\bibinfo {title} {Band structure of absorptive photonic crystals},\
  }\href {https://doi.org/10.1088/0305-4470/33/35/311} {\bibfield  {journal}
  {\bibinfo  {journal} {Journal of Physics A: Mathematical and General}\
  }\textbf {\bibinfo {volume} {33}},\ \bibinfo {pages} {6223} (\bibinfo {year}
  {2000})}\BibitemShut {NoStop}%
\bibitem [{\citenamefont {Klugkist}\ \emph {et~al.}(2006)\citenamefont
  {Klugkist}, \citenamefont {Mostovoy},\ and\ \citenamefont
  {Knoester}}]{Klugkist06}%
  \BibitemOpen
  \bibfield  {author} {\bibinfo {author} {\bibfnamefont {J.~A.}\ \bibnamefont
  {Klugkist}}, \bibinfo {author} {\bibfnamefont {M.}~\bibnamefont {Mostovoy}},\
  and\ \bibinfo {author} {\bibfnamefont {J.}~\bibnamefont {Knoester}},\
  }\bibfield  {title} {\bibinfo {title} {Mode softening, ferroelectric
  transition, and tunable photonic band structures in a point-dipole crystal},\
  }\href {https://doi.org/10.1103/PhysRevLett.96.163903} {\bibfield  {journal}
  {\bibinfo  {journal} {Phys. Rev. Lett.}\ }\textbf {\bibinfo {volume} {96}},\
  \bibinfo {pages} {163903} (\bibinfo {year} {2006})}\BibitemShut {NoStop}%
\bibitem [{\citenamefont {Antezza}\ and\ \citenamefont
  {Castin}(2009{\natexlab{a}})}]{Antezza2009}%
  \BibitemOpen
  \bibfield  {author} {\bibinfo {author} {\bibfnamefont {M.}~\bibnamefont
  {Antezza}}\ and\ \bibinfo {author} {\bibfnamefont {Y.}~\bibnamefont
  {Castin}},\ }\bibfield  {title} {\bibinfo {title} {{Spectrum of Light in a
  Quantum Fluctuating Periodic Structure}},\ }\href
  {https://doi.org/10.1103/PhysRevLett.103.123903} {\bibfield  {journal}
  {\bibinfo  {journal} {Phys. Rev. Lett.}\ }\textbf {\bibinfo {volume} {103}},\
  \bibinfo {pages} {123903} (\bibinfo {year} {2009}{\natexlab{a}})}\BibitemShut
  {NoStop}%
\bibitem [{\citenamefont {Antezza}\ and\ \citenamefont
  {Castin}(2009{\natexlab{b}})}]{Antezza09b}%
  \BibitemOpen
  \bibfield  {author} {\bibinfo {author} {\bibfnamefont {M.}~\bibnamefont
  {Antezza}}\ and\ \bibinfo {author} {\bibfnamefont {Y.}~\bibnamefont
  {Castin}},\ }\bibfield  {title} {\bibinfo {title} {Fano-hopfield model and
  photonic band gaps for an arbitrary atomic lattice},\ }\href
  {https://doi.org/10.1103/PhysRevA.80.013816} {\bibfield  {journal} {\bibinfo
  {journal} {Phys. Rev. A}\ }\textbf {\bibinfo {volume} {80}},\ \bibinfo
  {pages} {013816} (\bibinfo {year} {2009}{\natexlab{b}})}\BibitemShut
  {NoStop}%
\bibitem [{\citenamefont {Antezza}\ and\ \citenamefont
  {Castin}(2013)}]{Castin13}%
  \BibitemOpen
  \bibfield  {author} {\bibinfo {author} {\bibfnamefont {M.}~\bibnamefont
  {Antezza}}\ and\ \bibinfo {author} {\bibfnamefont {Y.}~\bibnamefont
  {Castin}},\ }\bibfield  {title} {\bibinfo {title} {Photonic band gap in an
  imperfect atomic diamond lattice: Penetration depth and effects of finite
  size and vacancies},\ }\href {https://doi.org/10.1103/PhysRevA.88.033844}
  {\bibfield  {journal} {\bibinfo  {journal} {Phys. Rev. A}\ }\textbf {\bibinfo
  {volume} {88}},\ \bibinfo {pages} {033844} (\bibinfo {year}
  {2013})}\BibitemShut {NoStop}%
\bibitem [{\citenamefont {Parmee}\ \emph {et~al.}(2022)\citenamefont {Parmee},
  \citenamefont {Ballantine},\ and\ \citenamefont {Ruostekoski}}]{Parmee22b}%
  \BibitemOpen
  \bibfield  {author} {\bibinfo {author} {\bibfnamefont {C.~D.}\ \bibnamefont
  {Parmee}}, \bibinfo {author} {\bibfnamefont {K.~E.}\ \bibnamefont
  {Ballantine}},\ and\ \bibinfo {author} {\bibfnamefont {J.}~\bibnamefont
  {Ruostekoski}},\ }\bibfield  {title} {\bibinfo {title} {Spontaneous symmetry
  breaking in frustrated triangular atom arrays due to cooperative light
  scattering},\ }\href {https://doi.org/10.1103/PhysRevResearch.4.043039}
  {\bibfield  {journal} {\bibinfo  {journal} {Phys. Rev. Res.}\ }\textbf
  {\bibinfo {volume} {4}},\ \bibinfo {pages} {043039} (\bibinfo {year}
  {2022})}\BibitemShut {NoStop}%
\bibitem [{\citenamefont {Bettles}\ \emph {et~al.}(2015)\citenamefont
  {Bettles}, \citenamefont {Gardiner},\ and\ \citenamefont
  {Adams}}]{Bettles_lattice}%
  \BibitemOpen
  \bibfield  {author} {\bibinfo {author} {\bibfnamefont {R.~J.}\ \bibnamefont
  {Bettles}}, \bibinfo {author} {\bibfnamefont {S.~A.}\ \bibnamefont
  {Gardiner}},\ and\ \bibinfo {author} {\bibfnamefont {C.~S.}\ \bibnamefont
  {Adams}},\ }\bibfield  {title} {\bibinfo {title} {Cooperative ordering in
  lattices of interacting two-level dipoles},\ }\href
  {https://doi.org/10.1103/PhysRevA.92.063822} {\bibfield  {journal} {\bibinfo
  {journal} {Phys. Rev. A}\ }\textbf {\bibinfo {volume} {92}},\ \bibinfo
  {pages} {063822} (\bibinfo {year} {2015})}\BibitemShut {NoStop}%
\bibitem [{\citenamefont {Yoo}\ and\ \citenamefont {Paik}(2016)}]{Yoo2016}%
  \BibitemOpen
  \bibfield  {author} {\bibinfo {author} {\bibfnamefont {S.-M.}\ \bibnamefont
  {Yoo}}\ and\ \bibinfo {author} {\bibfnamefont {S.~M.}\ \bibnamefont {Paik}},\
  }\bibfield  {title} {\bibinfo {title} {{Cooperative optical response of 2D
  dense lattices with strongly correlated dipoles}},\ }\href
  {https://doi.org/10.1364/OE.24.002156} {\bibfield  {journal} {\bibinfo
  {journal} {Opt. Express}\ }\textbf {\bibinfo {volume} {24}},\ \bibinfo
  {pages} {2156} (\bibinfo {year} {2016})}\BibitemShut {NoStop}%
\bibitem [{\citenamefont {Perczel}\ \emph
  {et~al.}(2017{\natexlab{a}})\citenamefont {Perczel}, \citenamefont
  {Borregaard}, \citenamefont {Chang}, \citenamefont {Pichler}, \citenamefont
  {Yelin}, \citenamefont {Zoller},\ and\ \citenamefont {Lukin}}]{Perczel2017a}%
  \BibitemOpen
  \bibfield  {author} {\bibinfo {author} {\bibfnamefont {J.}~\bibnamefont
  {Perczel}}, \bibinfo {author} {\bibfnamefont {J.}~\bibnamefont {Borregaard}},
  \bibinfo {author} {\bibfnamefont {D.~E.}\ \bibnamefont {Chang}}, \bibinfo
  {author} {\bibfnamefont {H.}~\bibnamefont {Pichler}}, \bibinfo {author}
  {\bibfnamefont {S.~F.}\ \bibnamefont {Yelin}}, \bibinfo {author}
  {\bibfnamefont {P.}~\bibnamefont {Zoller}},\ and\ \bibinfo {author}
  {\bibfnamefont {M.~D.}\ \bibnamefont {Lukin}},\ }\bibfield  {title} {\bibinfo
  {title} {{Photonic band structure of two-dimensional atomic lattices}},\
  }\href {https://doi.org/10.1103/PhysRevA.96.063801} {\bibfield  {journal}
  {\bibinfo  {journal} {Phys. Rev. A}\ }\textbf {\bibinfo {volume} {96}},\
  \bibinfo {pages} {063801} (\bibinfo {year} {2017}{\natexlab{a}})}\BibitemShut
  {NoStop}%
\bibitem [{\citenamefont {Ozawa}\ \emph {et~al.}(2019)\citenamefont {Ozawa},
  \citenamefont {Price}, \citenamefont {Amo}, \citenamefont {Goldman},
  \citenamefont {Hafezi}, \citenamefont {Lu}, \citenamefont {Rechtsman},
  \citenamefont {Schuster}, \citenamefont {Simon}, \citenamefont {Zilberberg},\
  and\ \citenamefont {Carusotto}}]{Ozawa_review}%
  \BibitemOpen
  \bibfield  {author} {\bibinfo {author} {\bibfnamefont {T.}~\bibnamefont
  {Ozawa}}, \bibinfo {author} {\bibfnamefont {H.~M.}\ \bibnamefont {Price}},
  \bibinfo {author} {\bibfnamefont {A.}~\bibnamefont {Amo}}, \bibinfo {author}
  {\bibfnamefont {N.}~\bibnamefont {Goldman}}, \bibinfo {author} {\bibfnamefont
  {M.}~\bibnamefont {Hafezi}}, \bibinfo {author} {\bibfnamefont
  {L.}~\bibnamefont {Lu}}, \bibinfo {author} {\bibfnamefont {M.~C.}\
  \bibnamefont {Rechtsman}}, \bibinfo {author} {\bibfnamefont {D.}~\bibnamefont
  {Schuster}}, \bibinfo {author} {\bibfnamefont {J.}~\bibnamefont {Simon}},
  \bibinfo {author} {\bibfnamefont {O.}~\bibnamefont {Zilberberg}},\ and\
  \bibinfo {author} {\bibfnamefont {I.}~\bibnamefont {Carusotto}},\ }\bibfield
  {title} {\bibinfo {title} {Topological photonics},\ }\href
  {https://doi.org/10.1103/RevModPhys.91.015006} {\bibfield  {journal}
  {\bibinfo  {journal} {Rev. Mod. Phys.}\ }\textbf {\bibinfo {volume} {91}},\
  \bibinfo {pages} {015006} (\bibinfo {year} {2019})}\BibitemShut {NoStop}%
\bibitem [{\citenamefont {Khanikaev}\ and\ \citenamefont
  {Shvets}(2017)}]{Khanikaev17}%
  \BibitemOpen
  \bibfield  {author} {\bibinfo {author} {\bibfnamefont {A.~B.}\ \bibnamefont
  {Khanikaev}}\ and\ \bibinfo {author} {\bibfnamefont {G.}~\bibnamefont
  {Shvets}},\ }\bibfield  {title} {\bibinfo {title} {Two-dimensional
  topological photonics},\ }\href {https://doi.org/10.1038/s41566-017-0048-5}
  {\bibfield  {journal} {\bibinfo  {journal} {Nature Photonics}\ }\textbf
  {\bibinfo {volume} {11}},\ \bibinfo {pages} {763} (\bibinfo {year}
  {2017})}\BibitemShut {NoStop}%
\bibitem [{\citenamefont {Haldane}(2017)}]{Haldane17}%
  \BibitemOpen
  \bibfield  {author} {\bibinfo {author} {\bibfnamefont {F.~D.~M.}\
  \bibnamefont {Haldane}},\ }\bibfield  {title} {\bibinfo {title} {Nobel
  lecture: Topological quantum matter},\ }\href
  {https://doi.org/10.1103/RevModPhys.89.040502} {\bibfield  {journal}
  {\bibinfo  {journal} {Rev. Mod. Phys.}\ }\textbf {\bibinfo {volume} {89}},\
  \bibinfo {pages} {040502} (\bibinfo {year} {2017})}\BibitemShut {NoStop}%
\bibitem [{\citenamefont {Cooper}\ \emph {et~al.}(2019)\citenamefont {Cooper},
  \citenamefont {Dalibard},\ and\ \citenamefont {Spielman}}]{Cooper19}%
  \BibitemOpen
  \bibfield  {author} {\bibinfo {author} {\bibfnamefont {N.~R.}\ \bibnamefont
  {Cooper}}, \bibinfo {author} {\bibfnamefont {J.}~\bibnamefont {Dalibard}},\
  and\ \bibinfo {author} {\bibfnamefont {I.~B.}\ \bibnamefont {Spielman}},\
  }\bibfield  {title} {\bibinfo {title} {Topological bands for ultracold
  atoms},\ }\href {https://doi.org/10.1103/RevModPhys.91.015005} {\bibfield
  {journal} {\bibinfo  {journal} {Rev. Mod. Phys.}\ }\textbf {\bibinfo {volume}
  {91}},\ \bibinfo {pages} {015005} (\bibinfo {year} {2019})}\BibitemShut
  {NoStop}%
\bibitem [{\citenamefont {Bettles}\ \emph {et~al.}(2017)\citenamefont
  {Bettles}, \citenamefont {Min\'a\ifmmode~\check{r}\else \v{r}\fi{}},
  \citenamefont {Adams}, \citenamefont {Lesanovsky},\ and\ \citenamefont
  {Olmos}}]{Bettles_17topo}%
  \BibitemOpen
  \bibfield  {author} {\bibinfo {author} {\bibfnamefont {R.~J.}\ \bibnamefont
  {Bettles}}, \bibinfo {author} {\bibfnamefont {J.}~\bibnamefont
  {Min\'a\ifmmode~\check{r}\else \v{r}\fi{}}}, \bibinfo {author} {\bibfnamefont
  {C.~S.}\ \bibnamefont {Adams}}, \bibinfo {author} {\bibfnamefont
  {I.}~\bibnamefont {Lesanovsky}},\ and\ \bibinfo {author} {\bibfnamefont
  {B.}~\bibnamefont {Olmos}},\ }\bibfield  {title} {\bibinfo {title}
  {Topological properties of a dense atomic lattice gas},\ }\href
  {https://doi.org/10.1103/PhysRevA.96.041603} {\bibfield  {journal} {\bibinfo
  {journal} {Phys. Rev. A}\ }\textbf {\bibinfo {volume} {96}},\ \bibinfo
  {pages} {041603} (\bibinfo {year} {2017})}\BibitemShut {NoStop}%
\bibitem [{\citenamefont {Perczel}\ \emph
  {et~al.}(2017{\natexlab{b}})\citenamefont {Perczel}, \citenamefont
  {Borregaard}, \citenamefont {Chang}, \citenamefont {Pichler}, \citenamefont
  {Yelin}, \citenamefont {Zoller},\ and\ \citenamefont {Lukin}}]{Perczel2017b}%
  \BibitemOpen
  \bibfield  {author} {\bibinfo {author} {\bibfnamefont {J.}~\bibnamefont
  {Perczel}}, \bibinfo {author} {\bibfnamefont {J.}~\bibnamefont {Borregaard}},
  \bibinfo {author} {\bibfnamefont {D.~E.}\ \bibnamefont {Chang}}, \bibinfo
  {author} {\bibfnamefont {H.}~\bibnamefont {Pichler}}, \bibinfo {author}
  {\bibfnamefont {S.~F.}\ \bibnamefont {Yelin}}, \bibinfo {author}
  {\bibfnamefont {P.}~\bibnamefont {Zoller}},\ and\ \bibinfo {author}
  {\bibfnamefont {M.~D.}\ \bibnamefont {Lukin}},\ }\bibfield  {title} {\bibinfo
  {title} {{Topological Quantum Optics in Two-Dimensional Atomic Arrays}},\
  }\href {https://doi.org/10.1103/PhysRevLett.119.023603} {\bibfield  {journal}
  {\bibinfo  {journal} {Phys. Rev. Lett.}\ }\textbf {\bibinfo {volume} {119}},\
  \bibinfo {pages} {023603} (\bibinfo {year} {2017}{\natexlab{b}})}\BibitemShut
  {NoStop}%
\bibitem [{\citenamefont {Holzinger}\ \emph {et~al.}(2021)\citenamefont
  {Holzinger}, \citenamefont {Moreno-Cardoner},\ and\ \citenamefont
  {Ritsch}}]{Holzinger21}%
  \BibitemOpen
  \bibfield  {author} {\bibinfo {author} {\bibfnamefont {R.}~\bibnamefont
  {Holzinger}}, \bibinfo {author} {\bibfnamefont {M.}~\bibnamefont
  {Moreno-Cardoner}},\ and\ \bibinfo {author} {\bibfnamefont {H.}~\bibnamefont
  {Ritsch}},\ }\bibfield  {title} {\bibinfo {title} {Nanoscale continuous
  quantum light sources based on driven dipole emitter arrays},\ }\href
  {https://doi.org/10.1063/5.0049270} {\bibfield  {journal} {\bibinfo
  {journal} {Applied Physics Letters}\ }\textbf {\bibinfo {volume} {119}},\
  \bibinfo {pages} {024002} (\bibinfo {year} {2021})}\BibitemShut {NoStop}%
\bibitem [{\citenamefont {Plankensteiner}\ \emph {et~al.}(2015)\citenamefont
  {Plankensteiner}, \citenamefont {Ostermann}, \citenamefont {Ritsch},\ and\
  \citenamefont {Genes}}]{Plankensteiner2015}%
  \BibitemOpen
  \bibfield  {author} {\bibinfo {author} {\bibfnamefont {D.}~\bibnamefont
  {Plankensteiner}}, \bibinfo {author} {\bibfnamefont {L.}~\bibnamefont
  {Ostermann}}, \bibinfo {author} {\bibfnamefont {H.}~\bibnamefont {Ritsch}},\
  and\ \bibinfo {author} {\bibfnamefont {C.}~\bibnamefont {Genes}},\ }\bibfield
   {title} {\bibinfo {title} {Selective protected state preparation of coupled
  dissipative quantum emitters},\ }\href {https://doi.org/10.1038/srep16231}
  {\bibfield  {journal} {\bibinfo  {journal} {Scientific Reports}\ }\textbf
  {\bibinfo {volume} {5}},\ \bibinfo {pages} {16231 EP } (\bibinfo {year}
  {2015})},\ \bibinfo {note} {article}\BibitemShut {NoStop}%
\bibitem [{\citenamefont {Jen}\ \emph {et~al.}(2016)\citenamefont {Jen},
  \citenamefont {Chang},\ and\ \citenamefont {Chen}}]{Jen2016}%
  \BibitemOpen
  \bibfield  {author} {\bibinfo {author} {\bibfnamefont {H.~H.}\ \bibnamefont
  {Jen}}, \bibinfo {author} {\bibfnamefont {M.-S.}\ \bibnamefont {Chang}},\
  and\ \bibinfo {author} {\bibfnamefont {Y.-C.}\ \bibnamefont {Chen}},\
  }\bibfield  {title} {\bibinfo {title} {Cooperative single-photon subradiant
  states},\ }\href {https://doi.org/10.1103/PhysRevA.94.013803} {\bibfield
  {journal} {\bibinfo  {journal} {Phys. Rev. A}\ }\textbf {\bibinfo {volume}
  {94}},\ \bibinfo {pages} {013803} (\bibinfo {year} {2016})}\BibitemShut
  {NoStop}%
\bibitem [{\citenamefont {Ferioli}\ \emph {et~al.}(2021)\citenamefont
  {Ferioli}, \citenamefont {Glicenstein}, \citenamefont {Henriet},
  \citenamefont {Ferrier-Barbut},\ and\ \citenamefont {Browaeys}}]{Ferioli21}%
  \BibitemOpen
  \bibfield  {author} {\bibinfo {author} {\bibfnamefont {G.}~\bibnamefont
  {Ferioli}}, \bibinfo {author} {\bibfnamefont {A.}~\bibnamefont
  {Glicenstein}}, \bibinfo {author} {\bibfnamefont {L.}~\bibnamefont
  {Henriet}}, \bibinfo {author} {\bibfnamefont {I.}~\bibnamefont
  {Ferrier-Barbut}},\ and\ \bibinfo {author} {\bibfnamefont {A.}~\bibnamefont
  {Browaeys}},\ }\bibfield  {title} {\bibinfo {title} {Storage and release of
  subradiant excitations in a dense atomic cloud},\ }\href
  {https://doi.org/10.1103/PhysRevX.11.021031} {\bibfield  {journal} {\bibinfo
  {journal} {Phys. Rev. X}\ }\textbf {\bibinfo {volume} {11}},\ \bibinfo
  {pages} {021031} (\bibinfo {year} {2021})}\BibitemShut {NoStop}%
\bibitem [{Note2()}]{Note2}%
  \BibitemOpen
  \bibinfo {note} {Classical analogy is an electron that starts rotating on a
  circular orbit under the Lorentz force.}\BibitemShut {Stop}%
\bibitem [{\citenamefont {Asenjo-Garcia}\ \emph {et~al.}(2017)\citenamefont
  {Asenjo-Garcia}, \citenamefont {Moreno-Cardoner}, \citenamefont {Albrecht},
  \citenamefont {Kimble},\ and\ \citenamefont {Chang}}]{Asenjo-Garcia2017a}%
  \BibitemOpen
  \bibfield  {author} {\bibinfo {author} {\bibfnamefont {A.}~\bibnamefont
  {Asenjo-Garcia}}, \bibinfo {author} {\bibfnamefont {M.}~\bibnamefont
  {Moreno-Cardoner}}, \bibinfo {author} {\bibfnamefont {A.}~\bibnamefont
  {Albrecht}}, \bibinfo {author} {\bibfnamefont {H.~J.}\ \bibnamefont
  {Kimble}},\ and\ \bibinfo {author} {\bibfnamefont {D.~E.}\ \bibnamefont
  {Chang}},\ }\bibfield  {title} {\bibinfo {title} {{Exponential Improvement in
  Photon Storage Fidelities Using Subradiance and “Selective Radiance” in
  Atomic Arrays}},\ }\href {https://doi.org/10.1103/PhysRevX.7.031024}
  {\bibfield  {journal} {\bibinfo  {journal} {Phys. Rev. X}\ }\textbf {\bibinfo
  {volume} {7}},\ \bibinfo {pages} {031024} (\bibinfo {year}
  {2017})}\BibitemShut {NoStop}%
\bibitem [{\citenamefont {Zhang}\ and\ \citenamefont
  {M\o{}lmer}(2020)}]{Zhang20b}%
  \BibitemOpen
  \bibfield  {author} {\bibinfo {author} {\bibfnamefont {Y.-X.}\ \bibnamefont
  {Zhang}}\ and\ \bibinfo {author} {\bibfnamefont {K.}~\bibnamefont
  {M\o{}lmer}},\ }\bibfield  {title} {\bibinfo {title} {Subradiant emission
  from regular atomic arrays: Universal scaling of decay rates from the
  generalized bloch theorem},\ }\href
  {https://doi.org/10.1103/PhysRevLett.125.253601} {\bibfield  {journal}
  {\bibinfo  {journal} {Phys. Rev. Lett.}\ }\textbf {\bibinfo {volume} {125}},\
  \bibinfo {pages} {253601} (\bibinfo {year} {2020})}\BibitemShut {NoStop}%
\bibitem [{\citenamefont {Gerbier}\ \emph {et~al.}(2006)\citenamefont
  {Gerbier}, \citenamefont {Widera}, \citenamefont {F\"olling}, \citenamefont
  {Mandel},\ and\ \citenamefont {Bloch}}]{gerbier_pra_2006}%
  \BibitemOpen
  \bibfield  {author} {\bibinfo {author} {\bibfnamefont {F.}~\bibnamefont
  {Gerbier}}, \bibinfo {author} {\bibfnamefont {A.}~\bibnamefont {Widera}},
  \bibinfo {author} {\bibfnamefont {S.}~\bibnamefont {F\"olling}}, \bibinfo
  {author} {\bibfnamefont {O.}~\bibnamefont {Mandel}},\ and\ \bibinfo {author}
  {\bibfnamefont {I.}~\bibnamefont {Bloch}},\ }\bibfield  {title} {\bibinfo
  {title} {Resonant control of spin dynamics in ultracold quantum gases by
  microwave dressing},\ }\href {https://doi.org/10.1103/PhysRevA.73.041602}
  {\bibfield  {journal} {\bibinfo  {journal} {Phys. Rev. A}\ }\textbf {\bibinfo
  {volume} {73}},\ \bibinfo {pages} {041602} (\bibinfo {year}
  {2006})}\BibitemShut {NoStop}%
\bibitem [{\citenamefont {Ballantine}\ \emph {et~al.}(2022)\citenamefont
  {Ballantine}, \citenamefont {Wilkowski},\ and\ \citenamefont
  {Ruostekoski}}]{Ballantine22str}%
  \BibitemOpen
  \bibfield  {author} {\bibinfo {author} {\bibfnamefont {K.~E.}\ \bibnamefont
  {Ballantine}}, \bibinfo {author} {\bibfnamefont {D.}~\bibnamefont
  {Wilkowski}},\ and\ \bibinfo {author} {\bibfnamefont {J.}~\bibnamefont
  {Ruostekoski}},\ }\bibfield  {title} {\bibinfo {title} {Optical magnetism and
  wavefront control by arrays of strontium atoms},\ }\href
  {https://doi.org/10.1103/PhysRevResearch.4.033242} {\bibfield  {journal}
  {\bibinfo  {journal} {Phys. Rev. Res.}\ }\textbf {\bibinfo {volume} {4}},\
  \bibinfo {pages} {033242} (\bibinfo {year} {2022})}\BibitemShut {NoStop}%
\bibitem [{\citenamefont {Fleischhauer}\ \emph {et~al.}(2005)\citenamefont
  {Fleischhauer}, \citenamefont {Imamoglu},\ and\ \citenamefont
  {Marangos}}]{FleischhauerEtAlRMP2005}%
  \BibitemOpen
  \bibfield  {author} {\bibinfo {author} {\bibfnamefont {M.}~\bibnamefont
  {Fleischhauer}}, \bibinfo {author} {\bibfnamefont {A.}~\bibnamefont
  {Imamoglu}},\ and\ \bibinfo {author} {\bibfnamefont {J.~P.}\ \bibnamefont
  {Marangos}},\ }\bibfield  {title} {\bibinfo {title} {Electromagnetically
  induced transparency: Optics in coherent media},\ }\href
  {https://doi.org/10.1103/RevModPhys.77.633} {\bibfield  {journal} {\bibinfo
  {journal} {Rev. Mod. Phys.}\ }\textbf {\bibinfo {volume} {77}},\ \bibinfo
  {pages} {633} (\bibinfo {year} {2005})}\BibitemShut {NoStop}%
\bibitem [{\citenamefont {Liu}\ \emph {et~al.}(2001)\citenamefont {Liu},
  \citenamefont {Dutton}, \citenamefont {Behroozi},\ and\ \citenamefont
  {Hau}}]{LiuEtAlNature2001}%
  \BibitemOpen
  \bibfield  {author} {\bibinfo {author} {\bibfnamefont {C.}~\bibnamefont
  {Liu}}, \bibinfo {author} {\bibfnamefont {Z.}~\bibnamefont {Dutton}},
  \bibinfo {author} {\bibfnamefont {C.~H.}\ \bibnamefont {Behroozi}},\ and\
  \bibinfo {author} {\bibfnamefont {L.~H.}\ \bibnamefont {Hau}},\ }\bibfield
  {title} {\bibinfo {title} {Observation of coherent optical information
  storage in and atomic medium using halted light pulses},\ }\href
  {https://doi.org/10.1038/35054017} {\bibfield  {journal} {\bibinfo  {journal}
  {Nature}\ }\textbf {\bibinfo {volume} {409}},\ \bibinfo {pages} {490}
  (\bibinfo {year} {2001})}\BibitemShut {NoStop}%
\bibitem [{\citenamefont {Fleischhauer}\ \emph {et~al.}(2000)\citenamefont
  {Fleischhauer}, \citenamefont {Matsko},\ and\ \citenamefont
  {Scully}}]{Fleisch_magneto}%
  \BibitemOpen
  \bibfield  {author} {\bibinfo {author} {\bibfnamefont {M.}~\bibnamefont
  {Fleischhauer}}, \bibinfo {author} {\bibfnamefont {A.~B.}\ \bibnamefont
  {Matsko}},\ and\ \bibinfo {author} {\bibfnamefont {M.~O.}\ \bibnamefont
  {Scully}},\ }\bibfield  {title} {\bibinfo {title} {Quantum limit of optical
  magnetometry in the presence of ac stark shifts},\ }\href
  {https://doi.org/10.1103/PhysRevA.62.013808} {\bibfield  {journal} {\bibinfo
  {journal} {Phys. Rev. A}\ }\textbf {\bibinfo {volume} {62}},\ \bibinfo
  {pages} {013808} (\bibinfo {year} {2000})}\BibitemShut {NoStop}%
\bibitem [{\citenamefont {Javanainen}(2020)}]{Javanainen20}%
  \BibitemOpen
  \bibfield  {author} {\bibinfo {author} {\bibfnamefont {J.}~\bibnamefont
  {Javanainen}},\ }\bibfield  {title} {\bibinfo {title} {Cooperative band-stop
  filters},\ }\href
  {https://doi.org/https://doi.org/10.1016/j.ijleo.2020.164792} {\bibfield
  {journal} {\bibinfo  {journal} {Optik}\ }\textbf {\bibinfo {volume} {216}},\
  \bibinfo {pages} {164792} (\bibinfo {year} {2020})}\BibitemShut {NoStop}%
\bibitem [{\citenamefont {Ba{\ss}ler}\ \emph {et~al.}(2023)\citenamefont
  {Ba{\ss}ler}, \citenamefont {Reitz}, \citenamefont {Schmidt},\ and\
  \citenamefont {Genes}}]{Bassler23}%
  \BibitemOpen
  \bibfield  {author} {\bibinfo {author} {\bibfnamefont {N.~S.}\ \bibnamefont
  {Ba{\ss}ler}}, \bibinfo {author} {\bibfnamefont {M.}~\bibnamefont {Reitz}},
  \bibinfo {author} {\bibfnamefont {K.~P.}\ \bibnamefont {Schmidt}},\ and\
  \bibinfo {author} {\bibfnamefont {C.}~\bibnamefont {Genes}},\ }\bibfield
  {title} {\bibinfo {title} {Linear optical elements based on cooperative
  subwavelength emitter arrays},\ }\href {https://doi.org/10.1364/OE.476830}
  {\bibfield  {journal} {\bibinfo  {journal} {Opt. Express}\ }\textbf {\bibinfo
  {volume} {31}},\ \bibinfo {pages} {6003} (\bibinfo {year}
  {2023})}\BibitemShut {NoStop}%
\bibitem [{\citenamefont {Ballantine}\ and\ \citenamefont
  {Ruostekoski}(2020{\natexlab{b}})}]{Ballantine20Huygens}%
  \BibitemOpen
  \bibfield  {author} {\bibinfo {author} {\bibfnamefont {K.~E.}\ \bibnamefont
  {Ballantine}}\ and\ \bibinfo {author} {\bibfnamefont {J.}~\bibnamefont
  {Ruostekoski}},\ }\bibfield  {title} {\bibinfo {title} {Optical magnetism and
  {Huygens'} surfaces in arrays of atoms induced by cooperative responses},\
  }\href {https://doi.org/10.1103/PhysRevLett.125.143604} {\bibfield  {journal}
  {\bibinfo  {journal} {Phys. Rev. Lett.}\ }\textbf {\bibinfo {volume} {125}},\
  \bibinfo {pages} {143604} (\bibinfo {year} {2020}{\natexlab{b}})}\BibitemShut
  {NoStop}%
\bibitem [{\citenamefont {Alaee}\ \emph {et~al.}(2020)\citenamefont {Alaee},
  \citenamefont {Gurlek}, \citenamefont {Albooyeh}, \citenamefont
  {Mart\'{\i}n-Cano},\ and\ \citenamefont {Sandoghdar}}]{Alaee20}%
  \BibitemOpen
  \bibfield  {author} {\bibinfo {author} {\bibfnamefont {R.}~\bibnamefont
  {Alaee}}, \bibinfo {author} {\bibfnamefont {B.}~\bibnamefont {Gurlek}},
  \bibinfo {author} {\bibfnamefont {M.}~\bibnamefont {Albooyeh}}, \bibinfo
  {author} {\bibfnamefont {D.}~\bibnamefont {Mart\'{\i}n-Cano}},\ and\ \bibinfo
  {author} {\bibfnamefont {V.}~\bibnamefont {Sandoghdar}},\ }\bibfield  {title}
  {\bibinfo {title} {Quantum metamaterials with magnetic response at optical
  frequencies},\ }\href {https://doi.org/10.1103/PhysRevLett.125.063601}
  {\bibfield  {journal} {\bibinfo  {journal} {Phys. Rev. Lett.}\ }\textbf
  {\bibinfo {volume} {125}},\ \bibinfo {pages} {063601} (\bibinfo {year}
  {2020})}\BibitemShut {NoStop}%
\bibitem [{\citenamefont {Ballantine}\ and\ \citenamefont
  {Ruostekoski}(2021{\natexlab{c}})}]{Ballantine21wavefront}%
  \BibitemOpen
  \bibfield  {author} {\bibinfo {author} {\bibfnamefont {K.~E.}\ \bibnamefont
  {Ballantine}}\ and\ \bibinfo {author} {\bibfnamefont {J.}~\bibnamefont
  {Ruostekoski}},\ }\bibfield  {title} {\bibinfo {title} {Cooperative optical
  wavefront engineering with atomic arrays},\ }\href
  {https://doi.org/doi:10.1515/nanoph-2021-0059} {\bibfield  {journal}
  {\bibinfo  {journal} {Nanophotonics}\ }\textbf {\bibinfo {volume} {10}},\
  \bibinfo {pages} {1901} (\bibinfo {year} {2021}{\natexlab{c}})}\BibitemShut
  {NoStop}%
\bibitem [{\citenamefont {Huygens}(1690)}]{Huygens}%
  \BibitemOpen
  \bibfield  {author} {\bibinfo {author} {\bibfnamefont {C.}~\bibnamefont
  {Huygens}},\ }\href@noop {} {\emph {\bibinfo {title} {Trait{\'e} de la
  Lumi{\'e}re}}}\ (\bibinfo  {publisher} {Pieter van der Aa},\ \bibinfo
  {address} {Leyden},\ \bibinfo {year} {1690})\BibitemShut {NoStop}%
\bibitem [{\citenamefont {Love}(1901)}]{Love1901}%
  \BibitemOpen
  \bibfield  {author} {\bibinfo {author} {\bibfnamefont {A.~E.~H.}\
  \bibnamefont {Love}},\ }\bibfield  {title} {\bibinfo {title} {The integration
  of the equations of propagation of electric waves},\ }\href
  {https://doi.org/10.1098/rsta.1901.0013} {\bibfield  {journal} {\bibinfo
  {journal} {Phil. Trans. R. Soc. London A}\ }\textbf {\bibinfo {volume}
  {197}},\ \bibinfo {pages} {1} (\bibinfo {year} {1901})}\BibitemShut {NoStop}%
\bibitem [{\citenamefont {Pfeiffer}\ and\ \citenamefont
  {Grbic}(2013)}]{Pfeiffer13}%
  \BibitemOpen
  \bibfield  {author} {\bibinfo {author} {\bibfnamefont {C.}~\bibnamefont
  {Pfeiffer}}\ and\ \bibinfo {author} {\bibfnamefont {A.}~\bibnamefont
  {Grbic}},\ }\bibfield  {title} {\bibinfo {title} {Metamaterial {H}uygens'
  surfaces: Tailoring wave fronts with reflectionless sheets},\ }\href
  {https://doi.org/10.1103/PhysRevLett.110.197401} {\bibfield  {journal}
  {\bibinfo  {journal} {Phys. Rev. Lett.}\ }\textbf {\bibinfo {volume} {110}},\
  \bibinfo {pages} {197401} (\bibinfo {year} {2013})}\BibitemShut {NoStop}%
\bibitem [{\citenamefont {Decker}\ \emph {et~al.}(2015)\citenamefont {Decker},
  \citenamefont {Staude}, \citenamefont {Falkner}, \citenamefont {Dominguez},
  \citenamefont {Neshev}, \citenamefont {Brener}, \citenamefont {Pertsch},\
  and\ \citenamefont {Kivshar}}]{Decker15}%
  \BibitemOpen
  \bibfield  {author} {\bibinfo {author} {\bibfnamefont {M.}~\bibnamefont
  {Decker}}, \bibinfo {author} {\bibfnamefont {I.}~\bibnamefont {Staude}},
  \bibinfo {author} {\bibfnamefont {M.}~\bibnamefont {Falkner}}, \bibinfo
  {author} {\bibfnamefont {J.}~\bibnamefont {Dominguez}}, \bibinfo {author}
  {\bibfnamefont {D.~N.}\ \bibnamefont {Neshev}}, \bibinfo {author}
  {\bibfnamefont {I.}~\bibnamefont {Brener}}, \bibinfo {author} {\bibfnamefont
  {T.}~\bibnamefont {Pertsch}},\ and\ \bibinfo {author} {\bibfnamefont {Y.~S.}\
  \bibnamefont {Kivshar}},\ }\bibfield  {title} {\bibinfo {title}
  {High-efficiency dielectric huygens’ surfaces},\ }\href
  {https://doi.org/10.1002/adom.201400584} {\bibfield  {journal} {\bibinfo
  {journal} {Advanced Optical Materials}\ }\textbf {\bibinfo {volume} {3}},\
  \bibinfo {pages} {813} (\bibinfo {year} {2015})}\BibitemShut {NoStop}%
\bibitem [{\citenamefont {Yu}\ \emph {et~al.}(2015)\citenamefont {Yu},
  \citenamefont {Zhu}, \citenamefont {Paniagua-Domínguez}, \citenamefont {Fu},
  \citenamefont {Luk'yanchuk},\ and\ \citenamefont {Kuznetsov}}]{Yu15}%
  \BibitemOpen
  \bibfield  {author} {\bibinfo {author} {\bibfnamefont {Y.~F.}\ \bibnamefont
  {Yu}}, \bibinfo {author} {\bibfnamefont {A.~Y.}\ \bibnamefont {Zhu}},
  \bibinfo {author} {\bibfnamefont {R.}~\bibnamefont {Paniagua-Domínguez}},
  \bibinfo {author} {\bibfnamefont {Y.~H.}\ \bibnamefont {Fu}}, \bibinfo
  {author} {\bibfnamefont {B.}~\bibnamefont {Luk'yanchuk}},\ and\ \bibinfo
  {author} {\bibfnamefont {A.~I.}\ \bibnamefont {Kuznetsov}},\ }\bibfield
  {title} {\bibinfo {title} {High-transmission dielectric metasurface with
  {$2\pi$} phase control at visible wavelengths},\ }\href
  {https://onlinelibrary.wiley.com/doi/abs/10.1002/lpor.201500041} {\bibfield
  {journal} {\bibinfo  {journal} {Laser \& Photonics Reviews}\ }\textbf
  {\bibinfo {volume} {9}},\ \bibinfo {pages} {412} (\bibinfo {year}
  {2015})}\BibitemShut {NoStop}%
\bibitem [{\citenamefont {{Sievenpiper}}\ \emph {et~al.}(1999)\citenamefont
  {{Sievenpiper}}, \citenamefont {{Lijun Zhang}}, \citenamefont {{Broas}},
  \citenamefont {{Alexopolous}},\ and\ \citenamefont
  {{Yablonovitch}}}]{Sievenpiper99}%
  \BibitemOpen
  \bibfield  {author} {\bibinfo {author} {\bibfnamefont {D.}~\bibnamefont
  {{Sievenpiper}}}, \bibinfo {author} {\bibnamefont {{Lijun Zhang}}}, \bibinfo
  {author} {\bibfnamefont {R.~F.~J.}\ \bibnamefont {{Broas}}}, \bibinfo
  {author} {\bibfnamefont {N.~G.}\ \bibnamefont {{Alexopolous}}},\ and\
  \bibinfo {author} {\bibfnamefont {E.}~\bibnamefont {{Yablonovitch}}},\
  }\bibfield  {title} {\bibinfo {title} {High-impedance electromagnetic
  surfaces with a forbidden frequency band},\ }\href
  {https://doi.org/10.1109/22.798001} {\bibfield  {journal} {\bibinfo
  {journal} {IEEE Transactions on Microwave Theory and Techniques}\ }\textbf
  {\bibinfo {volume} {47}},\ \bibinfo {pages} {2059} (\bibinfo {year}
  {1999})}\BibitemShut {NoStop}%
\bibitem [{\citenamefont {Schwanecke}\ \emph {et~al.}(2006)\citenamefont
  {Schwanecke}, \citenamefont {Fedotov}, \citenamefont {Khardikov},
  \citenamefont {Prosvirnin}, \citenamefont {Chen},\ and\ \citenamefont
  {Zheludev}}]{Schwanecke06}%
  \BibitemOpen
  \bibfield  {author} {\bibinfo {author} {\bibfnamefont {A.~S.}\ \bibnamefont
  {Schwanecke}}, \bibinfo {author} {\bibfnamefont {V.~A.}\ \bibnamefont
  {Fedotov}}, \bibinfo {author} {\bibfnamefont {V.~V.}\ \bibnamefont
  {Khardikov}}, \bibinfo {author} {\bibfnamefont {S.~L.}\ \bibnamefont
  {Prosvirnin}}, \bibinfo {author} {\bibfnamefont {Y.}~\bibnamefont {Chen}},\
  and\ \bibinfo {author} {\bibfnamefont {N.~I.}\ \bibnamefont {Zheludev}},\
  }\bibfield  {title} {\bibinfo {title} {Optical magnetic mirrors},\ }\href
  {https://doi.org/10.1088/1464-4258/9/1/l01} {\bibfield  {journal} {\bibinfo
  {journal} {Journal of Optics A: Pure and Applied Optics}\ }\textbf {\bibinfo
  {volume} {9}},\ \bibinfo {pages} {L1} (\bibinfo {year} {2006})}\BibitemShut
  {NoStop}%
\bibitem [{\citenamefont {Liu}\ \emph {et~al.}(2014)\citenamefont {Liu},
  \citenamefont {Sinclair}, \citenamefont {Mahony}, \citenamefont {Jun},
  \citenamefont {Campione}, \citenamefont {Ginn}, \citenamefont {Bender},
  \citenamefont {Wendt}, \citenamefont {Ihlefeld}, \citenamefont {Clem},
  \citenamefont {Wright},\ and\ \citenamefont {Brener}}]{Liu14}%
  \BibitemOpen
  \bibfield  {author} {\bibinfo {author} {\bibfnamefont {S.}~\bibnamefont
  {Liu}}, \bibinfo {author} {\bibfnamefont {M.~B.}\ \bibnamefont {Sinclair}},
  \bibinfo {author} {\bibfnamefont {T.~S.}\ \bibnamefont {Mahony}}, \bibinfo
  {author} {\bibfnamefont {Y.~C.}\ \bibnamefont {Jun}}, \bibinfo {author}
  {\bibfnamefont {S.}~\bibnamefont {Campione}}, \bibinfo {author}
  {\bibfnamefont {J.}~\bibnamefont {Ginn}}, \bibinfo {author} {\bibfnamefont
  {D.~A.}\ \bibnamefont {Bender}}, \bibinfo {author} {\bibfnamefont {J.~R.}\
  \bibnamefont {Wendt}}, \bibinfo {author} {\bibfnamefont {J.~F.}\ \bibnamefont
  {Ihlefeld}}, \bibinfo {author} {\bibfnamefont {P.~G.}\ \bibnamefont {Clem}},
  \bibinfo {author} {\bibfnamefont {J.~B.}\ \bibnamefont {Wright}},\ and\
  \bibinfo {author} {\bibfnamefont {I.}~\bibnamefont {Brener}},\ }\bibfield
  {title} {\bibinfo {title} {Optical magnetic mirrors without metals},\ }\href
  {https://doi.org/10.1364/OPTICA.1.000250} {\bibfield  {journal} {\bibinfo
  {journal} {Optica}\ }\textbf {\bibinfo {volume} {1}},\ \bibinfo {pages} {250}
  (\bibinfo {year} {2014})}\BibitemShut {NoStop}%
\bibitem [{\citenamefont {Lin}\ \emph {et~al.}(2016)\citenamefont {Lin},
  \citenamefont {Jiang}, \citenamefont {Ma}, \citenamefont {Yun}, \citenamefont
  {Liu}, \citenamefont {Werner},\ and\ \citenamefont {Mayer}}]{Lin16}%
  \BibitemOpen
  \bibfield  {author} {\bibinfo {author} {\bibfnamefont {L.}~\bibnamefont
  {Lin}}, \bibinfo {author} {\bibfnamefont {Z.~H.}\ \bibnamefont {Jiang}},
  \bibinfo {author} {\bibfnamefont {D.}~\bibnamefont {Ma}}, \bibinfo {author}
  {\bibfnamefont {S.}~\bibnamefont {Yun}}, \bibinfo {author} {\bibfnamefont
  {Z.}~\bibnamefont {Liu}}, \bibinfo {author} {\bibfnamefont {D.~H.}\
  \bibnamefont {Werner}},\ and\ \bibinfo {author} {\bibfnamefont {T.~S.}\
  \bibnamefont {Mayer}},\ }\bibfield  {title} {\bibinfo {title} {Dielectric
  nanoresonator based lossless optical perfect magnetic mirror with near-zero
  reflection phase},\ }\href {https://doi.org/10.1063/1.4947274} {\bibfield
  {journal} {\bibinfo  {journal} {Applied Physics Letters}\ }\textbf {\bibinfo
  {volume} {108}},\ \bibinfo {pages} {171902} (\bibinfo {year}
  {2016})}\BibitemShut {NoStop}%
\bibitem [{\citenamefont {Ballantine}\ and\ \citenamefont
  {Ruostekoski}(2023)}]{Ballantine23}%
  \BibitemOpen
  \bibfield  {author} {\bibinfo {author} {\bibfnamefont {K.~E.}\ \bibnamefont
  {Ballantine}}\ and\ \bibinfo {author} {\bibfnamefont {J.}~\bibnamefont
  {Ruostekoski}},\ }\href@noop {} {\bibinfo {title} {unpublished}} (\bibinfo
  {year} {2023})\BibitemShut {NoStop}%
\bibitem [{\citenamefont {Grankin}\ \emph {et~al.}(2018)\citenamefont
  {Grankin}, \citenamefont {Guimond}, \citenamefont {Vasilyev}, \citenamefont
  {Vermersch},\ and\ \citenamefont {Zoller}}]{Grankin18}%
  \BibitemOpen
  \bibfield  {author} {\bibinfo {author} {\bibfnamefont {A.}~\bibnamefont
  {Grankin}}, \bibinfo {author} {\bibfnamefont {P.~O.}\ \bibnamefont
  {Guimond}}, \bibinfo {author} {\bibfnamefont {D.~V.}\ \bibnamefont
  {Vasilyev}}, \bibinfo {author} {\bibfnamefont {B.}~\bibnamefont
  {Vermersch}},\ and\ \bibinfo {author} {\bibfnamefont {P.}~\bibnamefont
  {Zoller}},\ }\bibfield  {title} {\bibinfo {title} {Free-space photonic
  quantum link and chiral quantum optics},\ }\href
  {https://doi.org/10.1103/PhysRevA.98.043825} {\bibfield  {journal} {\bibinfo
  {journal} {Phys. Rev. A}\ }\textbf {\bibinfo {volume} {98}},\ \bibinfo
  {pages} {043825} (\bibinfo {year} {2018})}\BibitemShut {NoStop}%
\bibitem [{\citenamefont {Kimble}(2008)}]{Kimble08}%
  \BibitemOpen
  \bibfield  {author} {\bibinfo {author} {\bibfnamefont {H.~J.}\ \bibnamefont
  {Kimble}},\ }\bibfield  {title} {\bibinfo {title} {The quantum internet},\
  }\href {https://doi.org/10.1038/nature07127} {\bibfield  {journal} {\bibinfo
  {journal} {Nature}\ }\textbf {\bibinfo {volume} {453}},\ \bibinfo {pages}
  {1023} (\bibinfo {year} {2008})}\BibitemShut {NoStop}%
\bibitem [{\citenamefont {Ritter}\ \emph {et~al.}(2012)\citenamefont {Ritter},
  \citenamefont {N{\"o}lleke}, \citenamefont {Hahn}, \citenamefont {Reiserer},
  \citenamefont {Neuzner}, \citenamefont {Uphoff}, \citenamefont {M{\"u}cke},
  \citenamefont {Figueroa}, \citenamefont {Bochmann},\ and\ \citenamefont
  {Rempe}}]{Ritter12}%
  \BibitemOpen
  \bibfield  {author} {\bibinfo {author} {\bibfnamefont {S.}~\bibnamefont
  {Ritter}}, \bibinfo {author} {\bibfnamefont {C.}~\bibnamefont {N{\"o}lleke}},
  \bibinfo {author} {\bibfnamefont {C.}~\bibnamefont {Hahn}}, \bibinfo {author}
  {\bibfnamefont {A.}~\bibnamefont {Reiserer}}, \bibinfo {author}
  {\bibfnamefont {A.}~\bibnamefont {Neuzner}}, \bibinfo {author} {\bibfnamefont
  {M.}~\bibnamefont {Uphoff}}, \bibinfo {author} {\bibfnamefont
  {M.}~\bibnamefont {M{\"u}cke}}, \bibinfo {author} {\bibfnamefont
  {E.}~\bibnamefont {Figueroa}}, \bibinfo {author} {\bibfnamefont
  {J.}~\bibnamefont {Bochmann}},\ and\ \bibinfo {author} {\bibfnamefont
  {G.}~\bibnamefont {Rempe}},\ }\bibfield  {title} {\bibinfo {title} {An
  elementary quantum network of single atoms in optical cavities},\ }\href
  {https://doi.org/10.1038/nature11023} {\bibfield  {journal} {\bibinfo
  {journal} {Nature}\ }\textbf {\bibinfo {volume} {484}},\ \bibinfo {pages}
  {195} (\bibinfo {year} {2012})}\BibitemShut {NoStop}%
\bibitem [{\citenamefont {Ballantine}\ and\ \citenamefont
  {Ruostekoski}(2022)}]{Ballantine22bilayer}%
  \BibitemOpen
  \bibfield  {author} {\bibinfo {author} {\bibfnamefont {K.~E.}\ \bibnamefont
  {Ballantine}}\ and\ \bibinfo {author} {\bibfnamefont {J.}~\bibnamefont
  {Ruostekoski}},\ }\bibfield  {title} {\bibinfo {title} {Unidirectional
  absorption, storage, and emission of single photons in a collectively
  responding bilayer atomic array},\ }\href
  {https://doi.org/10.1103/PhysRevResearch.4.033200} {\bibfield  {journal}
  {\bibinfo  {journal} {Phys. Rev. Res.}\ }\textbf {\bibinfo {volume} {4}},\
  \bibinfo {pages} {033200} (\bibinfo {year} {2022})}\BibitemShut {NoStop}%
\bibitem [{\citenamefont {Gorshkov}\ \emph {et~al.}(2007)\citenamefont
  {Gorshkov}, \citenamefont {Andr\'e}, \citenamefont {Fleischhauer},
  \citenamefont {S\o{}rensen},\ and\ \citenamefont {Lukin}}]{Gorshkov07}%
  \BibitemOpen
  \bibfield  {author} {\bibinfo {author} {\bibfnamefont {A.~V.}\ \bibnamefont
  {Gorshkov}}, \bibinfo {author} {\bibfnamefont {A.}~\bibnamefont {Andr\'e}},
  \bibinfo {author} {\bibfnamefont {M.}~\bibnamefont {Fleischhauer}}, \bibinfo
  {author} {\bibfnamefont {A.~S.}\ \bibnamefont {S\o{}rensen}},\ and\ \bibinfo
  {author} {\bibfnamefont {M.~D.}\ \bibnamefont {Lukin}},\ }\bibfield  {title}
  {\bibinfo {title} {Universal approach to optimal photon storage in atomic
  media},\ }\href {https://doi.org/10.1103/PhysRevLett.98.123601} {\bibfield
  {journal} {\bibinfo  {journal} {Phys. Rev. Lett.}\ }\textbf {\bibinfo
  {volume} {98}},\ \bibinfo {pages} {123601} (\bibinfo {year}
  {2007})}\BibitemShut {NoStop}%
\bibitem [{\citenamefont {Dudin}\ and\ \citenamefont
  {Kuzmich}(2012)}]{Dudin12}%
  \BibitemOpen
  \bibfield  {author} {\bibinfo {author} {\bibfnamefont {Y.~O.}\ \bibnamefont
  {Dudin}}\ and\ \bibinfo {author} {\bibfnamefont {A.}~\bibnamefont
  {Kuzmich}},\ }\bibfield  {title} {\bibinfo {title} {Strongly interacting
  rydberg excitations of a cold atomic gas},\ }\href
  {https://doi.org/10.1126/science.1217901} {\bibfield  {journal} {\bibinfo
  {journal} {Science}\ }\textbf {\bibinfo {volume} {336}},\ \bibinfo {pages}
  {887} (\bibinfo {year} {2012})}\BibitemShut {NoStop}%
\bibitem [{\citenamefont {Bekenstein}\ \emph {et~al.}(2020)\citenamefont
  {Bekenstein}, \citenamefont {Pikovski}, \citenamefont {Pichler},
  \citenamefont {Shahmoon}, \citenamefont {Yelin},\ and\ \citenamefont
  {Lukin}}]{Bekenstein2020}%
  \BibitemOpen
  \bibfield  {author} {\bibinfo {author} {\bibfnamefont {R.}~\bibnamefont
  {Bekenstein}}, \bibinfo {author} {\bibfnamefont {I.}~\bibnamefont
  {Pikovski}}, \bibinfo {author} {\bibfnamefont {H.}~\bibnamefont {Pichler}},
  \bibinfo {author} {\bibfnamefont {E.}~\bibnamefont {Shahmoon}}, \bibinfo
  {author} {\bibfnamefont {S.~F.}\ \bibnamefont {Yelin}},\ and\ \bibinfo
  {author} {\bibfnamefont {M.~D.}\ \bibnamefont {Lukin}},\ }\bibfield  {title}
  {\bibinfo {title} {Quantum metasurfaces with atom arrays},\ }\href
  {https://doi.org/10.1038/s41567-020-0845-5} {\bibfield  {journal} {\bibinfo
  {journal} {Nature Physics}\ }\textbf {\bibinfo {volume} {16}},\ \bibinfo
  {pages} {676} (\bibinfo {year} {2020})}\BibitemShut {NoStop}%
\bibitem [{\citenamefont {Moreno-Cardoner}\ \emph {et~al.}(2021)\citenamefont
  {Moreno-Cardoner}, \citenamefont {Goncalves},\ and\ \citenamefont
  {Chang}}]{Moreno2021}%
  \BibitemOpen
  \bibfield  {author} {\bibinfo {author} {\bibfnamefont {M.}~\bibnamefont
  {Moreno-Cardoner}}, \bibinfo {author} {\bibfnamefont {D.}~\bibnamefont
  {Goncalves}},\ and\ \bibinfo {author} {\bibfnamefont {D.~E.}\ \bibnamefont
  {Chang}},\ }\bibfield  {title} {\bibinfo {title} {Quantum nonlinear optics
  based on two-dimensional {Rydberg} atom arrays},\ }\href
  {https://doi.org/10.1103/PhysRevLett.127.263602} {\bibfield  {journal}
  {\bibinfo  {journal} {Phys. Rev. Lett.}\ }\textbf {\bibinfo {volume} {127}},\
  \bibinfo {pages} {263602} (\bibinfo {year} {2021})}\BibitemShut {NoStop}%
\bibitem [{\citenamefont {Zhang}\ \emph {et~al.}(2022)\citenamefont {Zhang},
  \citenamefont {Walther}, \citenamefont {M{\o{}}lmer},\ and\ \citenamefont
  {Pohl}}]{Zhang2022}%
  \BibitemOpen
  \bibfield  {author} {\bibinfo {author} {\bibfnamefont {L.}~\bibnamefont
  {Zhang}}, \bibinfo {author} {\bibfnamefont {V.}~\bibnamefont {Walther}},
  \bibinfo {author} {\bibfnamefont {K.}~\bibnamefont {M{\o{}}lmer}},\ and\
  \bibinfo {author} {\bibfnamefont {T.}~\bibnamefont {Pohl}},\ }\bibfield
  {title} {\bibinfo {title} {Photon-photon interactions in {R}ydberg-atom
  arrays},\ }\href {https://doi.org/10.22331/q-2022-03-30-674} {\bibfield
  {journal} {\bibinfo  {journal} {{Quantum}}\ }\textbf {\bibinfo {volume}
  {6}},\ \bibinfo {pages} {674} (\bibinfo {year} {2022})}\BibitemShut {NoStop}%
\bibitem [{\citenamefont {Petrosyan}\ and\ \citenamefont
  {M\o{}lmer}(2018)}]{Petrosyan18}%
  \BibitemOpen
  \bibfield  {author} {\bibinfo {author} {\bibfnamefont {D.}~\bibnamefont
  {Petrosyan}}\ and\ \bibinfo {author} {\bibfnamefont {K.}~\bibnamefont
  {M\o{}lmer}},\ }\bibfield  {title} {\bibinfo {title} {Deterministic
  free-space source of single photons using rydberg atoms},\ }\href
  {https://doi.org/10.1103/PhysRevLett.121.123605} {\bibfield  {journal}
  {\bibinfo  {journal} {Phys. Rev. Lett.}\ }\textbf {\bibinfo {volume} {121}},\
  \bibinfo {pages} {123605} (\bibinfo {year} {2018})}\BibitemShut {NoStop}%
\bibitem [{\citenamefont {Rubies-Bigorda}\ \emph {et~al.}(2022)\citenamefont
  {Rubies-Bigorda}, \citenamefont {Walther}, \citenamefont {Patti},\ and\
  \citenamefont {Yelin}}]{Rubies22}%
  \BibitemOpen
  \bibfield  {author} {\bibinfo {author} {\bibfnamefont {O.}~\bibnamefont
  {Rubies-Bigorda}}, \bibinfo {author} {\bibfnamefont {V.}~\bibnamefont
  {Walther}}, \bibinfo {author} {\bibfnamefont {T.~L.}\ \bibnamefont {Patti}},\
  and\ \bibinfo {author} {\bibfnamefont {S.~F.}\ \bibnamefont {Yelin}},\
  }\bibfield  {title} {\bibinfo {title} {Photon control and coherent
  interactions via lattice dark states in atomic arrays},\ }\href
  {https://doi.org/10.1103/PhysRevResearch.4.013110} {\bibfield  {journal}
  {\bibinfo  {journal} {Phys. Rev. Research}\ }\textbf {\bibinfo {volume}
  {4}},\ \bibinfo {pages} {013110} (\bibinfo {year} {2022})}\BibitemShut
  {NoStop}%
\bibitem [{\citenamefont {Fayard}\ \emph {et~al.}(2023)\citenamefont {Fayard},
  \citenamefont {Ferrier-Barbut}, \citenamefont {Browaeys},\ and\ \citenamefont
  {Greffet}}]{Fayard2023}%
  \BibitemOpen
  \bibfield  {author} {\bibinfo {author} {\bibfnamefont {N.}~\bibnamefont
  {Fayard}}, \bibinfo {author} {\bibfnamefont {I.}~\bibnamefont
  {Ferrier-Barbut}}, \bibinfo {author} {\bibfnamefont {A.}~\bibnamefont
  {Browaeys}},\ and\ \bibinfo {author} {\bibfnamefont {J.-J.}\ \bibnamefont
  {Greffet}},\ }\href@noop {} {\bibinfo {title} {Optical control of collective
  states in 1d ordered atomic chains beyond the linear regime}} (\bibinfo
  {year} {2023}),\ \Eprint {https://arxiv.org/abs/2212.13022} {arXiv:2212.13022
  [quant-ph]} \BibitemShut {NoStop}%
\bibitem [{\citenamefont {Cidrim}\ \emph {et~al.}(2020)\citenamefont {Cidrim},
  \citenamefont {{do Espirito Santo}}, \citenamefont {Schachenmayer},
  \citenamefont {Kaiser},\ and\ \citenamefont {Bachelard}}]{Cidrim20}%
  \BibitemOpen
  \bibfield  {author} {\bibinfo {author} {\bibfnamefont {A.}~\bibnamefont
  {Cidrim}}, \bibinfo {author} {\bibfnamefont {T.~S.}\ \bibnamefont {{do
  Espirito Santo}}}, \bibinfo {author} {\bibfnamefont {J.}~\bibnamefont
  {Schachenmayer}}, \bibinfo {author} {\bibfnamefont {R.}~\bibnamefont
  {Kaiser}},\ and\ \bibinfo {author} {\bibfnamefont {R.}~\bibnamefont
  {Bachelard}},\ }\bibfield  {title} {\bibinfo {title} {Photon blockade with
  ground-state neutral atoms},\ }\href
  {https://doi.org/10.1103/PhysRevLett.125.073601} {\bibfield  {journal}
  {\bibinfo  {journal} {Phys. Rev. Lett.}\ }\textbf {\bibinfo {volume} {125}},\
  \bibinfo {pages} {073601} (\bibinfo {year} {2020})}\BibitemShut {NoStop}%
\bibitem [{\citenamefont {Williamson}\ \emph {et~al.}(2020)\citenamefont
  {Williamson}, \citenamefont {Borgh},\ and\ \citenamefont
  {Ruostekoski}}]{Williamson2020b}%
  \BibitemOpen
  \bibfield  {author} {\bibinfo {author} {\bibfnamefont {L.~A.}\ \bibnamefont
  {Williamson}}, \bibinfo {author} {\bibfnamefont {M.~O.}\ \bibnamefont
  {Borgh}},\ and\ \bibinfo {author} {\bibfnamefont {J.}~\bibnamefont
  {Ruostekoski}},\ }\bibfield  {title} {\bibinfo {title} {Superatom picture of
  collective nonclassical light emission and dipole blockade in atom arrays},\
  }\href {https://doi.org/10.1103/PhysRevLett.125.073602} {\bibfield  {journal}
  {\bibinfo  {journal} {Phys. Rev. Lett.}\ }\textbf {\bibinfo {volume} {125}},\
  \bibinfo {pages} {073602} (\bibinfo {year} {2020})}\BibitemShut {NoStop}%
\bibitem [{\citenamefont {Carmichael}\ and\ \citenamefont
  {Walls}(1976{\natexlab{a}})}]{Carmichael_1976}%
  \BibitemOpen
  \bibfield  {author} {\bibinfo {author} {\bibfnamefont {H.~J.}\ \bibnamefont
  {Carmichael}}\ and\ \bibinfo {author} {\bibfnamefont {D.~F.}\ \bibnamefont
  {Walls}},\ }\bibfield  {title} {\bibinfo {title} {Proposal for the
  measurement of the resonant {Stark} effect by photon correlation
  techniques},\ }\href {https://doi.org/10.1088/0022-3700/9/4/001} {\bibfield
  {journal} {\bibinfo  {journal} {Journal of Physics B: Atomic and Molecular
  Physics}\ }\textbf {\bibinfo {volume} {9}},\ \bibinfo {pages} {L43} (\bibinfo
  {year} {1976}{\natexlab{a}})}\BibitemShut {NoStop}%
\bibitem [{\citenamefont {Kimble}\ \emph {et~al.}(1977)\citenamefont {Kimble},
  \citenamefont {Dagenais},\ and\ \citenamefont {Mandel}}]{kimble1977}%
  \BibitemOpen
  \bibfield  {author} {\bibinfo {author} {\bibfnamefont {H.~J.}\ \bibnamefont
  {Kimble}}, \bibinfo {author} {\bibfnamefont {M.}~\bibnamefont {Dagenais}},\
  and\ \bibinfo {author} {\bibfnamefont {L.}~\bibnamefont {Mandel}},\
  }\bibfield  {title} {\bibinfo {title} {Photon antibunching in resonance
  fluorescence},\ }\href {https://doi.org/10.1103/PhysRevLett.39.691}
  {\bibfield  {journal} {\bibinfo  {journal} {Phys. Rev. Lett.}\ }\textbf
  {\bibinfo {volume} {39}},\ \bibinfo {pages} {691} (\bibinfo {year}
  {1977})}\BibitemShut {NoStop}%
\bibitem [{\citenamefont {Kimble}\ \emph {et~al.}(1978)\citenamefont {Kimble},
  \citenamefont {Dagenais},\ and\ \citenamefont {Mandel}}]{kimble1978}%
  \BibitemOpen
  \bibfield  {author} {\bibinfo {author} {\bibfnamefont {H.~J.}\ \bibnamefont
  {Kimble}}, \bibinfo {author} {\bibfnamefont {M.}~\bibnamefont {Dagenais}},\
  and\ \bibinfo {author} {\bibfnamefont {L.}~\bibnamefont {Mandel}},\
  }\bibfield  {title} {\bibinfo {title} {Multiatom and transit-time effects on
  photon-correlation measurements in resonance fluorescence},\ }\href
  {https://doi.org/10.1103/PhysRevA.18.201} {\bibfield  {journal} {\bibinfo
  {journal} {Phys. Rev. A}\ }\textbf {\bibinfo {volume} {18}},\ \bibinfo
  {pages} {201} (\bibinfo {year} {1978})}\BibitemShut {NoStop}%
\bibitem [{\citenamefont {Dagenais}\ and\ \citenamefont
  {Mandel}(1978)}]{dagenais1978}%
  \BibitemOpen
  \bibfield  {author} {\bibinfo {author} {\bibfnamefont {M.}~\bibnamefont
  {Dagenais}}\ and\ \bibinfo {author} {\bibfnamefont {L.}~\bibnamefont
  {Mandel}},\ }\bibfield  {title} {\bibinfo {title} {Investigation of two-time
  correlations in photon emissions from a single atom},\ }\href
  {https://doi.org/10.1103/PhysRevA.18.2217} {\bibfield  {journal} {\bibinfo
  {journal} {Phys. Rev. A}\ }\textbf {\bibinfo {volume} {18}},\ \bibinfo
  {pages} {2217} (\bibinfo {year} {1978})}\BibitemShut {NoStop}%
\bibitem [{\citenamefont {Walls}(1979)}]{walls1979}%
  \BibitemOpen
  \bibfield  {author} {\bibinfo {author} {\bibfnamefont {D.~F.}\ \bibnamefont
  {Walls}},\ }\bibfield  {title} {\bibinfo {title} {Evidence for the quantum
  nature of light},\ }\href {https://doi.org/10.1038/280451a0} {\bibfield
  {journal} {\bibinfo  {journal} {Nature}\ }\textbf {\bibinfo {volume} {280}},\
  \bibinfo {pages} {451} (\bibinfo {year} {1979})}\BibitemShut {NoStop}%
\bibitem [{\citenamefont {Jaksch}\ \emph {et~al.}(2000)\citenamefont {Jaksch},
  \citenamefont {Cirac}, \citenamefont {Zoller}, \citenamefont {Rolston},
  \citenamefont {C\^ot\'e},\ and\ \citenamefont {Lukin}}]{Jaksch00}%
  \BibitemOpen
  \bibfield  {author} {\bibinfo {author} {\bibfnamefont {D.}~\bibnamefont
  {Jaksch}}, \bibinfo {author} {\bibfnamefont {J.~I.}\ \bibnamefont {Cirac}},
  \bibinfo {author} {\bibfnamefont {P.}~\bibnamefont {Zoller}}, \bibinfo
  {author} {\bibfnamefont {S.~L.}\ \bibnamefont {Rolston}}, \bibinfo {author}
  {\bibfnamefont {R.}~\bibnamefont {C\^ot\'e}},\ and\ \bibinfo {author}
  {\bibfnamefont {M.~D.}\ \bibnamefont {Lukin}},\ }\bibfield  {title} {\bibinfo
  {title} {Fast quantum gates for neutral atoms},\ }\href
  {https://doi.org/10.1103/PhysRevLett.85.2208} {\bibfield  {journal} {\bibinfo
   {journal} {Phys. Rev. Lett.}\ }\textbf {\bibinfo {volume} {85}},\ \bibinfo
  {pages} {2208} (\bibinfo {year} {2000})}\BibitemShut {NoStop}%
\bibitem [{\citenamefont {Lukin}\ \emph {et~al.}(2001)\citenamefont {Lukin},
  \citenamefont {Fleischhauer}, \citenamefont {Cote}, \citenamefont {Duan},
  \citenamefont {Jaksch}, \citenamefont {Cirac},\ and\ \citenamefont
  {Zoller}}]{Lukin01}%
  \BibitemOpen
  \bibfield  {author} {\bibinfo {author} {\bibfnamefont {M.~D.}\ \bibnamefont
  {Lukin}}, \bibinfo {author} {\bibfnamefont {M.}~\bibnamefont {Fleischhauer}},
  \bibinfo {author} {\bibfnamefont {R.}~\bibnamefont {Cote}}, \bibinfo {author}
  {\bibfnamefont {L.~M.}\ \bibnamefont {Duan}}, \bibinfo {author}
  {\bibfnamefont {D.}~\bibnamefont {Jaksch}}, \bibinfo {author} {\bibfnamefont
  {J.~I.}\ \bibnamefont {Cirac}},\ and\ \bibinfo {author} {\bibfnamefont
  {P.}~\bibnamefont {Zoller}},\ }\bibfield  {title} {\bibinfo {title} {Dipole
  blockade and quantum information processing in mesoscopic atomic ensembles},\
  }\href {https://doi.org/10.1103/PhysRevLett.87.037901} {\bibfield  {journal}
  {\bibinfo  {journal} {Phys. Rev. Lett.}\ }\textbf {\bibinfo {volume} {87}},\
  \bibinfo {pages} {037901} (\bibinfo {year} {2001})}\BibitemShut {NoStop}%
\bibitem [{\citenamefont {Urban}\ \emph {et~al.}(2009)\citenamefont {Urban},
  \citenamefont {Johnson}, \citenamefont {Henage}, \citenamefont {Isenhower},
  \citenamefont {Yavuz}, \citenamefont {Walker},\ and\ \citenamefont
  {Saffman}}]{Urban09}%
  \BibitemOpen
  \bibfield  {author} {\bibinfo {author} {\bibfnamefont {E.}~\bibnamefont
  {Urban}}, \bibinfo {author} {\bibfnamefont {T.~A.}\ \bibnamefont {Johnson}},
  \bibinfo {author} {\bibfnamefont {T.}~\bibnamefont {Henage}}, \bibinfo
  {author} {\bibfnamefont {L.}~\bibnamefont {Isenhower}}, \bibinfo {author}
  {\bibfnamefont {D.~D.}\ \bibnamefont {Yavuz}}, \bibinfo {author}
  {\bibfnamefont {T.~G.}\ \bibnamefont {Walker}},\ and\ \bibinfo {author}
  {\bibfnamefont {M.}~\bibnamefont {Saffman}},\ }\bibfield  {title} {\bibinfo
  {title} {Observation of rydberg blockade between two atoms},\ }\href
  {https://doi.org/10.1038/nphys1178} {\bibfield  {journal} {\bibinfo
  {journal} {Nature Physics}\ }\textbf {\bibinfo {volume} {5}},\ \bibinfo
  {pages} {110} (\bibinfo {year} {2009})}\BibitemShut {NoStop}%
\bibitem [{\citenamefont {Ga{\"e}tan}\ \emph {et~al.}(2009)\citenamefont
  {Ga{\"e}tan}, \citenamefont {Miroshnychenko}, \citenamefont {Wilk},
  \citenamefont {Chotia}, \citenamefont {Viteau}, \citenamefont {Comparat},
  \citenamefont {Pillet}, \citenamefont {Browaeys},\ and\ \citenamefont
  {Grangier}}]{Grangier09}%
  \BibitemOpen
  \bibfield  {author} {\bibinfo {author} {\bibfnamefont {A.}~\bibnamefont
  {Ga{\"e}tan}}, \bibinfo {author} {\bibfnamefont {Y.}~\bibnamefont
  {Miroshnychenko}}, \bibinfo {author} {\bibfnamefont {T.}~\bibnamefont
  {Wilk}}, \bibinfo {author} {\bibfnamefont {A.}~\bibnamefont {Chotia}},
  \bibinfo {author} {\bibfnamefont {M.}~\bibnamefont {Viteau}}, \bibinfo
  {author} {\bibfnamefont {D.}~\bibnamefont {Comparat}}, \bibinfo {author}
  {\bibfnamefont {P.}~\bibnamefont {Pillet}}, \bibinfo {author} {\bibfnamefont
  {A.}~\bibnamefont {Browaeys}},\ and\ \bibinfo {author} {\bibfnamefont
  {P.}~\bibnamefont {Grangier}},\ }\bibfield  {title} {\bibinfo {title}
  {Observation of collective excitation of two individual atoms in the rydberg
  blockade regime},\ }\href {https://doi.org/10.1038/nphys1183} {\bibfield
  {journal} {\bibinfo  {journal} {Nature Physics}\ }\textbf {\bibinfo {volume}
  {5}},\ \bibinfo {pages} {115} (\bibinfo {year} {2009})}\BibitemShut {NoStop}%
\bibitem [{\citenamefont {Saffman}\ \emph {et~al.}(2010)\citenamefont
  {Saffman}, \citenamefont {Walker},\ and\ \citenamefont
  {M\o{}lmer}}]{Saffman10}%
  \BibitemOpen
  \bibfield  {author} {\bibinfo {author} {\bibfnamefont {M.}~\bibnamefont
  {Saffman}}, \bibinfo {author} {\bibfnamefont {T.~G.}\ \bibnamefont
  {Walker}},\ and\ \bibinfo {author} {\bibfnamefont {K.}~\bibnamefont
  {M\o{}lmer}},\ }\bibfield  {title} {\bibinfo {title} {Quantum information
  with rydberg atoms},\ }\href {https://doi.org/10.1103/RevModPhys.82.2313}
  {\bibfield  {journal} {\bibinfo  {journal} {Rev. Mod. Phys.}\ }\textbf
  {\bibinfo {volume} {82}},\ \bibinfo {pages} {2313} (\bibinfo {year}
  {2010})}\BibitemShut {NoStop}%
\bibitem [{\citenamefont {Schau{\ss}}\ \emph {et~al.}(2012)\citenamefont
  {Schau{\ss}}, \citenamefont {Cheneau}, \citenamefont {Endres}, \citenamefont
  {Fukuhara}, \citenamefont {Hild}, \citenamefont {Omran}, \citenamefont
  {Pohl}, \citenamefont {Gross}, \citenamefont {Kuhr},\ and\ \citenamefont
  {Bloch}}]{Schauss12}%
  \BibitemOpen
  \bibfield  {author} {\bibinfo {author} {\bibfnamefont {P.}~\bibnamefont
  {Schau{\ss}}}, \bibinfo {author} {\bibfnamefont {M.}~\bibnamefont {Cheneau}},
  \bibinfo {author} {\bibfnamefont {M.}~\bibnamefont {Endres}}, \bibinfo
  {author} {\bibfnamefont {T.}~\bibnamefont {Fukuhara}}, \bibinfo {author}
  {\bibfnamefont {S.}~\bibnamefont {Hild}}, \bibinfo {author} {\bibfnamefont
  {A.}~\bibnamefont {Omran}}, \bibinfo {author} {\bibfnamefont
  {T.}~\bibnamefont {Pohl}}, \bibinfo {author} {\bibfnamefont {C.}~\bibnamefont
  {Gross}}, \bibinfo {author} {\bibfnamefont {S.}~\bibnamefont {Kuhr}},\ and\
  \bibinfo {author} {\bibfnamefont {I.}~\bibnamefont {Bloch}},\ }\bibfield
  {title} {\bibinfo {title} {Observation of spatially ordered structures in a
  two-dimensional rydberg gas},\ }\href {https://doi.org/10.1038/nature11596}
  {\bibfield  {journal} {\bibinfo  {journal} {Nature}\ }\textbf {\bibinfo
  {volume} {491}},\ \bibinfo {pages} {87} (\bibinfo {year} {2012})}\BibitemShut
  {NoStop}%
\bibitem [{\citenamefont {Ripka}\ \emph {et~al.}(2018)\citenamefont {Ripka},
  \citenamefont {K{\"u}bler}, \citenamefont {L{\"o}w},\ and\ \citenamefont
  {Pfau}}]{ripka2018}%
  \BibitemOpen
  \bibfield  {author} {\bibinfo {author} {\bibfnamefont {F.}~\bibnamefont
  {Ripka}}, \bibinfo {author} {\bibfnamefont {H.}~\bibnamefont {K{\"u}bler}},
  \bibinfo {author} {\bibfnamefont {R.}~\bibnamefont {L{\"o}w}},\ and\ \bibinfo
  {author} {\bibfnamefont {T.}~\bibnamefont {Pfau}},\ }\bibfield  {title}
  {\bibinfo {title} {A room-temperature single-photon source based on strongly
  interacting {Rydberg} atoms},\ }\href
  {https://doi.org/10.1126/science.aau1949} {\bibfield  {journal} {\bibinfo
  {journal} {Science}\ }\textbf {\bibinfo {volume} {362}},\ \bibinfo {pages}
  {446} (\bibinfo {year} {2018})}\BibitemShut {NoStop}%
\bibitem [{\citenamefont {Carmichael}\ and\ \citenamefont
  {Walls}(1976{\natexlab{b}})}]{carmichael1976}%
  \BibitemOpen
  \bibfield  {author} {\bibinfo {author} {\bibfnamefont {H.}~\bibnamefont
  {Carmichael}}\ and\ \bibinfo {author} {\bibfnamefont {D.}~\bibnamefont
  {Walls}},\ }\bibfield  {title} {\bibinfo {title} {A quantum-mechanical master
  equation treatment of the dynamical {S}tark effect},\ }\href
  {https://doi.org/10.1088/0022-3700/9/8/007} {\bibfield  {journal} {\bibinfo
  {journal} {J. Phys. B: At. Mol. Phys.}\ }\textbf {\bibinfo {volume} {9}},\
  \bibinfo {pages} {1199} (\bibinfo {year} {1976}{\natexlab{b}})}\BibitemShut
  {NoStop}%
\bibitem [{\citenamefont {Pedersen}\ \emph {et~al.}(2023)\citenamefont
  {Pedersen}, \citenamefont {Zhang},\ and\ \citenamefont {Pohl}}]{Pedersen23}%
  \BibitemOpen
  \bibfield  {author} {\bibinfo {author} {\bibfnamefont {S.~P.}\ \bibnamefont
  {Pedersen}}, \bibinfo {author} {\bibfnamefont {L.}~\bibnamefont {Zhang}},\
  and\ \bibinfo {author} {\bibfnamefont {T.}~\bibnamefont {Pohl}},\ }\bibfield
  {title} {\bibinfo {title} {Quantum nonlinear metasurfaces from dual arrays of
  ultracold atoms},\ }\href {https://doi.org/10.1103/PhysRevResearch.5.L012047}
  {\bibfield  {journal} {\bibinfo  {journal} {Phys. Rev. Res.}\ }\textbf
  {\bibinfo {volume} {5}},\ \bibinfo {pages} {L012047} (\bibinfo {year}
  {2023})}\BibitemShut {NoStop}%
\bibitem [{\citenamefont {Cano}(2021)}]{Cano21}%
  \BibitemOpen
  \bibfield  {author} {\bibinfo {author} {\bibfnamefont {D.}~\bibnamefont
  {Cano}},\ }\bibfield  {title} {\bibinfo {title} {Photon statistics of the
  light transmitted and reflected by a two-dimensional atomic array},\ }\href
  {https://doi.org/10.1103/PhysRevA.104.053709} {\bibfield  {journal} {\bibinfo
   {journal} {Phys. Rev. A}\ }\textbf {\bibinfo {volume} {104}},\ \bibinfo
  {pages} {053709} (\bibinfo {year} {2021})}\BibitemShut {NoStop}%
\bibitem [{\citenamefont {Rusconi}\ \emph {et~al.}(2021)\citenamefont
  {Rusconi}, \citenamefont {Shi},\ and\ \citenamefont {Cirac}}]{Rusconi2021}%
  \BibitemOpen
  \bibfield  {author} {\bibinfo {author} {\bibfnamefont {C.~C.}\ \bibnamefont
  {Rusconi}}, \bibinfo {author} {\bibfnamefont {T.}~\bibnamefont {Shi}},\ and\
  \bibinfo {author} {\bibfnamefont {J.~I.}\ \bibnamefont {Cirac}},\ }\bibfield
  {title} {\bibinfo {title} {Exploiting the photonic nonlinearity of free-space
  subwavelength arrays of atoms},\ }\href
  {https://doi.org/10.1103/PhysRevA.104.033718} {\bibfield  {journal} {\bibinfo
   {journal} {Phys. Rev. A}\ }\textbf {\bibinfo {volume} {104}},\ \bibinfo
  {pages} {033718} (\bibinfo {year} {2021})}\BibitemShut {NoStop}%
\bibitem [{\citenamefont {Jen}(2018)}]{Jen2018}%
  \BibitemOpen
  \bibfield  {author} {\bibinfo {author} {\bibfnamefont {H.~H.}\ \bibnamefont
  {Jen}},\ }\bibfield  {title} {\bibinfo {title} {{Directional subradiance from
  helical-phase-imprinted multiphoton states}},\ }\href
  {https://doi.org/10.1038/s41598-018-25592-5} {\bibfield  {journal} {\bibinfo
  {journal} {Sci. Rep.}\ }\textbf {\bibinfo {volume} {8}},\ \bibinfo {pages}
  {7163} (\bibinfo {year} {2018})}\BibitemShut {NoStop}%
\bibitem [{\citenamefont {Moreno-Cardoner}\ \emph {et~al.}(2019)\citenamefont
  {Moreno-Cardoner}, \citenamefont {Plankensteiner}, \citenamefont {Ostermann},
  \citenamefont {Chang},\ and\ \citenamefont {Ritsch}}]{Moreno-Cardoner19}%
  \BibitemOpen
  \bibfield  {author} {\bibinfo {author} {\bibfnamefont {M.}~\bibnamefont
  {Moreno-Cardoner}}, \bibinfo {author} {\bibfnamefont {D.}~\bibnamefont
  {Plankensteiner}}, \bibinfo {author} {\bibfnamefont {L.}~\bibnamefont
  {Ostermann}}, \bibinfo {author} {\bibfnamefont {D.~E.}\ \bibnamefont
  {Chang}},\ and\ \bibinfo {author} {\bibfnamefont {H.}~\bibnamefont
  {Ritsch}},\ }\bibfield  {title} {\bibinfo {title} {Subradiance-enhanced
  excitation transfer between dipole-coupled nanorings of quantum emitters},\
  }\href {https://doi.org/10.1103/PhysRevA.100.023806} {\bibfield  {journal}
  {\bibinfo  {journal} {Phys. Rev. A}\ }\textbf {\bibinfo {volume} {100}},\
  \bibinfo {pages} {023806} (\bibinfo {year} {2019})}\BibitemShut {NoStop}%
\bibitem [{\citenamefont {Needham}\ \emph {et~al.}(2019)\citenamefont
  {Needham}, \citenamefont {Lesanovsky},\ and\ \citenamefont
  {Olmos}}]{Needham19}%
  \BibitemOpen
  \bibfield  {author} {\bibinfo {author} {\bibfnamefont {J.~A.}\ \bibnamefont
  {Needham}}, \bibinfo {author} {\bibfnamefont {I.}~\bibnamefont
  {Lesanovsky}},\ and\ \bibinfo {author} {\bibfnamefont {B.}~\bibnamefont
  {Olmos}},\ }\bibfield  {title} {\bibinfo {title} {Subradiance-protected
  excitation transport},\ }\href {https://doi.org/10.1088/1367-2630/ab31e8}
  {\bibfield  {journal} {\bibinfo  {journal} {New Journal of Physics}\ }\textbf
  {\bibinfo {volume} {21}},\ \bibinfo {pages} {073061} (\bibinfo {year}
  {2019})}\BibitemShut {NoStop}%
\bibitem [{\citenamefont {Holzinger}\ \emph {et~al.}(2020)\citenamefont
  {Holzinger}, \citenamefont {Plankensteiner}, \citenamefont {Ostermann},\ and\
  \citenamefont {Ritsch}}]{Holzinger20}%
  \BibitemOpen
  \bibfield  {author} {\bibinfo {author} {\bibfnamefont {R.}~\bibnamefont
  {Holzinger}}, \bibinfo {author} {\bibfnamefont {D.}~\bibnamefont
  {Plankensteiner}}, \bibinfo {author} {\bibfnamefont {L.}~\bibnamefont
  {Ostermann}},\ and\ \bibinfo {author} {\bibfnamefont {H.}~\bibnamefont
  {Ritsch}},\ }\bibfield  {title} {\bibinfo {title} {Nanoscale coherent light
  source},\ }\href {https://doi.org/10.1103/PhysRevLett.124.253603} {\bibfield
  {journal} {\bibinfo  {journal} {Phys. Rev. Lett.}\ }\textbf {\bibinfo
  {volume} {124}},\ \bibinfo {pages} {253603} (\bibinfo {year}
  {2020})}\BibitemShut {NoStop}%
\bibitem [{\citenamefont {Moreno-Cardoner}\ \emph {et~al.}(2022)\citenamefont
  {Moreno-Cardoner}, \citenamefont {Holzinger},\ and\ \citenamefont
  {Ritsch}}]{Moreno-Cardoner22}%
  \BibitemOpen
  \bibfield  {author} {\bibinfo {author} {\bibfnamefont {M.}~\bibnamefont
  {Moreno-Cardoner}}, \bibinfo {author} {\bibfnamefont {R.}~\bibnamefont
  {Holzinger}},\ and\ \bibinfo {author} {\bibfnamefont {H.}~\bibnamefont
  {Ritsch}},\ }\bibfield  {title} {\bibinfo {title} {Efficient nano-photonic
  antennas based on dark states in quantum emitter rings},\ }\href
  {https://doi.org/10.1364/OE.437396} {\bibfield  {journal} {\bibinfo
  {journal} {Opt. Express}\ }\textbf {\bibinfo {volume} {30}},\ \bibinfo
  {pages} {10779} (\bibinfo {year} {2022})}\BibitemShut {NoStop}%
\bibitem [{\citenamefont {Duan}\ \emph {et~al.}(2001)\citenamefont {Duan},
  \citenamefont {Lukin}, \citenamefont {Cirac},\ and\ \citenamefont
  {Zoller}}]{Duan01}%
  \BibitemOpen
  \bibfield  {author} {\bibinfo {author} {\bibfnamefont {L.~M.}\ \bibnamefont
  {Duan}}, \bibinfo {author} {\bibfnamefont {M.~D.}\ \bibnamefont {Lukin}},
  \bibinfo {author} {\bibfnamefont {J.~I.}\ \bibnamefont {Cirac}},\ and\
  \bibinfo {author} {\bibfnamefont {P.}~\bibnamefont {Zoller}},\ }\bibfield
  {title} {\bibinfo {title} {Long-distance quantum communication with atomic
  ensembles and linear optics},\ }\href {https://doi.org/10.1038/35106500}
  {\bibfield  {journal} {\bibinfo  {journal} {Nature}\ }\textbf {\bibinfo
  {volume} {414}},\ \bibinfo {pages} {413} (\bibinfo {year}
  {2001})}\BibitemShut {NoStop}%
\bibitem [{\citenamefont {van Enk}\ \emph {et~al.}(1998)\citenamefont {van
  Enk}, \citenamefont {Cirac},\ and\ \citenamefont {Zoller}}]{Enk98}%
  \BibitemOpen
  \bibfield  {author} {\bibinfo {author} {\bibfnamefont {S.~J.}\ \bibnamefont
  {van Enk}}, \bibinfo {author} {\bibfnamefont {J.~I.}\ \bibnamefont {Cirac}},\
  and\ \bibinfo {author} {\bibfnamefont {P.}~\bibnamefont {Zoller}},\
  }\bibfield  {title} {\bibinfo {title} {Photonic channels for quantum
  communication},\ }\href {https://doi.org/10.1126/science.279.5348.205}
  {\bibfield  {journal} {\bibinfo  {journal} {Science}\ }\textbf {\bibinfo
  {volume} {279}},\ \bibinfo {pages} {205} (\bibinfo {year}
  {1998})}\BibitemShut {NoStop}%
\bibitem [{\citenamefont {Duan}\ and\ \citenamefont {Kimble}(2003)}]{Duan03}%
  \BibitemOpen
  \bibfield  {author} {\bibinfo {author} {\bibfnamefont {L.-M.}\ \bibnamefont
  {Duan}}\ and\ \bibinfo {author} {\bibfnamefont {H.~J.}\ \bibnamefont
  {Kimble}},\ }\bibfield  {title} {\bibinfo {title} {Efficient engineering of
  multiatom entanglement through single-photon detections},\ }\href
  {https://doi.org/10.1103/PhysRevLett.90.253601} {\bibfield  {journal}
  {\bibinfo  {journal} {Phys. Rev. Lett.}\ }\textbf {\bibinfo {volume} {90}},\
  \bibinfo {pages} {253601} (\bibinfo {year} {2003})}\BibitemShut {NoStop}%
\bibitem [{\citenamefont {Duan}\ and\ \citenamefont {Kimble}(2004)}]{Duan04}%
  \BibitemOpen
  \bibfield  {author} {\bibinfo {author} {\bibfnamefont {L.-M.}\ \bibnamefont
  {Duan}}\ and\ \bibinfo {author} {\bibfnamefont {H.~J.}\ \bibnamefont
  {Kimble}},\ }\bibfield  {title} {\bibinfo {title} {Scalable photonic quantum
  computation through cavity-assisted interactions},\ }\href
  {https://doi.org/10.1103/PhysRevLett.92.127902} {\bibfield  {journal}
  {\bibinfo  {journal} {Phys. Rev. Lett.}\ }\textbf {\bibinfo {volume} {92}},\
  \bibinfo {pages} {127902} (\bibinfo {year} {2004})}\BibitemShut {NoStop}%
\bibitem [{\citenamefont {Thompson}\ \emph {et~al.}(1992)\citenamefont
  {Thompson}, \citenamefont {Rempe},\ and\ \citenamefont
  {Kimble}}]{Thompson92}%
  \BibitemOpen
  \bibfield  {author} {\bibinfo {author} {\bibfnamefont {R.~J.}\ \bibnamefont
  {Thompson}}, \bibinfo {author} {\bibfnamefont {G.}~\bibnamefont {Rempe}},\
  and\ \bibinfo {author} {\bibfnamefont {H.~J.}\ \bibnamefont {Kimble}},\
  }\bibfield  {title} {\bibinfo {title} {Observation of normal-mode splitting
  for an atom in an optical cavity},\ }\href
  {https://doi.org/10.1103/PhysRevLett.68.1132} {\bibfield  {journal} {\bibinfo
   {journal} {Phys. Rev. Lett.}\ }\textbf {\bibinfo {volume} {68}},\ \bibinfo
  {pages} {1132} (\bibinfo {year} {1992})}\BibitemShut {NoStop}%
\bibitem [{\citenamefont {Khitrova}\ \emph {et~al.}(2006)\citenamefont
  {Khitrova}, \citenamefont {Gibbs}, \citenamefont {Kira}, \citenamefont
  {Koch},\ and\ \citenamefont {Scherer}}]{Khitrova2006}%
  \BibitemOpen
  \bibfield  {author} {\bibinfo {author} {\bibfnamefont {G.}~\bibnamefont
  {Khitrova}}, \bibinfo {author} {\bibfnamefont {H.~M.}\ \bibnamefont {Gibbs}},
  \bibinfo {author} {\bibfnamefont {M.}~\bibnamefont {Kira}}, \bibinfo {author}
  {\bibfnamefont {S.~W.}\ \bibnamefont {Koch}},\ and\ \bibinfo {author}
  {\bibfnamefont {A.}~\bibnamefont {Scherer}},\ }\bibfield  {title} {\bibinfo
  {title} {{Vacuum Rabi splitting in semiconductors}},\ }\href
  {https://doi.org/10.1038/nphys227} {\bibfield  {journal} {\bibinfo  {journal}
  {Nat. Phys.}\ }\textbf {\bibinfo {volume} {2}},\ \bibinfo {pages} {81}
  (\bibinfo {year} {2006})}\BibitemShut {NoStop}%
\bibitem [{\citenamefont {Mkhitaryan}\ \emph {et~al.}(2018)\citenamefont
  {Mkhitaryan}, \citenamefont {Meng}, \citenamefont {Marini},\ and\
  \citenamefont {de~Abajo}}]{Mkhitaryan18}%
  \BibitemOpen
  \bibfield  {author} {\bibinfo {author} {\bibfnamefont {V.}~\bibnamefont
  {Mkhitaryan}}, \bibinfo {author} {\bibfnamefont {L.}~\bibnamefont {Meng}},
  \bibinfo {author} {\bibfnamefont {A.}~\bibnamefont {Marini}},\ and\ \bibinfo
  {author} {\bibfnamefont {F.~J.~G.}\ \bibnamefont {de~Abajo}},\ }\bibfield
  {title} {\bibinfo {title} {Lasing and amplification from two-dimensional atom
  arrays},\ }\href {https://doi.org/10.1103/PhysRevLett.121.163602} {\bibfield
  {journal} {\bibinfo  {journal} {Phys. Rev. Lett.}\ }\textbf {\bibinfo
  {volume} {121}},\ \bibinfo {pages} {163602} (\bibinfo {year}
  {2018})}\BibitemShut {NoStop}%
\bibitem [{\citenamefont {Shahmoon}\ \emph {et~al.}(2019)\citenamefont
  {Shahmoon}, \citenamefont {Lukin},\ and\ \citenamefont
  {Yelin}}]{Shahmoon2019_opto}%
  \BibitemOpen
  \bibfield  {author} {\bibinfo {author} {\bibfnamefont {E.}~\bibnamefont
  {Shahmoon}}, \bibinfo {author} {\bibfnamefont {M.~D.}\ \bibnamefont
  {Lukin}},\ and\ \bibinfo {author} {\bibfnamefont {S.~F.}\ \bibnamefont
  {Yelin}},\ }\bibfield  {title} {\bibinfo {title} {Chapter one - collective
  motion of an atom array under laser illumination},\ }in\ \href
  {https://doi.org/https://doi.org/10.1016/bs.aamop.2019.03.001} {\emph
  {\bibinfo {booktitle} {Advances in Atomic, Molecular, and Optical
  Physics}}},\ Vol.~\bibinfo {volume} {68},\ \bibinfo {editor} {edited by\
  \bibinfo {editor} {\bibfnamefont {L.~F.}\ \bibnamefont {Dimauro}}, \bibinfo
  {editor} {\bibfnamefont {H.}~\bibnamefont {Perrin}},\ and\ \bibinfo {editor}
  {\bibfnamefont {S.~F.}\ \bibnamefont {Yelin}}}\ (\bibinfo  {publisher}
  {Academic Press},\ \bibinfo {year} {2019})\ pp.\ \bibinfo {pages}
  {1--38}\BibitemShut {NoStop}%
\bibitem [{\citenamefont {Vaneecloo}\ \emph {et~al.}(2022)\citenamefont
  {Vaneecloo}, \citenamefont {Garcia},\ and\ \citenamefont
  {Ourjoumtsev}}]{Vaneecloo22}%
  \BibitemOpen
  \bibfield  {author} {\bibinfo {author} {\bibfnamefont {J.}~\bibnamefont
  {Vaneecloo}}, \bibinfo {author} {\bibfnamefont {S.}~\bibnamefont {Garcia}},\
  and\ \bibinfo {author} {\bibfnamefont {A.}~\bibnamefont {Ourjoumtsev}},\
  }\bibfield  {title} {\bibinfo {title} {Intracavity rydberg superatom for
  optical quantum engineering: Coherent control, single-shot detection, and
  optical $\ensuremath{\pi}$ phase shift},\ }\href
  {https://doi.org/10.1103/PhysRevX.12.021034} {\bibfield  {journal} {\bibinfo
  {journal} {Phys. Rev. X}\ }\textbf {\bibinfo {volume} {12}},\ \bibinfo
  {pages} {021034} (\bibinfo {year} {2022})}\BibitemShut {NoStop}%
\bibitem [{\citenamefont {Gross}\ and\ \citenamefont
  {Bloch}(2017)}]{Gross2017}%
  \BibitemOpen
  \bibfield  {author} {\bibinfo {author} {\bibfnamefont {C.}~\bibnamefont
  {Gross}}\ and\ \bibinfo {author} {\bibfnamefont {I.}~\bibnamefont {Bloch}},\
  }\bibfield  {title} {\bibinfo {title} {{Quantum simulations with ultracold
  atoms in optical lattices}},\ }\href
  {https://doi.org/10.1126/science.aal3837} {\bibfield  {journal} {\bibinfo
  {journal} {Science}\ }\textbf {\bibinfo {volume} {357}},\ \bibinfo {pages}
  {995} (\bibinfo {year} {2017})}\BibitemShut {NoStop}%
\bibitem [{\citenamefont {Domokos}\ and\ \citenamefont
  {Ritsch}(2002)}]{Domokos2002}%
  \BibitemOpen
  \bibfield  {author} {\bibinfo {author} {\bibfnamefont {P.}~\bibnamefont
  {Domokos}}\ and\ \bibinfo {author} {\bibfnamefont {H.}~\bibnamefont
  {Ritsch}},\ }\bibfield  {title} {\bibinfo {title} {Collective cooling and
  self-organization of atoms in a cavity},\ }\href
  {https://doi.org/10.1103/PhysRevLett.89.253003} {\bibfield  {journal}
  {\bibinfo  {journal} {Phys. Rev. Lett.}\ }\textbf {\bibinfo {volume} {89}},\
  \bibinfo {pages} {253003} (\bibinfo {year} {2002})}\BibitemShut {NoStop}%
\bibitem [{\citenamefont {Baumann}\ \emph {et~al.}(2010)\citenamefont
  {Baumann}, \citenamefont {Guerlin}, \citenamefont {Brennecke},\ and\
  \citenamefont {Esslinger}}]{Baumann2010}%
  \BibitemOpen
  \bibfield  {author} {\bibinfo {author} {\bibfnamefont {K.}~\bibnamefont
  {Baumann}}, \bibinfo {author} {\bibfnamefont {C.}~\bibnamefont {Guerlin}},
  \bibinfo {author} {\bibfnamefont {F.}~\bibnamefont {Brennecke}},\ and\
  \bibinfo {author} {\bibfnamefont {T.}~\bibnamefont {Esslinger}},\ }\bibfield
  {title} {\bibinfo {title} {{Dicke quantum phase transition with a superfluid
  gas in an optical cavity}},\ }\href {https://doi.org/10.1038/nature09009}
  {\bibfield  {journal} {\bibinfo  {journal} {Nature}\ }\textbf {\bibinfo
  {volume} {464}},\ \bibinfo {pages} {1301} (\bibinfo {year}
  {2010})}\BibitemShut {NoStop}%
\bibitem [{\citenamefont {Gopalakrishnan}\ \emph {et~al.}(2009)\citenamefont
  {Gopalakrishnan}, \citenamefont {Lev},\ and\ \citenamefont
  {Goldbart}}]{Gopalakrishnan09}%
  \BibitemOpen
  \bibfield  {author} {\bibinfo {author} {\bibfnamefont {S.}~\bibnamefont
  {Gopalakrishnan}}, \bibinfo {author} {\bibfnamefont {B.~L.}\ \bibnamefont
  {Lev}},\ and\ \bibinfo {author} {\bibfnamefont {P.~M.}\ \bibnamefont
  {Goldbart}},\ }\bibfield  {title} {\bibinfo {title} {Emergent crystallinity
  and frustration with {Bose}--{Einstein} condensates in multimode cavities},\
  }\href {https://doi.org/10.1038/nphys1403} {\bibfield  {journal} {\bibinfo
  {journal} {Nature Physics}\ }\textbf {\bibinfo {volume} {5}},\ \bibinfo
  {pages} {845} (\bibinfo {year} {2009})}\BibitemShut {NoStop}%
\bibitem [{\citenamefont {Strack}\ and\ \citenamefont
  {Sachdev}(2011)}]{Strack11}%
  \BibitemOpen
  \bibfield  {author} {\bibinfo {author} {\bibfnamefont {P.}~\bibnamefont
  {Strack}}\ and\ \bibinfo {author} {\bibfnamefont {S.}~\bibnamefont
  {Sachdev}},\ }\bibfield  {title} {\bibinfo {title} {Dicke quantum spin glass
  of atoms and photons},\ }\href
  {https://doi.org/10.1103/PhysRevLett.107.277202} {\bibfield  {journal}
  {\bibinfo  {journal} {Phys. Rev. Lett.}\ }\textbf {\bibinfo {volume} {107}},\
  \bibinfo {pages} {277202} (\bibinfo {year} {2011})}\BibitemShut {NoStop}%
\bibitem [{\citenamefont {Daley}\ \emph {et~al.}(2008)\citenamefont {Daley},
  \citenamefont {Boyd}, \citenamefont {Ye},\ and\ \citenamefont
  {Zoller}}]{Daley08}%
  \BibitemOpen
  \bibfield  {author} {\bibinfo {author} {\bibfnamefont {A.~J.}\ \bibnamefont
  {Daley}}, \bibinfo {author} {\bibfnamefont {M.~M.}\ \bibnamefont {Boyd}},
  \bibinfo {author} {\bibfnamefont {J.}~\bibnamefont {Ye}},\ and\ \bibinfo
  {author} {\bibfnamefont {P.}~\bibnamefont {Zoller}},\ }\bibfield  {title}
  {\bibinfo {title} {Quantum computing with alkaline-earth-metal atoms},\
  }\href {https://doi.org/10.1103/PhysRevLett.101.170504} {\bibfield  {journal}
  {\bibinfo  {journal} {Phys. Rev. Lett.}\ }\textbf {\bibinfo {volume} {101}},\
  \bibinfo {pages} {170504} (\bibinfo {year} {2008})}\BibitemShut {NoStop}%
\bibitem [{\citenamefont {Fukuhara}\ \emph {et~al.}(2009)\citenamefont
  {Fukuhara}, \citenamefont {Sugawa}, \citenamefont {Sugimoto}, \citenamefont
  {Taie},\ and\ \citenamefont {Takahashi}}]{Fukuhara09}%
  \BibitemOpen
  \bibfield  {author} {\bibinfo {author} {\bibfnamefont {T.}~\bibnamefont
  {Fukuhara}}, \bibinfo {author} {\bibfnamefont {S.}~\bibnamefont {Sugawa}},
  \bibinfo {author} {\bibfnamefont {M.}~\bibnamefont {Sugimoto}}, \bibinfo
  {author} {\bibfnamefont {S.}~\bibnamefont {Taie}},\ and\ \bibinfo {author}
  {\bibfnamefont {Y.}~\bibnamefont {Takahashi}},\ }\bibfield  {title} {\bibinfo
  {title} {Mott insulator of ultracold alkaline-earth-metal-like atoms},\
  }\href {https://doi.org/10.1103/PhysRevA.79.041604} {\bibfield  {journal}
  {\bibinfo  {journal} {Phys. Rev. A}\ }\textbf {\bibinfo {volume} {79}},\
  \bibinfo {pages} {041604} (\bibinfo {year} {2009})}\BibitemShut {NoStop}%
\bibitem [{\citenamefont {Ye}\ \emph {et~al.}(2008)\citenamefont {Ye},
  \citenamefont {Kimble},\ and\ \citenamefont {Katori}}]{Ye08}%
  \BibitemOpen
  \bibfield  {author} {\bibinfo {author} {\bibfnamefont {J.}~\bibnamefont
  {Ye}}, \bibinfo {author} {\bibfnamefont {H.~J.}\ \bibnamefont {Kimble}},\
  and\ \bibinfo {author} {\bibfnamefont {H.}~\bibnamefont {Katori}},\
  }\bibfield  {title} {\bibinfo {title} {Quantum state engineering and
  precision metrology using state-insensitive light traps},\ }\href
  {https://doi.org/10.1126/science.1148259} {\bibfield  {journal} {\bibinfo
  {journal} {Science}\ }\textbf {\bibinfo {volume} {320}},\ \bibinfo {pages}
  {1734} (\bibinfo {year} {2008})}\BibitemShut {NoStop}%
\bibitem [{\citenamefont {Stellmer}\ \emph {et~al.}(2012)\citenamefont
  {Stellmer}, \citenamefont {Pasquiou}, \citenamefont {Grimm},\ and\
  \citenamefont {Schreck}}]{Stellmer12}%
  \BibitemOpen
  \bibfield  {author} {\bibinfo {author} {\bibfnamefont {S.}~\bibnamefont
  {Stellmer}}, \bibinfo {author} {\bibfnamefont {B.}~\bibnamefont {Pasquiou}},
  \bibinfo {author} {\bibfnamefont {R.}~\bibnamefont {Grimm}},\ and\ \bibinfo
  {author} {\bibfnamefont {F.}~\bibnamefont {Schreck}},\ }\bibfield  {title}
  {\bibinfo {title} {Creation of ultracold ${\mathrm{sr}}_{2}$ molecules in the
  electronic ground state},\ }\href
  {https://doi.org/10.1103/PhysRevLett.109.115302} {\bibfield  {journal}
  {\bibinfo  {journal} {Phys. Rev. Lett.}\ }\textbf {\bibinfo {volume} {109}},\
  \bibinfo {pages} {115302} (\bibinfo {year} {2012})}\BibitemShut {NoStop}%
\bibitem [{\citenamefont {Kr{\"{a}}mer}\ \emph {et~al.}(2016)\citenamefont
  {Kr{\"{a}}mer}, \citenamefont {Ostermann},\ and\ \citenamefont
  {Ritsch}}]{Kramer2016}%
  \BibitemOpen
  \bibfield  {author} {\bibinfo {author} {\bibfnamefont {S.}~\bibnamefont
  {Kr{\"{a}}mer}}, \bibinfo {author} {\bibfnamefont {L.}~\bibnamefont
  {Ostermann}},\ and\ \bibinfo {author} {\bibfnamefont {H.}~\bibnamefont
  {Ritsch}},\ }\bibfield  {title} {\bibinfo {title} {{Optimized geometries for
  future generation optical lattice clocks}},\ }\href
  {https://doi.org/10.1209/0295-5075/114/14003} {\bibfield  {journal} {\bibinfo
   {journal} {Europhys. Lett.}\ }\textbf {\bibinfo {volume} {114}},\ \bibinfo
  {pages} {14003} (\bibinfo {year} {2016})}\BibitemShut {NoStop}%
\bibitem [{\citenamefont {Qu}\ and\ \citenamefont {Rey}(2019)}]{Qu19}%
  \BibitemOpen
  \bibfield  {author} {\bibinfo {author} {\bibfnamefont {C.}~\bibnamefont
  {Qu}}\ and\ \bibinfo {author} {\bibfnamefont {A.~M.}\ \bibnamefont {Rey}},\
  }\bibfield  {title} {\bibinfo {title} {Spin squeezing and many-body dipolar
  dynamics in optical lattice clocks},\ }\href
  {https://doi.org/10.1103/PhysRevA.100.041602} {\bibfield  {journal} {\bibinfo
   {journal} {Phys. Rev. A}\ }\textbf {\bibinfo {volume} {100}},\ \bibinfo
  {pages} {041602} (\bibinfo {year} {2019})}\BibitemShut {NoStop}%
\bibitem [{\citenamefont {Robicheaux}\ and\ \citenamefont
  {Huang}(2019)}]{Robicheaux19}%
  \BibitemOpen
  \bibfield  {author} {\bibinfo {author} {\bibfnamefont {F.}~\bibnamefont
  {Robicheaux}}\ and\ \bibinfo {author} {\bibfnamefont {S.}~\bibnamefont
  {Huang}},\ }\bibfield  {title} {\bibinfo {title} {Atom recoil during coherent
  light scattering from many atoms},\ }\href
  {https://doi.org/10.1103/PhysRevA.99.013410} {\bibfield  {journal} {\bibinfo
  {journal} {Phys. Rev. A}\ }\textbf {\bibinfo {volume} {99}},\ \bibinfo
  {pages} {013410} (\bibinfo {year} {2019})}\BibitemShut {NoStop}%
\bibitem [{\citenamefont {Olmos}\ \emph {et~al.}(2013)\citenamefont {Olmos},
  \citenamefont {Yu}, \citenamefont {Singh}, \citenamefont {Schreck},
  \citenamefont {Bongs},\ and\ \citenamefont {Lesanovsky}}]{Olmos13}%
  \BibitemOpen
  \bibfield  {author} {\bibinfo {author} {\bibfnamefont {B.}~\bibnamefont
  {Olmos}}, \bibinfo {author} {\bibfnamefont {D.}~\bibnamefont {Yu}}, \bibinfo
  {author} {\bibfnamefont {Y.}~\bibnamefont {Singh}}, \bibinfo {author}
  {\bibfnamefont {F.}~\bibnamefont {Schreck}}, \bibinfo {author} {\bibfnamefont
  {K.}~\bibnamefont {Bongs}},\ and\ \bibinfo {author} {\bibfnamefont
  {I.}~\bibnamefont {Lesanovsky}},\ }\bibfield  {title} {\bibinfo {title}
  {Long-range interacting many-body systems with alkaline-earth-metal atoms},\
  }\href {https://doi.org/10.1103/PhysRevLett.110.143602} {\bibfield  {journal}
  {\bibinfo  {journal} {Phys. Rev. Lett.}\ }\textbf {\bibinfo {volume} {110}},\
  \bibinfo {pages} {143602} (\bibinfo {year} {2013})}\BibitemShut {NoStop}%
\bibitem [{\citenamefont {Zhou}\ \emph {et~al.}(2010)\citenamefont {Zhou},
  \citenamefont {Xu}, \citenamefont {Chen},\ and\ \citenamefont
  {Chen}}]{Zhou10}%
  \BibitemOpen
  \bibfield  {author} {\bibinfo {author} {\bibfnamefont {X.}~\bibnamefont
  {Zhou}}, \bibinfo {author} {\bibfnamefont {X.}~\bibnamefont {Xu}}, \bibinfo
  {author} {\bibfnamefont {X.}~\bibnamefont {Chen}},\ and\ \bibinfo {author}
  {\bibfnamefont {J.}~\bibnamefont {Chen}},\ }\bibfield  {title} {\bibinfo
  {title} {Magic wavelengths for terahertz clock transitions},\ }\href
  {https://doi.org/10.1103/PhysRevA.81.012115} {\bibfield  {journal} {\bibinfo
  {journal} {Phys. Rev. A}\ }\textbf {\bibinfo {volume} {81}},\ \bibinfo
  {pages} {012115} (\bibinfo {year} {2010})}\BibitemShut {NoStop}%
\bibitem [{\citenamefont {Werij}\ \emph {et~al.}(1992)\citenamefont {Werij},
  \citenamefont {Greene}, \citenamefont {Theodosiou},\ and\ \citenamefont
  {Gallagher}}]{Werij92}%
  \BibitemOpen
  \bibfield  {author} {\bibinfo {author} {\bibfnamefont {H.~G.~C.}\
  \bibnamefont {Werij}}, \bibinfo {author} {\bibfnamefont {C.~H.}\ \bibnamefont
  {Greene}}, \bibinfo {author} {\bibfnamefont {C.~E.}\ \bibnamefont
  {Theodosiou}},\ and\ \bibinfo {author} {\bibfnamefont {A.}~\bibnamefont
  {Gallagher}},\ }\bibfield  {title} {\bibinfo {title} {Oscillator strengths
  and radiative branching ratios in atomic sr},\ }\href
  {https://doi.org/10.1103/PhysRevA.46.1248} {\bibfield  {journal} {\bibinfo
  {journal} {Phys. Rev. A}\ }\textbf {\bibinfo {volume} {46}},\ \bibinfo
  {pages} {1248} (\bibinfo {year} {1992})}\BibitemShut {NoStop}%
\bibitem [{\citenamefont {Beloy}\ \emph {et~al.}(2012)\citenamefont {Beloy},
  \citenamefont {Sherman}, \citenamefont {Lemke}, \citenamefont {Hinkley},
  \citenamefont {Oates},\ and\ \citenamefont {Ludlow}}]{Beloy12}%
  \BibitemOpen
  \bibfield  {author} {\bibinfo {author} {\bibfnamefont {K.}~\bibnamefont
  {Beloy}}, \bibinfo {author} {\bibfnamefont {J.~A.}\ \bibnamefont {Sherman}},
  \bibinfo {author} {\bibfnamefont {N.~D.}\ \bibnamefont {Lemke}}, \bibinfo
  {author} {\bibfnamefont {N.}~\bibnamefont {Hinkley}}, \bibinfo {author}
  {\bibfnamefont {C.~W.}\ \bibnamefont {Oates}},\ and\ \bibinfo {author}
  {\bibfnamefont {A.~D.}\ \bibnamefont {Ludlow}},\ }\bibfield  {title}
  {\bibinfo {title} {Determination of the $5d6s$ ${}^{3}{D}_{1}$ state lifetime
  and blackbody-radiation clock shift in yb},\ }\href
  {https://doi.org/10.1103/PhysRevA.86.051404} {\bibfield  {journal} {\bibinfo
  {journal} {Phys. Rev. A}\ }\textbf {\bibinfo {volume} {86}},\ \bibinfo
  {pages} {051404} (\bibinfo {year} {2012})}\BibitemShut {NoStop}%
\bibitem [{\citenamefont {Covey}\ \emph {et~al.}(2019)\citenamefont {Covey},
  \citenamefont {Sipahigil}, \citenamefont {Szoke}, \citenamefont {Sinclair},
  \citenamefont {Endres},\ and\ \citenamefont {Painter}}]{Covey19}%
  \BibitemOpen
  \bibfield  {author} {\bibinfo {author} {\bibfnamefont {J.~P.}\ \bibnamefont
  {Covey}}, \bibinfo {author} {\bibfnamefont {A.}~\bibnamefont {Sipahigil}},
  \bibinfo {author} {\bibfnamefont {S.}~\bibnamefont {Szoke}}, \bibinfo
  {author} {\bibfnamefont {N.}~\bibnamefont {Sinclair}}, \bibinfo {author}
  {\bibfnamefont {M.}~\bibnamefont {Endres}},\ and\ \bibinfo {author}
  {\bibfnamefont {O.}~\bibnamefont {Painter}},\ }\bibfield  {title} {\bibinfo
  {title} {Telecom-band quantum optics with ytterbium atoms and silicon
  nanophotonics},\ }\href {https://doi.org/10.1103/PhysRevApplied.11.034044}
  {\bibfield  {journal} {\bibinfo  {journal} {Phys. Rev. Appl.}\ }\textbf
  {\bibinfo {volume} {11}},\ \bibinfo {pages} {034044} (\bibinfo {year}
  {2019})}\BibitemShut {NoStop}%
\bibitem [{\citenamefont {Yamamoto}\ \emph {et~al.}(2016)\citenamefont
  {Yamamoto}, \citenamefont {Kobayashi}, \citenamefont {Kuno}, \citenamefont
  {Kato},\ and\ \citenamefont {Takahashi}}]{Yamamoto2016}%
  \BibitemOpen
  \bibfield  {author} {\bibinfo {author} {\bibfnamefont {R.}~\bibnamefont
  {Yamamoto}}, \bibinfo {author} {\bibfnamefont {J.}~\bibnamefont {Kobayashi}},
  \bibinfo {author} {\bibfnamefont {T.}~\bibnamefont {Kuno}}, \bibinfo {author}
  {\bibfnamefont {K.}~\bibnamefont {Kato}},\ and\ \bibinfo {author}
  {\bibfnamefont {Y.}~\bibnamefont {Takahashi}},\ }\bibfield  {title} {\bibinfo
  {title} {An ytterbium quantum gas microscope with narrow-line laser
  cooling},\ }\href {https://doi.org/10.1088/1367-2630/18/2/023016} {\bibfield
  {journal} {\bibinfo  {journal} {New Journal of Physics}\ }\textbf {\bibinfo
  {volume} {18}},\ \bibinfo {pages} {023016} (\bibinfo {year}
  {2016})}\BibitemShut {NoStop}%
\bibitem [{\citenamefont {Kim}\ \emph {et~al.}(2016)\citenamefont {Kim},
  \citenamefont {Lee}, \citenamefont {Lee}, \citenamefont {Jo}, \citenamefont
  {Song},\ and\ \citenamefont {Ahn}}]{Kim16}%
  \BibitemOpen
  \bibfield  {author} {\bibinfo {author} {\bibfnamefont {H.}~\bibnamefont
  {Kim}}, \bibinfo {author} {\bibfnamefont {W.}~\bibnamefont {Lee}}, \bibinfo
  {author} {\bibfnamefont {H.-g.}\ \bibnamefont {Lee}}, \bibinfo {author}
  {\bibfnamefont {H.}~\bibnamefont {Jo}}, \bibinfo {author} {\bibfnamefont
  {Y.}~\bibnamefont {Song}},\ and\ \bibinfo {author} {\bibfnamefont
  {J.}~\bibnamefont {Ahn}},\ }\bibfield  {title} {\bibinfo {title} {In situ
  single-atom array synthesis using dynamic holographic optical tweezers},\
  }\href {https://doi.org/10.1038/ncomms13317} {\bibfield  {journal} {\bibinfo
  {journal} {Nature Communications}\ }\textbf {\bibinfo {volume} {7}},\
  \bibinfo {pages} {13317} (\bibinfo {year} {2016})}\BibitemShut {NoStop}%
\bibitem [{\citenamefont {Endres}\ \emph {et~al.}(2016)\citenamefont {Endres},
  \citenamefont {Bernien}, \citenamefont {Keesling}, \citenamefont {Levine},
  \citenamefont {Anschuetz}, \citenamefont {Krajenbrink}, \citenamefont
  {Senko}, \citenamefont {Vuletic}, \citenamefont {Greiner},\ and\
  \citenamefont {Lukin}}]{Endres16}%
  \BibitemOpen
  \bibfield  {author} {\bibinfo {author} {\bibfnamefont {M.}~\bibnamefont
  {Endres}}, \bibinfo {author} {\bibfnamefont {H.}~\bibnamefont {Bernien}},
  \bibinfo {author} {\bibfnamefont {A.}~\bibnamefont {Keesling}}, \bibinfo
  {author} {\bibfnamefont {H.}~\bibnamefont {Levine}}, \bibinfo {author}
  {\bibfnamefont {E.~R.}\ \bibnamefont {Anschuetz}}, \bibinfo {author}
  {\bibfnamefont {A.}~\bibnamefont {Krajenbrink}}, \bibinfo {author}
  {\bibfnamefont {C.}~\bibnamefont {Senko}}, \bibinfo {author} {\bibfnamefont
  {V.}~\bibnamefont {Vuletic}}, \bibinfo {author} {\bibfnamefont
  {M.}~\bibnamefont {Greiner}},\ and\ \bibinfo {author} {\bibfnamefont {M.~D.}\
  \bibnamefont {Lukin}},\ }\bibfield  {title} {\bibinfo {title} {Atom-by-atom
  assembly of defect-free one-dimensional cold atom arrays},\ }\href
  {https://doi.org/10.1126/science.aah3752} {\bibfield  {journal} {\bibinfo
  {journal} {Science}\ }\textbf {\bibinfo {volume} {354}},\ \bibinfo {pages}
  {1024} (\bibinfo {year} {2016})}\BibitemShut {NoStop}%
\bibitem [{\citenamefont {Barredo}\ \emph {et~al.}(2018)\citenamefont
  {Barredo}, \citenamefont {Lienhard}, \citenamefont {de~L{\'e}s{\'e}leuc},
  \citenamefont {Lahaye},\ and\ \citenamefont {Browaeys}}]{Barredo18}%
  \BibitemOpen
  \bibfield  {author} {\bibinfo {author} {\bibfnamefont {D.}~\bibnamefont
  {Barredo}}, \bibinfo {author} {\bibfnamefont {V.}~\bibnamefont {Lienhard}},
  \bibinfo {author} {\bibfnamefont {S.}~\bibnamefont {de~L{\'e}s{\'e}leuc}},
  \bibinfo {author} {\bibfnamefont {T.}~\bibnamefont {Lahaye}},\ and\ \bibinfo
  {author} {\bibfnamefont {A.}~\bibnamefont {Browaeys}},\ }\bibfield  {title}
  {\bibinfo {title} {Synthetic three-dimensional atomic structures assembled
  atom by atom},\ }\href {https://doi.org/10.1038/s41586-018-0450-2} {\bibfield
   {journal} {\bibinfo  {journal} {Nature}\ }\textbf {\bibinfo {volume}
  {561}},\ \bibinfo {pages} {79} (\bibinfo {year} {2018})}\BibitemShut
  {NoStop}%
\bibitem [{\citenamefont {Cooper}\ \emph {et~al.}(2018)\citenamefont {Cooper},
  \citenamefont {Covey}, \citenamefont {Madjarov}, \citenamefont {Porsev},
  \citenamefont {Safronova},\ and\ \citenamefont {Endres}}]{Cooper18}%
  \BibitemOpen
  \bibfield  {author} {\bibinfo {author} {\bibfnamefont {A.}~\bibnamefont
  {Cooper}}, \bibinfo {author} {\bibfnamefont {J.~P.}\ \bibnamefont {Covey}},
  \bibinfo {author} {\bibfnamefont {I.~S.}\ \bibnamefont {Madjarov}}, \bibinfo
  {author} {\bibfnamefont {S.~G.}\ \bibnamefont {Porsev}}, \bibinfo {author}
  {\bibfnamefont {M.~S.}\ \bibnamefont {Safronova}},\ and\ \bibinfo {author}
  {\bibfnamefont {M.}~\bibnamefont {Endres}},\ }\bibfield  {title} {\bibinfo
  {title} {Alkaline-earth atoms in optical tweezers},\ }\href
  {https://doi.org/10.1103/PhysRevX.8.041055} {\bibfield  {journal} {\bibinfo
  {journal} {Phys. Rev. X}\ }\textbf {\bibinfo {volume} {8}},\ \bibinfo {pages}
  {041055} (\bibinfo {year} {2018})}\BibitemShut {NoStop}%
\bibitem [{\citenamefont {Saskin}\ \emph {et~al.}(2019)\citenamefont {Saskin},
  \citenamefont {Wilson}, \citenamefont {Grinkemeyer},\ and\ \citenamefont
  {Thompson}}]{Saskin19}%
  \BibitemOpen
  \bibfield  {author} {\bibinfo {author} {\bibfnamefont {S.}~\bibnamefont
  {Saskin}}, \bibinfo {author} {\bibfnamefont {J.~T.}\ \bibnamefont {Wilson}},
  \bibinfo {author} {\bibfnamefont {B.}~\bibnamefont {Grinkemeyer}},\ and\
  \bibinfo {author} {\bibfnamefont {J.~D.}\ \bibnamefont {Thompson}},\
  }\bibfield  {title} {\bibinfo {title} {Narrow-line cooling and imaging of
  ytterbium atoms in an optical tweezer array},\ }\href
  {https://doi.org/10.1103/PhysRevLett.122.143002} {\bibfield  {journal}
  {\bibinfo  {journal} {Phys. Rev. Lett.}\ }\textbf {\bibinfo {volume} {122}},\
  \bibinfo {pages} {143002} (\bibinfo {year} {2019})}\BibitemShut {NoStop}%
\bibitem [{\citenamefont {Schymik}\ \emph {et~al.}(2022)\citenamefont
  {Schymik}, \citenamefont {Ximenez}, \citenamefont {Bloch}, \citenamefont
  {Dreon}, \citenamefont {Signoles}, \citenamefont {Nogrette}, \citenamefont
  {Barredo}, \citenamefont {Browaeys},\ and\ \citenamefont
  {Lahaye}}]{Schymik22}%
  \BibitemOpen
  \bibfield  {author} {\bibinfo {author} {\bibfnamefont {K.-N.}\ \bibnamefont
  {Schymik}}, \bibinfo {author} {\bibfnamefont {B.}~\bibnamefont {Ximenez}},
  \bibinfo {author} {\bibfnamefont {E.}~\bibnamefont {Bloch}}, \bibinfo
  {author} {\bibfnamefont {D.}~\bibnamefont {Dreon}}, \bibinfo {author}
  {\bibfnamefont {A.}~\bibnamefont {Signoles}}, \bibinfo {author}
  {\bibfnamefont {F.}~\bibnamefont {Nogrette}}, \bibinfo {author}
  {\bibfnamefont {D.}~\bibnamefont {Barredo}}, \bibinfo {author} {\bibfnamefont
  {A.}~\bibnamefont {Browaeys}},\ and\ \bibinfo {author} {\bibfnamefont
  {T.}~\bibnamefont {Lahaye}},\ }\bibfield  {title} {\bibinfo {title} {In situ
  equalization of single-atom loading in large-scale optical tweezer arrays},\
  }\href {https://doi.org/10.1103/PhysRevA.106.022611} {\bibfield  {journal}
  {\bibinfo  {journal} {Phys. Rev. A}\ }\textbf {\bibinfo {volume} {106}},\
  \bibinfo {pages} {022611} (\bibinfo {year} {2022})}\BibitemShut {NoStop}%
\bibitem [{\citenamefont {Wang}\ \emph {et~al.}(2018)\citenamefont {Wang},
  \citenamefont {Subhankar}, \citenamefont {Bienias}, \citenamefont
  {\L{}k{a}cki}, \citenamefont {Tsui}, \citenamefont {Baranov}, \citenamefont
  {Gorshkov}, \citenamefont {Zoller}, \citenamefont {Porto},\ and\
  \citenamefont {Rolston}}]{Wang18}%
  \BibitemOpen
  \bibfield  {author} {\bibinfo {author} {\bibfnamefont {Y.}~\bibnamefont
  {Wang}}, \bibinfo {author} {\bibfnamefont {S.}~\bibnamefont {Subhankar}},
  \bibinfo {author} {\bibfnamefont {P.}~\bibnamefont {Bienias}}, \bibinfo
  {author} {\bibfnamefont {M.}~\bibnamefont {\L{}k{a}cki}}, \bibinfo {author}
  {\bibfnamefont {T.-C.}\ \bibnamefont {Tsui}}, \bibinfo {author}
  {\bibfnamefont {M.~A.}\ \bibnamefont {Baranov}}, \bibinfo {author}
  {\bibfnamefont {A.~V.}\ \bibnamefont {Gorshkov}}, \bibinfo {author}
  {\bibfnamefont {P.}~\bibnamefont {Zoller}}, \bibinfo {author} {\bibfnamefont
  {J.~V.}\ \bibnamefont {Porto}},\ and\ \bibinfo {author} {\bibfnamefont
  {S.~L.}\ \bibnamefont {Rolston}},\ }\bibfield  {title} {\bibinfo {title}
  {Dark state optical lattice with a subwavelength spatial structure},\ }\href
  {https://doi.org/10.1103/PhysRevLett.120.083601} {\bibfield  {journal}
  {\bibinfo  {journal} {Phys. Rev. Lett.}\ }\textbf {\bibinfo {volume} {120}},\
  \bibinfo {pages} {083601} (\bibinfo {year} {2018})}\BibitemShut {NoStop}%
\bibitem [{\citenamefont {Bienias}\ \emph {et~al.}(2020)\citenamefont
  {Bienias}, \citenamefont {Subhankar}, \citenamefont {Wang}, \citenamefont
  {Tsui}, \citenamefont {Jendrzejewski}, \citenamefont {Tiecke}, \citenamefont
  {Juzeli\ifmmode~\bar{u}\else \={u}\fi{}nas}, \citenamefont {Jiang},
  \citenamefont {Rolston}, \citenamefont {Porto},\ and\ \citenamefont
  {Gorshkov}}]{Bienias20}%
  \BibitemOpen
  \bibfield  {author} {\bibinfo {author} {\bibfnamefont {P.}~\bibnamefont
  {Bienias}}, \bibinfo {author} {\bibfnamefont {S.}~\bibnamefont {Subhankar}},
  \bibinfo {author} {\bibfnamefont {Y.}~\bibnamefont {Wang}}, \bibinfo {author}
  {\bibfnamefont {T.-C.}\ \bibnamefont {Tsui}}, \bibinfo {author}
  {\bibfnamefont {F.}~\bibnamefont {Jendrzejewski}}, \bibinfo {author}
  {\bibfnamefont {T.}~\bibnamefont {Tiecke}}, \bibinfo {author} {\bibfnamefont
  {G.}~\bibnamefont {Juzeli\ifmmode~\bar{u}\else \={u}\fi{}nas}}, \bibinfo
  {author} {\bibfnamefont {L.}~\bibnamefont {Jiang}}, \bibinfo {author}
  {\bibfnamefont {S.~L.}\ \bibnamefont {Rolston}}, \bibinfo {author}
  {\bibfnamefont {J.~V.}\ \bibnamefont {Porto}},\ and\ \bibinfo {author}
  {\bibfnamefont {A.~V.}\ \bibnamefont {Gorshkov}},\ }\bibfield  {title}
  {\bibinfo {title} {Coherent optical nanotweezers for ultracold atoms},\
  }\href {https://doi.org/10.1103/PhysRevA.102.013306} {\bibfield  {journal}
  {\bibinfo  {journal} {Phys. Rev. A}\ }\textbf {\bibinfo {volume} {102}},\
  \bibinfo {pages} {013306} (\bibinfo {year} {2020})}\BibitemShut {NoStop}%
\bibitem [{\citenamefont {Tsui}\ \emph {et~al.}(2020)\citenamefont {Tsui},
  \citenamefont {Wang}, \citenamefont {Subhankar}, \citenamefont {Porto},\ and\
  \citenamefont {Rolston}}]{Tsui20}%
  \BibitemOpen
  \bibfield  {author} {\bibinfo {author} {\bibfnamefont {T.-C.}\ \bibnamefont
  {Tsui}}, \bibinfo {author} {\bibfnamefont {Y.}~\bibnamefont {Wang}}, \bibinfo
  {author} {\bibfnamefont {S.}~\bibnamefont {Subhankar}}, \bibinfo {author}
  {\bibfnamefont {J.~V.}\ \bibnamefont {Porto}},\ and\ \bibinfo {author}
  {\bibfnamefont {S.~L.}\ \bibnamefont {Rolston}},\ }\bibfield  {title}
  {\bibinfo {title} {Realization of a stroboscopic optical lattice for cold
  atoms with subwavelength spacing},\ }\href
  {https://doi.org/10.1103/PhysRevA.101.041603} {\bibfield  {journal} {\bibinfo
   {journal} {Phys. Rev. A}\ }\textbf {\bibinfo {volume} {101}},\ \bibinfo
  {pages} {041603} (\bibinfo {year} {2020})}\BibitemShut {NoStop}%
\bibitem [{\citenamefont {Anderson}\ \emph {et~al.}(2020)\citenamefont
  {Anderson}, \citenamefont {Trypogeorgos}, \citenamefont {Vald\'es-Curiel},
  \citenamefont {Liang}, \citenamefont {Tao}, \citenamefont {Zhao},
  \citenamefont {Andrijauskas}, \citenamefont {Juzeli\ifmmode~\bar{u}\else
  \={u}\fi{}nas},\ and\ \citenamefont {Spielman}}]{Anderson20}%
  \BibitemOpen
  \bibfield  {author} {\bibinfo {author} {\bibfnamefont {R.~P.}\ \bibnamefont
  {Anderson}}, \bibinfo {author} {\bibfnamefont {D.}~\bibnamefont
  {Trypogeorgos}}, \bibinfo {author} {\bibfnamefont {A.}~\bibnamefont
  {Vald\'es-Curiel}}, \bibinfo {author} {\bibfnamefont {Q.-Y.}\ \bibnamefont
  {Liang}}, \bibinfo {author} {\bibfnamefont {J.}~\bibnamefont {Tao}}, \bibinfo
  {author} {\bibfnamefont {M.}~\bibnamefont {Zhao}}, \bibinfo {author}
  {\bibfnamefont {T.}~\bibnamefont {Andrijauskas}}, \bibinfo {author}
  {\bibfnamefont {G.}~\bibnamefont {Juzeli\ifmmode~\bar{u}\else
  \={u}\fi{}nas}},\ and\ \bibinfo {author} {\bibfnamefont {I.~B.}\ \bibnamefont
  {Spielman}},\ }\bibfield  {title} {\bibinfo {title} {Realization of a deeply
  subwavelength adiabatic optical lattice},\ }\href
  {https://doi.org/10.1103/PhysRevResearch.2.013149} {\bibfield  {journal}
  {\bibinfo  {journal} {Phys. Rev. Res.}\ }\textbf {\bibinfo {volume} {2}},\
  \bibinfo {pages} {013149} (\bibinfo {year} {2020})}\BibitemShut {NoStop}%
\bibitem [{\citenamefont {Kubala}\ \emph {et~al.}(2021)\citenamefont {Kubala},
  \citenamefont {Zakrzewski},\ and\ \citenamefont {\L{}acki}}]{Kubala21}%
  \BibitemOpen
  \bibfield  {author} {\bibinfo {author} {\bibfnamefont {P.}~\bibnamefont
  {Kubala}}, \bibinfo {author} {\bibfnamefont {J.}~\bibnamefont {Zakrzewski}},\
  and\ \bibinfo {author} {\bibfnamefont {M.}~\bibnamefont {\L{}acki}},\
  }\bibfield  {title} {\bibinfo {title} {Optical lattice for a tripodlike
  atomic level structure},\ }\href
  {https://doi.org/10.1103/PhysRevA.104.053312} {\bibfield  {journal} {\bibinfo
   {journal} {Phys. Rev. A}\ }\textbf {\bibinfo {volume} {104}},\ \bibinfo
  {pages} {053312} (\bibinfo {year} {2021})}\BibitemShut {NoStop}%
\bibitem [{\citenamefont {Chomaz}\ \emph {et~al.}(2012)\citenamefont {Chomaz},
  \citenamefont {Corman}, \citenamefont {Yefsah}, \citenamefont {Desbuquois},\
  and\ \citenamefont {Dalibard}}]{dalibardexp}%
  \BibitemOpen
  \bibfield  {author} {\bibinfo {author} {\bibfnamefont {L.}~\bibnamefont
  {Chomaz}}, \bibinfo {author} {\bibfnamefont {L.}~\bibnamefont {Corman}},
  \bibinfo {author} {\bibfnamefont {T.}~\bibnamefont {Yefsah}}, \bibinfo
  {author} {\bibfnamefont {R.}~\bibnamefont {Desbuquois}},\ and\ \bibinfo
  {author} {\bibfnamefont {J.}~\bibnamefont {Dalibard}},\ }\bibfield  {title}
  {\bibinfo {title} {Absorption imaging of a quasi-two-dimensional gas: a
  multiple scattering analysis},\ }\href
  {https://doi.org/10.1088/1367-2630/14/5/055001} {\bibfield  {journal}
  {\bibinfo  {journal} {New Journal of Physics}\ }\textbf {\bibinfo {volume}
  {14}},\ \bibinfo {pages} {005501} (\bibinfo {year} {2012})}\BibitemShut
  {NoStop}%
\bibitem [{\citenamefont {Javanainen}\ \emph {et~al.}(2017)\citenamefont
  {Javanainen}, \citenamefont {Ruostekoski}, \citenamefont {Li},\ and\
  \citenamefont {Yoo}}]{Javanainen17}%
  \BibitemOpen
  \bibfield  {author} {\bibinfo {author} {\bibfnamefont {J.}~\bibnamefont
  {Javanainen}}, \bibinfo {author} {\bibfnamefont {J.}~\bibnamefont
  {Ruostekoski}}, \bibinfo {author} {\bibfnamefont {Y.}~\bibnamefont {Li}},\
  and\ \bibinfo {author} {\bibfnamefont {S.-M.}\ \bibnamefont {Yoo}},\
  }\bibfield  {title} {\bibinfo {title} {Exact electrodynamics versus standard
  optics for a slab of cold dense gas},\ }\href
  {https://doi.org/10.1103/PhysRevA.96.033835} {\bibfield  {journal} {\bibinfo
  {journal} {Phys. Rev. A}\ }\textbf {\bibinfo {volume} {96}},\ \bibinfo
  {pages} {033835} (\bibinfo {year} {2017})}\BibitemShut {NoStop}%
\bibitem [{\citenamefont {Belov}\ and\ \citenamefont
  {Simovski}(2005)}]{Belov05}%
  \BibitemOpen
  \bibfield  {author} {\bibinfo {author} {\bibfnamefont {P.~A.}\ \bibnamefont
  {Belov}}\ and\ \bibinfo {author} {\bibfnamefont {C.~R.}\ \bibnamefont
  {Simovski}},\ }\bibfield  {title} {\bibinfo {title} {Homogenization of
  electromagnetic crystals formed by uniaxial resonant scatterers},\ }\href
  {https://doi.org/10.1103/PhysRevE.72.026615} {\bibfield  {journal} {\bibinfo
  {journal} {Phys. Rev. E}\ }\textbf {\bibinfo {volume} {72}},\ \bibinfo
  {pages} {026615} (\bibinfo {year} {2005})}\BibitemShut {NoStop}%
\bibitem [{\citenamefont {Novotny}\ and\ \citenamefont
  {Hecht}(2012)}]{Novotny_nanooptics}%
  \BibitemOpen
  \bibfield  {author} {\bibinfo {author} {\bibfnamefont {L.}~\bibnamefont
  {Novotny}}\ and\ \bibinfo {author} {\bibfnamefont {B.}~\bibnamefont
  {Hecht}},\ }\href@noop {} {\emph {\bibinfo {title} {Principles of
  Nano-Optics}}},\ \bibinfo {edition} {2nd}\ ed.\ (\bibinfo  {publisher}
  {Cambridge University Press, Cambridge},\ \bibinfo {year} {2012})\BibitemShut
  {NoStop}%
\bibitem [{\citenamefont {Benisty}\ \emph {et~al.}(2022)\citenamefont
  {Benisty}, \citenamefont {Greffet},\ and\ \citenamefont
  {Lalanne}}]{Greffet_nanophotonics}%
  \BibitemOpen
  \bibfield  {author} {\bibinfo {author} {\bibfnamefont {H.}~\bibnamefont
  {Benisty}}, \bibinfo {author} {\bibfnamefont {J.-J.}\ \bibnamefont
  {Greffet}},\ and\ \bibinfo {author} {\bibfnamefont {P.}~\bibnamefont
  {Lalanne}},\ }\href@noop {} {\emph {\bibinfo {title} {Introduction to
  Nanophotonics}}},\ \bibinfo {edition} {1st}\ ed.\ (\bibinfo  {publisher}
  {Oxford University Press, Oxford},\ \bibinfo {year} {2022})\BibitemShut
  {NoStop}%
\end{thebibliography}
\end{document}